\documentclass[a4paper,11pt]{article}

\usepackage{jcappub}
\usepackage{amsmath}
\usepackage{graphicx}
\usepackage{latexsym}
\usepackage{xspace}
\usepackage[table,usenames,dvipsnames]{xcolor}
\usepackage{color}
\usepackage{hyperref} 
\usepackage{bm}
\usepackage{relsize}
\usepackage{multirow}
\usepackage{array}
\usepackage{tabularx}

\makeatletter
\gdef\@fpheader{}
\makeatother

\newcommand{\ie}{i.e.\xspace}

\newcommand{\order}[1]{\mathcal{O}\!\left(#1\right)}

\newcommand{\ASPIC}{\texttt{ASPIC}\xspace}

\newcommand{\dd}{\mathrm{d}}

\newcommand{\phim}{\varphi_\um}
\newcommand{\phir}{\varphi_\ur}

\newcommand{\sss}[1]{{\scriptscriptstyle{#1}}}

\newcommand{\uPl}{\mathrm{Pl}}

\newcommand{\uend}{\mathrm{end}}
\newcommand{\udec}{\mathrm{dec}}
\newcommand{\ueq}{\mathrm{eq}}

\newcommand{\ureh}{\mathrm{reh}}
\newcommand{\urad}{\mathrm{rad}}

\newcommand{\um}{\mathrm{m}}
\newcommand{\ur}{\mathrm{r}}

\newcommand{\uS}{\mathrm{S}}

\newcommand{\usssS}{\sss{\uS}}

\newcommand{\usssPl}{\sss{\uPl}}

\newcommand{\nS}{n_\usssS}

\newcommand{\uNL}{\mathrm{NL}}

\newcommand{\calP}{\mathcal{P}}

\newcommand{\Mp}{M_\usssPl}

\newcommand{\vev}{\textit{vev}~}

\newcommand{\fnl}{f_\uNL}

\newcommand{\efolds}{$e$-folds~}

\newcommand{\beq}{\begin{equation}}
\newcommand{\eeq}{\end{equation}}
\newcommand{\bea}{\begin{eqnarray}}
\newcommand{\eea}{\end{eqnarray}}

\newlength{\wsingfig}
\newlength{\wdblefig}
\newlength{\wappfig}
\newlength{\wquadfig}
\newlength{\wtriplefig}
\newcommand{\Eq}[1]{Eq.~(\ref{#1})}
\newcommand{\Eqs}[1]{Eqs.~(\ref{#1})}
\newcommand{\Fig}[1]{Fig.~{\ref{#1}}}

\newcommand{\Ref}[1]{Ref.~{\cite{#1}}}
\newcommand{\Refs}[1]{Refs.~{\cite{#1}}}

\newcommand{\SF}{\star} 
\newcommand{\matter}{\mathrm{m}}
\newcommand{\radiation}{\mathrm{r}}

\newcommand{\MYhref}[3][blue]{\href{#2}{\color{#1}{#3}}}%

\subheader{}

\title{Encyclop{\ae}dia Curvatonis}

\author{Vincent Vennin,}
\author{Kazuya Koyama}
\author{and David Wands}

\affiliation{Institute of Cosmology \& Gravitation, University of Portsmouth, Dennis Sciama Building, Burnaby Road, Portsmouth, PO1 3FX, United Kingdom}

\emailAdd{vincent.vennin@port.ac.uk}
\emailAdd{kazuya.koyama@port.ac.uk}
\emailAdd{david.wands@port.ac.uk}

\date{today}

\begin{document}

\abstract{We investigate whether the predictions of single-field models of inflation are robust under the introduction of additional scalar degrees of freedom, and whether these extra fields change the potentials for which the data show the strongest preference.
We study the situation where an extra light scalar field contributes both to the total curvature perturbations and to the reheating kinematic properties. Ten reheating scenarios are identified, and all necessary formulas allowing a systematic computation of the predictions for this class of models are derived. They are implemented in the public library \ASPIC, which contains more than $75$ single-field potentials. This paves the way for a forthcoming full Bayesian analysis of the problem. A few representative examples are displayed and discussed.}



\maketitle
\pagebreak

\section{Introduction and Summary of the Main Results}
Inflation is the leading paradigm for explaining the physical conditions that prevailed in the very early Universe~\cite{Starobinsky:1980te,Sato:1980yn,Guth:1980zm,Linde:1981mu,Albrecht:1982wi,Linde:1983gd}. It consists of a phase of accelerated expansion that solves the standard hot big bang model problems, and provides a causal mechanism for generating inhomogeneities on cosmological scales~\cite{Mukhanov:1981xt,Mukhanov:1982nu,Starobinsky:1982ee,Guth:1982ec,Hawking:1982cz,Bardeen:1983qw}. 
The most recent Planck measurements~\cite{Adam:2015rua,Ade:2015lrj,Ade:2015ava} of the Cosmic Microwave Background (CMB) indicate that these cosmological perturbations are almost scale invariant, with no evidence of non-Gaussianities or isocurvature components. At this stage, the full set of observations can therefore be accounted for~\cite{Giannantonio:2014rva} in a minimal setup, where inflation is driven by a single scalar inflaton field $\phi$ with canonical kinetic term, minimally coupled to gravity, and evolving in a flat potential $V(\phi)$ in the slow-roll regime. When testing these models one by one in a systematic and Bayesian way~\cite{Martin:2014vha,Martin:2013nzq,Martin:2014rqa}, the best potentials, mostly of the ``Plateau'' type, can be identified.

From the theoretical point of view, inflation proceeds at energy scales where particle physics is not known and has not been tested in accelerators. The physical details of how the inflaton is connected with the standard model of particle physics and its extensions therefore remain elusive. In particular, most physical setups that have been proposed to embed inflation contain extra scalar fields that can play a role either during inflation or afterwards. A natural question is therefore whether single-field model predictions are robust under the introduction of these extra fields. 

From the observational point of view, one should also focus on the extensions to the minimal scenarios that the coming generation of cosmological and astrophysical surveys will be able to rule out. As far as non-Gaussianities are concerned for example, the accuracy of currently proposed experiments is typically of the order~\cite{Amendola:2012ys,Giannantonio:2011ya,Camera:2014bwa,Dore:2014cca,Leistedt:2014zqa,Munoz:2015eqa} $\sigma(\fnl)\sim 1$. Since models predicting $\vert\fnl\vert\gg 1$ are already excluded by Planck while models yielding $\vert\fnl\vert \ll 1$ will not be constrained in the short future, models that should draw our attention are those giving $\vert\fnl\vert\sim 1$.

For these reasons, in this paper we begin a systematic analysis of single-field slow-roll models of inflation when an extra light massive scalar field $\sigma$ is introduced and can play a role both during inflation and afterwards.\footnote{In many (notably string theory) high energy setups, a larger number of light scalar degrees of freedom are usually to be considered~\cite{Assadullahi:2007uw,Baumann:2014nda,Price:2014xpa}. Here, for simplicity, we only implement a single additional light scalar field but the method we present can be extended to such scenarios.} The potentials under scrutiny are thus of the form $V(\phi)+m_\sigma^2\sigma^2/2$, where $\sigma$ is taken to be lighter than $\phi$ at the end of inflation. Both fields are assumed to be slowly rolling during inflation, and eventually decay into radiation fluids with decay rates denoted $\Gamma_\phi$ and $\Gamma_\sigma$ respectively, during reheating. In the limit where the extra field $\sigma$ is entirely responsible for the observed primordial curvature perturbations, the class of models studied here is essentially the curvaton scenario~\cite{Linde:1996gt,Enqvist:2001zp,Lyth:2001nq,Moroi:2001ct}, for which a natural value of the local non-Gaussianity parameter is~\cite{Valiviita:2006mz} $\fnl = -5/4$, hence of order one.

However, in the present work, we address the generic setup where both $\phi$ and $\sigma$ can a priori contribute to curvature perturbations~\cite{Dimopoulos:2003az,Langlois:2004nn,Lazarides:2004we,Moroi:2005np}. When $V(\phi)$ is quadratic, these ``mixed'' scenarios have recently been studied in \Refs{Bartolo:2002vf,Ellis:2013iea,Byrnes:2014xua} and it has been shown~\cite{Enqvist:2013paa} that the fit of quartic chaotic inflation can be significantly improved in the curvaton limit. In \Ref{Hardwick:2015tma}, a Bayesian analysis was carried out for the quadratic inflaton + curvaton models assuming instantaneous reheating, and these models were found not to be disfavored with respect to standard quadratic inflation. In this paper however, we aim at addressing {\em all} single-field potentials as recently mapped in \Ref{Martin:2014vha}. Importantly, we do not make any assumption as to the ordering of the three events: $\sigma$ becomes massive, $\phi$ decays and $\sigma$ decays. Nor do we restrict the epochs during which $\sigma$ can dominate the energy content of the Universe. Assuming that the inflaton energy density decreases as matter after inflation and before it decays into radiation, this leaves us with 10 possible cases (including situations where $\sigma$ drives a secondary phase of inflation~\cite{Langlois:2004nn, Moroi:2005kz, Ichikawa:2008iq, Dimopoulos:2011gb}), see \Fig{fig:rho}, that we study one by one. Furthermore, reheating kinematic effects are properly taken into account so that the number of \efolds elapsed between the Hubble exit time of the CMB pivot scale and the end of inflation is not a free parameter but is given by an explicit function of the potential $V(\phi)$ parameters, $m_\sigma$, $\Gamma_\phi$, $\Gamma_\sigma$ and $\sigma_\uend$, the \vev of $\sigma$ at the end of inflation. This effect is particularly important for curvaton-like scenarios, since in these cases the same parameters determine the statistical properties of perturbations and the kinematics of reheating. It is therefore crucial to properly account for the interplay between these two physical effects and the suppression of degeneracies it yields.

In practice, we make use of the $\delta N$ formalism to relate observables to variations in the energy densities of both fields at decay time of the last field. This allows us to calculate all relevant physical quantities by only keeping track of the background energy densities. The main steps employed are described in section~\ref{sec:Method}, while detailed formulas are provided in appendices~\ref{sec:inflation:formulas} and \ref{sec:reheating:formulas}. The corresponding numerical routines have been implemented in the publicly available \ASPIC code~\cite{aspic}. In this manner, one can now easily compute the predictions of the $\sim 75$ inflationary potentials originally contained in this library, when a massive scalar field is added, in all 10 cases mentioned above.

In this paper, we discuss the results only for a few representative models: Large Field Inflation, Higgs Inflation (the Starobinsky model) and Natural Inflation. In appendix~\ref{sec:predictions} we display their predictions in all 10 cases, in terms of scalar spectral index $\nS$, tensor-to-scalar ratio $r$, and local non-Gaussianity parameter $\fnl$. In general, isocurvature perturbations can also be produced~\cite{Moroi:2001ct,Lyth:2002my}. Their amplitudes depend on the details of the couplings between the fields $\phi$ and $\sigma$ and the other species that here we leave unspecified. Indeed, they do not survive if either the field that last decays dominates the energy budget before decaying and/or the decay products fully thermalize after the last field decay. In principle however, non-adiabatic perturbations can give rise to extra constraints~\cite{Easson:2010uw}, and we leave this for future work.

\begin{figure}[t]
\begin{center}
\includegraphics[width=\wdblefig,clip=true]{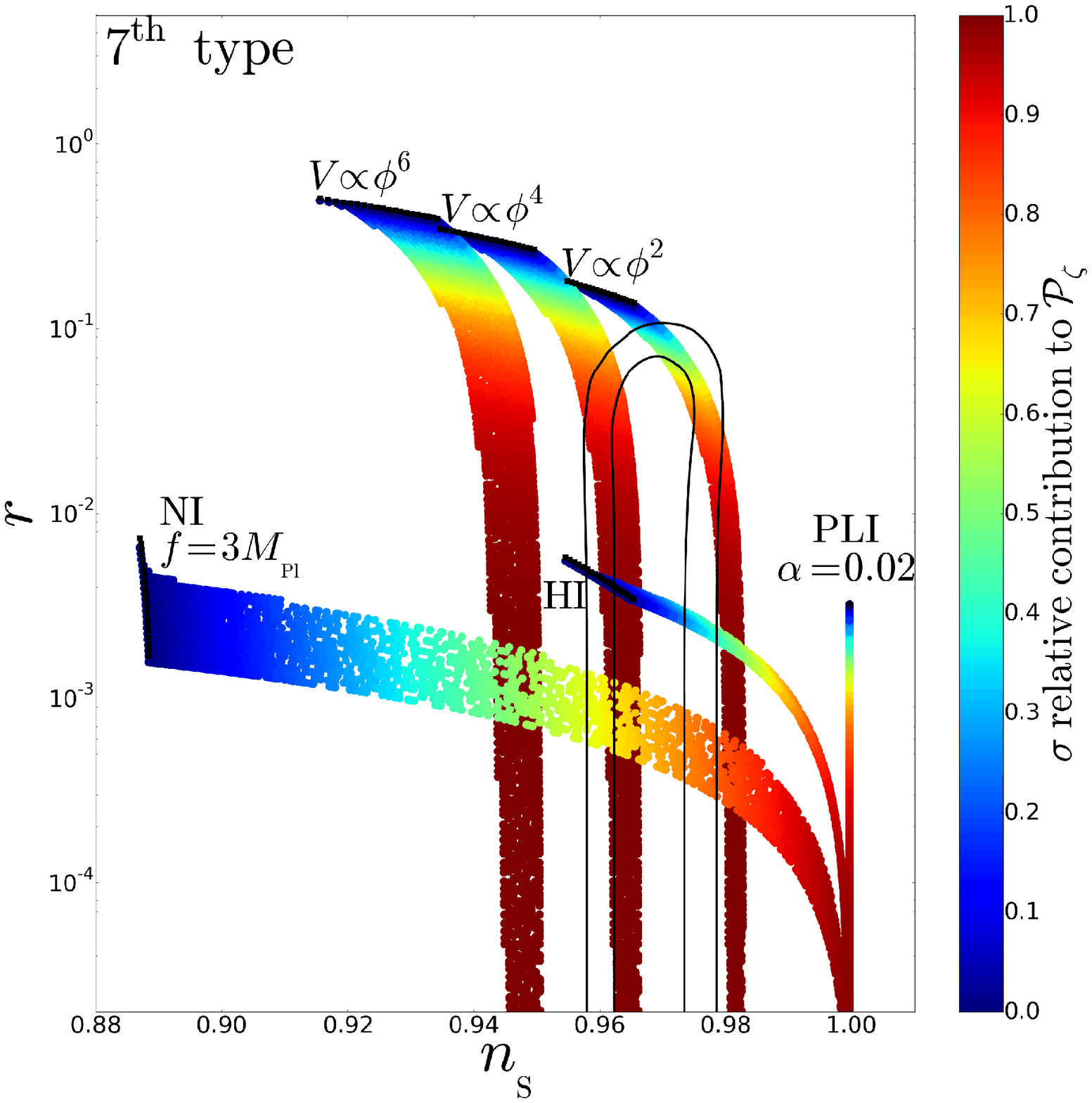}
\includegraphics[width=\wdblefig,clip=true]{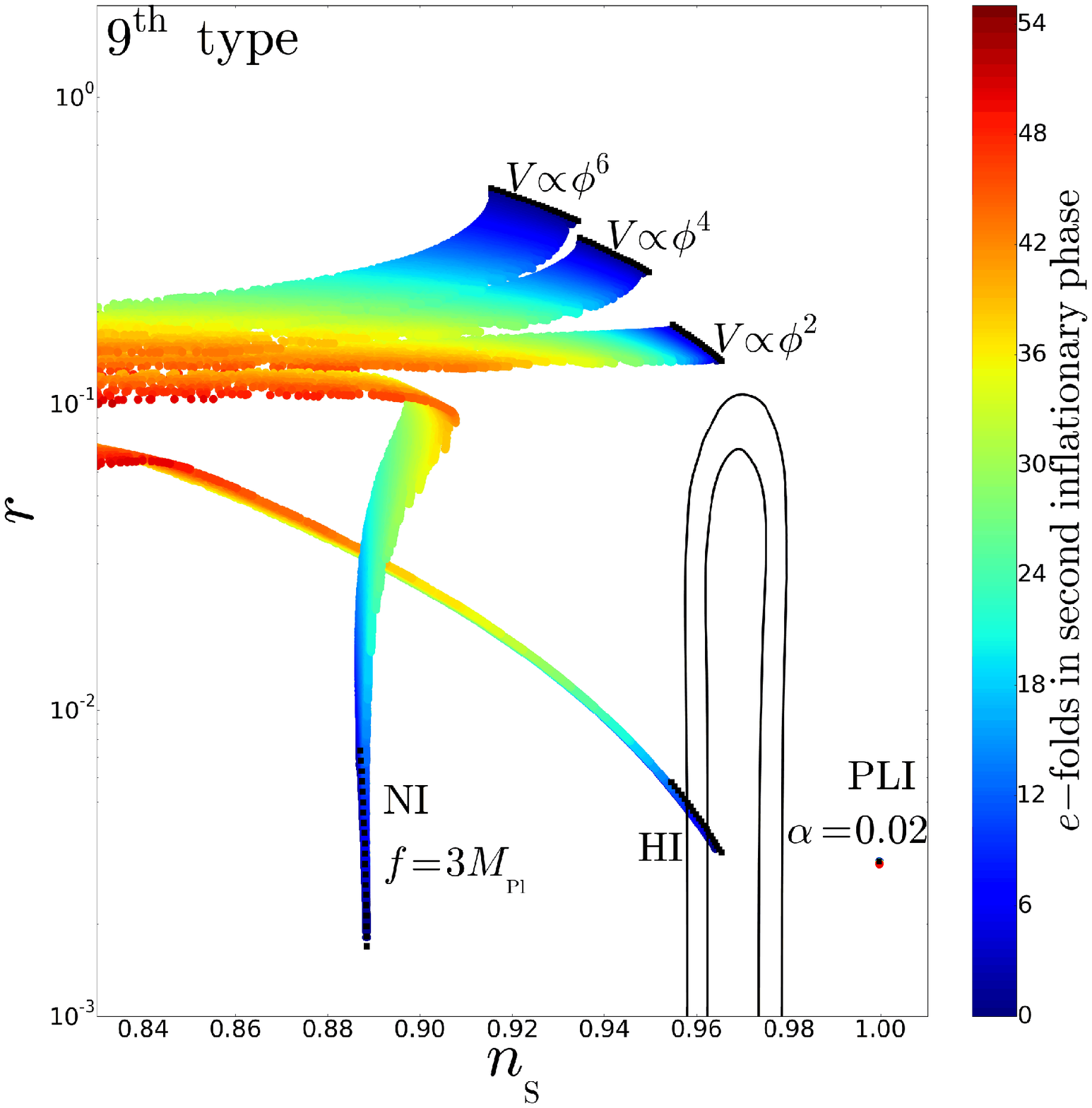}
\caption{Predictions of a few inflationary models (Large Field Inflation $V\propto\phi^p$ with $p=2$, $4$ and $6$, Higgs Inflation $V\propto [1-\exp(-\sqrt{2/3}\phi/\Mp)]^2$ - the Starobinsky model -, Natural Inflation $V\propto 1+\cos(\phi/f)$ with $f=3\Mp$ and Power Law Inflation $V\propto \exp(-\alpha\phi/\Mp)$ with $\alpha=0.02$), when a massive field $\sigma$ is added, and when the reheating scenario is of the $7^\mathrm{th}$ (left panel, corresponding to the standard ``curvaton scenario'') and $9^\mathrm{th}$ (right panel, corresponding to a typical situation where the added massive field drives a secondary phase of inflation) type. In the left panel, the color encodes the relative contribution from $\sigma$ to the total curvature power spectrum. In the right panel, the color encodes the number of \efolds realized during the second phase of inflation, taking place during reheating. The black lines are the one and two sigma Planck 2015 contours~\cite{Ade:2015lrj}, and the black squares stand for the predictions of the single-field versions of the models.}
\label{fig:summarynsR}
\end{center}
\end{figure}
We describe and analyze our main results in section~\ref{sec:Discussion}. However, in order to provide a quick picture of the situation, in \Fig{fig:summarynsR}, the predictions for $\nS$ and $r$ are superimposed for the models mentioned above, when reheating is of the $7^\mathrm{th}$ (left panel) and $9^\mathrm{th}$ (right panel) types (see \Fig{fig:rho}): 

$\bullet$ In the $7^\mathrm{th}$ type, the inflaton decays before the $\sigma$ field becomes massive and then decays. This can be viewed as a standard ``curvaton'' scenario as originally proposed~\cite{Linde:1996gt,Enqvist:2001zp,Lyth:2001nq,Moroi:2001ct}. The tensor-to-scalar ratio $r$ is driven to zero when the relative contribution from $\sigma$ to scalar perturbations increases. For small-field models (Natural Inflation, Higgs Inflation or Power Law Inflation in the figure), this does not make a big difference since those models already predict a value for $r$ that is currently unconstrained. For large-field models however (Large Field Inflation in the figure), the observational upper bound on $r$ strongly favors the model parameters for which $r$ is reduced. On the other hand, the spectral index $\nS$ increases as the extra field contribution to the scalar power spectrum increases, reaching the asymptotic value $\nS\simeq 1-2\epsilon_{1*}$, where $\epsilon_{1*}$ is the first slow-roll parameter evaluated at Hubble exit time of the pivot scale. For small field models, one effectively has $\nS\sim 1$ in this limit and three cases can therefore be distinguished. If the single-field version of the model predicts a too red value for $\nS$ (for example, Natural Inflation in the figure), a small fraction of the mixed scenarios can provide a good fit to the data, otherwise $\nS$ is either too red or too blue. If the single-field version of the model already predicts the right value for $\nS$ (for example, Higgs Inflation in the figure), adding a massive field can only make the fit to the data worse. The same conclusion holds if the single-field version of the model predicts a too blue value for $\nS$ (for example, Power Law Inflation in the figure). For Large Field Inflation on the other hand, in the limit where only the extra field contributes to scalar perturbations, one has $\nS\sim 1-2p/[p+4(N_\uend- N_*)]$. This means that $p=6$ predicts values of $\nS$ that are too red and $p=2$ predicts values of $\nS$ that are too blue in this limit, while $p=4$ provides a good fit. 

$\bullet$ In the $9^\mathrm{th}$ type (right panel of \Fig{fig:summarynsR}), $\sigma$ drives a secondary phase of inflation. For small-field models, the main effect is that $N_\uend - N_*$, the number of \efolds elapsed between the Hubble exit time of the pivot scale and the end of the \emph{first} phase of inflation, decreases. (In particular, one can check that Power Law Inflation, which has an inflaton shift symmetry, has its predictions essentially unchanged). As a consequence, predictions are computed closer to the ending point of the first inflationary phase, where the potential $V(\phi)$ is less flat. This is why $r$ typically increases and $\nS$ is shifted away from scale invariance. The situation is therefore opposite to the $7^\mathrm{th}$ kind displayed in the left panel: models predicting too red values of $\nS$ provide even worse fit to the data once $\sigma$ is included, models predicting the right value of $\nS$ are also made worse, while models yielding too blue values of $\nS$ (but still red) can provide a better fit for some parameters. For large-field models, this effect almost compensates with the reduction in $r$ arising from the increase in the contribution from $\sigma$ to the total perturbations, so that $r$ does not change much. In particular, $r$ does not decrease sufficiently so that, even if one starts from a large single-field model with a too blue value of $\nS$, a good fit to the data can never be obtained.

In order to further investigate how the observational status of a given single-field model of inflation changes when a massive scalar field is added, one needs to incorporate non-Gaussianities to the discussion. This is done in detail in the rest of the paper. The aim of the above discussion is only to provide the reader with a taste of the physical effects we will encounter later. Finally, let us note that in some reheating scenarios and for some potentials, the best fit to the data can be improved when adding a massive scalar field, but this might be realized only for limited ranges of parameters. The immediate question to ask is of course whether the increase in the fit and the corresponding volume in parameter space could lead to a substantial increase in the Bayesian evidence of such models. In a future publication~\cite{Vennin:2016}, we will address this issue by computing the Bayesian evidence of the models present in the \ASPIC library~\cite{aspic}, when a massive scalar field is added. This will allow us to quantitatively discuss the observational status of curvaton-type models of inflation, as well as assessing the robustness of single-field models of inflation when massive scalar fields are added.
\section{Method}
\label{sec:Method}
In this paper, we investigate the situation where inflation is mainly driven by an inflaton scalar field $\phi$ with potential $V(\phi)$, in the presence of a light free scalar field $\sigma$. Here, ``light'' must be understood in the sense that the mass of this extra field is smaller than the Hubble scale during the whole inflationary period, \ie $m_\sigma/\sqrt{2}\ll H_\uend$, where $H_\uend$ denotes the Hubble factor at the end of inflation. The two fields follow Klein-Gordon equations in Friedmann-Lema\^itre-Robertson-Walker space-times given by
\bea
\ddot{\phi}+3H\dot{\phi}+V_\phi&=&0 \, ,\\
\ddot{\sigma}+3H\dot{\sigma}+m_\sigma^2\sigma &=&0\, ,
\label{eq:KG:sigma}
\eea
where the Hubble factor $H$ obeys the Friedmann equation
\beq
3\Mp^2 H^2=\dfrac{\dot{\phi}^2}{2}+V\left(\phi\right)+\dfrac{\dot{\sigma}^2}{2}+\frac{m_\sigma^2}{2}\sigma^2\, .
\label{eq:Friedmann}
\eeq
During inflation, we assume that both fields are slowly rolling. When inflation is over, we work under the sudden decay approximation, since it has been shown to be a good approximation to the full numerical results in the usual curvaton scenarios~\cite{Malik:2006pm,Sasaki:2006kq}. In practice, this means that when $H$ drops below $\Gamma_\phi$ (respectively $\Gamma_\sigma$), the energy contained in $\phi$ (respectively $\sigma$) instantaneously decays to radiation.\footnote{Here, $\Gamma_\phi$ (respectively $\Gamma_\sigma$) are effective values for which assuming instantaneous decay at $H=\Gamma_\phi$ (respectively $H=\Gamma_\sigma$) provides a good description of the full decay dynamics. They do not exactly match the ``decay rates'' as would be defined in a dynamical way. However, since $\Gamma_\phi$ (respectively $\Gamma_\sigma$) are varied across orders of magnitudes, this plays a minor role.} We also assume that close to its minimum, the inflationary potential is quadratic so that the energy density contained in $\phi$ behaves as matter between the end of inflation and the decay of $\phi$. 

On top of the parameters describing the inflationary potential $V(\phi)$, the system therefore contains four free parameters: $m_\sigma$, $\Gamma_\phi$, $\Gamma_\sigma$ and $\sigma_\uend$. The kinematic parameters of reheating (energy density at the onset of the radiation era and mean equation of state parameter, see section~\ref{sec:reheating}) being entirely fixed by these quantities, there is no other free parameter.
\begin{figure*}[t]
\begin{center}
\includegraphics[width=0.9\textwidth]{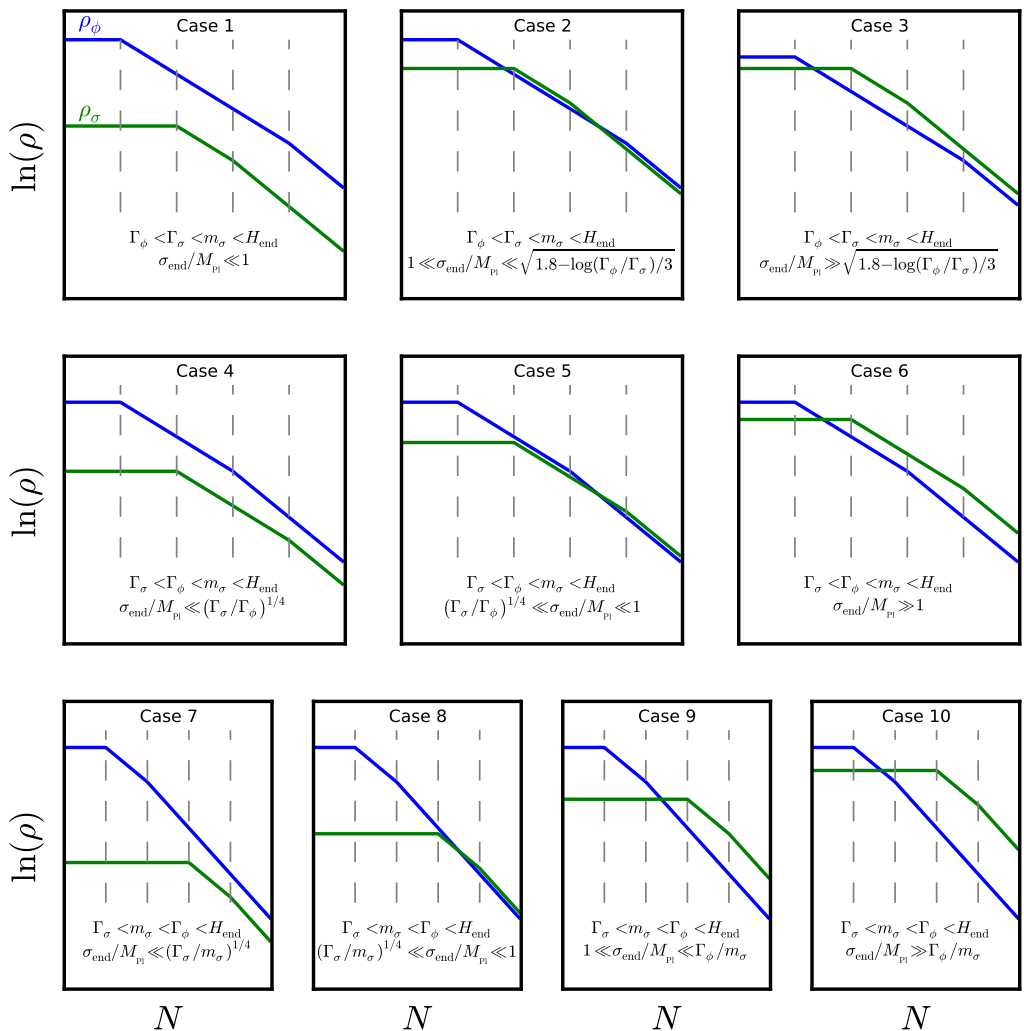}
\caption{Different possible reheating scenarios, depending on the values taken by $\Gamma_\sigma$, $m_\sigma$, $\Gamma_\phi$, $H_\uend$ and $\sigma_\uend$. Cases 1, 2 and 3 correspond to $\Gamma_\phi<\Gamma_\sigma<m_\sigma<H_\uend$; cases 4, 5 and 6 correspond to $\Gamma_\sigma<\Gamma_\phi<m_\sigma<H_\uend$; cases 7, 8, 9 and 10 correspond to $\Gamma_\sigma<m_\sigma<\Gamma_\phi<H_\uend$. Within each row, different cases are distinguished by $\sigma_\uend/\Mp$ which controls when $\sigma$ dominates the total energy density. The blue curves stand for the energy density of $\phi$ while the green ones are for $\sigma$.}
\label{fig:rho}
\end{center}
\end{figure*}

Let us now see what the different scenarios for the reheating phase are. Here, by ``reheating'', we refer to anything happening between the end of the phase of inflation driven by $\phi$ and the onset of the final radiation dominated era. Four energy scales are present in the model, $\Gamma_\sigma$, $m_\sigma$, $\Gamma_\phi$ and $H_\uend$. Since $m_\sigma<H_\uend$ by definition, and since $m_\sigma>\Gamma_\sigma$ (\ie we assume perturbative decay) and $H_\uend>\Gamma_\phi$ ($\phi$ decays after inflation ends), these values must be ordered according to one of the three following possibilities: $\Gamma_\phi<\Gamma_\sigma<m_\sigma<H_\uend$, $\Gamma_\sigma<\Gamma_\phi<m_\sigma<H_\uend$ or $\Gamma_\sigma<m_\sigma<\Gamma_\phi<H_\uend$. The time evolution of the energy densities associated with $\phi$ and $\sigma$ are schematically represented in \Fig{fig:rho}. Depending on the value of $\sigma_\uend$, hence on the energy density contained in $\sigma$, one can see that 10 sub-cases must be distinguished.
\subsection{Observable Quantities as Energy Density Variations}
Since we want to address a large number of inflationary potentials $V(\phi)$ and describe all $10$ possible cases in a systematic way, it is necessary to design a calculational framework that allows us to obtain the predictions of these models in a standardized and methodical way. A numerically-oriented well-optimized solution is to make use of the ``quasi-isotropic''~\cite{Lifshitz:1960,Starobinsky:1982mr,Khalatnikov:2002kn} or ``separate universe''~\cite{Sasaki:1998ug,Wands:2000dp,Lyth:2003im,Lyth:2004gb} approach and the corresponding $\delta N$ formalism~\cite{Starobinsky:1986fxa,Sasaki:1995aw,Wands:2000dp,Lyth:2005fi} to express all observable quantities in terms of the variations of the energy densities at the last decay time, with respect to the fields values at first Hubble crossing time during inflation. Let us see how this can be done.

When the last field decays, since the other field has already decayed, one has two fluids: a matter fluid that corresponds to the field $\phim$ oscillating  at the bottom of its potential, and a radiation fluid that corresponds to the decay products of the other field $\phir$. In cases 1-3, one has $\phim=\phi$ and $\phir=\sigma$ since the field that decays last is $\phi$, while in cases 4-10, one has $\phim=\sigma$ and $\phir=\phi$ since the field that last decays is $\sigma$.

The non-linear curvature perturbation of the matter fluid on uniform-matter density hypersurfaces is given by~\cite{Lyth:2004gb}
\bea
\zeta_\matter\left(t,\bm{x}\right)&=&\delta N\left(t,\bm{x}\right)+\int_{\bar{\rho}_\matter(t)}^{\rho_\matter(t,\bm{x})}\dfrac{\dd\tilde{\rho}_\matter}{3\tilde{\rho}_\matter}
=
\zeta\left(t,\bm{x}\right) + \dfrac{1}{3}\ln\left[\dfrac{\rho_\matter(t,\bm{x})}{\bar{\rho}_\matter(t)}\right]\, ,
\label{eq:zetam:deltarho}
\eea
where in the last equality $\rho_\matter(t,\bm{x})$ is defined on uniform total energy density hypersurfaces. In the same manner, the non-linear curvature perturbation of the radiation fluid on uniform-radiation density hypersurfaces is given by,
\bea
\zeta_\radiation\left(t,\bm{x}\right)&=&\zeta\left(t,\bm{x}\right) + \dfrac{1}{4}\ln\left[\dfrac{\rho_\radiation(t,\bm{x})}{\bar{\rho}_\radiation(t)}\right]\, ,
\label{eq:zetar:deltarho}
\eea
where similarly, $\rho_\radiation(t,\bm{x})$ is defined on uniform total energy density hypersurfaces. The total density is uniform on the last decay surface, parametrized by $H=\Gamma$, so one has $\rho_\matter\left(t_\udec,\bm{x}\right)+\rho_\radiation\left(t_\udec,\bm{x}\right)=\bar{\rho}\left(t_\udec\right)$, where a bar denotes background, homogeneous quantities. This gives rise to~\cite{Sasaki:2006kq}
\beq
\bar{\rho}_\matter\left(t_\udec\right)e^{3\left(\zeta_\matter-\zeta\right)}+\bar{\rho}_\radiation\left(t_\udec\right)e^{4\left(\zeta_\radiation-\zeta\right)}=\bar{\rho}_\matter\left(t_\udec\right)+\bar{\rho}_\radiation\left(t_\udec\right)\, .
\label{eq:decaysurface}
\eeq
This non-linear relation allows us to relate $\zeta$ with $\zeta_\matter$ and $\zeta_\radiation$ at any order in perturbation theory. 
\subsubsection{Power Spectra}
\label{sec:powerspectrum}
Let us now expand \Eq{eq:decaysurface} at linear order to work out the scalar power spectrum. Defining
\beq
\label{eq:rdec:def}
r_\mathrm{dec}=\dfrac{3\bar{\rho}_\matter\left(t_\udec\right)}{3\bar{\rho}_\matter\left(t_\udec\right)+4\bar{\rho}_\radiation\left(t_\udec\right)}\, ,
\eeq
one obtains $\zeta=r_\mathrm{dec}\zeta_\matter+\left(1-r_\mathrm{dec}\right)\zeta_\radiation$. We thus need to specify $\zeta_\matter$ and $\zeta_\radiation$. Since the two corresponding calculations proceed in exactly the same way, let us give the details only for $\zeta_\matter$. A first remark is that, from \Eq{eq:zetam:deltarho}, one can write
\beq
\zeta_\matter=\dfrac{1}{3}\ln\left[\dfrac{\rho_\matter(t,\bm{x})}{\bar{\rho}_\matter(t)}\right]_{\delta N=0}\, .
\label{eq:zetam1:rhom}
\eeq

On the other hand, after Fourier expanding the density field $\rho_\matter$, one has, on super-Hubble scales,
\beq
\left.\delta \rho_\matter(t,\bm{k})\right\vert_{\delta N=0}=\left.\dfrac{\partial\bar{\rho}_\matter(t)}{\partial\phi_{\matter,k}}\right\vert_{\delta N=0}\delta\varphi_{\matter,\bm{k}}+\left.\dfrac{\partial\bar{\rho}_\matter(t)}{\partial\phi_{\radiation,k}}\right\vert_{\delta N=0}\delta\varphi_{\radiation,\bm{k}}\, .
\label{eq:rhom1:Fourrier}
\eeq
Here, $\delta\varphi_{\matter,\bm{k}}$ and $\delta\varphi_{\radiation,\bm{k}}$ are the field fluctuations at the Hubble exit time of the wavenumber $\bm{k}$. At leading order in slow roll, they are uncorrelated Gaussian fluctuations with amplitude $H(k)/(2\pi)$, where $H(k)$ is the value of the Hubble parameter when $\bm{k}$ exits the Hubble radius (see \Ref{Byrnes:2006fr} for inclusion of higher order slow-roll corrections). Here also, $\vert_{\delta N=0}$ means that the derivatives with respect to $\varphi_k$, the values of the fields when $\bm{k}$ exits the Hubble radius, are to be evaluated under the constraint that the number of \efolds elapsed between the time $N_k$ when the mode $\bm{k}$ crosses the Hubble radius during inflation and the time $N_\udec$ of the last decay, is fixed. Plugging \Eq{eq:rhom1:Fourrier} into \Eq{eq:zetam1:rhom}, one obtains, at linear order,
\beq
\label{eq:zetam1:deltafieldstar}
\zeta_\matter\left(t,\bm{k}\right)=\dfrac{1}{3\bar{\rho}_\matter(t)}\left[\left.\dfrac{\partial\bar{\rho}_\matter(t)}{\partial\phi_{\matter,k}}\right\vert_{\delta N=0}\delta\varphi_{\matter,\bm{k}}+\left.\dfrac{\partial\bar{\rho}_\matter(t)}{\partial\phi_{\radiation,k}}\right\vert_{\delta N=0}\delta\varphi_{\radiation,\bm{k}}\right]\, .
\eeq
A similar expression for $\zeta_\radiation$ can be derived, where one just needs to exchange matter and radiation indices, and $1/4$ replaces the overall $1/3$ factor. One then obtains, at the last field decay time,
\bea
\zeta\left(\bm{k}\right) &=& \left[\dfrac{r_\mathrm{dec}}{3\bar{\rho}_\matter(t_\mathrm{dec})}\left.\dfrac{\partial\bar{\rho}_\matter(t_\mathrm{dec})}{\partial\phi_{\matter,k}}\right\vert_{\delta\left(N_\udec-N_k\right)=0} +\dfrac{1-r_\mathrm{dec}}{4\bar{\rho}_\radiation(t_\mathrm{dec})}\left.\dfrac{\partial\bar{\rho}_\radiation(t_\mathrm{dec})}{\partial\phi_{\matter,k}}\right\vert_{\delta\left(N_\udec-N_k\right)=0}\right]\delta\varphi_{\matter,\bm{k}}
\nonumber\\ &+&
 \left[\dfrac{r_\mathrm{dec}}{3\bar{\rho}_\matter(t_\mathrm{dec})}\left.\dfrac{\partial\bar{\rho}_\matter(t_\mathrm{dec})}{\partial\phi_{\radiation,k}}\right\vert_{\delta\left(N_\udec-N_k\right)=0} +\dfrac{1-r_\mathrm{dec}}{4\bar{\rho}_\radiation(t_\mathrm{dec})}\left.\dfrac{\partial\bar{\rho}_\radiation(t_\mathrm{dec})}{\partial\phi_{\radiation,k}}\right\vert_{\delta\left(N_\udec-N_k\right)=0}\right]\delta\varphi_{\radiation,\bm{k}}\, .
\nonumber\\
\label{eq:zeta1}
\eea
In order to make notations more compact, let us introduce the quantities
\beq
\label{eq:Aalphabeta:def}
A^\alpha_\beta\left(k\right)\equiv\left.\frac{\partial\ln\bar{\rho}_\alpha\left(t_\udec\right)}{\partial\varphi_{\beta,k}}\right\vert_{\delta\left(N_\udec-N_k\right)=0}\, ,
\eeq
where $\alpha,\beta=\matter,\radiation$. The scalar curvature power spectrum is then given by $\calP_\zeta(k)=\left\langle \zeta^2 (k)\right\rangle$, that is to say
\bea
\calP_\zeta(k) &=&\left[
\left(\dfrac{r_\mathrm{dec}}{3} A^\matter_\matter +\dfrac{1-r_\mathrm{dec}}{4} A^\radiation_\matter\right)^2
+\left(\dfrac{r_\mathrm{dec}}{3}A^\matter_\radiation +\dfrac{1-r_\mathrm{dec}}{4}A^\radiation_\radiation\right)^2
\right] \left[\dfrac{H\left(k\right)}{2\pi}\right]^2\, .
\label{eq:powerspectrum:rho}
\eea
From here, the spectral index and the running can be obtained as simple expressions of the potential derivatives at first~\cite{Wands:2002bn} and second~\cite{Byrnes:2006fr} order in slow roll. For simplicity, here we do not reproduce these formulas since they are explicit. From \Eq{eq:powerspectrum:rho}, one can also identify the relative contribution from the matter field to the total scalar power spectrum (first term in the brackets) or, equivalently, the relative contribution from the radiation field (second term). These quantities are not directly observable by themselves but they can be useful to study as they indicate which field mostly contributes to scalar perturbations.  

Let us now consider the gravitational waves power spectrum. The tensor perturbations, in contrast to the scalar perturbations, remain frozen on large scales and decouple from the scalar perturbations at linear order. Thus the primordial perturbation spectrum for gravitational waves is given by the usual formula~\cite{Starobinsky:1979ty},
\beq
\calP_h(k)= \frac{2 H^2\left(k\right)}{\pi^2\Mp^2}\, .
\label{eq:powerspectrum:h}
\eeq
\subsubsection{Non-Gaussianities}
Due to the presence of non-adiabatic field perturbations on large scales after inflation, curvaton-like scenarios can produce local type non-Gaussianities~\cite{Sasaki:2006kq}, which is also the observationally most constrained type of non-Gaussianities~\cite{Ade:2015ava}. This is why in this paper, we characterize non-Gaussianities predictions by the value of the local $\fnl$ parameter. It can be calculated by expanding perturbations up to second order, $\zeta=\zeta_1+\zeta_2/2$, where $\zeta_1$ is the quantity we already calculated for the scalar power spectrum in \Eq{eq:zeta1} in section~\ref{sec:powerspectrum}. The local non-Gaussianities parameter $\fnl$ is defined through the ratio of the bispectrum to the power spectrum squared, 
\beq
\fnl=\frac{5}{6}\frac{\left\langle \zeta^3 \right\rangle}{3\left\langle \zeta^2\right\rangle^2}=\frac{5}{6}\frac{\left\langle \zeta_1^2 \zeta_2 \right\rangle}{\left\langle \zeta_1^2\right\rangle^2}\, ,
\eeq
where the second expression is valid at leading order in the perturbations. Let us see how it can be calculated.

Similarly to what was done for the scalar power spectrum, one can first expand \Eq{eq:decaysurface} at second order in the perturbations to relate $\zeta_2$ at the last decay time to its matter and radiation components. One obtains
\beq
\zeta_2=r_\udec\zeta_{\matter2}+\left(1-r_\udec\right)\zeta_{\radiation2}+r_\udec\left(1-r_\udec\right)\left(3+r_\udec\right)\left(\zeta_{\matter1}-\zeta_{\radiation1}\right)^2\, .
\eeq
The first order curvature perturbations $\zeta_{\matter1}$ and $\zeta_{\radiation1}$ have already been expressed in terms of the scalar field fluctuations at Hubble exit time in section~\ref{sec:powerspectrum}, see \Eq{eq:zetam1:deltafieldstar}. One needs to work out similar expressions for $\zeta_{\matter2}$ and $\zeta_{\radiation2}$. Expanding \Eqs{eq:zetam:deltarho} and ~(\ref{eq:zetar:deltarho}) at second order in the perturbations, one obtains
\beq
\zeta_{\matter2}\left(\bm{k}\right)=\frac{1}{6}
\left.\frac{\partial^2\ln\bar{\rho}_\matter}{\partial\phi_{\matter,k}^2}\right\vert_{\delta N=0}
\delta\varphi_{\matter,\bm{k}}^2
+\frac{1}{6}\left.\frac{\partial^2\ln\bar{\rho}_\matter}{\partial\phi_{\radiation,k}^2}\right\vert_{\delta N=0}
\delta\varphi_{\radiation,\bm{k}}^2
+\frac{1}{3}\left.\frac{\partial^2\ln\bar{\rho}_\matter}{\partial\phi_{\radiation,k}\partial\phi_{\matter,k}}\right\vert_{\delta N=0}
\delta\varphi_{\matter,\bm{k}}\delta\varphi_{\radiation,\bm{k}}\, ,
\eeq
and a similar expression for $\zeta_{\matter2}$. Recalling that, at Hubble crossing time, one has~\cite{Byrnes:2006vq} $\langle\delta\varphi_{\matter,\bm{k}}^2\rangle=\langle\delta\varphi_{\radiation,\bm{k}}^2\rangle=H(k)^2/(2\pi)^2$,  $\langle\delta\varphi_{\matter,\bm{k}}^4\rangle=\langle\delta\varphi_{\radiation,\bm{k}}^4\rangle=2 H(k)^4/(2\pi)^4$ and $\langle\delta\varphi_{\matter,\bm{k}}\delta\varphi_{\radiation,\bm{k}}\rangle=0$, the previous results give rise to
\bea
\fnl &=& \frac{5}{3} \left\lbrace 81 \left({A^\radiation_{\radiation}}\right)^2 \left({A^\radiation_{\mathrm{mm}}} + 2 {A^\radiation_{\mathrm{rr}}}\right) +
27 {A^\radiation_{\radiation}} \left\lbrace 8 {A^\matter_{\matter}} {A^\radiation_{\mathrm{mr}}} + 8 {A^\matter_{\matter}}^2 {A^\radiation_{\radiation}}
+ 16 \left({A^\matter_{\radiation}}\right)^2 {A^\radiation_{\radiation}} +
{A^\radiation_{\radiation}} \left[ 4 {A^\matter_{\mathrm{mm}}}
\right. \right. \right. \nonumber \\ & & \left. \left. \left. 
+ 8 {A^\matter_{\mathrm{rr}}} 
- 9 {A^\radiation_{\mathrm{mm}}} + 9 \left({A^\radiation_{\radiation}}\right)^2 - 18 {A^\radiation_{\mathrm{rr}}}\right]
+ 8 {A^\matter_{\radiation}} \left[{A^\radiation_{\mathrm{mm}}} - 3 \left({A^\radiation_{\radiation}}\right)^2
+ 2 {A^\radiation_{\mathrm{rr}}}\right]\right\rbrace r_\udec 
\right. \nonumber \\ & & \left.
+ 9 \left(16 {A^\matter_{\matter}} \left(2 {A^\matter_{\radiation}} {A^\radiation_{\mathrm{mr}}}
+ 4 {A^\matter_{\mathrm{mr}}} {A^\radiation_{\radiation}} - 3 {A^\radiation_{\mathrm{mr}}}
{A^\radiation_{\radiation}}\right) +
16 \left({A^\matter_{\matter}}\right)^2 \left[2 {A^\radiation_{\mathrm{mm}}} 
+ 2 \left(6 {A^\matter_{\radiation}} - 5 {A^\radiation_{\radiation}}\right)
\right. \right. \right. \nonumber \\ & & \left.\left.\left.
{A^\radiation_{\radiation}} + {A^\radiation_{\mathrm{rr}}}\right] + \left(4 {A^\matter_{\radiation}}
- 3 {A^\radiation_{\radiation}}\right) \left\lbrace 32 \left({A^\matter_{\radiation}}\right)^2 {A^\radiation_{\radiation}} +
{A^\radiation_{\radiation}} \left[8 {A^\matter_{\mathrm{mm}}}
+ 16 {A^\matter_{\mathrm{rr}}} - 9 {A^\radiation_{\mathrm{mm}}} 
\right. \right. \right. \right. \nonumber \\ & & \left. \left. \left. \left.
+ 24 \left({A^\radiation_{\radiation}}\right)^2
- 18 {A^\radiation_{\mathrm{rr}}}\right] +
4 {A^\matter_{\radiation}} \left[{A^\radiation_{\mathrm{mm}}} - 14 \left({A^\radiation_{\radiation}}\right)^2
+ 2 {A^\radiation_{\mathrm{rr}}}\right]\right\rbrace\right) r_\udec^2 +
3 \left\lbrace 256 \left({A^\matter_{\matter}}\right)^4
\right. \right. \nonumber \\ & & \left.\left.
+8 {A^\matter_{\matter}} \left(8 {A^\matter_{\mathrm{mr}}}
- 3 {A^\radiation_{\mathrm{mr}}}\right) \left(4 {A^\matter_{\radiation}}
- 3 {A^\radiation_{\radiation}}\right) + \left(4 {A^\matter_{\radiation}} -
3 {A^\radiation_{\radiation}}\right)^2
\left[4 {A^\matter_{\mathrm{mm}}} + 16 \left({A^\matter_{\radiation}}\right)^2 + 8 {A^\matter_{\mathrm{rr}}} 
\right. \right. \right. \nonumber \\ & & \left. \left. \left.
- 3 {A^\radiation_{\mathrm{mm}}} -
40 {A^\matter_{\radiation}} {A^\radiation_{\radiation}} + 18 \left({A^\radiation_{\radiation}}\right)^2
- 6 {A^\radiation_{\mathrm{rr}}}\right] +
8 \left({A^\matter_{\matter}}\right)^2 \left[16 {A^\matter_{\mathrm{mm}}} + 96 \left({A^\matter_{\radiation}}\right)^2 + 8 {A^\matter_{\mathrm{rr}}}
\right. \right. \right. \nonumber \\ & & \left. \left. \left.
- 192 {A^\matter_{\radiation}} 
{A^\radiation_{\radiation}} +
87 \left({A^\radiation_{\radiation}}\right)^2 - 6 \left(2 {A^\radiation_{\mathrm{mm}}} + {A^\radiation_{\mathrm{rr}}}\right)\right]\right\rbrace r_\udec^3 -
8 \left[64 \left({A^\matter_{\matter}}\right)^4 + {A^\matter_{\radiation}}
\left(4 {A^\matter_{\radiation}} - 3 {A^\radiation_{\radiation}}\right)^3
\right.\right. \nonumber \\ & & \left. \left. 
 + 6 \left({A^\matter_{\matter}}\right)^2 \left(4 {A^\matter_{\radiation}} - 3 {A^\radiation_{\radiation}}\right) \left(8 {A^\matter_{\radiation}} - 3 {A^\radiation_{\radiation}}\right)\right] r_\udec^4 - \left[256
\left({A^\matter_{\matter}}\right)^4 +
48 \left({A^\matter_{\matter}}\right)^2
\left(4 {A^\matter_{\radiation}} - 3 {A^\radiation_{\radiation}}\right)^2 
\right. \right. \nonumber \\ & & \left. \left.
+ \left(4 {A^\matter_{\radiation}} - 3 {A^\radiation_{\radiation}}\right)^4\right] r_\udec^5 -
81 \left({A^\radiation_{\matter}}\right)^4
\left(r_\udec-1\right)^3 r_\udec \left(3 + r_\udec\right) +
216 {A^\matter_{\matter}} \left({A^\radiation_{\matter}}\right)^3
\right. \nonumber \\ & & \left.
 \left(r_\udec-1\right)^2 r_\udec \left(3 + r_\udec\right) \left(2 r_\udec-1\right)
+6 {A^\radiation_{\matter}} \left(r_\udec-1\right) \left[-9 {A^\radiation_{\mathrm{mr}}}
\left(r_\udec-1\right) 
\right. \right. \nonumber \\ & & \left.\left.
\left[3 {A^\radiation_{\radiation}} \left(r_\udec-1\right) - 4 {A^\matter_{\radiation}} r_\udec\right] +
2 r_\udec
\left(32 \left({A^\matter_{\matter}}\right)^3 r_\udec \left(3 + r_\udec\right) \left(2 r_\udec-1\right)
\right.\right.\right. \nonumber \\ & & \left. \left. \left.
+ 12 {A^\matter_{\mathrm{mr}}} \left[3 {A^\radiation_{\radiation}}
\left(r_\udec-1\right) - 4 {A^\matter_{\radiation}} r_\udec\right]
+ 3 {A^\matter_{\matter}} \left\lbrace 12 {A^\radiation_{\mathrm{mm}}}
\left(r_\udec-1\right)-6 {A^\radiation_{\mathrm{rr}}}
\right.\right.\right.\right. \nonumber \\ & & \left.\left.\left.\left.
- 8 \left(2 {A^\matter_{\mathrm{mm}}} + {A^\matter_{\mathrm{rr}}}\right) r_\udec +
16 \left({A^\matter_{\radiation}}\right)^2
r_\udec \left(3 + r_\udec\right) \left(2 r_\udec-1\right)
- 8 {A^\matter_{\radiation}} {A^\radiation_{\radiation}} \left(3 + r_\udec\right)
\right.\right.\right.\right. \nonumber \\ & & \left.\left.\left.\left.
\left[1 + 6 \left(r_\udec-1\right) r_\udec\right]
+  3 \left[2 {A^\radiation_{\mathrm{rr}}} r_\udec
+ 3  \left({A^\radiation_{\radiation}}\right)^2 \left(r_\udec-1\right)
\left(3 + r_\udec\right) \left(2 r_\udec -1\right) \right] \right\rbrace \right) \right] 
\right. \nonumber \\ & & \left.
-9 \left({A^\radiation_{\matter}}\right)^2 \left(r_\udec-1\right) \left[18 {A^\radiation_{\mathrm{mm}}} \left(r_\udec-1\right)^2
+ 9 {A^\radiation_{\mathrm{rr}}}
\left(r_\udec-1\right)^2 +
r_\udec 
\right.\right. \nonumber \\ & & \left.\left.
\left(3 \left[8 {A^\matter_{\mathrm{mm}}} + 8 \left({A^\matter_{\radiation}}\right)^2 + 4 {A^\matter_{\mathrm{rr}}} - 36 {A^\matter_{\radiation}} {A^\radiation_{\radiation}}
+ 27 \left({A^\radiation_{\radiation}}\right)^2\right]
+ 16 \left({A^\matter_{\matter}}\right)^2 \left(3 + r_\udec\right) 
\right.\right.\right. \nonumber \\ & & \left.\left.\left.
\left[1 + 6 \left(r_\udec-1\right) r_\udec\right] +
r_\udec \left\lbrace 27 \left({A^\radiation_{\radiation}}\right)^2
\left(r_\udec^2 + r_\udec -5\right)-24 {A^\matter_{\mathrm{mm}}} - 12 {A^\matter_{\mathrm{rr}}}
\right.\right.\right.\right. \nonumber \\ & & \left.\left.\left.\left.
 +8 \left({A^\matter_{\radiation}}\right)^2 \left[(6 r_\udec \left(2 + r_\udec\right)-17\right] -
36 {A^\matter_{\radiation}} {A^\radiation_{\radiation}} \left[r_\udec
\left(3 + 2 r_\udec\right)-8\right]\right\rbrace\right)\right]\right\rbrace/
\nonumber \\ & &
\left\lbrace 9 \left[\left({A^\radiation_{\matter}}\right)^2+\left({A^\radiation_{\radiation}}\right)^2\right] \left(r_\udec-1 \right)^2  - 24  \left( r_\udec -1\right)
 r_\udec \left({A^\matter_{\matter}} {A^\radiation_{\matter}}+{A^\matter_{\radiation}} {A^\radiation_{\radiation}}\right)
\right. \nonumber \\ & & \left.
 + 16 \left[\left({A^\matter_{\matter}}\right)^2 + \left({A^\matter_{\radiation}}\right)^2\right] r_\udec^2\right\rbrace^2\, .
\label{eq:fnl}
\eea
Here, similarly to \Eq{eq:Aalphabeta:def}, for display convenience, we have introduced the short notation
\beq
\label{eq:Aalphabetagamma:def}
A^\alpha_{\beta\gamma}\left(k\right)\equiv\left.\frac{\partial\ln\bar{\rho}_\alpha\left(t_\udec\right)}{\partial\varphi_{\beta,k}\partial\varphi_{\gamma,k}}\right\vert_{\delta\left(N_\udec-N_k\right)=0}\, .
\eeq

This formalism might seem a little heavy (in particular, in light of \Eq{eq:fnl} or considering the various equations given in appendix~\ref{sec:reheating:formulas}), but it allows us to design a computational strategy that is fully generic, since it only relies on tracking down energy density components during the different stages of the scenarios under consideration. When evaluated at the last field decay time indeed, these give rise to $r_\mathrm{dec}$ through \Eq{eq:rdec:def}. Then, fixing $N_k$ and $N_\mathrm{dec}$ to their background values, we just need to take the derivative of these quantities with respect to $\phi_k$ and $\sigma_k$ up to second order in order to calculate $A^\alpha_\beta$ and $A^\alpha_{\beta\gamma}$ defined in \Eqs{eq:Aalphabeta:def} and~(\ref{eq:Aalphabetagamma:def}). Observable quantities are finally expressed in terms of these parameters only through \Eqs{eq:powerspectrum:rho}, (\ref{eq:powerspectrum:h}) and~(\ref{eq:fnl}). This is why in the next sections, we calculate how energy densities evolve throughout inflation and reheating.
\subsection{Inflation}
\label{sec:inflation}
During inflation, we assume that both fields $\phi$ and $\sigma$ are slowly rolling.  In terms of the number of \efolds $N\equiv \ln(a)$, \Eqs{eq:KG:sigma} and~(\ref{eq:Friedmann}) then give rise to
\bea
\dfrac{\dd\phi}{\dd N}=-\Mp^2\dfrac{V^\prime}{V+m_\sigma^2\sigma^2/2}
\label{eq:kg:sr:phi}
\, ,\quad
\dfrac{\dd\sigma}{\dd N}=-\Mp^2\dfrac{m_\sigma^2 \sigma}{V+m_\sigma^2\sigma^2/2}\, ,
\label{eq:kg:sr:sigma}
\eea
where a prime denotes derivative with respect to $\phi$. Combined together, these equations yield $\dd\sigma/\dd\phi=m_\sigma^2\sigma/V^\prime$. This relation can be integrated, and one obtains
\beq
\sigma_*=\sigma_\uend\exp\left[m_\sigma^2\int_{\phi_\uend}^{\phi_*}\frac{\dd\phi}{V^\prime}\right]\, .
\label{eq:phiofsigma}
\eeq
On the other hand, combining \Eqs{eq:kg:sr:phi} and~(\ref{eq:kg:sr:sigma}) also gives rise to $V/V^\prime\dd\phi + \sigma/2\dd\sigma=-\Mp^2\dd N $, so that one has
\beq
\int_{\phi_\uend}^{\phi_*}\frac{V}{V^\prime}\frac{\dd\phi}{\Mp^2}+\frac{\sigma_*^2-\sigma_\uend^2}{4\Mp^2}=N_\uend-N_*\, .
\label{eq:sr:trajintegrated:both}
\eeq
Plugging \Eq{eq:phiofsigma} into this relation leads to the number of \efolds realized between $\phi_*$ and $\phi_\uend$,
\beq
N_\uend-N_*=\int_{\phi_\uend}^{\phi_*}\frac{V}{V^\prime}\frac{\dd\phi}{\Mp^2}+\dfrac{\sigma_\uend^2}{4\Mp^2}\left[\exp\left(2m_\sigma^2\int_{\phi_\uend}^{\phi_*}\frac{\dd\phi}{V^\prime}\right)-1\right]\, .
\label{eq:sr:trajintegrated}
\eeq
When $N_\uend-N_*$ is known, $\phi_*$ can then be computed (often numerically) from inverting \Eq{eq:sr:trajintegrated} and $\sigma_*$ can then be derived from \Eq{eq:phiofsigma}, provided $\phi_\uend$ is specified. This can be obtained from requiring that the first slow-roll parameter,
\beq
\epsilon_1\simeq \frac{\Mp^2}{2}\frac{{V^\prime}^2+m_\sigma^4\sigma^2}{\left(V+\frac{m_\sigma^2}{2}\sigma^2\right)^2}\, ,
\label{eq:epsilon1}
\eeq
equals one at the end of inflation. One then has $\dot{\phi}^2/2+\dot{\sigma}^2/2=V(\phi)/2+m_\sigma^2/2$, so that $\rho_\uend=(3/2)V(\phi_\uend)+3m_\sigma^2\sigma_\uend^2/4$ and $H_\uend$ can be computed through the Friedmann equation~(\ref{eq:Friedmann}). From here, variations with respect to $\phi_*$ and $\sigma_*$ of $\sigma_\uend$, $N_\uend$ and the energy densities contained in both fields at the end of inflation can be calculated, and we display the corresponding formulas in appendix~\ref{sec:inflation:formulas}.
\subsection{Reheating}
\label{sec:reheating}
During reheating, the way energy densities evolve is different for the ten cases under consideration. In appendix~\ref{sec:reheating:formulas}, we detail the calculation of $r_\udec$ and the energy density variations case by case, and we derive under which conditions on the parameters $m_\sigma$, $\Gamma_\phi$, $\Gamma_\sigma$ and $\sigma_\uend$ each scenario takes place. 

It is also important to mention that the parameters of the model completely fix $N_\uend - N_*$, the number of \efolds elapsed between the Hubble exit time of the pivot scale $k_*$ around which observable quantities are expressed, and the end of inflation. This can be seen introducing the reheating kinematic parameter,
\beq
\ln R_\urad = \dfrac{1-3\bar{w}_\ureh}{12\left(1+\bar{w}_\ureh\right)}\ln\left(\dfrac{\rho_\ureh}{\rho_\uend}\right)\, .
\eeq
Here, the averaged equation of state $\bar{w}_\ureh$ is defined as
\beq
\bar{w}_\ureh=\dfrac{1}{\Delta N_\ureh}\int_\ureh w\left(N\right)\dd N\, .
\label{eq:wbar:def}
\eeq
As explained in \Refs{Martin:2010kz,Easther:2011yq}, the reheating parameter is related to $N_\uend - N_*$ through
\beq
\label{eq:DeltaNstar:reheating}
N_\uend -  N_*=\ln R_\urad+\frac{1}{4}\ln\left(\frac{2}{27}\frac{\rho_*^2}{\Mp^4\rho_\uend}\frac{1}{1-\epsilon_{1*}/3}\right)-\ln\left(\frac{k_*/a_\mathrm{now}}{\tilde{\rho}_{\radiation,\mathrm{now}}^{1/4}}\right)\, ,
\eeq
where $\tilde{\rho}_{\radiation,\mathrm{now}}$ is the energy density of radiation today rescaled by the number of relativistic degrees of freedom. The reason why $\ln R_\urad$ is convenient is that it follows a simple additivity rule.\footnote{Indeed, let us consider the case where reheating is made of two phases. The first one drives the energy density from $\rho_0$ to $\rho_1$ with an average equation of state $\bar{w}_1$, and lasts $N_1=\ln(\rho_0/\rho_1)/[3(1+\bar{w}_1)]$ $e$-folds. Similarly, the second phase drives the energy density from $\rho_1$ to $\rho_2$ with an average equation of state $\bar{w}_2$, and lasts $N_2=\ln(\rho_1/\rho_2)/[3(1+\bar{w}_2)]$ $e$-folds. From here, one can calculate the averaged equation of state parameter $\bar{w}_\ureh=(N_1\bar{w}1+N_2\bar{w}2)/(N_1+N_2)$, and the reheating parameter
\beq
\ln R_\urad=\dfrac{1-3\bar{w}_1}{12\left(1+\bar{w}_1\right)}\ln\left(\dfrac{\rho_1}{\rho_0}\right)+\dfrac{1-3\bar{w}_2}{12\left(1+\bar{w}_2\right)}\ln\left(\dfrac{\rho_2}{\rho_1}\right)=\dfrac{1-3\bar{w}_\ureh}{12\left(1+\bar{w}_\ureh\right)}\ln\left(\frac{\rho_2}{\rho_0}\right)
\eeq
which is simply given by the sum of the reheating parameters during each phase.} In particular, this means that any radiation dominated phase will not contribute to $\ln R_\urad$ (since, in this case, $1-3w_\ureh=0$). This is consistent with the fact that reheating is, by definition, the connection between the end of inflation and the beginning of the radiation era. 

In all ten cases detailed in appendix~\ref{sec:reheating:formulas}, reheating is comprised of three kinds of sub-phases: matter phase, radiation phase, and inflating phase. For matter dominated phases occurring between $\rho_1$ and $\rho_2$, $\bar{w}=0$ and one simply has
\beq
\ln R_\urad^{\matter} = \dfrac{1}{12}\ln\left(\dfrac{\rho_2}{\rho_1}\right)\, .
\label{eq:Rrad:mat}
\eeq
When an extra phase of inflation occurs during $\Delta N_{\mathrm{inf}}$ \efolds and proceeds between $\rho_1$ and $\rho_2$, one has\footnote{The integral in \Eq{eq:wbar:def} can be expressed as
\bea
\int w\dd N &=& \int \left(-1+\dfrac{2}{3}\epsilon_1\right)\dd N 
= - \Delta N_-\dfrac{2}{3}\int\dfrac{\dd H}{H}
=  - \Delta N+\dfrac{1}{3}\ln \left(\dfrac{\rho_1}{\rho_2}\right)\, ,
\eea
where we have used the Friedmann equation in the last equality. This gives rise to $\bar{w}=-1+\ln \left(\rho_1/\rho_2\right)/(3 \Delta N)$.}
\beq
\ln R_\urad^{\mathrm{inf}}=-\Delta N_{\mathrm{inf}}+\dfrac{1}{4}\ln\left(\dfrac{\rho_1}{\rho_2}\right)\, .
\label{eq:Rrad:inf}
\eeq
Finally, as already noticed, during a radiation phase, $\ln R_\urad^\radiation=0$. For each reheating scenario, one then just add to sum the contributions of all sub-phases to compute $\ln R_\urad$, hence $N_\uend -  N_*$.
\section{Classification and Discussion}
\label{sec:Discussion}
The numerical routines corresponding to the procedure detailed in section~\ref{sec:Method}, with relevant formulas given in appendices~\ref{sec:inflation:formulas} and~\ref{sec:predictions}, have been added to the publicly available \ASPIC library~\cite{aspic} and can be downloaded from the \MYhref[violet]{http://cp3.irmp.ucl.ac.be/~ringeval/aspic.html}{\ASPIC website}. This means that one can now easily compute the predictions of the $\sim 75$ inflationary potentials originally contained in this library, when a massive scalar field is added, in all 10 reheating scenarios. For illustrative purposes, in appendix~\ref{sec:predictions}, we display these predictions for three prototypical examples: 
\begin{itemize}
\item Large-field inflation, for which $V=M^4(\phi/\Mp)^p$, is presented in appendix~\ref{sec:plots:lfi}. It is a typical example yielding a value for $r$ which is too large in its single-field version.
\item Higgs inflation (the Starobinsky model) is presented in appendix~\ref{sec:plots:hi}, and its potential is given by $V=M^4[1-\exp(-\sqrt{2/3}\phi/\Mp)]^2$. It is a typical example for which the single-field version of the model provides a good fit to the data, both in terms of $\nS$ and $r$.
\item Natural inflation is presented in appendix~\ref{sec:plots:ni}, and its potential is given by $V=M^4[1+\cos(\phi/f)]$. When $f$ is not super-Planckian, it is a typical example that yields a value of $\nS$ which is too small in the single-field version of the model. 
\end{itemize}
In all $10$ cases, predictions are plotted in the $(\nS,r)$, $(\nS,\fnl)$ and $(\fnl,r)$ planes. They are compared to the predictions of the single-field versions of the models, if one takes the averaged equation of state parameter for the reheating epoch of these models to be $\bar{w}_\ureh=0$. Indeed, in the present work, let us recall that we assume that close to its minimum, $V(\phi)$ is quadratic so that the energy density contained in $\phi$ behaves as matter between the end of inflation and the decay of $\phi$. For fair comparison purpose, we therefore make the same assumption regarding the single-field versions of the models. Finally, the Planck observational constraints are also superimposed.

Let us describe the behaviors obtained in all ten cases. In order to gain some intuition on the order of magnitude of the predicted amount of non-Gaussianities, a useful formula is the estimate~\cite{Lyth:2002my,Enqvist:2013paa,Sasaki:2006kq}
\beq
\label{eq:fnl:approx}
\fnl\sim \frac{5}{12}\left(\frac{\calP_\zeta^\sigma}{\calP_\zeta}\right)^2\left(\frac{3}{r_\udec}-4-2 r_\udec\right)\, ,
\eeq
where $\calP_\zeta^\sigma/\calP_\zeta$ is the fractional contribution from $\sigma$ to the total curvature power spectrum, defined below \Eq{eq:powerspectrum:rho}. In particular, one can see that the only way to get large non-Gaussianities, is to have $\sigma$ providing a substantial contribution to the total amount of scalar perturbations while the field that first decays remains dominant afterwards (so that $r_\udec\ll 1$). We now review the different cases, and schematically summarize the main observed trends in table~\ref{tab:summary}.
\begin{table}[t]
\begin{center}
\newcolumntype{L}[1]{>{\raggedright\let\newline\\\arraybackslash\hspace{0pt}}m{#1}}
\newcolumntype{C}[1]{>{\centering}m{#1}}
\newcolumntype{R}[1]{>{\raggedleft\let\newline\\\arraybackslash\hspace{0pt}}m{#1}}
\begin{tabular}{{|c||c|c|c|c|c|c|c|c|c|}}
\hline
Reheating & Extra & \multirow{2}{*}{ $r_\udec$} & \multicolumn{2}{c|}{$\calP_\zeta^\sigma/\calP_\zeta(\sigma_\uend) $} & \multicolumn{2}{c|}{$\nS(\sigma_\uend) $}& \multicolumn{2}{c|}{$r(\sigma_\uend) $}& \multirow{2}{*}{ $\fnl$}\\
\cline{4-9}
Scenario  & Inflation & & SF & LF &  $\, \quad$ SF $\, \quad$ & LF& SF & LF& \\
\hline
1 & no & $\scriptstyle{\sim 1}$ & \multicolumn{2}{c|}{$ {\scriptstyle{0}}\nearrow {\scriptstyle{0}} $} & \multicolumn{2}{c|}{$\scriptstyle{\SF} $} & \multicolumn{2}{c|}{$\scriptstyle{\SF} $} & $\scriptstyle{\ll 1}$\\
\hline
2 & yes &$\scriptstyle{\sim 1}$ & \multicolumn{2}{c|}{${\scriptstyle{0}} \nearrow {\scriptstyle{0}} $} & \multicolumn{2}{c|}{$ \scriptstyle{\SF}  $} & \multicolumn{2}{c|}{$\scriptstyle{\SF}   $} & $\scriptstyle{<\order{0.1}}$\\
\hline
3 & yes &$\scriptstyle{\ll 1}$ & ${\scriptstyle{0}} \nearrow {\scriptstyle{0}} $ & ${\scriptstyle{0}} \nearrow {\scriptstyle{1}} $ & \multicolumn{2}{c|}{${\scriptstyle{\SF }} \searrow $} & $ {\scriptstyle{\SF}} \nearrow$ & $\scriptstyle{\SF} $ & $\scriptstyle{\ll 1}$\\
\hline
4 & no &$\scriptstyle{\ll 1}$ & \multicolumn{2}{c|}{${\scriptstyle{0}}\nearrow {\scriptstyle{1}} $}  & ${\scriptstyle{\SF}} \nearrow {\scriptstyle{1}} $& ${\scriptstyle{\SF}}\nearrow {\scriptstyle{1-2\epsilon_{1*}}} $ &  \multicolumn{2}{c|}{$ {\scriptstyle{\SF}}\searrow {\scriptstyle{0}}$}  & $\scriptstyle{\gg 1}$\\
\hline
5 & no &$\scriptstyle{\sim 1}$ & \multicolumn{2}{c|}{${\scriptstyle{1}} \searrow {\scriptstyle{0}} $}  & ${\scriptstyle{1}} \searrow {\scriptstyle{\SF}} $& ${\scriptstyle{1-2\epsilon_{1*}}} \searrow {\scriptstyle{\SF}} $ &  \multicolumn{2}{c|}{${\scriptstyle{0}}\nearrow {\scriptstyle{\SF}}$}  & $\scriptstyle{\mathcal{O}(1)}$\\
\hline
6 & yes &$\scriptstyle{\sim 1}$ & ${\scriptstyle{0}} \nearrow {\scriptstyle{0}} $ & ${\scriptstyle{0 }}\nearrow {\scriptstyle{1}} $ & \multicolumn{2}{c|}{$ {\scriptstyle{\SF}}\searrow $} & $ {\scriptstyle{\SF}} \nearrow$ & $\scriptstyle{\SF} $ & $\scriptstyle{\ll 1}$\\
\hline
7 & no &$\scriptstyle{\ll 1}$ & \multicolumn{2}{c|}{${\scriptstyle{0}}\nearrow {\scriptstyle{1}} $}  & ${\scriptstyle{\SF}} \nearrow {\scriptstyle{1}} $& ${\scriptstyle{ \SF}}\nearrow {\scriptstyle{ 1-2\epsilon_{1*}}} $ &  \multicolumn{2}{c|}{${\scriptstyle{\SF}}\searrow {\scriptstyle{0}}$}  & $\scriptstyle{\gg 1}$\\
\hline
8 & no &$\scriptstyle{\sim 1}$ & \multicolumn{2}{c|}{${\scriptstyle{1}}\searrow {\scriptstyle{0}} $}  & ${\scriptstyle{1}}\searrow {\scriptstyle{\SF}} $& ${\scriptstyle{1-2\epsilon_{1*} }}  \searrow {\scriptstyle{\SF}} $ &  \multicolumn{2}{c|}{${\scriptstyle{0}}\nearrow {\scriptstyle{\SF}}$}  & $\scriptstyle{\mathcal{O}(1)}$\\
\hline
9 & yes &$\scriptstyle{\sim 1}$ & ${\scriptstyle{0}} \nearrow {\scriptstyle{0}} $ & ${\scriptstyle{ 0}} \nearrow {\scriptstyle{1}} $ & \multicolumn{2}{c|}{${\scriptstyle{\SF}} \searrow $} & $ {\scriptstyle{\SF}} \nearrow$ & $\scriptstyle{\SF} $ & $\scriptstyle{\ll 1}$\\
\hline
10 & yes &$\scriptstyle{\sim 1}$ & ${\scriptstyle{0}}\nearrow {\scriptstyle{0}} $ & ${\scriptstyle{ 0}} \nearrow {\scriptstyle{1}} $ & \multicolumn{2}{c|}{${\scriptstyle{\SF}} \searrow $} & $ {\scriptstyle{\SF}} \nearrow$ & $\scriptstyle{\SF} $ & $\scriptstyle{\ll 1}$\\
\hline
\hline
\multicolumn{2}{l}{$\scriptstyle{\SF}$: single-field predictions}
\end{tabular}
\end{center}
\caption[]{Schematic trends observed for the 10 reheating scenarios. The second column shows whether an extra phase of inflation takes place during reheating or not. The schematic value given for the ratio $r_\udec$ defined in \Eq{eq:rdec:def} indicates whether the matter field dominates the energy content of the Universe at the last field decay time ($r_\udec\sim 1$) or whether the radiation field does ($r_\udec\ll 1$). The ratio $\calP_\zeta^\sigma/\calP_\zeta$ shows how the relative contribution from $\sigma$ to the total scalar power spectrum evolves with $\sigma_\uend$, and between which typical values it varies. In some cases, the trends depends on whether the  single-field version of the model under consideration is of the small-field (``SF'') or large-field (``LF'') type (where small and large field must be understood in terms of $\epsilon_{1*}$). The same trends are shown for $\nS$ and $r$. Here, the symbol $\SF$ refers to the single-field version of the model predictions. Finally, we report whether large non-Gaussianities can be produced.} 
\label{tab:summary} 
\end{table}

Case 1 is in practice indistinguishable from the single-field version of the model since $\sigma$ never dominates the energy content of the Universe and never provides the dominant contribution to the scalar perturbations. It decays before $\phi$ and is a purely spectator field.

Case 2 is the only situation where $\sigma$ dominates the energy content of the Universe for a transient phase only. It even drives a short second phase of inflation. It decays before $\phi$ and is subdominant at the last field decay, which means that, again, it does not contribute significantly to scalar perturbations. This is why the predicted values for $\nS$ and $r$ are again almost the same as the ones for the single-field version of the model. However, non-Gaussianities are slightly enhanced, and can typically reach up to $\order{0.1}$ values (which, though small, are significantly larger than the single-field predictions).

Cases 3, 6, 9 and 10 have similar phenomenologies. In these cases, a secondary phase of inflation takes place during reheating. The difference with case 2 is that this phase does not necessarily need to be short, and importantly, at the last field decay time, $\sigma$ dominates the energy budget of the Universe. This explains why non-gaussianities remain small. However, the behavior of $\nS$ and $r$ depends on whether the single-field version of the model is of the small or large field type. For small-field models, the main effect of increasing $\sigma_\uend$ is that more \efolds are realized in the second phase of inflation. As a consequence, $N_\uend - N_*$ decreases, and the predictions are calculated at a field location $\phi_*$ that gets closer and closer to $\phi_\uend$, where the potential is steeper. This means that $\nS$ is shifted away from $1$ and $r$ increases. Let us stress that the interplay between $\sigma_\uend$, the number of \efolds realized in the second phase of inflation and $N_\uend -  N_*$, is described by the reheating kinematic analysis of section~\ref{sec:reheating}. A proper inclusion of these effects is therefore  crucial to correctly account for these regimes, where reheating kinematic effects dominate. When $\sigma_\uend$ continues to increase, the contribution from $\sigma$ to the total amount of scalar perturbations starts to increase as well. However, before it reaches non vanishingly small values, the number of \efolds realized in the second phase of inflation is too large so that the scales that we observe in the CMB would not have exited the Hubble radius during the first phase of inflation. For large-field versions of the single-field model, on the contrary, both effects have the same order of magnitude: adding more \efolds during reheating drives $\phi_*$ closer to $\phi_\uend$ (hence $\nS$ is shifted away from one, as in the small-field case), but increasing $\sigma_\uend$ also enlarges the contribution from $\sigma$ to the total amount of scalar perturbations. As far as $r$ is concerned, both effects act in opposite directions and roughly cancel out (especially if the potential is quadratic, where, up to mild logarithmic terms, they exactly cancel out), which implies that $r$ is almost unchanged.

Cases 4 and 7 give comparable phenomenologies. The extra field $\sigma$ remains subdominant throughout the entire scenario but contrary to case 1, it decays after $\phi$, which means that as $\sigma_\uend$ increases, its contribution to the total amount of scalar perturbations becomes significant. In this regime, large levels of non-Gaussianities can therefore be produced. As far as $\nS$ and $r$ are concerned, one can show that they are well approximated by~\cite{Wands:2002bn}
\bea
\nS-1 &\sim&\left(1-\frac{\calP_\zeta^\sigma}{\calP_\zeta}\right)\left(\nS-1\right)_{\mathrm{single-field}}-2\epsilon_{1*}\frac{\calP_\zeta^\sigma}{\calP_\zeta}\, ,\\
r&\sim & \left(1-\frac{\calP_\zeta^\sigma}{\calP_\zeta}\right) r_{\mathrm{single-field}}\, .
\eea
When $\sigma_\uend$ is small, $\calP_\zeta^\sigma$ represents only a small fraction of $\calP_\zeta$ and one can check that the same predictions as in the single-field version of the model are obtained. In the opposite regime, $\calP_\zeta^\sigma\sim \calP_\zeta$, $r$ takes vanishingly small values and $\nS$ reaches $1-2\epsilon_{1*}$. For small-field models, this is essentially not distinguishable from $\nS=1$. For large-field models however, $\nS$ can be substantially different from $1$ and one can see that, for example, $V\propto\phi^4$ provides a good fit to the measured value of $\nS$ in this regime.

Cases 5 and 8 are similar to these two previous cases except that $\sigma$ dominates the energy budget of the Universe when it decays. This implies that large non-Gaussianities cannot be produced since $r_\udec\sim 1$. In fact, when $\sigma_\uend$ is not too large, $\sigma$ provides the main contribution to the scalar perturbations and one has $\fnl\sim -5/4$, see \Eq{eq:fnl:approx}. When $\sigma_\uend$ increases, its contribution to $\calP_\zeta$ decreases, $\fnl$ takes $\order{1}$ values and eventually becomes much smaller than one. As far as $\nS$ and $r$ are concerned, they interpolate between the same values as for cases 4 and 7. The only difference is that the relative contribution from $\sigma$ to $\calP_\zeta$ now decreases with $\sigma_\uend$, hence $\sigma_\uend$ varies in the opposite direction between the two asymptotic values.

\section{Conclusion}
\label{sec:Conclusion}
The ongoing quest for high accuracy astrophysical and cosmological data offers an unprecedented opportunity to constrain the inflationary theory. So far, observations are compatible with the simplest framework, where inflation is driven by a single scalar field, minimally coupled to gravity and slowly rolling down a flat potential with a canonical kinetic term. However, most physical setups that have been proposed to embed inflation contain extra scalar degrees of freedom that can play a role either during inflation or afterwards. This is why it is important to ask the two following questions: Are single-field models predictions robust under the introduction of these extra fields? Do the data show preference for the same single-field potentials once these extra fields are included?

This paper is the first of a series which tackles these two issues. It investigates the situation where an extra light scalar field is added to single-field scenarios, and contributes both to the total curvature perturbations and to the reheating kinematic properties. Since it has been shown~\cite{Martin:2014nya} that the later plays a crucial role when constraining models from measuring the former, it is indeed necessary to properly describe these two effects simultaneously. In practice, ten different types of reheating scenarios have been identified and studied separately for each single-field potential. Making use of the $\delta N$ formalism to relate observable quantities to variations in the energy densities contained in both fields, we have derived all necessary formulas to implement a systematic and ``industrial'' treatment of this class of models. The corresponding routines have been added to the publicly available runtime library \ASPIC~\cite{aspic}. It now allows the fast computation of the predictions of the $\sim 75$ inflationary potentials originally contained, when an extra massive scalar field is added and in all $10$ reheating scenarios.

For illustrative purpose, we have displayed the predictions obtained for a few potentials. The few generic trends we have thus identified have a number of observational consequences for the models we considered:

$\bullet$ In their single-field versions, large-field models are plagued with values of $r$ that are too large. We saw that scenarios producing a second phase of inflation during reheating do not help to solve this issue. The only scenarios where $r$ is decreased are $4$, $5$, $7$ and $8$. However, in cases $4$ and $7$, large non-Gaussianities are produced and one concludes that these models are safe only in scenarios $5$ and $8$. The question then is how much parameter space allows one to obtain the correct predictions and how much these models are improved. 

$\bullet$ Models that already provide a good fit to the data in their single-field version, such as Higgs Inflation (the Starobinsky model), can only perform worse once an extra scalar field is introduced and shifts the predictions away from the  preferred values. The question is again how much parameter space is wasted in these disfavored regions and how it affects the likelihood of these models. 

$\bullet$ For small-field models producing values of $\nS$ that are too small, such as Natural Inflation, the only way to increase $\nS$ is to consider cases $4$, $5$, $7$ or $8$. However, cases $4$ and $7$ yield large non-Gaussianities and one is left with cases $5$ and $8$. There, in the asymptotic regime, $\nS$ is driven to $1$ so there exists only an intermediate range of values for the model parameters such that $\nS$ acquires the right amplitude. The question is of course how fine tuned this range of values is.

To address these issues, one needs to resort to a complete Bayesian analysis of the problem. More generally, this is the only way to make accurate statements about the observational status of the models we considered, the robustness of single-field models of inflation under the introduction of extra scalar degrees of freedom, and the constraining power of the current and future data to prove or disprove such models. Such an analysis will be presented in a separate publication~\cite{Vennin:2016}.

\section*{Acknowledgments}
It is a pleasure to thank Chris Byrnes and Robert Hardwick for enjoyable discussions and useful comments. This work is supported by STFC grants ST/K00090X/1 and ST/L005573/1.

\appendix
\section{Energy Density Variations During Inflation}
\label{sec:inflation:formulas}
The slow-roll trajectory~(\ref{eq:phiofsigma}) gives rise to
\bea
\label{eq:dsigmaenddfieldstar:start}
\frac{\partial\sigma_\uend}{\partial\sigma_*}&=&\left[\frac{\sigma_*}{\sigma_\uend}-\dfrac{m_\sigma^2\sigma_*}{V^\prime\left(\phi_\uend\right)}\dfrac{\dd\phi_\uend}{\dd\sigma_\uend}\right]^{-1}\\
\frac{\partial^2\sigma_\uend}{{\partial\sigma_*}^2}&=&\left(\frac{\partial\sigma_\uend}{\partial\sigma_*}\right)^2\left[\frac{\sigma_*}{\sigma_\uend^2}\frac{\partial\sigma_\uend}{\partial\sigma_*}+2\frac{m_\sigma^2\sigma_*}{V^\prime\left(\phi_\uend\right)}\frac{\dd\phi_\uend}{\dd\sigma_\uend}-m_\sigma^2\sigma_*^2\frac{V^{\prime\prime}\left(\phi_\uend\right)}{{V^\prime}^2\left(\phi_\uend\right)}\left(\frac{\dd\phi_\uend}{\dd\sigma_\uend}\right)^2\frac{\partial\sigma_\uend}{\partial\sigma_*}
\right.\nonumber\\ & &\left.
+\frac{m_\sigma^2\sigma_*^2}{V^\prime\left(\phi_\uend\right)}\frac{\dd^2\phi_\uend}{\partial\sigma_\uend^2}\frac{\partial\sigma_\uend}{\partial\sigma_*}-\frac{1}{\sigma_\uend} \right]\\
\frac{\partial\sigma_\uend}{\partial\phi_*}&=&\left[\frac{V^\prime\left(\phi_*\right)}{V^\prime\left(\phi_\uend\right)}\frac{\dd\phi_\uend}{\dd\sigma_\uend}-\frac{V^\prime\left(\phi_*\right)}{m_\sigma^2\sigma_\uend}\right]^{-1}\\
\frac{\partial^2\sigma_\uend}{{\partial\phi_*}^2}&=&\left(\frac{\partial\sigma_\uend}{\partial\phi_*}\right)^2\left[\frac{V^\prime\left(\phi_*\right)}{{V^\prime}^2\left(\phi_\uend\right)}V^{\prime\prime}\left(\phi_\uend\right)\left(\frac{\dd\phi_\uend}{\dd\sigma_\uend}\right)^2\frac{\partial\sigma_\uend}{\partial\phi_*}-\frac{V^\prime\left(\phi_*\right)}{V^\prime\left(\phi_\uend\right)}\frac{\dd^2\phi_\uend}{\dd\sigma_\uend^2}\frac{\partial\sigma_\uend}{\partial\phi_*}
\right.\nonumber\\ & & \left.
+\frac{V^{\prime\prime}\left(\phi_*\right)}{m_\sigma^2\sigma_\uend}-\frac{V^\prime\left(\phi_*\right)}{m_\sigma^2\sigma_\uend}\frac{\partial\sigma_\uend}{\partial\phi_*}-\frac{V^{\prime\prime}\left(\phi_*\right)}{V^\prime\left(\phi_\uend\right)}\frac{\dd\phi_\uend}{\dd\sigma_\uend}\right]\\
\frac{\partial^2\sigma_\uend}{\partial\sigma_*\partial\phi_*}&=&\left(\frac{\partial\sigma_\uend}{\partial\phi_*}\right)^2\left[\frac{V^\prime\left(\phi_*\right)}{{V^\prime}^2\left(\phi_\uend\right)}V^{\prime\prime}\left(\phi_\uend\right)\left(\frac{\dd\phi_\uend}{\dd\sigma_\uend}\right)^2\frac{\partial\sigma_\uend}{\partial\sigma_*}-\frac{V^\prime\left(\phi_*\right)}{V^\prime\left(\phi_\uend\right)}\frac{\dd^2\phi_\uend}{\dd\sigma_\uend^2}\frac{\partial\sigma_\uend}{\partial\sigma_*}
\right. \nonumber\\ & & \left.
-\frac{V^\prime\left(\phi_*\right)}{m_\sigma^2\sigma_\uend^2}\frac{\partial\sigma_\uend}{\partial\sigma_*}\right]\, ,
\label{eq:dsigmaenddfieldstar:end}
\eea
where the derivatives of $\phi_\uend(\sigma_\uend)$ are given by varying $\phi_\uend$ and $\sigma_\uend$ in the condition $\epsilon_{1\uend}=1$ and making use of \Eq{eq:epsilon1}. Combined with \Eqs{eq:dsigmaenddfieldstar:start}-(\ref{eq:dsigmaenddfieldstar:end}), they also give rise to the derivatives of $\phi_\uend(\phi_*,\sigma_*)$. Making use of the Friedmann equation $H_\uend^2=[2V(\phi_\uend)+3m_\sigma^2\sigma_\uend^2]/(4\Mp^2)$, the derivatives of $H_\uend$ with respect to $\phi_*$ and $\sigma_*$ can also be obtained.  Then, noting that
\bea
\rho_\uend^\phi & = & 3\Mp^2H_\uend^2-\rho^\sigma_\uend \simeq 3\Mp^2H_\uend^2 - \dfrac{m_\sigma^2}{2}\sigma^2_\uend\left(1+\dfrac{m_\sigma^2}{9H_\uend^2}\right)\, ,
\label{eq:rhophiend}
\eea 
the derivatives of $\rho_\uend^\phi$ with respect to $\phi_*$ and $\sigma_*$ can be derived. We do not reproduce them here since they are straightforward. Finally, the number of inflationary $e$-folds~(\ref{eq:sr:trajintegrated:both}) obeys (fixing $N_*$)
\bea
\frac{\partial N_\uend}{\partial\phi_*}&=&\frac{V\left(\phi_*\right)}{\Mp^2V^\prime\left(\phi_*\right)}-\frac{\partial\phi_\uend}{\partial\phi_*}\frac{V\left(\phi_\uend\right)}{\Mp^2V^\prime\left(\phi_\uend\right)}-\frac{\sigma_\uend}{2\Mp^2}\frac{\partial\sigma_\uend}{\partial\phi_*}\\
\frac{\partial^2 N_\uend}{{\partial\phi_*}^2}&=&\frac{1}{\Mp^2}\left[1-\frac{V\left(\phi_*\right)V^{\prime\prime}\left(\phi_*\right)}{{V^\prime}^2\left(\phi_*\right)}\right]-\frac{1}{\Mp^2}\frac{\partial^2\phi_\uend}{\partial{\phi_*}^2}\frac{V\left(\phi_\uend\right)}{V^\prime\left(\phi_\uend\right)}-\frac{1}{2\Mp^2}\left(\frac{\partial\sigma_\uend}{\partial\phi_*}\right)^2
\nonumber\\ & &
-\frac{1}{\Mp^2}\left(\frac{\partial\phi_\uend}{\partial\phi_*}\right)^2\left[1-\frac{V\left(\phi_\uend\right)V^{\prime\prime}\left(\phi_\uend\right)}{{V^\prime}^2\left(\phi_\uend\right)}\right]-\frac{\sigma_\uend}{2\Mp^2}\frac{\partial^2\sigma_\uend}{\partial\phi_*^2}
\\
\frac{\partial N_\uend}{\partial\sigma_*}&=&\frac{\sigma_*}{2\Mp^2}-\frac{\sigma_\uend}{2\Mp^2}\frac{\partial\sigma_\uend}{\partial\sigma_*}-\frac{V\left(\phi_\uend\right)}{V^\prime\left(\phi_\uend\right)}\frac{\partial\phi_\uend}{\partial\sigma_*}\\
\frac{\partial^2 N_\uend}{{\partial\sigma_*}^2}&=&\frac{1}{2\Mp^2}-\frac{1}{2\Mp^2}\left(\frac{\partial\sigma_\uend}{\partial\sigma_*}\right)^2-\left[1-\frac{V\left(\phi_\uend\right)V^{\prime\prime}\left(\phi_\uend\right)}{{V^\prime}^2\left(\phi_\uend\right)}\right]\left(\frac{\partial\phi_\uend}{\partial\sigma_*}\right)^2
\nonumber\\ & &
-\frac{\sigma_\uend}{2\Mp^2}\frac{\partial^2\sigma_\uend}{\partial{\sigma_*}^2}-\frac{V\left(\phi_\uend\right)}{V^\prime\left(\phi_\uend\right)}\frac{\partial^2\phi_\uend}{\partial{\sigma_*}^2}\\
\frac{\partial^2 N_\uend}{\partial\sigma_*\partial\phi_*}&=&-\frac{1}{2\Mp^2}\left\lbrace 2\frac{\partial\phi_\uend}{\partial\sigma_*}\frac{\partial\phi_\uend}{\partial\phi_*}\left[1-\frac{V\left(\phi_\uend\right)V^{\prime\prime}\left(\phi_\uend\right)}{{V^\prime}^2\left(\phi_\uend\right)}\right]+\frac{\partial\sigma_\uend}{\partial\sigma_*}\frac{\partial\sigma_\uend}{\partial\phi_*}
\right.\\ & & \left.
+2\frac{V\left(\phi_\uend\right)}{V^\prime\left(\phi_\uend\right)}\frac{\partial^2\phi_\uend}{\partial\sigma_*\partial\phi_*}+\sigma_\uend\frac{\partial^2\sigma_\uend}{\partial\sigma_*\partial\phi_*}\right\rbrace\, .
\eea
\section{Energy Density Variations During Reheating }
\label{sec:reheating:formulas}
\subsection{$\Gamma_\phi<\Gamma_\sigma<m_\sigma<H_\uend$}
In this case, the curvaton becomes massive and decays before the inflaton decays. This corresponds to cases 1, 2 and 3. Since the last field to decay is the inflaton, one has $\phim=\phi$ and $\phir=\sigma$.
\subsubsection{Case 1}
\label{sec:case1}
In this case, between the end of inflation and the decay of the inflaton, the Universe is dominated by the inflaton which behaves as matter, and one has
\beq
H=\dfrac{H_\uend}{1+\dfrac{3}{2}H_\uend\left(t-t_\uend\right)}\, .
\label{eq:H:matter}
\eeq 
The curvaton field $\sigma$ becomes massive when $H_{\sigma\mathrm{-mass}}=m_\sigma/\sqrt{2}$. Together with \Eq{eq:H:matter}, this gives rise to
\beq
t_{\sigma\mathrm{-mass}}-t_\uend=\frac{2}{3}\left(\frac{\sqrt{2}}{m_\sigma}-\frac{1}{H_\uend}\right)\, .
\label{eq:tsigmamass:1}
\eeq
Assuming that $\sigma$ is slowly rolling between the end of inflation and the moment it becomes massive, one has
\beq
\label{eq:sigmaSR}
\ln\left(\frac{\sigma}{\sigma_\uend}\right)=-\frac{m_\sigma^2}{3H_\uend}\left(t-t_\uend\right)\left[1+\frac{3}{4}H_\uend\left(t-t_\uend\right)\right]\, .
\eeq
Combined with \Eq{eq:tsigmamass:1}, this gives rise to
\beq
\label{eq:sigma:sigmamass:mat}
\sigma_{\sigma\mathrm{-mass}}=\sigma_\uend\exp\left(-\frac{2}{9}+\frac{m_\sigma^2}{9H_\uend^2}\right)\, .
\eeq
One then has $\rho^\sigma_{\sigma\mathrm{-mass}}=3m_\sigma^2\sigma_{\sigma\mathrm{-mass}}^2/4$. Then $\sigma$ simply behaves as matter, and one has
\bea
\rho^\sigma_{\sigma\mathrm{-dec}}&=&\rho^\sigma_{\sigma\mathrm{-mass}}\left(\frac{a_{\sigma\mathrm{-dec}}}{a_{\sigma\mathrm{-mass}}}\right)^{-3}=\rho^\sigma_{\sigma\mathrm{-mass}}\left(\frac{H_{\sigma\mathrm{-dec}}}{H_{\sigma\mathrm{-mass}}}\right)^{2}\\
&=&\frac{3}{2}\Gamma_\sigma^2\sigma_\uend^2\exp\left(-\frac{4}{9}+\frac{2m_\sigma^2}{9H_\uend^2}\right)\, .
\eea
Then $\sigma$ decays and the associated energy density evolves as radiation. At time of inflaton decaying, one then has
\bea
\rho^\radiation_{\phi\mathrm{-dec}}&=&\rho^\sigma_{\sigma\mathrm{-dec}}\left(\dfrac{a_{\phi\mathrm{-dec}}}{a_{\sigma\mathrm{-dec}}}\right)^{-4}=\rho^\sigma_{\sigma\mathrm{-dec}}\left(\dfrac{H_{\phi\mathrm{-dec}}}{H_{\sigma\mathrm{-dec}}}\right)^{8/3}=\rho^\sigma_{\sigma\mathrm{-dec}}\left(\dfrac{\Gamma_\phi}{\Gamma_\sigma}\right)^{8/3}\\
&=&\frac{3}{2}\Gamma_\sigma^2\sigma_\uend^2\left(\dfrac{\Gamma_\phi}{\Gamma_\sigma}\right)^{8/3}\exp\left(-\frac{4}{9}+\frac{2m_\sigma^2}{9H_\uend^2}\right)\, .
\label{eq:rhoraddec:1}
\eea
As for the inflaton, since it dominates the energy content of the Universe when it decays, its energy density is simply given by
\beq
\rho^{\matter}_{\phi\mathrm{-dec}}=3\Mp^2\Gamma_\phi^2-\rho^\radiation_{\phi\mathrm{-dec}}\, ,
\label{eq:rhomatdec:1}
\eeq
and $r_\udec$ can be computed making use of \Eqs{eq:rdec:def}, (\ref{eq:rhoraddec:1}) and~(\ref{eq:rhomatdec:1}).

Let us now work out the perturbed energy densities. They must be evaluated at the fixed background time $N_{\phi\mathrm{-dec}}$, so remember that this quantity must remain unperturbed. For the matter component, one has
\beq
\label{eq:rhogammadec:Nef:1}
\rho^{\matter}_{\phi\mathrm{-dec}}=\rho_\uend^\phi\exp\left[-3\left(N_{\phi\mathrm{-dec}}-N_\uend\right)\right]\, .
\eeq
This gives rise to
\bea
\label{eq:1:drhophidecdfields:start}
A^\matter_\alpha &=& \frac{\partial\ln\rho_\uend^\phi}{\partial\phi_{\alpha *}}+3\frac{\partial N_\uend}{\partial\phi_{\alpha *}}\\
A^\matter_{\alpha\beta} &=& \frac{\partial^2\ln\rho_\uend^\phi}{\partial\phi_{\alpha *}\partial\phi_{\beta *}}+3\frac{\partial N_\uend}{\partial\phi_{\alpha *}\partial\phi_{\beta *}}\, .
\label{eq:1:drhophidecdfields:end}
\eea
As for the radiation component, one has
\bea
\rho^\radiation_{\phi\mathrm{-dec}}&=&\rho^\sigma_{\sigma\mathrm{-dec}}\exp\left[-4\left(N_{\phi\mathrm{-dec}}-N_{\sigma\mathrm{-dec}}\right)\right]
\nonumber\\ &=&
\frac{3}{2}\Gamma_\sigma^2\sigma_\uend^2\left(\frac{H_\uend}{\Gamma_\sigma}\right)^{8/3}\exp\left(-\frac{4}{9}+\frac{2m_\sigma^2}{9H_\uend^2}\right)\exp\left[-4\left(N_{\phi\mathrm{-dec}}-N_\uend\right)\right]\, .
\eea
This gives rise to
\bea
A^\radiation_\alpha &=& 2\frac{\partial\ln\sigma_\uend}{\partial\phi_{\alpha *}}
+\frac{4}{3}\frac{\partial\ln H_\uend}{\partial\phi_{\alpha *}}\left(2-\frac{m_\sigma^2}{3H_\uend^2}\right)+4\frac{\partial N_\uend}{\partial\phi_{\alpha *}}\\
A^\radiation_{\alpha\beta} &=& 2\frac{\partial^2\ln\sigma_\uend}{\partial\phi_{\alpha *}\partial\phi_{\beta *}}
+\frac{4}{3}\frac{\partial^2\ln H_\uend}{\partial\phi_{\alpha *}\partial\phi_{\beta *}}\left(2-\frac{m_\sigma^2}{3H_\uend^2}\right)+\frac{8}{9}\frac{m_\sigma^2}{H_\uend^2}\frac{\partial\ln H_\uend}{\partial\phi_{\alpha *}}\frac{\partial\ln H_\uend}{\partial\phi_{\beta *}}
+4\frac{\partial^2 N_\uend}{\partial\phi_{\alpha *}\partial\phi_{\beta *}}\, .\nonumber\\
\eea

Let us now calculate the reheating parameter. Since the curvaton never dominates the energy budget of the Universe, and reheating is only made of a matter dominated phase taking place between $H_\uend$ and $\Gamma_\phi$, one simply has
\beq
\label{eq:Rrad:1-4-7}
\ln R_\urad= \dfrac{1}{6}\ln\left(\dfrac{\Gamma_\phi}{H_\uend}\right)\, .
\eeq

Finally, let us determine under which condition case 1 takes place. In the limit where $m_\sigma\ll H_\uend$, one has $\rho^\sigma_{\sigma\mathrm{-mass}}\simeq 3m_\sigma^2\sigma_\uend^2\exp(-4/9)/4$, while $\rho^\phi_{\sigma\mathrm{-mass}}\simeq 3\Mp^2 m_\sigma^2$. Requiring that $\rho^\sigma_{\sigma\mathrm{-mass}} < \rho^\phi_{\sigma\mathrm{-mass}}$ then leads to
\beq
\dfrac{\sigma_\uend}{\Mp}\ll \sqrt{2}e^{2/9}\, .
\label{eq:1:sigmaendmax}
\eeq
\subsubsection{Case 2}
\label{sec:case2}
In this case, the curvaton comes to dominate the energy content of Universe while it is still slow rolling, and let the inflaton dominate again after $\sigma$ decays. Let us denote $t_{\mathrm{eq}_1}$ and $t_{\mathrm{eq}_2}$ the two times when $\rho^\phi=\rho^\sigma$.
At time of first equality between $\rho_\sigma$ and $\rho_\phi$, one has
\beq
\label{eq:2:rhophieq1}
\rho_{\mathrm{eq}_1}^\phi=\rho^\phi_\mathrm{end}\left(\dfrac{a_{\mathrm{eq}_1}}{a_\uend}\right)^{-3}=\rho^\phi_\mathrm{end}\left(\dfrac{H_{\mathrm{eq}_1}}{H_\uend}\right)^{2}\, .
\eeq
On the other hand, one can combine \Eqs{eq:H:matter} and~(\ref{eq:sigmaSR}) to get
\beq
\label{eq:2:sigmaeq1}
\ln\left(\frac{\sigma_{\ueq_1}}{\sigma_\uend}\right)=\frac{m_\sigma^2}{9 H_{\ueq_1}^2}\left[\left(\frac{H_{\ueq_1}}{H_\uend}\right)^2-1\right]\, ,
\eeq
which gives rise to
\beq
\rho^\sigma_{\ueq_1}=\frac{m_\sigma^2}{2}\sigma_{\ueq_1}^2\left(1+\frac{m_\sigma^2}{9 H_{\ueq_1}^2}\right)=
\frac{m_\sigma^2}{2}\sigma_\uend^2\left(1+\frac{m_\sigma^2}{9 H_{\ueq_1}^2}\right)\exp\left[-\frac{2 m_\sigma^2}{9H_{\ueq_1}^2}\left(1-\frac{H_{\ueq_1}^2}{H_\uend^2}\right)\right]\, .
\label{eq:2:rhosigmaeq1}
\eeq
The Hubble parameter $H_{\mathrm{eq}_1}$ can then be obtained numerically equaling \Eqs{eq:2:rhophieq1} and~(\ref{eq:2:rhosigmaeq1}), and $\sigma_{\ueq_1}$ follows from \Eq{eq:2:sigmaeq1}. Then follows a second phase of inflation driven by $\sigma$, which ends when the curvaton becomes massive, that is when $\sigma_{\sigma\mathrm{-mass}}=\sqrt{2}\Mp$ and $H_{\sigma\mathrm{-mass}}=m_\sigma/\sqrt{2}$. The number of \efolds realized in this phase is given by $N_{\sigma\mathrm{-mass}}-N_{\mathrm{eq}_1}=(\sigma_{\ueq_1}^2-\sigma^2_{\sigma\mathrm{-mass}})/(4\Mp^2)=\sigma_{\ueq_1}^2/(4\Mp^2)-1/2$. As a consequence, we have
\beq
\rho_{\sigma\mathrm{-mass}}^\phi=\rho_{\ueq_1}^\phi\exp\left[-3\left(N_{\sigma\mathrm{-mass}}-N_{\mathrm{eq}_1}\right)\right]
=\rho^\phi_\mathrm{end}\left(\dfrac{H_{\mathrm{eq}_1}}{H_\uend}\right)^{2}\exp\left(\dfrac{3}{2}-\dfrac{3\sigma_{\mathrm{eq}_1}^2}{4\Mp^2}\right)\, .
\eeq
On the other hand, one simply has $\rho_{\sigma\mathrm{-mass}}^\sigma = 3/4 m_\sigma^2 \sigma_{\sigma\mathrm{-mass}}^2=3/2 m_\sigma^2 \Mp^2$. Then, the Universe is dominated by a matter fluid, until the curvaton decays. One then has
\bea
\rho_{\sigma\mathrm{-dec}}^\phi&=&\rho_{\sigma\mathrm{-mass}}^\phi\left(\dfrac{a_{\sigma\mathrm{-dec}}}{a_{\sigma\mathrm{-mass}}}\right)^{-3}=\rho_{\sigma\mathrm{-mass}}^\phi\left(\dfrac{H_{\sigma\mathrm{-dec}}}{H_{\sigma\mathrm{-mass}}}\right)^{2}
\nonumber\\ &=&
2\left(\dfrac{\Gamma_\sigma}{m_\sigma}\right)^2\rho_\uend^\phi\left(\dfrac{H_{\mathrm{eq}_1}}{H_\uend}\right)^{2}\exp\left(\dfrac{3}{2}-\dfrac{3\sigma_{\mathrm{eq}_1}^2}{4\Mp^2}\right)\, ,
\eea
while the energy density of the curvaton is simply $\rho_{\sigma\mathrm{-dec}}\simeq 3\Mp^2\Gamma_\sigma^2$. Then follows a phase where the Universe is dominated by the decay products of the curvaton, which behave as radiation, until the inflaton dominates the energy budget again. At this time, the energy density of the inflaton is given by
\bea
\rho_{\mathrm{eq}_2}^\phi&=&\rho_{\sigma\mathrm{-dec}}^\phi\left(\dfrac{a_{\mathrm{eq}_2}}{a_{\sigma\mathrm{-dec}}}\right)^{-3}=\rho_{\sigma\mathrm{-dec}}^\phi\left(\dfrac{H_{\mathrm{eq}_2}}{H_{\sigma\mathrm{-dec}}}\right)^{3/2}
\nonumber\\ &=&
2\left(\dfrac{H_{\mathrm{eq}_2}}{\Gamma_\sigma}\right)^{3/2}\left(\dfrac{\Gamma_\sigma}{m_\sigma}\right)^2\rho_\uend^\phi\left(\dfrac{H_{\mathrm{eq}_1}}{H_\uend}\right)^{2}\exp\left(\dfrac{3}{2}-\dfrac{3\sigma_{\mathrm{eq}_1}^2}{4\Mp^2}\right)
\eea
while the energy density contained in radiation is $\rho_{\mathrm{eq}_2}^\radiation\simeq 3\Mp^2 H_{\mathrm{eq}_2}^2-\rho_{\mathrm{eq}_2}^\phi$. By equating these two quantities, one obtains
\beq
H_{\mathrm{eq}_2}=\dfrac{16}{9}\Gamma_\sigma\dfrac{\left(\rho_\uend^\phi\right)^2}{\Mp^4 m_\sigma^4}\left(\dfrac{H_{\mathrm{eq}_1}}{H_\uend}\right)^4\exp\left(3-\dfrac{3\sigma_{\mathrm{eq}_1}^2}{2\Mp^2}\right)\, .
\label{eq:Heq2}
\eeq
Then, the Universe is dominated by the inflaton again, which behaves as matter, until it decays. The energy densities of the matter and radiation fluids at this time are respectively given by
\beq
\rho_{\phi\mathrm{-dec}}^{\matter}=\rho_{\mathrm{eq}_2}^\phi\left(\dfrac{a_{\phi\mathrm{-dec}}}{a_{\mathrm{eq}_2}}\right)^{-3}
=\rho_{\mathrm{eq}_2}^\phi\left(\dfrac{H_{\phi\mathrm{-dec}}}{H_{\mathrm{eq}_2}}\right)^{2}
=3\Mp^2\Gamma_\phi^2\, ,
\eeq
and
\beq
\rho_{\phi\mathrm{-dec}}^{\radiation}=\rho_{\mathrm{eq}_2}^\radiation\left(\dfrac{a_{\phi\mathrm{-dec}}}{a_{\mathrm{eq}_2}}\right)^{-4}
=\rho_{\mathrm{eq}_2}^\radiation\left(\dfrac{H_{\phi\mathrm{-dec}}}{H_{\mathrm{eq}_2}}\right)^{8/3}
=\frac{3}{2}\Mp^2\dfrac{\Gamma_\phi^{8/3}}{H_{\mathrm{eq}2}^{2/3}}\, ,
\eeq
where $H_{\mathrm{eq}2}^{2/3}$ is given by \Eq{eq:Heq2}, and $r_\udec$ can be computed making use of \Eq{eq:rdec:def}.

Let us now work out the perturbed energy densities. As far as matter is concerned, \Eq{eq:rhogammadec:Nef:1} is still valid, hence \Eqs{eq:1:drhophidecdfields:start}-(\ref{eq:1:drhophidecdfields:end}) apply. As for the radiation component, one has
\bea
\rho_{\phi\mathrm{-dec}}^{\radiation} &=&\rho_{\ueq_2}^{\radiation}\exp\left[-4\left(N_{\phi\mathrm{-dec}}-N_{\ueq_2}\right)\right]\\ 
&=&\rho_{\ueq_2}^{\radiation}\exp\left[-4\left(N_{\phi\mathrm{-dec}}-N_{\uend}\right)\right]\exp\left[-4\left(N_{\uend}-N_{\ueq_1}\right)\right]\exp\left[-4\left(N_{\ueq_1}-N_{\sigma\mathrm{-mass}}\right)\right]
\nonumber\\& & \exp\left[-4\left(N_{\sigma\mathrm{-mass}}-N_{\sigma\mathrm{-dec}}\right)\right]
\exp\left[-4\left(N_{\sigma\mathrm{-dec}}-N_{\ueq_2}\right)\right]\\
&=& 3\Mp^2\Gamma_\sigma^2\left(\frac{H_\uend}{H_{\ueq_1}}\right)^{8/3}\left(\frac{m_\sigma^2}{2\Gamma_\sigma^2}\right)^{4/3}\exp\left(\frac{\sigma_{\ueq_1}^2}{\Mp^2}-2\right)\exp\left[-4\left(N_{\phi\mathrm{-dec}}-N_\uend\right)\right]\, .
\eea
By varying this expression under the constraint that $N_{\phi\mathrm{-dec}}$ is fixed, one obtains
\bea
A^\radiation_\alpha&=&\frac{8}{3}\frac{\partial\ln H_\uend}{\partial\phi_{\alpha *}} - \frac{8}{3}\frac{\partial\ln H_{\ueq_1}}{\partial\phi_{\alpha *}}+2\frac{\sigma_{\ueq_1}^2}{\Mp^2}\frac{\partial\ln\sigma_{\ueq_1}}{\partial\phi_{\alpha *}}+4\frac{\partial N_\uend}{\partial\phi_{\alpha *}}\\
A^\radiation_{\alpha\beta}&=&\frac{8}{3}\frac{\partial^2\ln H_\uend}{{\partial\phi_{\alpha*}}{\partial\phi_{\beta*}}} - \frac{8}{3}\frac{\partial^2\ln H_{\ueq_1}}{{\partial\phi_{\alpha*}}{\partial\phi_{\beta*}}}+2\frac{\sigma_{\ueq_1}^2}{\Mp^2}\left(\frac{\partial^2\ln\sigma_{\ueq_1}}{{\partial\phi_{\alpha*}}{\partial\phi_{\beta*}}}+2\frac{\partial\ln\sigma_{\ueq_1}}{\partial\phi_{\alpha*}}\frac{\partial\ln\sigma_{\ueq_1}}{\partial\phi_{\beta*}}\right)+4\frac{\partial^2 N_\uend}{{\partial\phi_{\alpha*}}{\partial\phi_{\beta*}}}\nonumber\\
\eea
In these equations, the perturbations of $H_{\ueq_1}$ can be obtained by taking the derivatives of the equality between \Eqs{eq:2:rhophieq1} and~(\ref{eq:2:rhosigmaeq1}). One obtains

\bea
\label{eq:2:dHeq1dfields:start}
\frac{\partial\ln H_{\ueq_1}}{\partial\phi_{\alpha *}}&=&\left[\frac{\partial\ln\sigma_\uend}{\partial\phi_{\alpha *}}-\frac{1}{2}\frac{\partial\ln\rho_\uend^\phi}{\partial\phi_{\alpha *}}+\left(1-\frac{2m_\sigma^2}{9H_\uend^2}\right)\frac{\partial\ln H_\uend}{\partial\phi_{\alpha *}}\right]
/ \left(1+\frac{m_\sigma^2}{9H_{\ueq_1}^2+m_\sigma^2}-\frac{2m_\sigma^2}{9H_{\ueq_1}^2}\right)\nonumber\\ \\
\frac{\partial^2\ln H_{\ueq_1}}{\partial\phi_{\alpha *}\partial\phi_{\beta *}}&=&\left\lbrace\frac{\partial^2\ln\sigma_\uend}{{\partial\phi_{\alpha *}}{\partial\phi_{\beta *}}}-\frac{1}{2}\frac{\partial^2\ln\rho_\uend^\phi}{{\partial\phi_{\alpha *}}{\partial\phi_{\beta *}}}+\left(1-\frac{2m_\sigma^2}{9H_\uend^2}\right)\frac{\partial^2\ln H_\uend}{{\partial\phi_{\alpha *}}{\partial\phi_{\beta *}}}
\right. \nonumber\\ & & \left. +\frac{4m_\sigma^2}{9H_\uend^2}\frac{\partial\ln H_\uend}{\partial\phi_{\alpha *}}\frac{\partial\ln H_\uend}{\partial\phi_{\beta *}}-2\frac{\partial\ln H_{\ueq_1}}{\partial\phi_{\alpha *}}\frac{\partial\ln H_{\ueq_1}}{\partial\phi_{\beta *}}\left[\frac{2m_\sigma^2}{9H_{\ueq_1}^2}-\frac{9m_\sigma^2 H_{\ueq_1}^2}{\left(m_\sigma^2+9H_{\ueq_1}^2\right)^2}\right]\right\rbrace
\nonumber\\& &\times \left(1+\frac{m_\sigma^2}{9H_{\ueq_1}^2+m_\sigma^2}-\frac{2m_\sigma^2}{9H_{\ueq_1}^2}\right)^{-1}\, .
\label{eq:2:dHeq1dfields:end}
\eea
At last, the perturbations of $\sigma_{\ueq_1}$ can be obtained from \Eq{eq:2:sigmaeq1}. One has
\bea
\label{eq:2:dsigmaeq1dfields:start}
\frac{\partial\ln\sigma_{\ueq_1}}{\partial\phi_{\alpha*}}&=&\frac{\partial\ln\sigma_{\uend}}{\partial\phi_{\alpha*}}-\frac{2m_\sigma^2}{9H_\uend^2}\frac{\partial\ln H_\uend}{\partial\phi_{\alpha*}}+\frac{2m_\sigma^2}{9H_{\ueq_1}^2}\frac{\partial\ln H_{\ueq_1}}{\partial\phi_{\alpha*}}\\
\frac{\partial^2\ln\sigma_{\ueq_1}}{{\partial\phi_{\alpha*}}{\partial\phi_{\beta*}}}&=&\frac{\partial^2\ln\sigma_{\uend}}{{\partial\phi_{\alpha*}}{\partial\phi_{\beta*}}}-\frac{2m_\sigma^2}{9H_\uend^2}\left(\frac{\partial^2\ln H_\uend}{{\partial\phi_{\alpha*}}{\partial\phi_{\beta*}}}-2\frac{\partial\ln H_\uend}{\partial\phi_{\alpha*}}\frac{\partial\ln H_\uend}{\partial\phi_{\beta*}}\right)
\nonumber\\ & &+\frac{2m_\sigma^2}{9H_{\ueq_1}^2}\left(\frac{\partial^2\ln H_{\ueq_1}}{{\partial\phi_{\alpha*}}{\partial\phi_{\beta*}}}-2\frac{\partial\ln H_{\ueq_1}}{\partial\phi_{\alpha*}}\frac{\partial\ln H_{\ueq_1}}{\partial\phi_{\beta*}}\right)\, .
\label{eq:2:dsigmaeq1dfields:end}
\eea

Let us now calculate the reheating parameter. Reheating is made of a phase of matter (driven by $\phi$), inflation, matter (driven by $\sigma$), radiation, and matter (driven by $\phi$ again). Making use of the formulas derived above, one has
\beq
\ln R_\urad^{\matter,\phi,1}=\dfrac{1}{6}\ln\left(\dfrac{H_{\mathrm{eq}1}}{H_\uend}\right)\, ,
\eeq
\beq
\ln R_\urad^\mathrm{inf}=\dfrac{1}{2}-\dfrac{\sigma_{\ueq,1}^2}{4\Mp^2}+\dfrac{1}{4}\ln\left(\dfrac{2H_{\ueq,1}^2}{m_\sigma^2}\right)\, ,
\eeq
\beq
\ln R_\urad^{\matter,\sigma}=\dfrac{1}{12}\ln\left(\dfrac{2\Gamma_\sigma^2}{m_\sigma^2}\right)\, ,
\eeq
\beq
\ln R_\urad^{\matter,\phi,2}=\dfrac{1}{6}\ln\left(\dfrac{\Gamma_\phi}{H_{\ueq,2}}\right)\, ,
\eeq
The parameter $\ln R_\urad$ is given by the sum of all these contributions.

Finally, let us derive under which conditions case 2 takes place.  In the limit $\Gamma_\phi\ll \Gamma_\sigma\ll m_\sigma\ll H_\uend$, one has $\sigma_{\mathrm{eq}_1}\simeq \sigma_\uend$, $H_{\mathrm{eq}_1}^2\simeq m_\sigma^2 \sigma_\uend^2/(6 \Mp^2)$ and $\rho_\uend^\phi\simeq 3\Mp^2H_\uend^2$. This gives rise to 
\beq
H_{\mathrm{eq}_2}\simeq\dfrac{4}{9}\Gamma_\sigma\dfrac{\sigma_\uend^4}{\Mp^4}\exp\left(3-\dfrac{3}{2}\dfrac{\sigma_\uend^2}{\Mp^2}\right)\, .
\eeq
The condition $\Gamma_\phi<H_{\mathrm{eq}_2}<\Gamma_\sigma$ then gives rise to
\beq
\sqrt{-\dfrac{4}{3}W_{-1}\left(-\dfrac{9}{8}e^{-3/2}\right)}<\dfrac{\sigma_\uend}{\Mp}<\sqrt{-\dfrac{4}{3}W_{-1}\left(-\dfrac{9}{8}e^{-3/2}\sqrt{\dfrac{\Gamma_\phi}{\Gamma_\sigma}}\right)}\, ,
\eeq
where $W_{-1}$ is the $-1$ branch of the Lambert function. The lower bound is a numerical constant, $\simeq 1.69$, which is close to the upper bound~(\ref{eq:1:sigmaendmax}) $\sqrt{2}e^{2/9}$ derived in case 1. The upper bound can be approximated in the limit $\Gamma_\sigma\gg \Gamma_\phi$ and one obtains
\beq
1.7<\dfrac{\sigma_\uend}{\Mp}<\sqrt{2-\dfrac{4}{3}\log\left(\dfrac{9}{8}\right)-\dfrac{2}{3}\log\left(\dfrac{\Gamma_\phi}{\Gamma_\sigma}\right)}
\, .
\eeq
\subsubsection{Case 3}
\label{sec:case3}
In case 3, the curvaton dominates the energy content of the Universe while it is still slow rolling and while the inflaton has not decayed yet, but does not let the inflaton dominate again at later times. In order to determine the time $t_\mathrm{eq}$ when this occurs, one needs to solve the equation $\rho^\phi_\mathrm{eq}=\rho^\sigma_\mathrm{eq}$, where 
\beq
\rho^\phi_\mathrm{eq}=\rho_{\uend}^\phi\left(\dfrac{a_\mathrm{eq}}{a_\uend}\right)^{-3}=\rho_{\uend}^\phi\left(\dfrac{H_\mathrm{eq}}{H_\uend}\right)^{2}\, ,
\label{eq:3:rhophieq}
\eeq
and, combining \Eqs{eq:H:matter} and~(\ref{eq:sigmaSR}),
\beq
\rho^\sigma_\ueq=\frac{m_\sigma^2}{2}\sigma_\ueq^2\left(1+\frac{m_\sigma^2}{9 H_\ueq^2}\right)=
\frac{m_\sigma^2}{2}\sigma_\uend^2\left(1+\frac{m_\sigma^2}{9 H_\ueq^2}\right)\exp\left[-\frac{2 m_\sigma^2}{9H_\ueq^2}\left(1-\frac{H_\ueq^2}{H_\uend^2}\right)\right]\, .
\label{eq:3:rhosigmaeq}
\eeq
The equation $\rho^\phi_\mathrm{eq}=\rho^\sigma_\mathrm{eq}$ then needs to be solved numerically.
Then follows a phase of inflation driven by $\sigma$, between $\sigma_\ueq$ and $\sigma_{\sigma\mathrm{-mass}}\simeq\sqrt{2}\Mp$. The realized number of \efolds is $N_{\sigma\mathrm{-mass}}-N_\ueq = \sigma_\ueq^2/(4\Mp^2)-1/2$. During this epoch, the energy density of the inflaton field decreases according to
\beq
\label{eq:3:rhophi:sigmamass}
\rho_{\sigma\mathrm{-mass}}^\phi=\rho_\ueq^\phi\exp\left[-3\left(N_{\sigma\mathrm{-mass}}-N_\ueq\right)\right]=\rho_\uend^\phi\left(\dfrac{H_\ueq}{H_\uend}\right)^2\exp\left(\dfrac{3}{2}-\dfrac{3\sigma_\ueq^2}{4\Mp^2}\right)\, .
\eeq
Then, the Universe is matter dominated until the curvaton decays. One then has
\beq
\rho^\phi_{\sigma\mathrm{-dec}}=\rho_{\sigma\mathrm{-mass}}^\phi\left(\dfrac{H_{\sigma\mathrm{-dec}}}{H_{\sigma\mathrm{-mass}}}\right)^2=2\rho_{\sigma\mathrm{-mass}}^\phi\left(\dfrac{\Gamma_\sigma}{m_\sigma}\right)^2\, .
\eeq
From this moment, the Universe undergoes a radiation dominated era, and one then has
\beq
\rho^\matter_{\phi\mathrm{-dec}}=\rho^\phi_{\sigma\mathrm{-dec}}\left(\dfrac{H_{\phi\mathrm{-dec}}}{H_{\sigma\mathrm{-dec}}}\right)^{3/2}=\rho^\phi_{\sigma\mathrm{-dec}}\left(\dfrac{\Gamma_\phi}{\Gamma_\sigma}\right)^{3/2}\, .
\eeq
One the other hand, one simply has
\beq
\rho_{\phi\mathrm{-dec}}^\radiation=3\Mp^2\Gamma_\phi^2-\rho^\phi_{\phi\mathrm{-dec}}\, ,
\eeq
and $r_\udec$ is then given by \Eq{eq:rdec:def}.

Let us now work out the perturbed energy densities. As far as matter is concerned, \Eq{eq:rhogammadec:Nef:1} is still valid, hence \Eqs{eq:1:drhophidecdfields:start}-(\ref{eq:1:drhophidecdfields:end}) apply. As for the radiation component, one has
\bea
\rho_{\phi\mathrm{-dec}}^\matter &=&\rho_{\sigma\mathrm{-dec}}^\sigma\exp\left[-4\left(N_{\phi\mathrm{-dec}}-N_{\sigma\mathrm{-dec}}\right)\right]\\
&=&\rho_{\sigma\mathrm{-dec}}^\sigma\exp\left[-4\left(N_{\phi\mathrm{-dec}}-N_\uend\right)\right]\exp\left[-4\left(N_\uend-N_\ueq\right)\right]
\nonumber\\ & & \exp\left[-4\left(N_\ueq-N_{\sigma\mathrm{-mass}}\right)\right]\exp\left[-4\left(N_{\sigma\mathrm{-mass}}-N_{\sigma\mathrm{-dec}}\right)\right]
\nonumber\\ &=& \frac{3}{2}\Mp^2 m_\sigma^2 \left(\frac{H_\uend}{H_\ueq}\right)^{8/3}\exp\left(\frac{\sigma_\ueq^2}{\Mp^2}-2\right)\exp\left[-4\left(N_{\phi\mathrm{-dec}}-N_\uend\right)\right]\, .
\eea
By varying this expression under the constraint that $N_{\phi\mathrm{-dec}}$ is fixed, one obtains
\bea
A^\radiation_\alpha &=& \frac{8}{3}\left(\frac{\partial\ln H_\uend}{\partial\phi_{\alpha *}}-\frac{\partial\ln H_\ueq}{\partial\phi_{\alpha *}}\right)+2\frac{\sigma_\ueq^2}{\Mp^2}\frac{\partial\ln\sigma_\ueq}{\partial\phi_{\alpha *}}+4\frac{\partial N_\uend}{\partial\phi_{\alpha *}}\\
A^\radiation_{\alpha\beta} &=& \frac{8}{3}\left(\frac{\partial^2\ln H_\uend}{{\partial\phi_{\alpha *}}{\partial\phi_{\beta *}}}-\frac{\partial^2\ln H_\ueq}{{\partial\phi_{\alpha *}}{\partial\phi_{\beta *}}}\right)
+2\frac{\sigma_\ueq^2}{\Mp^2}\left(\frac{\partial^2\ln\sigma_\ueq}{\partial\phi_{\alpha *}\partial\phi_{\beta *}}+2\frac{\partial\ln\sigma_\ueq}{\partial\phi_{\alpha *}}\frac{\partial\ln\sigma_\ueq}{\partial\phi_{\beta *}}\right)
+4\frac{\partial^2 N_\uend}{{\partial\phi_{\alpha *}}{\partial\phi_{\beta *}}}\, .\nonumber\\
\eea
In these expressions, the derivatives of $H_\ueq$ can be obtained by Taylor expanding \Eqs{eq:3:rhophieq} and~(\ref{eq:3:rhosigmaeq}), which exactly match \Eqs{eq:2:rhophieq1} and~(\ref{eq:2:rhosigmaeq1}). As a consequence, one obtains the same relations as \Eqs{eq:2:dHeq1dfields:start}-(\ref{eq:2:dHeq1dfields:end}). In the same manner, the variations of $\sigma_\ueq$ follow from \Eq{eq:3:rhosigmaeq} and are given by \Eqs{eq:2:dsigmaeq1dfields:start}-(\ref{eq:2:dsigmaeq1dfields:end}).

Let us now calculate the reheating parameter. Reheating is made of a matter phase driven by $\phi$, a phase of inflation driven by $\sigma$, and a matter phase driven by $\sigma$ also. One has
\beq
\label{eq:3-6-10:Rradmatphi}
\ln R_\urad^{\matter,\phi}=\dfrac{1}{6}\ln\left(\dfrac{H_{\mathrm{eq}}}{H_\uend}\right)\, ,
\eeq
\beq
\label{eq:3-6-10:Rradinfl}
\ln R_\urad^\mathrm{inf}=\dfrac{1}{2}-\dfrac{\sigma_{\ueq}^2}{4\Mp^2}+\dfrac{1}{4}\ln\left(\dfrac{2H_{\ueq}^2}{m_\sigma^2}\right)\, ,
\eeq
and
\beq
\label{eq:3-6-10:Rradmatsigma}
\ln R_\urad^{\matter,\sigma}=\dfrac{1}{12}\ln\left(\dfrac{2\Gamma_\sigma^2}{m_\sigma^2}\right)\, .
\eeq
The parameter $\ln R_\urad$ is given by the sum of all these contributions.

Finally, let us see under which condition case 3 takes place.  In the limit $\Gamma_\phi\ll \Gamma_\sigma\ll m_\sigma\ll H_\uend$, one has $\sigma_{\mathrm{eq}}\simeq \sigma_\uend$, $H_{\mathrm{eq}}^2\simeq 2 m_\sigma^2 \sigma_\uend^2/(6 \Mp^2)$ and $\rho_\uend^\phi\simeq 3\Mp^2H_\uend^2$. This gives rise to 
\beq
\rho^\phi_{\phi\mathrm{-dec}}\simeq 2\left(\dfrac{\Gamma_\phi}{\Gamma_\sigma}\right)^{3/2}\Gamma_\sigma^2\sigma_\uend^2\exp\left(\dfrac{3}{2}-\dfrac{3}{4}\dfrac{\sigma_\uend^2}{\Mp^2}\right)\, .
\eeq
Requiring that $\rho^\phi_{\phi\mathrm{-dec}}<\rho^\radiation_{\phi\mathrm{-dec}}$ yields
\beq
\dfrac{\sigma_\uend}{\Mp}>\sqrt{-\dfrac{4}{3}W_{-1}\left(-\dfrac{9}{8}e^{-3/2}\sqrt{\dfrac{\Gamma_\phi}{\Gamma_\sigma}}\right)}\, ,
\eeq
which exactly corresponds to the upper bound of case 2.
\subsection{$\Gamma_\sigma<\Gamma_\phi<m_\sigma<H_\uend$}
In this case, the curvaton becomes massive before the inflaton decay, and it decays only afterwards. This corresponds to cases 4, 5 and 6. Since the last field to decay is the curvaton, one has $\phim=\sigma$ and $\phir=\phi$.
\subsubsection{Case 4}
\label{sec:case4}
In this case, the curvaton never dominates the energy budget of the Universe. After inflation stops, the Universe experiences a phase of matter dominated era, followed by a phase of radiation. In this sense, the situation is quite similar to case 1 and these two cases share common formulas.
Because the Universe is still dominated by the inflaton between $H_\uend$ and $\Gamma_\phi$, the oscillations of the inflaton at the bottom of its quadratic potential give rise an era of matter domination, and $H(t)$ is given by \Eq{eq:H:matter}. The decay of the inflaton occurs when $H=\Gamma_\phi$, that is at time given by $3/2(t_{\phi\mathrm{-dec}}-t_\uend)=1/\Gamma_\phi-1/H_\uend$. At this decay time, the energy density of the inflaton, which is instantaneously converted to radiation, is given by 
\beq
\label{eq:rhophi:phidec:4}
\rho^\phi_{\phi\mathrm{-dec}}=\rho_\uend^\phi\Gamma_\phi^2/H_\uend^2\, .
\eeq 
Then, the universe is radiation dominated and the decay of $\sigma$ occurs when $H=\Gamma_\sigma$. At this point, $\rho_{\sigma\mathrm{-dec}}^\radiation$ is simply given by 
\beq
\rho_{\sigma\mathrm{-dec}}^\radiation=\rho_\uend^\phi\frac{\Gamma_\sigma^2}{H_\uend^2}\, .
\eeq
As far as $\sigma$ is concerned, at the moment when it becomes massive, $\sigma_{\sigma\mathrm{-mass}}$ is given by \Eq{eq:sigma:sigmamass:mat}, and one has $\rho^\sigma_{\sigma\mathrm{-mass}}=3m_\sigma^2\sigma_{\sigma\mathrm{-mass}}^2/4$. Then, the curvaton behaves as a matter fluid in a matter dominated Universe, and one has
\bea
\rho^\sigma_{\phi\mathrm{-dec}}&=&\rho^\sigma_{\sigma\mathrm{-mass}}\left(\frac{a_{\phi\mathrm{-dec}}}{a_{\sigma\mathrm{-mass}}}\right)^{-3}
=\rho^\sigma_{\sigma\mathrm{-mass}}\left(\frac{H_{\phi\mathrm{-dec}}}{H_{\sigma\mathrm{-mass}}}\right)^{2}\\
&=&\frac{3}{2} \Gamma_\phi^2 \sigma_\uend^2\exp\left(-\frac{4}{9}+\frac{2m_\sigma^2}{9H_\uend^2}\right)\, .
\label{eq:rhosigma:phidec:4}
\eea
Then, the curvaton behaves as a matter fluid in a radiation dominated Universe, and one has
\bea
\rho^\matter_{\sigma\mathrm{-dec}}&=&\rho^\sigma_{\phi\mathrm{-dec}}\left(\frac{a_{\sigma\mathrm{-dec}}}{a_{\phi\mathrm{-dec}}}\right)^{-3}
=\rho^\sigma_{\phi\mathrm{-dec}}\left(\frac{H_{\sigma\mathrm{-dec}}}{H_{\phi\mathrm{-dec}}}\right)^{3/2}
\\
&=&\frac{3}{2} \Gamma_\phi^2\left(\frac{\Gamma_\sigma}{\Gamma_\phi}\right)^{3/2} \sigma_\uend^2\exp\left(-\frac{4}{9}+\frac{2m_\sigma^2}{9H_\uend^2}\right)\, .
\eea
From here, $r_\udec$ is given by \Eq{eq:rdec:def}.

Let us now work out the perturbed energy densities. They must be evaluated at the fixed background time $N_{\sigma\mathrm{-dec}}$, so this quantity must remain unperturbed. For the radiation component, one has
\bea
\rho_{\sigma\mathrm{-dec}}^\radiation&=&\rho_\uend^\phi\exp\left[-3\left(N_{\phi\mathrm{-dec}}-N_\uend\right)-4\left(N_{\sigma\mathrm{-dec}}-N_{\phi\mathrm{-dec}}\right)\right]\\&=&
\rho_\uend^\phi\left(\frac{\rho_\uend^{\phi}}{3\Mp^2\Gamma_\phi^2}\right)^{1/3}\exp\left[4\left(N_\uend-N_{\sigma\mathrm{-dec}}\right)\right]\, .
\label{eq:rhophi:sigmadec:Nef:4}
\eea
This gives rise to
\bea
\label{eq:drhophidecdfieldstar:4:start}
A^\radiation_\alpha &=& 4\frac{\partial N_\uend}{\partial\phi_{\alpha *}}+\frac{4}{3}\frac{\partial\ln\rho_\uend^\phi}{\partial\phi_{\alpha *}}\\
A^\radiation_{\alpha\beta} &=& 4\frac{\partial^2 N_\uend}{\partial\phi_{\alpha *}\partial\phi_{\beta *}}+\frac{4}{3}\frac{\partial^2\ln\rho_\uend^\phi}{\partial\phi_{\alpha *}\partial\phi_{\beta *}}\, .
\label{eq:drhophidecdfieldstar:4:end}
\eea
As for matter, one has
\bea
\rho_{\sigma\mathrm{-dec}}^\matter &=&\rho^\sigma_{\sigma\mathrm{-mass}}\exp\left[-3\left(N_{\sigma\mathrm{-dec}}-N_{\sigma\mathrm{-mass}}\right)\right]
\nonumber \\
&=&\frac{\rho_\uend^\phi}{2}\left(\frac{\sigma_\uend}{\Mp}\right)^2\exp\left(-\frac{4}{9}+\frac{2m_\sigma^2}{9H_\uend^2}\right)\exp\left[-3\left(N_{\sigma\mathrm{-dec}}-N_{\uend}\right)\right]\,.
\label{eq:4:rhosigmadec:Nef}
\eea
This gives rise to
\bea
\label{eq:drhosigmadecdfieldstar:4:start}
A^\matter_\alpha &=& 3 \frac{\partial N_\uend}{\partial\phi_{\alpha*}}+\frac{\partial\ln\rho_\uend^\phi}{\partial\phi_{\alpha*}}+2\frac{\partial\ln\sigma_\uend}{\partial\phi_{\alpha*}}-\frac{4m_\sigma^2}{9H_\uend^2}\frac{\partial\ln H_\uend}{\partial\phi_{\alpha*}}\\
A^\matter_{\alpha \beta} &=& 3 \frac{\partial^2 N_\uend}{{\partial\phi_{\alpha*}}{\partial\phi_{\beta*}}}+\frac{\partial^2\ln\rho_\uend^\phi}{{\partial\phi_{\alpha*}}{\partial\phi_{\beta*}}}+2\frac{\partial^2\ln\sigma_\uend}{{\partial\phi_{\alpha*}}{\partial\phi_{\beta*}}}+\frac{4m_\sigma^2}{9 H_\uend^2}\left(2\frac{\partial\ln H_\uend}{\partial\phi_{\alpha*}} \frac{\partial\ln H_\uend}{\partial\phi_{\beta*}}-\frac{\partial^2\ln H_\uend}{\partial\phi_{\alpha*} \partial\phi_{\beta*}} \right)\, .\nonumber\\
\label{eq:drhosigmadecdfieldstar:4:end}
\eea

In case 4, the reheating parameter $R_\urad$ is given by the same expression as in case 1, namely \Eq{eq:Rrad:1-4-7}.

Finally, let us derive under which conditions case 4 takes place. One simply needs to make sure that the Universe is dominated by radiation at the decay time of the curvaton. Since $\rho^\radiation_{\sigma\mathrm{-dec}}\simeq 3\Mp^2\Gamma_\sigma^2$, one has
\beq
\frac{\rho^\matter_{\sigma\mathrm{-dec}}}{\rho^\radiation_{\sigma\mathrm{-dec}}}\simeq
\frac{1}{2}\left(\frac{\sigma_\uend}{\Mp}\right)^2\sqrt{\frac{\Gamma_\phi}{\Gamma_\sigma}}\exp\left(-\frac{4}{9}+\frac{m_\sigma^2}{9H_\uend^2}\right)\, .
\eeq
Therefore, case 4 corresponds to the regime where
\beq
\frac{\sigma_\uend}{\Mp}\ll \sqrt{2}e^{2/9}\left(\frac{\Gamma_\sigma}{\Gamma_\phi}\right)^{1/4}\, .
\eeq	
\subsubsection{Case 5}
\label{sec:case5}
In this case, the curvaton comes to dominate again the energy content of the Universe, at some point between the decay of the inflaton and its own decay.
First, let us notice that until the decay time of the inflaton, the situation is similar to case 4, so that $\rho^\phi_{\phi\mathrm{-dec}}$ is given by \Eq{eq:rhophi:phidec:4} and $\rho^\sigma_{\phi\mathrm{-dec}}$ is given by \Eq{eq:rhosigma:phidec:4}. Then, $\sigma$ behaves as a matter fluid in a radiation dominated universe, and its energy density at the time of equality between $\phi$ and $\sigma$ is given by
\bea
\rho^\sigma_\mathrm{eq}&=&\rho^\sigma_{\phi\mathrm{-dec}}\left(\frac{a_\ueq}{a_{\phi\mathrm{-dec}}}\right)^{-3}=\rho^\sigma_{\phi\mathrm{-dec}}\left(\frac{H_\ueq}{H_{\phi\mathrm{-dec}}}\right)^{3/2}\\
&=&\frac{3}{2}\Gamma_\phi^2 \sigma_\uend^2\exp\left(-\frac{4}{9}+\frac{2m_\sigma^2}{9H_\uend^2}\right)\left(\frac{H_\ueq}{\Gamma_\phi}\right)^{3/2}\, .
\eea
In the same manner, one has
\beq
\rho^\radiation_\mathrm{eq}=\rho_{\phi\mathrm{-dec}}^\phi\left(\dfrac{a_\mathrm{eq}}{a_{\phi\mathrm{-dec}}}\right)^{-4}=\rho_{\phi\mathrm{-dec}}^\phi\left(\dfrac{H_\mathrm{eq}}{H_{\phi\mathrm{-dec}}}\right)^{2}=
\rho_\uend^\phi\left(\dfrac{H_\mathrm{eq}}{H_\uend}\right)^2\, .
\eeq
Requiring $\rho^\radiation_\mathrm{eq}=\rho^\sigma_\mathrm{eq}$, one then obtains
\beq
H_\ueq = \left(\frac{3}{2}\right)^{2}\exp\left(-\frac{8}{9}+\frac{4 m_\sigma^2}{9 H_\uend^2}\right)\Gamma_\phi\sigma_\uend^4\frac{H_\uend^4}{\left(\rho_\uend^\phi\right)^2}\, .
\label{eq:5:Heq}
\eeq
Then, the Universe is dominated by a matter fluid, and one can write
\beq
\rho_{\sigma\mathrm{-dec}}^\sigma=\rho_{\mathrm{eq}}^\sigma\left(\frac{a_{\sigma\mathrm{-dec}}}{a_\ueq}\right)^{-3}=\rho_{\mathrm{eq}}^\sigma\left(\frac{H_{\sigma\mathrm{-dec}}}{H_\ueq}\right)^{2}=\Gamma_\sigma^2\frac{\rho_\uend^\phi}{H_\uend^2}\, ,
\eeq
in agreement with the fact that the curvaton dominates the energy budget at this time. On the other hand, the energy density of radiation is given by
\bea
\rho_{\sigma\mathrm{-dec}}^\radiation &=&\rho_\mathrm{eq}^\radiation\left(\dfrac{a_{\sigma\mathrm{-dec}}}{a_\mathrm{eq}}\right)^{-4}
=\rho_\mathrm{eq}^\radiation\left(\dfrac{H_{\sigma\mathrm{-dec}}}{H_\mathrm{eq}}\right)^{8/3}
\nonumber\\ &=&
\left(\frac{2}{3}\right)^{2/3}\left(\frac{\Gamma_\sigma}{\sigma_\uend}\right)^{8/3}\exp\left(\frac{16}{27}-\frac{8m_\sigma^2}{27 H_\uend^2}\right)\frac{\left(\rho_\uend^\phi\right)^{7/3}}{H_\uend^{14/3}\Gamma_\phi^{2/3}}\, ,
\label{eq:rhogamma:sigmadec:8}
\eea
and $r_\udec$ can be computed through \Eq{eq:rdec:def}.

Let us now work out the perturbed energy densities. For the radiation component, \Eq{eq:rhophi:sigmadec:Nef:4} applies, and therefore \Eqs{eq:drhophidecdfieldstar:4:start}-(\ref{eq:drhophidecdfieldstar:4:end}) are valid. In the same manner, for the matter component, \Eq{eq:4:rhosigmadec:Nef} still applies, and therefore \Eqs{eq:drhosigmadecdfieldstar:4:start}-(\ref{eq:drhosigmadecdfieldstar:4:end}) can be used.

Let us now calculate the reheating parameter. In this case reheating is made of a matter phase driven by $\phi$, a radiation phase, and a matter phase driven by $\sigma$. One has
\beq
\label{eq:5-8:Rradmatphi}
\ln R_\urad^{\matter,\phi}=\dfrac{1}{6}\ln\left(\dfrac{\Gamma_\phi}{H_\uend}\right)\, ,
\eeq
\beq
\label{eq:5-8:Rradmatsigma}
\ln R_\urad^{\matter,\sigma}=\dfrac{1}{6}\ln\left(\dfrac{\Gamma_\sigma}{H_\ueq}\right)\, ,
\eeq
The parameter $\ln R_\urad$ is given by the sum of these two contributions.

Finally, let us see under which conditions on the parameters case 5 takes place. Case 5 corresponds to the situation when equality between inflaton and curvaton energy densities happens between the inflaton decay and the curvaton decay, that is $\Gamma_\sigma < H_\ueq < \Gamma_\phi$. Together with \Eq{eq:5:Heq}, this leads to
\beq
\sqrt{2}e^{2/9} \left(\frac{\Gamma_\sigma}{\Gamma_\phi}\right)^{1/4}<\frac{\sigma_\uend}{\Mp}<\sqrt{2}e^{2/9}\, .
\eeq
\subsubsection{Case 6}
\label{sec:case6}
In this case, the curvaton dominates the energy content of the Universe from some point on between the moment when the inflaton becomes massive and the moment when the curvaton becomes massive. In order to determine the value of the Hubble parameter $H_\ueq$ when this happens, one can proceed as in case 3 and solve the equality between \Eqs{eq:3:rhophieq} and~(\ref{eq:3:rhosigmaeq}).
Then begins a new phase of inflation driven by $\sigma$. It ends when the curvaton becomes massive, at $H_{\sigma\mathrm{-mass}}=m_\sigma/\sqrt{2}$. The energy density contained in $\phi$, which behaves as matter during this period, $\rho_{\sigma\mathrm{-mass}}^\phi$, is then given as in case 3 by \Eq{eq:3:rhophi:sigmamass}. Then, the Universe is matter dominated and the inflaton continues to behave as matter, until it decays. One then has
\beq
\rho^\phi_{\phi\mathrm{-dec}}=\rho_{\sigma\mathrm{-mass}}^\phi\left(\dfrac{H_{\phi\mathrm{-dec}}}{H_{\sigma\mathrm{-mass}}}\right)^2=2\rho_{\sigma\mathrm{-mass}}^\phi\left(\dfrac{\Gamma_\phi}{m_\sigma}\right)^2\, .
\eeq
After the inflaton decay, its decay products scale as radiation and one has
\bea
\rho^\radiation_{\sigma\mathrm{-dec}}&=&\rho^\phi_{\phi\mathrm{-dec}}\left(\dfrac{H_{\sigma\mathrm{-dec}}}{H_{\phi\mathrm{-dec}}}\right)^{8/3}=\rho^\phi_{\phi\mathrm{-dec}}\left(\dfrac{\Gamma_\sigma}{\Gamma_\phi}\right)^{8/3}
\nonumber\\ &=&
2 \left(\dfrac{\Gamma_\sigma}{\Gamma_\phi}\right)^{8/3}\left(\dfrac{\Gamma_\phi}{m_\sigma}\right)^2 \rho_\uend^\phi\left(\dfrac{H_\ueq}{H_\uend}\right)^2\exp\left(\dfrac{3}{2}-\dfrac{3\sigma_\ueq^2}{4\Mp^2}\right)\, .
\eea
One the other hand, one simply has
\beq
\rho_{\sigma\mathrm{-dec}}^\matter=3\Mp^2\Gamma_\sigma^2-\rho^\radiation_{\sigma\mathrm{-dec}}\, ,
\eeq
and $r_\udec$ can be computed through \Eq{eq:rdec:def}.

Let us now work out the perturbed energy densities. For radiation, one has
\bea
\rho_{\sigma\mathrm{-dec}}^\radiation &=&\rho_{\phi\mathrm{-dec}}^\phi\exp\left[-4\left(N_{\sigma\mathrm{-dec}}-N_{\phi\mathrm{-dec}}\right)\right]\\
&=&\rho_{\phi\mathrm{-dec}}^\phi\exp\left[-4\left(N_{\sigma\mathrm{-dec}}-N_\uend\right)\right]\exp\left[-4\left(N_\uend-N_\ueq\right)\right]
\nonumber\\ & & \exp\left[-4\left(N_\ueq-N_{\sigma\mathrm{-mass}}\right)\right]\exp\left[-4\left(N_{\sigma\mathrm{-mass}}-N_{\phi\mathrm{-dec}}\right)\right]
\nonumber\\ &=& \rho_\uend^\phi\left(\frac{H_\uend m_\sigma}{H_\ueq \Gamma_\phi \sqrt{2}}\right)^{2/3}\exp\left(\frac{\sigma_\ueq^2}{4\Mp^2}-\frac{1}{2}\right)\exp\left[-4\left(N_{\sigma\mathrm{-dec}}-N_\uend\right)\right]\, .
\label{eq:rhogammadec:Nef:6}
\eea
This gives rise to

\bea
\label{eq:drhogammadec:dfields:6:start}
A^\radiation_\alpha &=& 4\frac{\partial N_\uend}{\partial\phi_{\alpha*}}+\frac{\partial\ln\rho_\uend^\phi}{\partial\phi_{\alpha*}}+\frac{2}{3}\frac{\partial\ln H_\uend}{\partial\phi_{\alpha*}}-\frac{2}{3}\frac{\partial\ln H_\ueq}{\partial\phi_{\alpha*}}+\frac{\sigma_\ueq^2}{2\Mp^2}\frac{\partial\ln \sigma_\ueq}{\partial\phi_{\alpha*}}\\
A^\radiation_{\alpha \beta} &=& 4\frac{\partial^2 N_\uend}{{\partial\phi_{\alpha*}}{\partial\phi_{\beta*}}}+\frac{\partial^2\ln\rho_\uend^\phi}{{\partial\phi_{\alpha*}}{\partial\phi_{\beta*}}}+\frac{2}{3}\frac{\partial^2\ln H_\uend}{{\partial\phi_{\alpha*}}{\partial\phi_{\beta*}}}-\frac{2}{3}\frac{\partial^2\ln H_\ueq}{{\partial\phi_{\alpha*}}{\partial\phi_{\beta*}}}
\nonumber\\ & &
+\frac{\sigma_\ueq^2}{2\Mp^2}\left(2 \frac{\partial^2\ln \sigma_\ueq}{\partial\phi_{\alpha*}\partial\phi_{\beta*}}+\frac{\partial\ln \sigma_\ueq}{\partial\phi_{\beta*}}\frac{\partial\ln \sigma_\ueq}{\partial\phi_{\beta*}}\right)\, .
\label{eq:drhogammadec:dfields:6:end}
\eea
In these expressions, the derivatives of $H_\ueq$ are given by \Eqs{eq:2:dHeq1dfields:start}-(\ref{eq:2:dHeq1dfields:end}) and the ones of $\sigma_\ueq$ are given by \Eqs{eq:2:dsigmaeq1dfields:start}-(\ref{eq:2:dsigmaeq1dfields:end}). As for matter, one has
\bea
\rho_{\sigma\mathrm{-dec}}^\matter &=&\rho_{\sigma\mathrm{-mass}}^\sigma\exp\left[-3\left(N_{\sigma\mathrm{-dec}}-N_{\sigma\mathrm{-mass}}\right)\right]\nonumber \\
&=&\rho_{\sigma\mathrm{-mass}}^\sigma\exp\left[-3\left(N_{\sigma\mathrm{-dec}}-N_{\uend}\right)\right] \exp\left[-3\left(N_{\uend}-N_\ueq\right)\right] \exp\left[-3\left(N_\ueq-N_{\sigma\mathrm{-mass}}\right)\right]
\nonumber \\
& =&\frac{3}{2}\Mp^2m_\sigma^2\left(\frac{H_\uend}{H_\ueq}\right)^2\exp\left(\frac{3\sigma_\ueq^2}{4\Mp^2}-\frac{3}{2}\right) \exp\left[-3\left(N_{\sigma\mathrm{-dec}}-N_{\uend}\right)\right] \, .
\label{eq:rhomattdec:Nef:6}
\eea
This gives rise to
\bea
\label{eq:drhomattdec:dfields:6:start}
A^\matter_\alpha &=& 3\frac{\partial N_\uend}{\partial\phi_{\alpha*}}+2\frac{\partial\ln H_\uend}{\partial\phi_{\alpha*}}-2\frac{\partial\ln H_\ueq}{\partial\phi_{\alpha*}}+\frac{3\sigma_\ueq^2}{2\Mp^2}\frac{\partial\ln\sigma_\ueq}{\partial\phi_{\alpha*}}\\
A^\matter_{\alpha\beta} &=& 3\frac{\partial^2 N_\uend}{{\partial\phi_{\alpha*}}{\partial\phi_{\beta*}}}+2\frac{\partial^2\ln H_\uend}{{\partial\phi_{\alpha*}}{\partial\phi_{\beta*}}}-2\frac{\partial^2\ln H_\ueq}{{\partial\phi_{\alpha*}}{\partial\phi_{\beta*}}}+\frac{3\sigma_\ueq^2}{2\Mp^2}\left(\frac{\partial^2\ln\sigma_\ueq}{\partial\phi_{\alpha*}\partial\phi_{\beta*}}+2\frac{\partial\ln\sigma_\ueq}{\partial\phi_{\alpha*}}\frac{\partial\ln\sigma_\ueq}{\partial\phi_{\beta*}}\right)\, .\nonumber\\
\label{eq:drhomattdec:dfields:6:end}
\eea

In case 6, the reheating parameter $R_\urad$ is given by the same expression as in case 3, and \Eqs{eq:3-6-10:Rradmatphi}, (\ref{eq:3-6-10:Rradinfl}) and (\ref{eq:3-6-10:Rradmatsigma}) apply.

Finally, let us see under which conditions case 6 takes place. In the limit where the curvaton remains frozen at $\sigma_\uend$ until equality, one has $\sigma_\ueq\simeq \sigma_\uend$ and $H_\ueq\simeq m_\sigma\sigma_\uend/(\sqrt{6}\Mp)$. Case 6 corresponds to the situation where this happens between the end of inflation and the moment when $\sigma$ becomes massive, \ie $m_\sigma/\sqrt{2}<H_\ueq<H_\uend$. This leads to the condition
\beq
\sqrt{3} \ll \frac{\sigma_\uend}{\Mp}\ll \sqrt{6}\frac{H_\uend}{m_\sigma}\, .
\eeq
The lower bound $\sqrt{3}$, is numerically very close to the upper bound derived for case 5, $\sqrt{2}e^{2/9}$. 
\subsection{$\Gamma_\sigma<m_\sigma<\Gamma_\phi<H_\uend$}
In this case, the inflaton decays before the curvaton becomes massive. This corresponds to cases 7, 8, 9 and 10. Since the last field to decay is the curvaton, one has $\phim=\sigma$ and $\phir=\phi$.
\subsubsection{Case 7}
\label{sec:case7}
In this case, the curvaton is always subdominant. Between the end of inflation and the inflaton decay, the Universe is matter dominated and \Eqs{eq:H:matter} and~(\ref{eq:sigmaSR}) apply. Evaluated at the moment when $\phi$ decays, when $H=\Gamma_\phi$, this gives rise to
\beq
\label{eq:7:sigmaphidec}
\sigma_{\phi\mathrm{-dec}}=\sigma_\uend\exp\left[-\frac{m_\sigma^2}{9\Gamma_\phi^2}\left(1-\frac{\Gamma_\phi^2}{H_\uend^2}\right)\right]\, .
\eeq
Then, the Universe is radiation dominated, and one has
\beq
\label{eq:H:rad}
H=\frac{\Gamma_\phi}{1+2\Gamma_\phi\left(t-t_{\phi\mathrm{-dec}}\right)}\, .
\eeq
In the limit where $\sigma$ is slowly rolling until the moment when it becomes massive, one then has
\beq
\label{eq:7:sigmaSR:rad}
\ln\left(\frac{\sigma}{\sigma_{\phi\mathrm{-dec}}}\right)=-\frac{m_\sigma^2}{3\Gamma_\phi}\left(t-t_{\phi\mathrm{-dec}}\right)\left[1+\Gamma_\phi\left(t-t_{\phi\mathrm{-dec}}\right)\right]\, .
\eeq
The field $\sigma$ becomes massive when $H=m_\sigma/\sqrt{2}$, and at this time
\beq
\label{eq:7:sigmasigmamass}
\sigma_{\sigma\mathrm{-mass}}=\sigma_{\phi\mathrm{-dec}}\exp\left(-\frac{1}{3}+\frac{m_\sigma^2}{12\Gamma_\phi^2}\right)=\sigma_\uend\exp\left(-\frac{1}{3}-\frac{m_\sigma^2}{36\Gamma_\phi^2}+\frac{m_\sigma^2}{9H_\uend^2}\right)\, ,
\eeq
and $\rho^\sigma_{\sigma\mathrm{-mass}}=3 m_\sigma^2 \sigma_{\sigma\mathrm{-mass}}^2/4$. Then, $\sigma$ becomes massive, and one has
\bea
\rho_{\sigma\mathrm{-dec}}^\matter&=&\rho_{\sigma\mathrm{-mass}}^\sigma\left(\frac{\Gamma_\sigma\sqrt{2}}{m_\sigma}\right)^{3/2}\\ &=&
\frac{3}{4}m_\sigma^2\sigma_\uend^2\left(\frac{\Gamma_\sigma\sqrt{2}}{m_\sigma}\right)^{3/2}\exp\left(-\frac{2}{3}-\frac{m_\sigma^2}{18\Gamma_\phi^2}+\frac{2m_\sigma^2}{9H_\uend^2}\right)\, .
\label{eq:rhosigmadec:7}
\eea
One the other hand, one simply has $\rho_{\sigma\mathrm{-dec}}^\radiation=3\Mp^2\Gamma_\sigma^2-\rho_{\sigma\mathrm{-dec}}^\matter$ and $r_\udec$ is given by \Eq{eq:rdec:def}.

Let us now work out the perturbed energy densities. They must be evaluated at the fixed background time $N_{\sigma\mathrm{-dec}}$, so remember that this quantity must remain unperturbed. For the radiation component, one has 
\bea
\rho_{\sigma\mathrm{-dec}}^\radiation &=&\rho_{\phi\mathrm{-dec}}^\phi\exp\left[-4\left(N_{\sigma\mathrm{-dec}}-N_{\phi\mathrm{-dec}}\right)\right]\\
 &=&\rho_{\phi\mathrm{-dec}}^\phi\exp\left[-4\left(N_{\sigma\mathrm{-dec}}-N_\uend\right)\right]\exp\left[-4\left(N_\uend-N_{\phi\mathrm{-dec}}\right)\right]\\
 &=&3\Mp^2\Gamma_\phi^2\left(\frac{H_\uend}{\Gamma_\phi}\right)^{8/3}\exp\left[-4\left(N_{\sigma\mathrm{-dec}}-N_\uend\right)\right]\, .
\label{eq:7:rhogammadec:Nef}
\eea
This gives rise to
\bea
\label{eq:7:drhogammadecdfields:start}
A^\radiation_\alpha &=&4\frac{\partial N_\uend}{\partial\phi_{\alpha*}}+\frac{8}{3}\frac{\partial\ln H_\uend}{\partial\phi_{\alpha*}}\\
A^\radiation_{\alpha\beta} &=& 4\frac{\partial^2 N_\uend}{{\partial\phi_{\alpha*}}{\partial\phi_{\beta*}}}+\frac{8}{3}\frac{\partial^2\ln H_\uend}{{\partial\phi_{\alpha*}}{\partial\phi_{\beta*}}}\, .
\label{eq:7:drhogammadecdfields:end}
\eea

As for the matter energy density fluctuations, one has
\bea
\rho_{\sigma\mathrm{-dec}}^\matter &=&\rho^\sigma_{\sigma\mathrm{-mass}}\exp\left[-3\left(N_{\sigma\mathrm{-dec}}-N_{\sigma\mathrm{-mass}}\right)\right]
\nonumber \\
&=&\rho^\sigma_{\sigma\mathrm{-mass}}\exp\left[-3\left(N_{\sigma\mathrm{-dec}}-N_\uend\right)\right]\exp\left[-3\left(N_\uend-N_{\phi\mathrm{-dec}}\right)\right]\exp\left[-3\left(N_{\phi\mathrm{-dec}}-N_{\sigma\mathrm{-mass}}\right)\right]\nonumber\\
&=&
\frac{3}{4}m_\sigma^2\sigma_\uend^2\left(\frac{\Gamma_\phi\sqrt{2}}{m_\sigma}\right)^{3/2}\left(\frac{H_\uend}{\Gamma_\phi}\right)^2\exp\left(-\frac{2}{3}-\frac{m_\sigma^2}{18\Gamma_\phi^2}+\frac{2m_\sigma^2}{9H_\uend^2}\right)\exp\left[-3\left(N_{\sigma\mathrm{-dec}}-N_\uend\right)\right]\,.\nonumber\\
\label{eq:7:rhosigmadec:Nef}
\eea
This gives rise to
\bea
\label{eq:7:drhosigmadecdfields:start}
A^\matter_\alpha &=& 3\frac{\partial N_\uend}{\partial\phi_{\alpha*}}+2\frac{\partial\ln\sigma_\uend}{\partial\phi_{\alpha*}}+\left(2-\frac{4m_\sigma^2}{9H_\uend^2}\right)\frac{\partial\ln H_\uend}{\partial\phi_{\alpha*}}\\
A^\matter_{\alpha\beta} &=& 3\frac{\partial^2 N_\uend}{{\partial\phi_{\alpha*}}{\partial\phi_{\beta*}}}+2\frac{\partial^2\ln\sigma_\uend}{{\partial\phi_{\alpha*}}{\partial\phi_{\beta*}}}+\left(2-\frac{4m_\sigma^2}{9H_\uend^2}\right)\frac{\partial^2\ln H_\uend}{{\partial\phi_{\alpha*}}{\partial\phi_{\beta*}}}+\frac{8}{9}\frac{m_\sigma^2}{H_\uend^2}\frac{\partial\ln H_\uend}{\partial\phi_{\alpha*}}\frac{\partial\ln H_\uend}{\partial\phi_{\beta*}}\, .\nonumber\\
\label{eq:7:drhosigmadecdfields:end}
\eea

In case 7, the reheating parameter $R_\urad$ is given by the same expression as in cases 1 and 4, namely \Eq{eq:Rrad:1-4-7}.

Finally, let us check under which conditions the assumptions of case 7 are valid. \Eq{eq:rhosigmadec:7} gives rise to
\beq
\dfrac{\rho_{\sigma\mathrm{-dec}}^\sigma}{\rho_{\sigma\mathrm{-dec}}^\radiation}\simeq\frac{1}{2^{5/4}e^{2/3}}\sqrt{\frac{m_\sigma}{\Gamma_\sigma}}\left(\frac{\sigma_\uend}{\Mp}\right)\, .
\eeq
Therefore, case 7 corresponds to the regime where 
\beq
\frac{\sigma_\uend}{\Mp}\ll 2^{5/8}e^{1/3}\left(\frac{\Gamma_\sigma}{m_\sigma}\right)^{1/4}\, .
\eeq 
\subsubsection{Case 8}
\label{sec:case8}
In this case the curvaton comes to dominate the energy budget of the Universe, after it becomes massive and when the inflaton has already decayed. Until the time  when the curvaton becomes massive, the same analysis as in case 7 apply, and \Eq{eq:7:sigmasigmamass} giving $\sigma_{\sigma\mathrm{-mass}}$ can be used. At the time of equality, the energy density of $\sigma$ is then given by
\beq
\rho^\sigma_\mathrm{eq}=\rho^\sigma_{\sigma\mathrm{-mass}}\left(\frac{a_\ueq}{a_{\sigma\mathrm{-mass}}}\right)^{-3}=\frac{3}{4	}m_\sigma^2\sigma_{\sigma\mathrm{-mass}}^2\left(\frac{H_\ueq}{H_{\sigma\mathrm{-mass}}}\right)^{3/2}=\frac{3}{4	}m_\sigma^2\sigma_{\sigma\mathrm{-mass}}^2\left(\frac{H_\ueq\sqrt{2}}{m_{\sigma}}\right)^{3/2}\, .
\eeq
On the other hand, one simply has
\beq
\rho^{\radiation}_\mathrm{eq}=\rho_{\phi\mathrm{-dec}}^\phi\left(\dfrac{a_\mathrm{eq}}{a_{\phi\mathrm{-dec}}}\right)^{-4}=\rho_{\phi\mathrm{-dec}}^\phi\left(\dfrac{H_\mathrm{eq}}{H_{\phi\mathrm{-dec}}}\right)^{2}=
\rho_\uend^\phi\left(\dfrac{H_\mathrm{eq}}{H_\uend}\right)^2\, ,
\eeq
and solving $\rho^\radiation_\mathrm{eq}=\rho^\sigma_\mathrm{eq}$ gives rise to
\beq
H_\ueq = \frac{9}{2^{5/2}}\exp\left(-\frac{4}{3}-\frac{m_\sigma^2}{9\Gamma_\phi^2}+\frac{4m_\sigma^2}{9H_\uend^2}\right)m_\sigma\sigma_\uend^4\frac{H_\uend^4}{\left(\rho_\uend^\phi\right)^2}\, .
\label{eq:8:Heq}
\eeq
Then, the Universe is radiation dominated and one has
\beq
\rho_{\sigma\mathrm{-dec}}^\sigma=\rho_{\mathrm{eq}}^\sigma\left(\frac{a_{\sigma\mathrm{-dec}}}{a_\ueq}\right)^{-3}=\rho_{\mathrm{eq}}^\sigma\left(\frac{H_{\sigma\mathrm{-dec}}}{H_\ueq}\right)^{2}=\Gamma_\sigma^2\frac{\rho_\uend^\phi}{H_\uend^2}\, ,
\eeq
in agreement with the fact that the curvaton dominates the energy budget at this time. On the other hand, the energy density of radiation is given by
\bea
\rho_{\sigma\mathrm{-dec}}^\radiation &=&\rho_\mathrm{eq}^\radiation\left(\dfrac{a_{\sigma\mathrm{-dec}}}{a_\mathrm{eq}}\right)^{-4}
=\rho_\mathrm{eq}^\radiation\left(\dfrac{H_{\sigma\mathrm{-dec}}}{H_\mathrm{eq}}\right)^{8/3}
\nonumber\\ &=&
\frac{2^{5/3}}{3^{4/3}}\exp\left(\frac{8}{9}+\frac{2m_\sigma^2}{27\Gamma_\phi^2}-\frac{8}{27}\frac{m_\sigma^2}{H_\uend^2}\right)\left(\frac{\rho_\uend^\phi}{H_\uend^2}\right)^{7/3}\left(\frac{\Gamma_\sigma}{\sigma_\uend}\right)^{8/3}m_\sigma^{-2/3}\, ,
\label{eq:rhogamma:sigmadec:8}
\eea
and $r_\udec$ can be calculated thanks to \Eq{eq:rdec:def}.

In case 8, the reheating parameter $R_\urad$ is given by the same expression as in case 5, and \Eqs{eq:5-8:Rradmatphi} and~(\ref{eq:5-8:Rradmatsigma}) apply.

Let us now work out the perturbed energy densities. For the radiation component, \Eq{eq:7:rhogammadec:Nef} still applies, and therefore \Eqs{eq:7:drhogammadecdfields:start}-(\ref{eq:7:drhogammadecdfields:end}) are valid in case 8 too. For the matter component, \Eq{eq:7:rhosigmadec:Nef} still applies, and therefore \Eqs{eq:7:drhosigmadecdfields:start}-(\ref{eq:7:drhosigmadecdfields:end}) are valid in case 8 too.

Finally, let us see under which conditions on the parameters case 8 takes place. Case 8 corresponds to the situation when equality between inflaton and curvaton energy densities happens between the moment when the curvaton becomes massive, and the decay of the curvaton, that is $\Gamma_\sigma<H_\mathrm{eq}<m_\sigma/\sqrt{2}$. Together with \Eq{eq:8:Heq}, this leads to the condition
\beq
2^{5/8}e^{1/3}\left(\dfrac{\Gamma_\sigma}{m_\sigma}\right)^{1/4}\ll \dfrac{\sigma_\uend}{\Mp}\ll \sqrt{2}e^{1/3}\, .
\eeq
\subsubsection{Case 9}
\label{sec:case9}
In this case the curvaton comes to dominate the energy budget of the Universe before it becomes massive but after the inflaton decay. This means that there is an extra phase of inflation. Until the moment when the curvaton dominates the energy content, the situation is similar to case 7 and \Eqs{eq:7:sigmaphidec},~(\ref{eq:H:rad}) and ~(\ref{eq:7:sigmaSR:rad}) apply. This gives rise to
\beq
\label{eq:9:sigmaeq}
\sigma_\ueq = \sigma_\uend\exp\left[-\frac{m_\sigma^2}{3}\left(\frac{1}{4 H_\ueq^2}+\frac{1}{12\Gamma_\phi^2}-\frac{1}{3H_\uend^2}\right)\right]\, ,
\eeq
and
\beq
\rho^\sigma_\ueq=\frac{m_\sigma^2}{2}\sigma_\ueq^2\left(1+\frac{m_\sigma^2}{9 H_\ueq^2}\right)\, .
\label{eq:9:rhosigmaeq}
\eeq
One the other hand, one has
\beq
\rho^\radiation_\mathrm{eq}=\rho_{\phi\mathrm{-dec}}^\phi\left(\dfrac{a_\mathrm{eq}}{a_{\phi\mathrm{-dec}}}\right)^{-4}=\rho_{\phi\mathrm{-dec}}^\phi\left(\dfrac{H_\mathrm{eq}}{H_{\phi\mathrm{-dec}}}\right)^{2}=
\rho_\uend^\phi\left(\dfrac{H_\mathrm{eq}}{H_\uend}\right)^2\, .
\label{eq:9:rhogammaeq}
\eeq
The value of $H_\ueq$ is obtained by numerically solving $\rho^\sigma_\ueq=\rho^\radiation_\mathrm{eq}$. The new phase of inflation terminates when $\sigma_{\sigma\mathrm{-mass}}=\sqrt{2}\Mp$ and $H_{\sigma\mathrm{-mass}}=m_\sigma/\sqrt{2}$. The realized number of \efolds is given by $N_{\sigma\mathrm{-mass}}-N_\mathrm{eq}=(\sigma_\mathrm{eq}^2-\sigma_{\sigma\mathrm{-mass}}^2)/(4\Mp^2)$, and one has
\beq
\rho^\radiation_{\sigma\mathrm{-mass}}=\rho_\mathrm{eq}^\radiation\left(\dfrac{a_{\sigma\mathrm{-mass}}}{a_\mathrm{eq}}\right)^{-4}=\rho_\mathrm{eq}^\radiation\exp\left[-4\left(N_{\sigma\mathrm{-mass}}-N_\mathrm{eq}\right)\right]
=\rho_\uend^\phi\left(\dfrac{H_\mathrm{eq}}{H_\uend}\right)^2\exp\left(2-\dfrac{\sigma_\mathrm{eq}^2}{\Mp^2}\right)\, .
\eeq
After the curvaton becomes massive, the Universe is dominated by a matter fluid until the curvaton decays. One then has
\bea
\rho^\radiation_{\sigma\mathrm{-dec}}&=&\rho^\radiation_{\sigma\mathrm{-mass}}\left(\dfrac{a_{\sigma\mathrm{-dec}}}{a_{\sigma\mathrm{-mass}}}\right)^{-4}=\rho^\radiation_{\sigma\mathrm{-mass}}\left(\dfrac{H_{\sigma\mathrm{-dec}}}{H_{\sigma\mathrm{-mass}}}\right)^{8/3}
\nonumber\\&=&
\rho_\uend^\phi\left(\dfrac{H_\mathrm{eq}}{H_\uend}\right)^2\exp\left(2-\dfrac{\sigma_\mathrm{eq}^2}{\Mp^2}\right)\left(\dfrac{\Gamma_\sigma\sqrt{2}}{m_\sigma}\right)^{8/3}\, ,
\eea
while $\rho^\matter_{\sigma\mathrm{-dec}}=3\Mp^2\Gamma_\sigma^2-\rho^\radiation_{\sigma\mathrm{-dec}}$, and $r_\udec$ is given by \Eq{eq:rdec:def}.

Let us now derive the perturbed energy densities. For radiation, \Eq{eq:7:rhogammadec:Nef} is still valid, and \Eqs{eq:7:drhogammadecdfields:start}-(\ref{eq:7:drhogammadecdfields:end}) still apply. As for matter, one has
\bea
\rho_{\sigma\mathrm{-dec}}^\matter &=&\rho_{\sigma\mathrm{-mass}}^\sigma\exp\left[-3\left(N_{\sigma\mathrm{-dec}}-N_{\sigma\mathrm{-mass}}\right)\right]
\\&=&\rho_{\sigma\mathrm{-mass}}^\sigma\exp\left[-3\left(N_{\sigma\mathrm{-dec}}-N_{\uend}\right)\right]\exp\left[-3\left(N_\uend-N_{\phi\mathrm{-dec}}\right)\right]
\nonumber\\ & & \exp\left[-3\left(N_{\phi\mathrm{-dec}}-N_{\ueq}\right)\right]\exp\left[-3\left(N_\ueq-N_{\sigma\mathrm{-mass}}\right)\right]\\
&=&\frac{3}{2}\Mp^2m_\sigma^2\left(\frac{H_\uend}{\Gamma_\phi}\right)^2\left(\frac{\Gamma_\phi}{H_\ueq}\right)^{3/2}\exp\left(\frac{3\sigma_\ueq^2}{4\Mp^2}-\frac{3}{2}\right)\exp\left[-3\left(N_{\sigma\mathrm{-dec}}-N_{\uend}\right)\right]\, .\nonumber\\
\eea
This gives rise to
\bea
A^\matter_\alpha &=& 3\frac{\partial N_\uend}{\partial\phi_{\alpha*}}+2\frac{\partial\ln H_\uend}{\partial\phi_{\alpha*}}-\frac{3}{2}\frac{\partial\ln H_\ueq}{\partial\phi_{\alpha*}}+\frac{3\sigma_\ueq^2}{2\Mp^2}\frac{\partial\ln\sigma_\ueq}{\partial\phi_{\alpha*}}\\
A^\matter_{\alpha\beta} &=& 3\frac{\partial^2 N_\uend}{{\partial\phi_{\alpha*}}{\partial\phi_{\beta*}}}+2\frac{\partial^2\ln H_\uend}{{\partial\phi_{\alpha*}}{\partial\phi_{\beta*}}}-\frac{3}{2}\frac{\partial^2\ln H_\ueq}{{\partial\phi_{\alpha*}}{\partial\phi_{\beta*}}}+\frac{3\sigma_\ueq^2}{2\Mp^2}\left(\frac{\partial^2\ln\sigma_\ueq}{\partial\phi_{\alpha*}\partial\phi_{\beta*}}+2\frac{\partial\ln\sigma_\ueq}{\partial\phi_{\alpha*}}\frac{\partial\ln\sigma_\ueq}{\partial\phi_{\beta*}}\right)\, .\nonumber\\
\eea
As explained above, $H_\ueq$ has to be determined numerically. However, Taylor expanding \Eqs{eq:9:rhogammaeq} and~(\ref{eq:9:rhosigmaeq}), one can express its derivatives as
\bea
\label{eq:9:dHeqdfields:start}
\frac{\partial \ln H_\ueq}{\partial\phi_{\alpha*}}&=&\left[\frac{\partial\ln\sigma_\uend}{\partial\phi_{\alpha*}}+\left(1-\frac{2 m_\sigma^2}{9H_\uend^2}\right)\frac{\partial\ln H_\uend}{\partial\phi_{\alpha*}}-\frac{1}{2}\frac{\partial\ln\rho_\uend^\phi}{\partial\phi_{\alpha*}}\right]/\left(1+\frac{m_\sigma^2}{9H_\ueq^2+m_\sigma^2}-\frac{m_\sigma^2}{6H_\ueq^2}\right)\nonumber \\ \\
\frac{\partial^2 \ln H_\ueq}{\partial\phi_{\alpha*}\partial\phi_{\beta*}}&=&\left\lbrace\frac{\partial^2\ln\sigma_\uend}{{\partial\phi_{\alpha*}}{\partial\phi_{\beta*}}}-\frac{1}{2}\frac{\partial^2\ln\rho_\uend^\phi}{{\partial\phi_{\alpha*}}{\partial\phi_{\beta*}}}+\left(1-\frac{2m_\sigma^2}{9H_\uend^2}\right)\frac{\partial^2\ln H_\uend}{{\partial\phi_{\alpha*}}{\partial\phi_{\beta*}}}
\right. \nonumber\\ & & \left. +\frac{4m_\sigma^2}{9H_\uend^2}\frac{\partial\ln H_\uend}{\partial\phi_{\alpha*}}\frac{\partial\ln H_\uend}{\partial\phi_{\beta*}}-2\frac{\partial\ln H_\ueq}{\partial\phi_{\alpha*}}\frac{\partial\ln H_\ueq}{\partial\phi_{\beta*}}\left[\frac{m_\sigma^2}{6H_\ueq^2}-\frac{9 H_\ueq^2 m_\sigma^2}{\left(9H_\ueq^2+m_\sigma^2\right)^2}\right]\right\rbrace
\nonumber\\& &\times
\left[1+\frac{m_\sigma^2}{9H_\ueq^2+m_\sigma^2}-\frac{m_\sigma^2}{6H_\ueq^2}\right]^{-1}\, .
\label{eq:9:dHeqdfields:end}
\eea
On the other hand, derivatives of $\sigma_\ueq$ are derived from \Eq{eq:9:sigmaeq}, and read
\bea
\label{eq:9:dsigmadfields:start}
\frac{\partial\ln\sigma_\ueq}{\partial\phi_{\alpha*}}&=&\frac{\partial\ln\sigma_\uend}{\partial\phi_{\alpha*}}+\frac{m_\sigma^2}{6H_\ueq^2}\frac{\partial \ln H_\ueq}{\partial\phi_{\alpha*}}-\frac{2m_\sigma^2}{9H_\uend^2}\frac{\partial \ln H_\uend}{\partial\phi_{\alpha*}}\\
\frac{\partial^2\ln \sigma_\ueq}{{\partial\phi_{\alpha*}}{\partial\phi_{\beta*}}}&=&\frac{\partial^2\ln\sigma_\uend}{{\partial\phi_{\alpha*}}{\partial\phi_{\beta*}}}+\frac{m_\sigma^2}{6H_\ueq^2}\left(\frac{\partial^2 \ln H_\ueq}{{\partial\phi_{\alpha*}}{\partial\phi_{\beta*}}}-2\frac{\partial \ln H_\ueq}{\partial\phi_{\alpha*}}\frac{\partial \ln H_\ueq}{\partial\phi_{\beta*}}\right)
\nonumber\\& & -\frac{2m_\sigma^2}{9H_\uend^2}\left(\frac{\partial^2 \ln H_\uend}{{\partial\phi_{\alpha*}}{\partial\phi_{\beta*}}}-2\frac{\partial \ln H_\uend}{\partial\phi_{\alpha*}}\frac{\partial \ln H_\uend}{\partial\phi_{\beta*}}\right)\, .
\label{eq:9:dsigmadfields:end}
\eea

Let us now calculate the reheating parameter. In case 9, reheating is made of a matter phase driven by $\phi$, a radiation phase, and inflation phase driven by $\sigma$ and a matter phase driven by $\sigma$. One has
\beq
\label{eq:9:Rradmatphi}
\ln R_\urad^{\matter,\phi}= \dfrac{1}{6}\ln\left(\dfrac{\Gamma_\phi}{H_\uend}\right)\, ,
\eeq
\beq
\label{eq:9:Rradinfl}
\ln R_\urad^{\mathrm{inf}}=\dfrac{1}{2}-\dfrac{\sigma_{\ueq}^2}{4\Mp^2}+\dfrac{1}{4}\ln\left(2\dfrac{H_{\ueq}^2}{m_\sigma^2}\right)\, ,
\eeq
\beq
\label{eq:9:Rradmatsigma}
\ln R_\urad^{\matter,\sigma}= \dfrac{1}{12}\ln\left(\dfrac{2\Gamma_\sigma^2}{m_\sigma^2}\right)\, .
\eeq
The parameter $\ln R_\urad$ is given by the sum of these two contributions.

Finally, let us check under which conditions case 9 takes place. In case 9, equality between curvaton and radiation occurs while the curvaton is still slowly rolling, but after the inflaton decay. In the limit $\Gamma_\sigma\ll m_\sigma\ll \Gamma_\phi\ll H_\uend$, one then has
\beq
\rho^\sigma_\mathrm{eq}\simeq\rho^\sigma_\mathrm{end}\simeq\dfrac{m_\sigma^2}{2}\sigma_\uend^2\, ,
\eeq
while
\beq
\rho_{\mathrm{eq}}^\matter\simeq \rho_{\phi\mathrm{-dec}}^\phi\left(\dfrac{a_{\mathrm{eq}}}{a_{\phi\mathrm{-dec}}}\right)^{-4}\simeq  \rho_{\phi\mathrm{-dec}}^\phi\left(\dfrac{H_{\mathrm{eq}}}{H_{\phi\mathrm{-dec}}}\right)^{2}
\simeq  3\Mp^2 H_\mathrm{eq}^2\, .
\eeq
Equaling the two above relations gives rise to $H_\mathrm{eq}\simeq m_\sigma\sigma_\uend/(\sqrt{6}\Mp)$. Case 9 corresponds to the situation where this happens between the decay of the inflaton and the moment when the curvaton becomes massive, that is $m_\sigma/\sqrt{2}<H_\mathrm{eq}<\Gamma_\phi$. This leads to the condition
\beq
\sqrt{3}\ll \dfrac{\sigma_\uend}{\Mp}\ll \sqrt{6} \dfrac{\Gamma_\phi}{m_\sigma}\, .
\eeq
\subsubsection{Case 10}
\label{sec:case10}
In this case equality between curvaton and inflaton energy densities occur while the inflaton is still massive, and when the curvaton is still slowly rolling. In order to determine the value of the Hubble parameter $H_\ueq$ when this happens, one can proceeds as in case 3 and solve the equality between \Eqs{eq:3:rhophieq} and~(\ref{eq:3:rhosigmaeq}). From this moment on, the Universe experiences a new phase of inflation. At some point during this inflationary second epoch, the inflaton decays. Since, in the slow roll approximation, one has $H_{\phi\mathrm{-dec}}\simeq m_\sigma \sigma_{\phi\mathrm{-dec}}/(\sqrt{6}\Mp)$, this occurs when $\sigma_{\phi\mathrm{-dec}}=\sqrt{6}\Mp\Gamma_\phi/m_\sigma$. The number of \efolds elapsed at this point is then given by $N_{\phi\mathrm{-dec}}-N_\mathrm{eq}=\sigma_\mathrm{eq}^2/(4\Mp^2)-3\Gamma_\phi^2/(2 m_\sigma^2)$. Therefore, the energy density contained in the inflaton field when the inflaton decays is given by
\bea
\rho^\phi_{\phi\mathrm{-dec}}&=&\rho^\phi_\mathrm{eq}\left(\dfrac{a_{\phi\mathrm{-dec}}}{a_\mathrm{eq}}\right)^{-3}=\rho_\phi^\mathrm{end}\left(\dfrac{H_\mathrm{eq}}{H_\uend}\right)^{2}\exp\left[-3\left(N_{\phi\mathrm{-dec}}-N_\mathrm{eq}\right)\right]
\nonumber \\ &=&
\rho^\phi_\mathrm{end}\left(\dfrac{H_\mathrm{eq}}{H_\uend}\right)^{2}\exp\left(\dfrac{9\Gamma_\phi^2}{2m_\sigma^2}-\dfrac{3\sigma_\mathrm{eq}^2}{4\Mp^2}\right)\, .
\eea
After the inflaton decays, the Universe continues to inflate until the curvaton becomes massive and $H_{\sigma\mathrm{-mass}}=m_\sigma/\sqrt{2}$. The number of \efolds elapsed during this period is given by $N_{\sigma\mathrm{-mass}}-N_{\phi\mathrm{-dec}}=3\Gamma_\phi^2/(2 m_\sigma^2)-1/2$, so that the energy density of radiation becomes
\beq
\rho^\radiation_{\sigma\mathrm{-mass}}=\rho^\phi_{\phi\mathrm{-dec}}\exp\left[-4\left(N_{\sigma\mathrm{-mass}}-N_{\phi\mathrm{-dec}}\right)\right]=
\rho_\phi^\mathrm{end}\left(\dfrac{H_\mathrm{eq}}{H_\uend}\right)^{2}\exp\left(2-\dfrac{3\sigma_\ueq^2}{4\Mp^2}-\dfrac{3\Gamma_\phi^2}{2m_\sigma^2}\right)\, .
\eeq
After the curvaton becomes massive, the Universe is matter dominated, so that the energy density of radiation scales as
\bea
\rho^\radiation_{\sigma\mathrm{-dec}}&=&\rho^\radiation_{\sigma\mathrm{-mass}}\left(\dfrac{a_{\sigma\mathrm{-dec}}}{a_{\sigma\mathrm{-mass}}}\right)^{-4}=\rho^\radiation_{\sigma\mathrm{-mass}}\left(\dfrac{H_{\sigma\mathrm{-dec}}}{H_{\sigma\mathrm{-mass}}}\right)^{8/3}
\nonumber\\ &=&
\rho_\phi^\mathrm{end}\left(\dfrac{H_\mathrm{eq}}{H_\uend}\right)^{2}\left(\dfrac{\Gamma_\sigma\sqrt{2}}{m_\sigma}\right)^{8/3}\exp\left(2-\dfrac{3\sigma_\ueq^2}{4\Mp^2}-\dfrac{3\Gamma_\phi^2}{2m_\sigma^2}\right)\, .
\eea
On the other hand, the energy density of the curvaton at this time is simply given by $\rho^\matter_{\sigma\mathrm{-dec}}=3\Mp^2\Gamma_\sigma^2-\rho^\radiation_{\sigma\mathrm{-dec}}$, and $r_\udec$ can be calculated from \Eq{eq:rdec:def}.

Let us now work out the perturbed energy densities. They must be evaluated at the fixed background time $N_{\sigma\mathrm{-dec}}$, so remember that this quantity must remain unperturbed. For the radiation component, one has 
\bea
\rho_{\sigma\mathrm{-dec}}^\radiation &=&\rho_{\phi\mathrm{-dec}}^\phi\exp\left[-4\left(N_{\sigma\mathrm{-dec}}-N_{\phi\mathrm{-dec}}\right)\right]\\
 &=&\rho_{\phi\mathrm{-dec}}^\phi\exp\left[-4\left(N_{\sigma\mathrm{-dec}}-N_\uend\right)\right]\exp\left[-4\left(N_\uend-N_{\ueq}\right)\right]
\nonumber\\ & &\times  
 \exp\left[-4\left(N_\ueq-N_{\phi\mathrm{-dec}}\right)\right]\\
&=&\rho_\uend^\phi\left(\frac{H_\uend}{H_\ueq}\right)^{2/3}\exp\left(\frac{\sigma_\ueq^2}{4\Mp^2}-\frac{3\Gamma_\phi^2}{2m_\sigma^2}\right)\exp\left[-4\left(N_{\sigma\mathrm{-dec}}-N_\uend\right)\right]\, .
\label{eq:10:rhogammadec:Nef}
\eea
One can see that this expression only differs from \Eq{eq:rhogammadec:Nef:6} by an overall constant factor that disappears in the logarithmic derivative, and therefore, \Eqs{eq:drhogammadec:dfields:6:start}-(\ref{eq:drhogammadec:dfields:6:end}) can be used. As for matter, \Eq{eq:rhomattdec:Nef:6} applies, and \Eqs{eq:drhomattdec:dfields:6:start}-(\ref{eq:drhomattdec:dfields:6:end}) can be used.

In case 10, the reheating parameter $R_\urad$ is given by the same expression as in cases 3 and 6, and \Eqs{eq:3-6-10:Rradmatphi}, (\ref{eq:3-6-10:Rradinfl}) and (\ref{eq:3-6-10:Rradmatsigma}) apply.

Finally, let us check under which conditions case 10 takes place. In the limit where the curvaton remains frozen at $\sigma_\uend$ until equality, one has $H_\ueq\simeq m_\sigma\sigma_\uend/(\sqrt{6}\Mp)$. Case 10 corresponds to the situation where equality happens between the end of inflation and the decay of the inflaton, that is $\Gamma_\phi<H_\ueq<H_\uend$. This leads to the condition
\beq
\sqrt{6}\frac{\Gamma_\phi}{m_\sigma}\ll \frac{\sigma_\uend}{\Mp}\ll \sqrt{6}\frac{H_\uend}{m_\sigma}\, .
\eeq
\section{Predictions for a few models}
\label{sec:predictions}
In this appendix, we plot the predictions of three single-field inflationary potentials when an extra massive scalar field $\sigma$ is added: Large Field inflation for which $V=M^4\left(\phi/\Mp\right)^p$, Higgs inflation - the Starobinsky model - for which $V=M^4[1-\exp(-\sqrt{2/3}\phi/\Mp)]^2$, and Natural Inflation for which $V=M^4[1+\cos(\phi/f)]$. Large Field inflation is a prototypical example of single-field model that predicts a value for $r$ that is too large, Natural Inflation is a prototypical example that predicts a value for $\nS$ that is too small, and Higgs inflation is a prototypical example that predicts the right value for $\nS$ and a value for $r$ that is sufficiently small. 

The value of $M$ is set so that \Eq{eq:powerspectrum:rho} fits the measured amplitude of the scalar power spectrum. For each potential, we investigate the $10$ reheating scenarios and calculate $\nS$, $r$ and $\fnl$. The conventions used in the following series of figures are the following. 
In the upper left corner of each plot, the name of the model is written, where ``MC'' (for ``Massive Curvaton''), a number (referring to the reheating scenario under consideration) and the acronym of the single-field version of the model are appended. For example, ``MC7NI'' corresponds to the natural inflation potential, where a massive scalar field is added, in the 7${}^{\mathrm{th}}$ reheating scenario.
The predictions of the model are displayed as colored circles, where a color bar on the right indicates which parameter the color encodes.\footnote{Most of the time, the color encodes $\sigma_\uend/\Mp$. However, when $\sigma_\uend/\Mp$ is not single-valued in the three planes $(\nS,r)$, $(\nS,\fnl)$ and $(\fnl,r)$, another quantity, $\sigma_\uend^2\Gamma_\phi/(\Mp^2\Gamma_\sigma)$, reflecting the relative contribution of $\sigma$ to $\calP_\zeta$, is displayed.} For comparison purpose, black squares stand for the predictions of the single-field version of the model, when $\bar{w}_\ureh=0$. For a generic reheating scenario (that is to say, when $-1<\bar{w}_\ureh<1/3$), those predictions would span a much larger region in the observable space. Finally, the black solid lines stand for the marginalized joint $68$\% and $95$\% confidence level regions of the Planck 2015 data~\cite{Ade:2015lrj} combined with BICEP2 and Keck Array. When $\fnl$ is displayed, we simply show the $95$\% bounds for the parameters on the axes.

In passing, let us notice that when going from one reheating scenario the other, the predictions evolve continuously with the models parameters. However, because all four parameters $m_\sigma$, $\Gamma_\phi$, $\Gamma_\sigma$ and $\sigma_\uend$ are varied simultaneously, this is not always obvious from only looking at how the color changes at the border between two adjacent reheating scenarios. Besides, in principle, our calculation is valid under the approximation that all events are well separated in time. It should therefore not be trusted close to the boundary between two reheating scenarios. In a Bayesian perspective in any case, this only represents a tiny region in parameter space and should play a negligible role in the final results.

\clearpage
\subsection{Large Field Inflation + Massive Scalar Field (MCLFI)}
\label{sec:plots:lfi}
\begin{figure}[!ht]
\begin{center}
\includegraphics[width=\wappfig,clip=true]{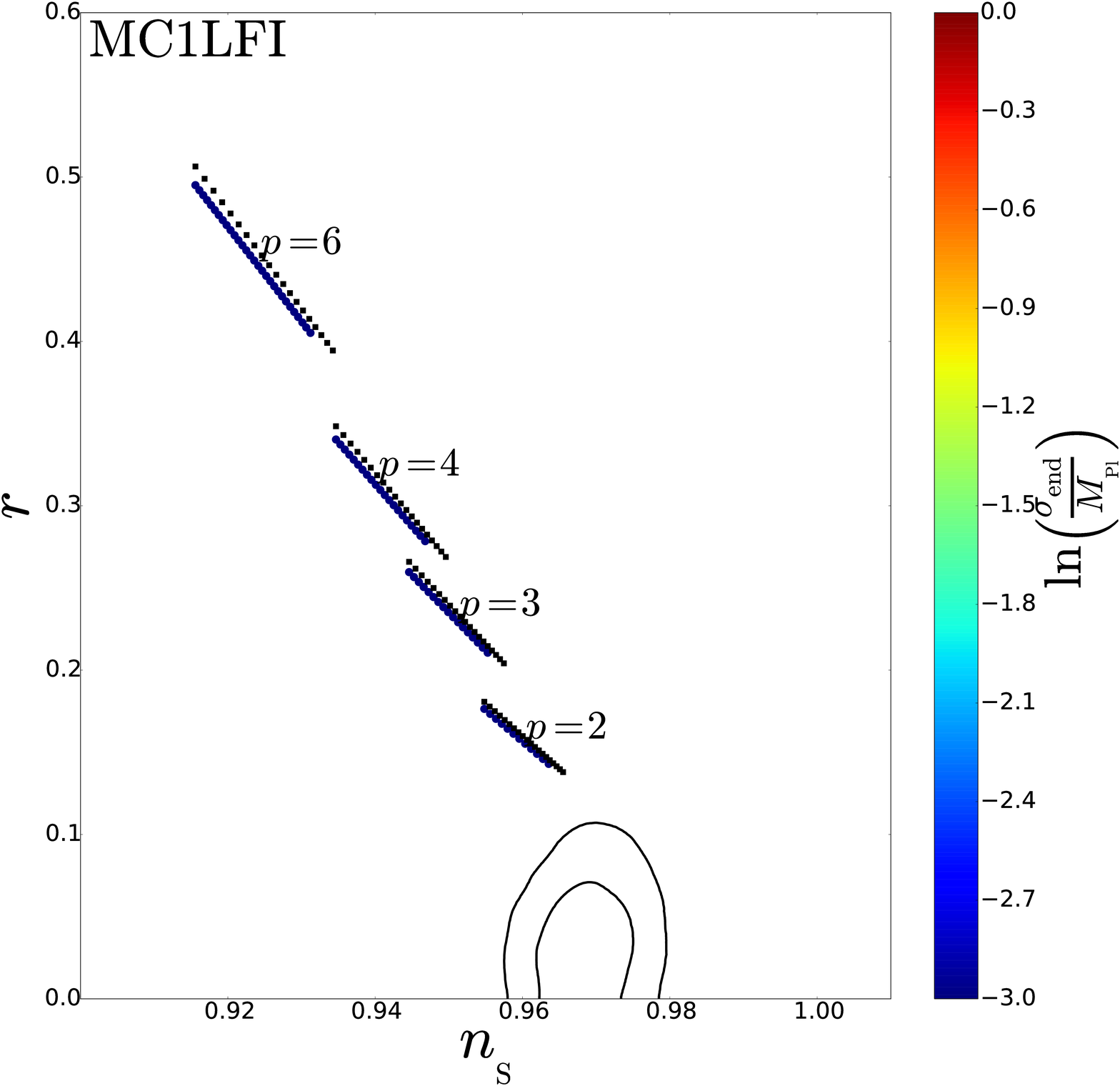}
\includegraphics[width=\wappfig,clip=true]{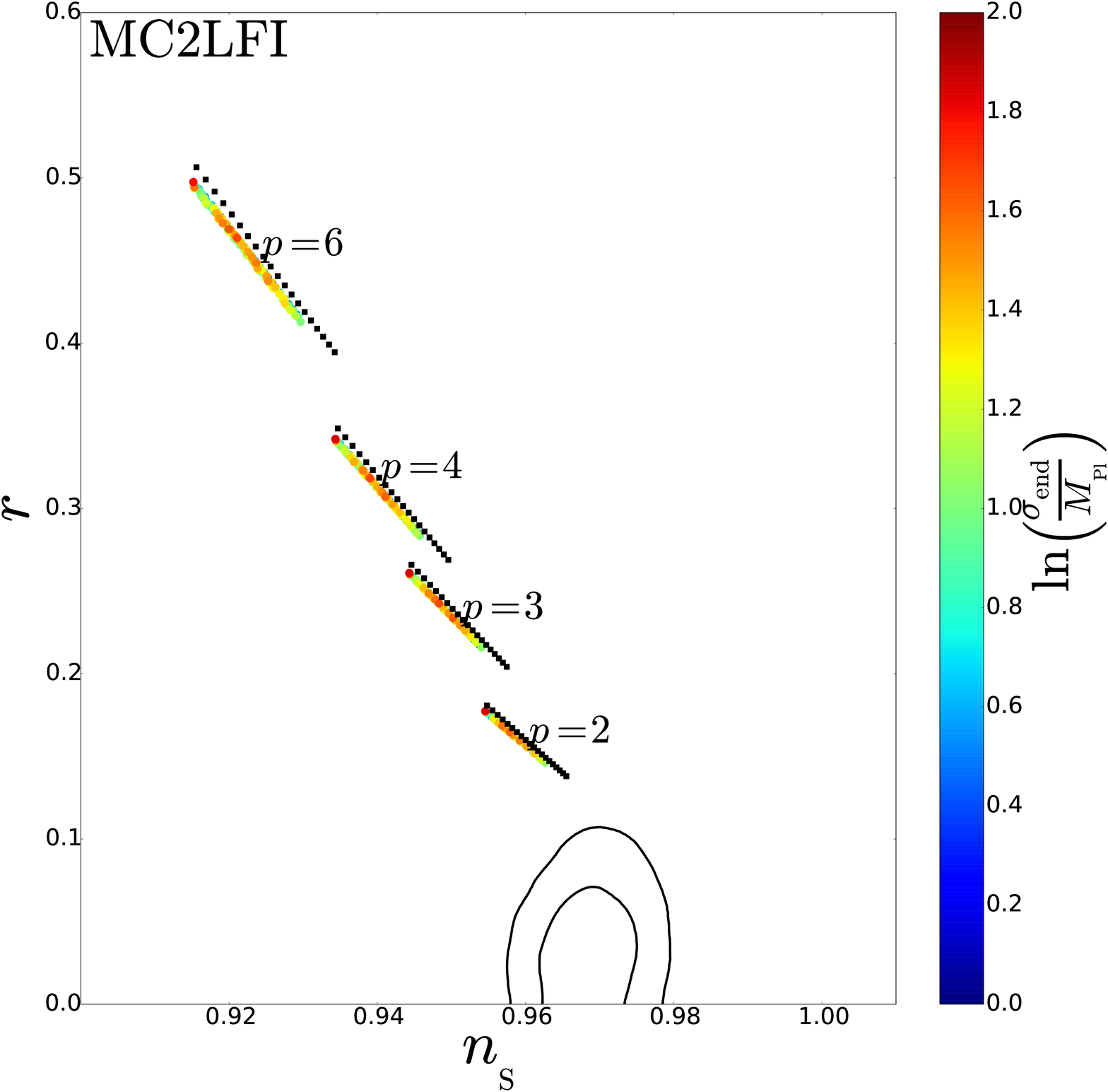}
\includegraphics[width=\wappfig,clip=true]{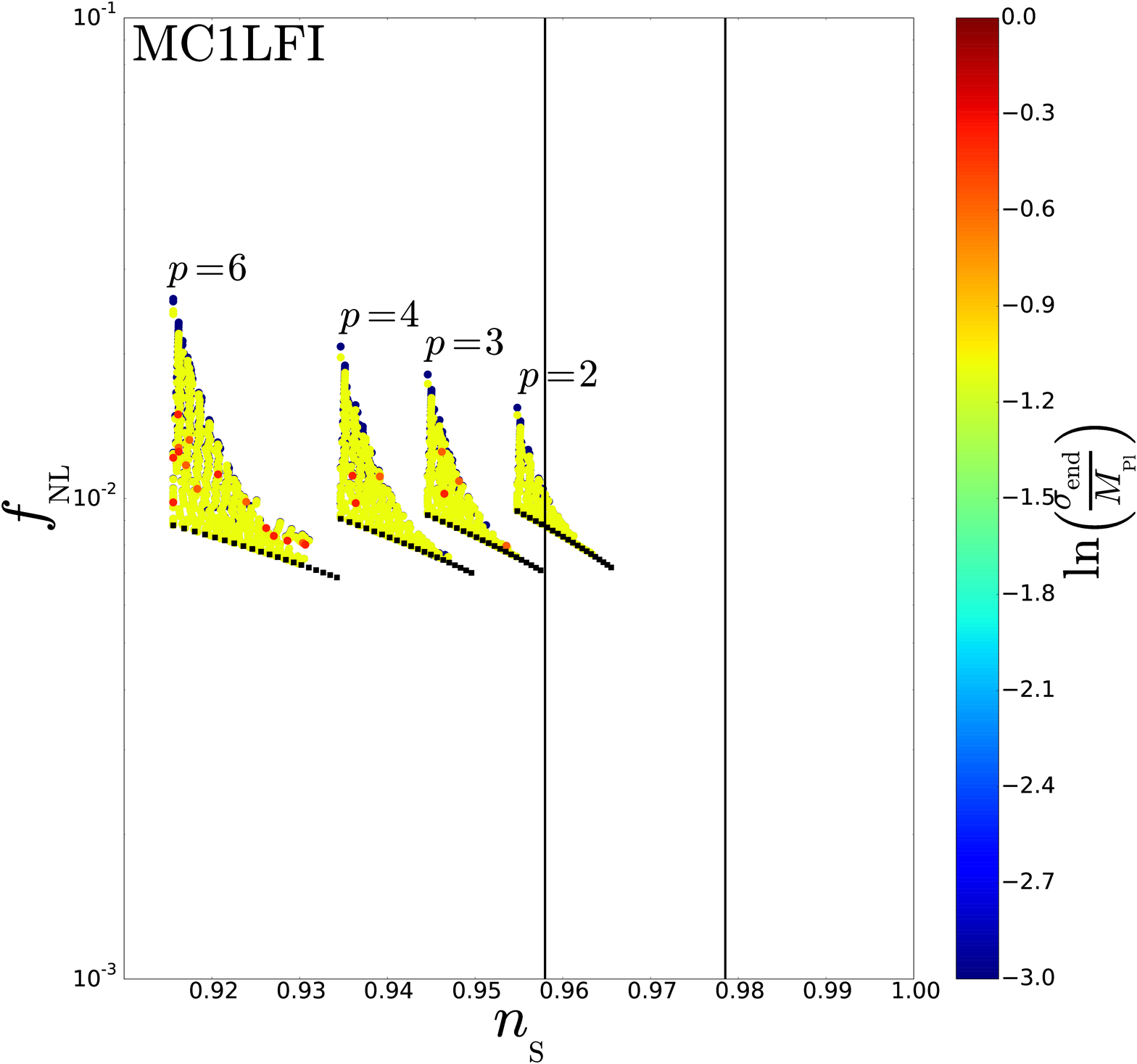}
\includegraphics[width=\wappfig,clip=true]{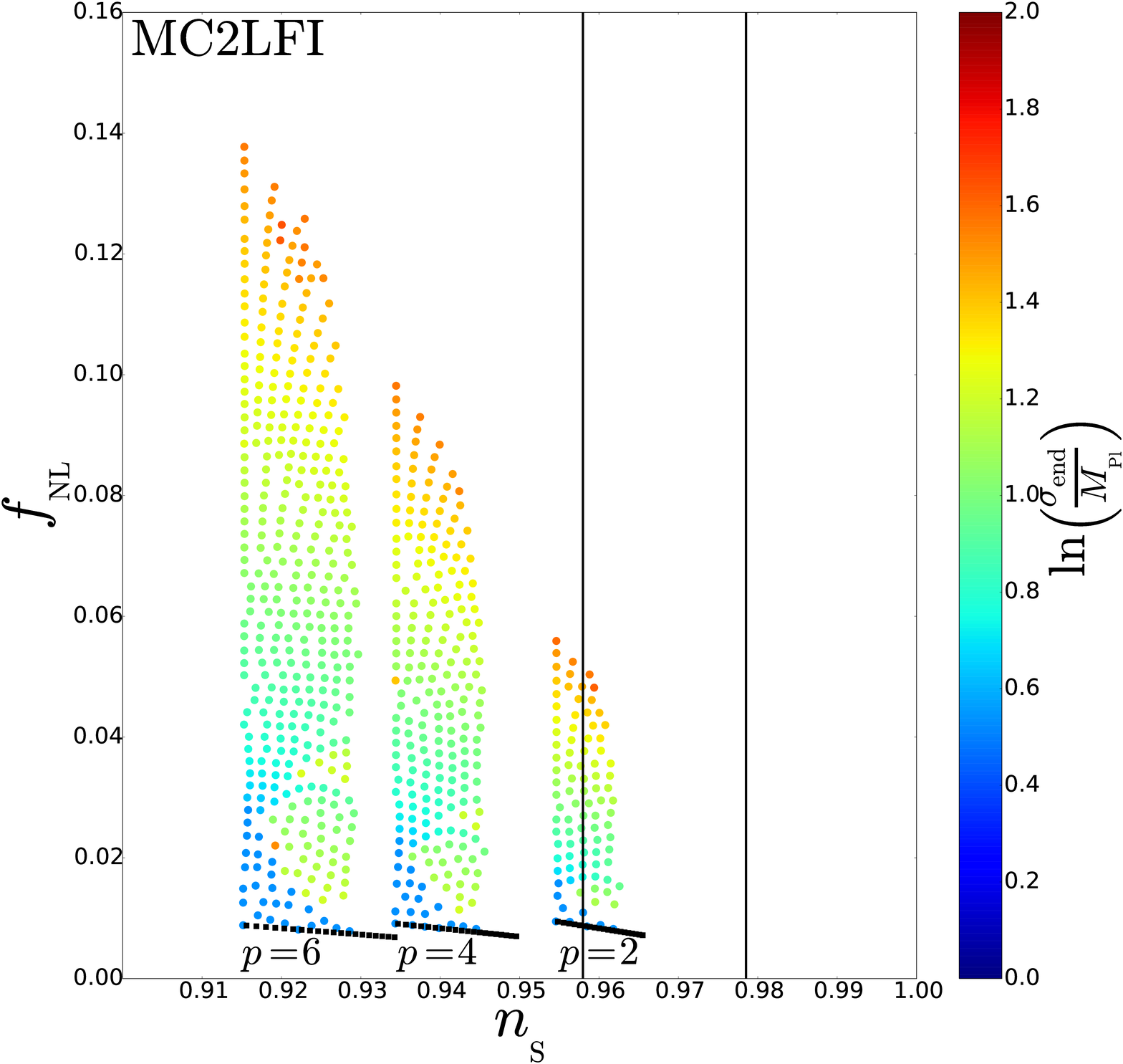}
\includegraphics[width=\wappfig,clip=true]{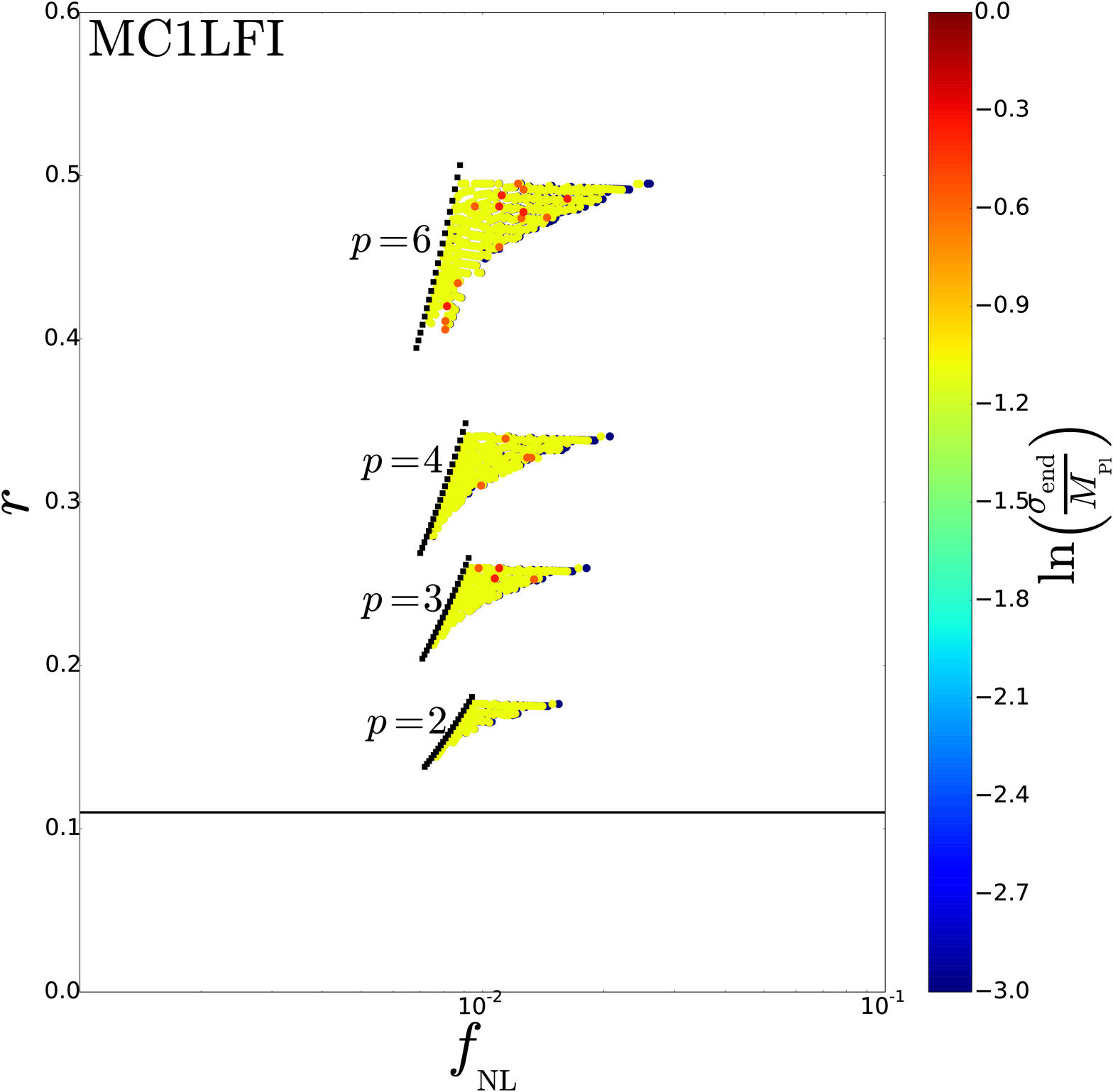}
\includegraphics[width=\wappfig,clip=true]{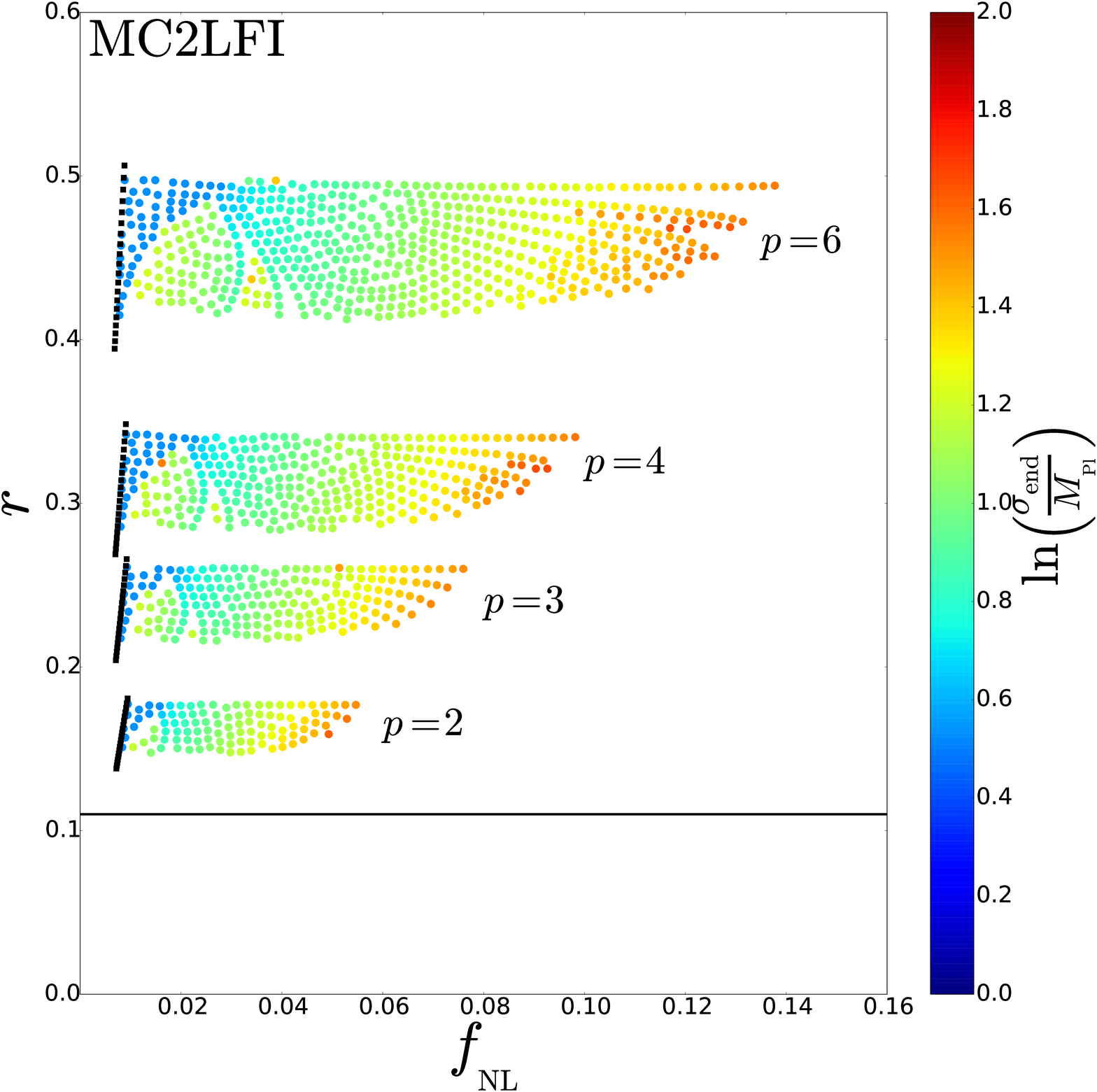}
\caption{Reheating consistent slow-roll predictions for the large field inflation models with a massive curvaton field, when reheating scenario is of the first (left panels) and second (right panels) type.}
\label{fig:CMBMCLFI12}
\end{center}
\end{figure}
\begin{figure}[!ht]
\begin{center}
\includegraphics[width=\wappfig,clip=true]{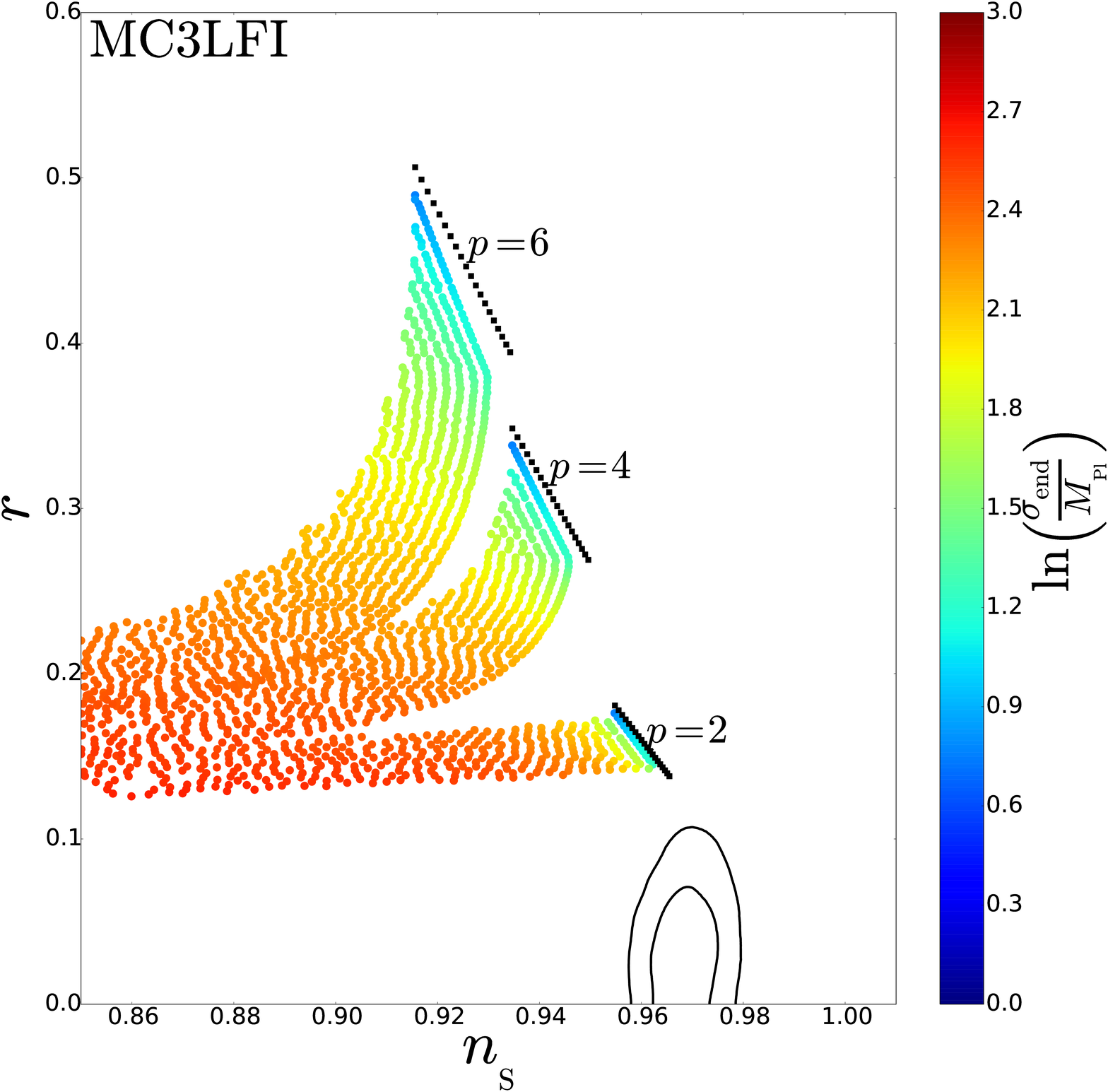}
\includegraphics[width=\wappfig,clip=true]{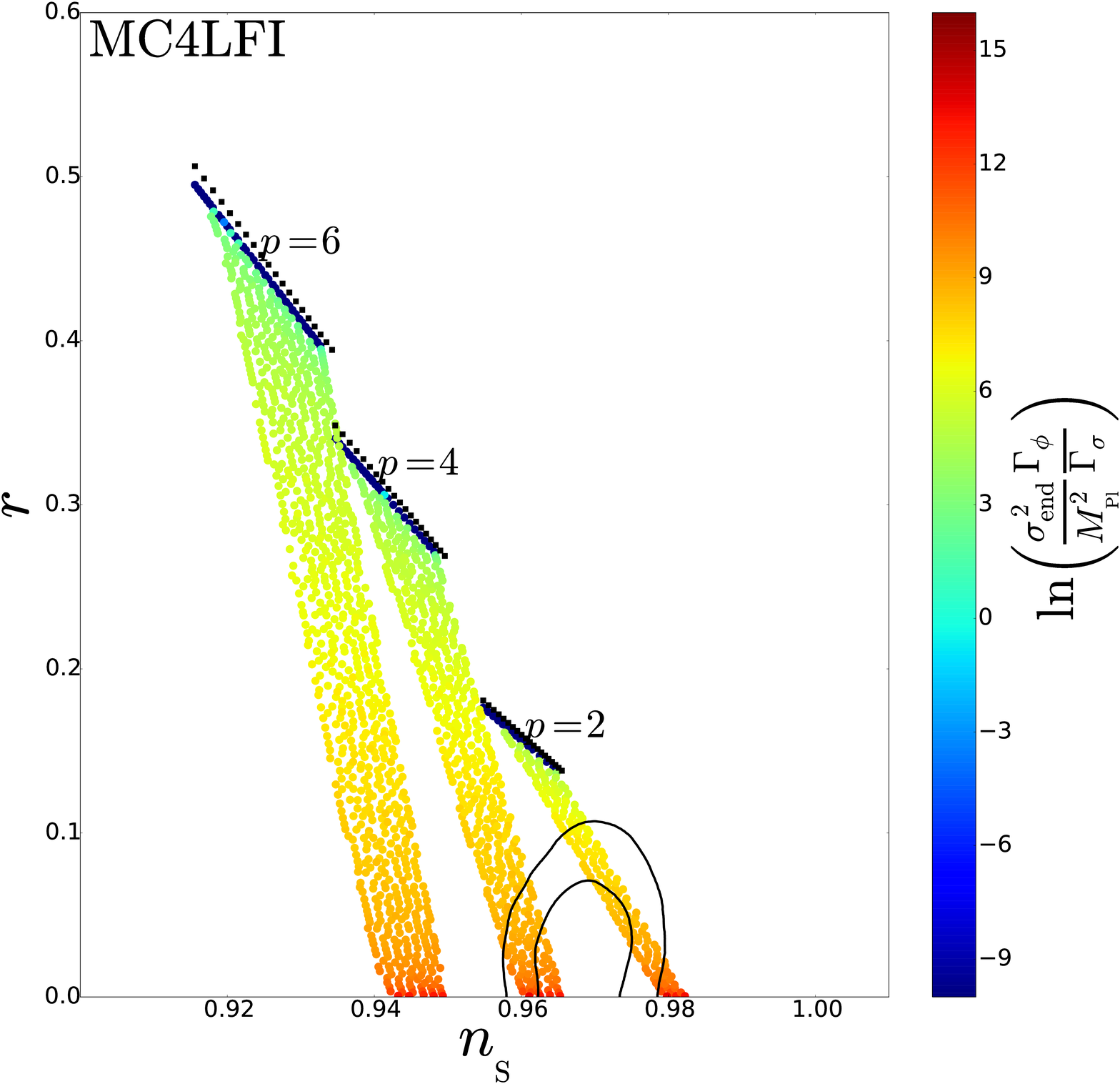}
\includegraphics[width=\wappfig,clip=true]{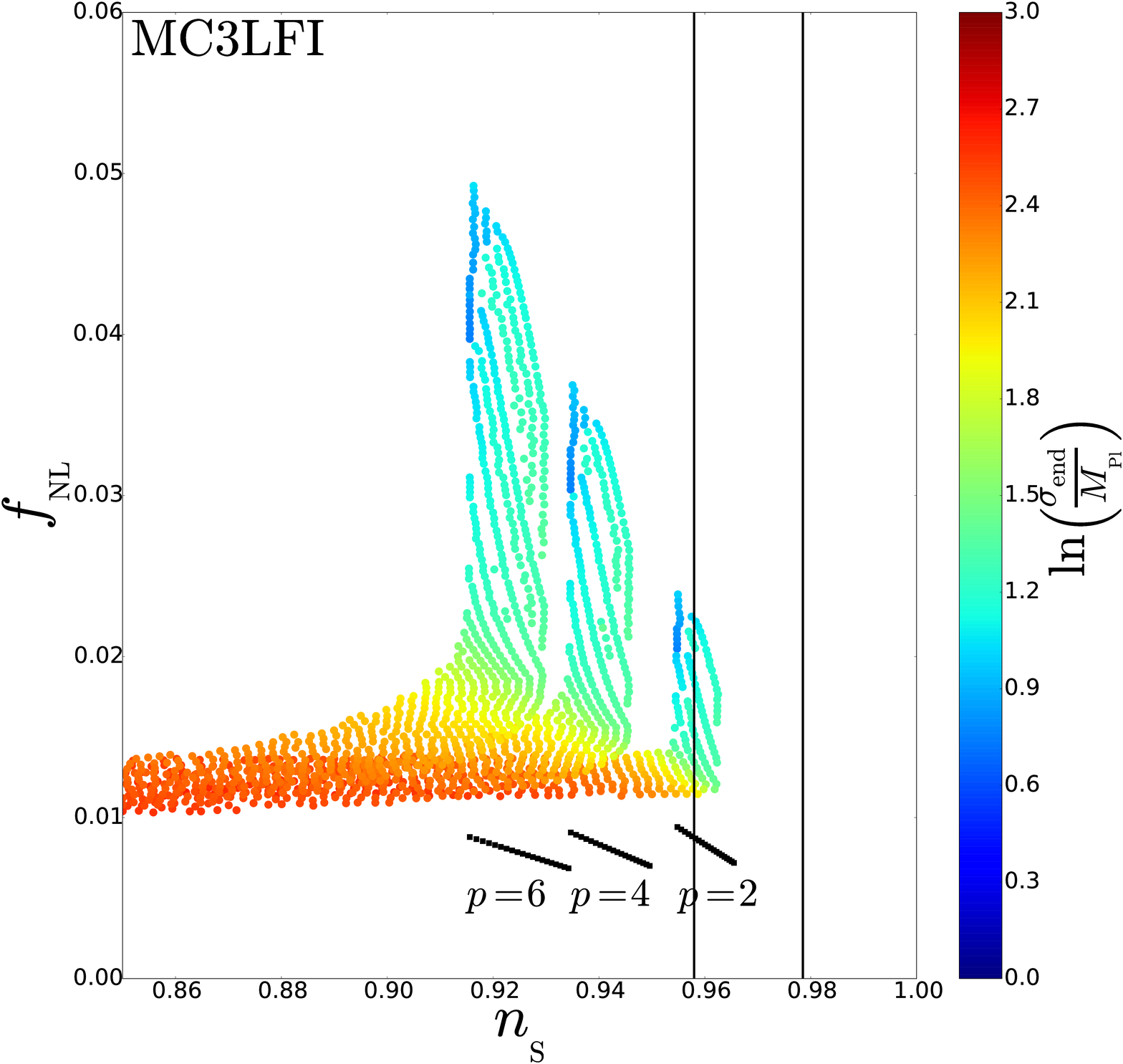}
\includegraphics[width=\wappfig,clip=true]{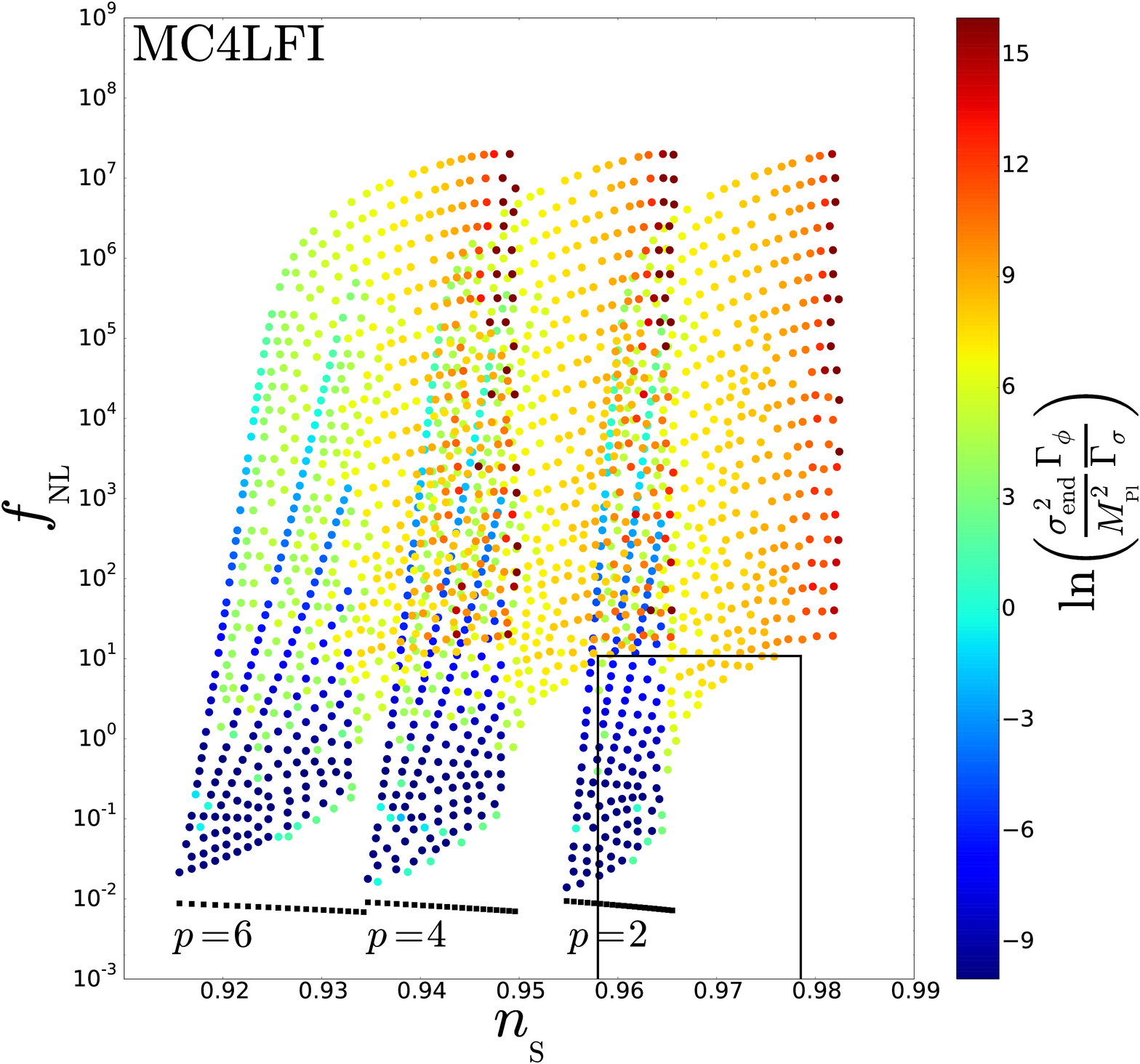}
\includegraphics[width=\wappfig,clip=true]{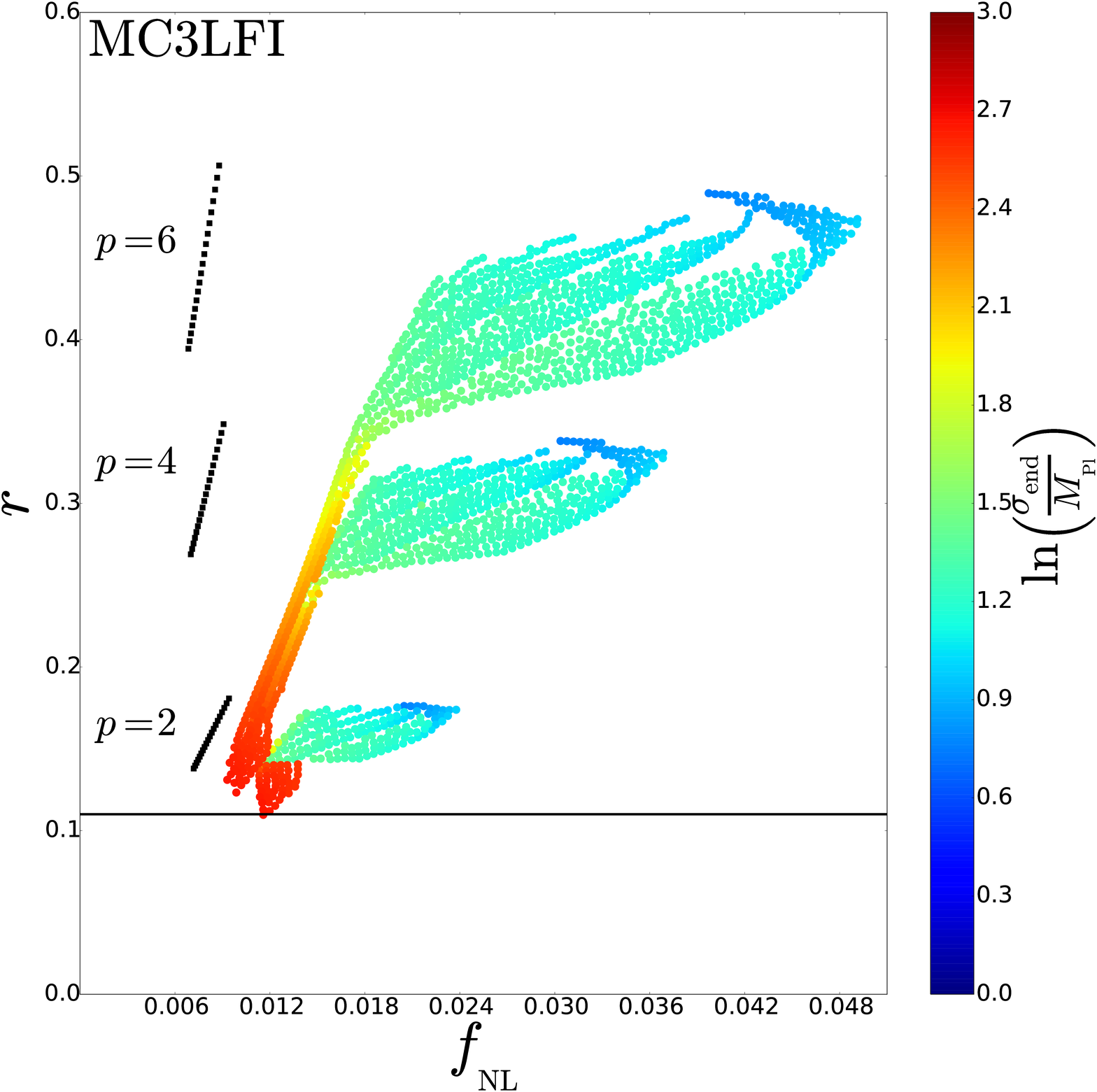}
\includegraphics[width=\wappfig,clip=true]{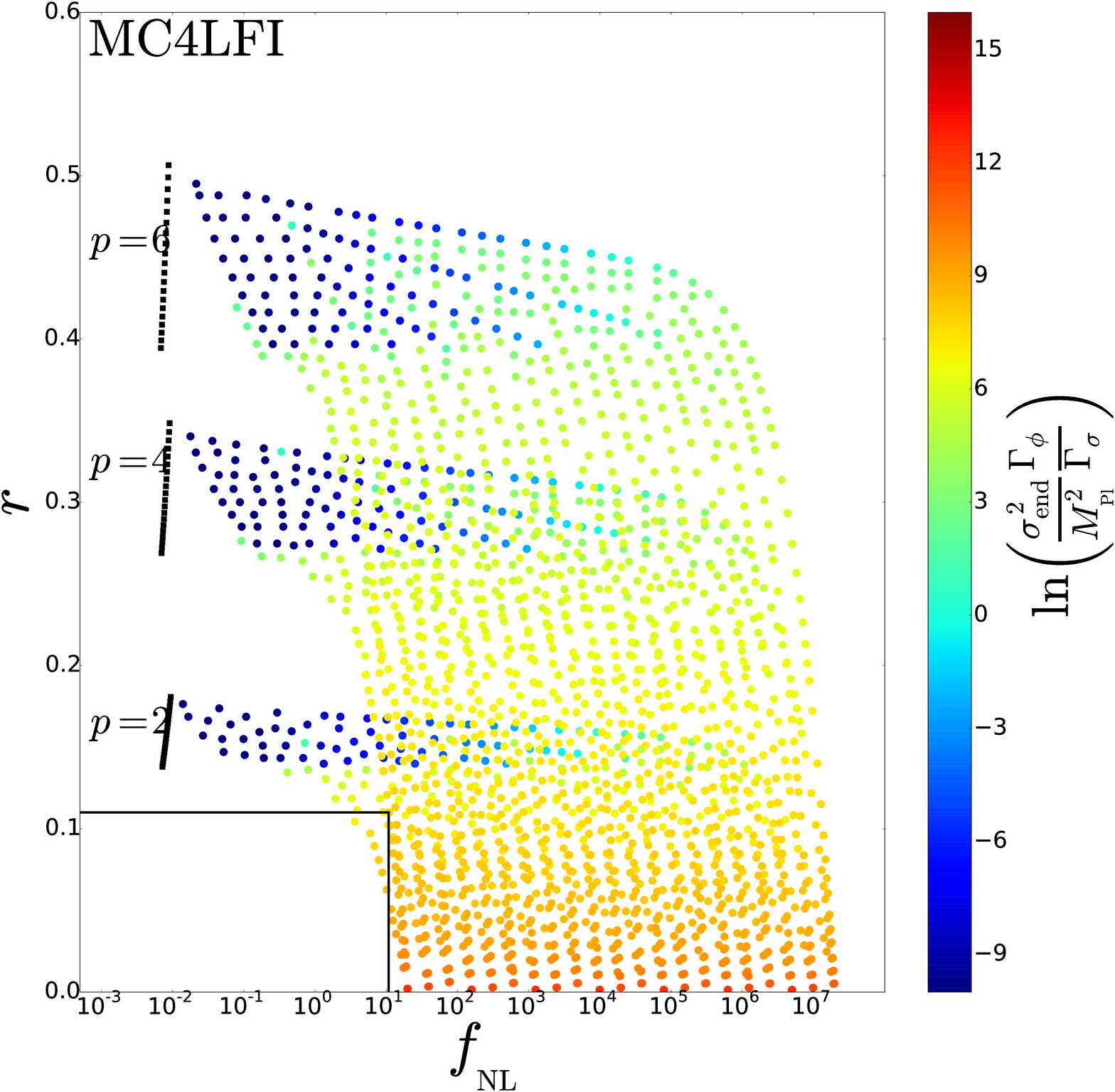}
\caption{Reheating consistent slow-roll predictions for the large field inflation models with a massive curvaton field, when reheating scenario is of the third (left panels) and fourth (right panels) type.}
\label{fig:CMBMCLFI34}
\end{center}
\end{figure}
\begin{figure}[!ht]
\begin{center}
\includegraphics[width=\wappfig,clip=true]{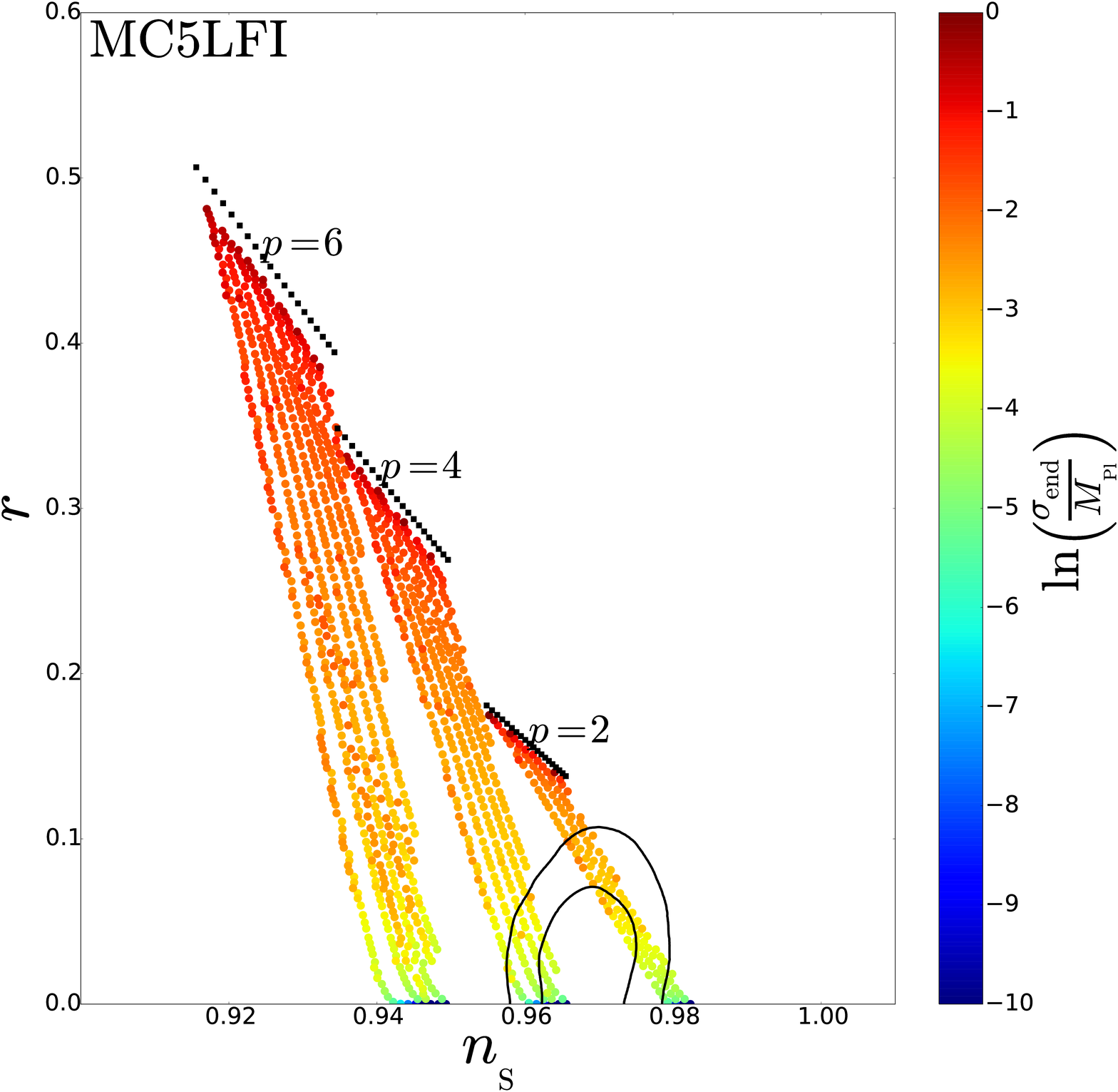}
\includegraphics[width=\wappfig,clip=true]{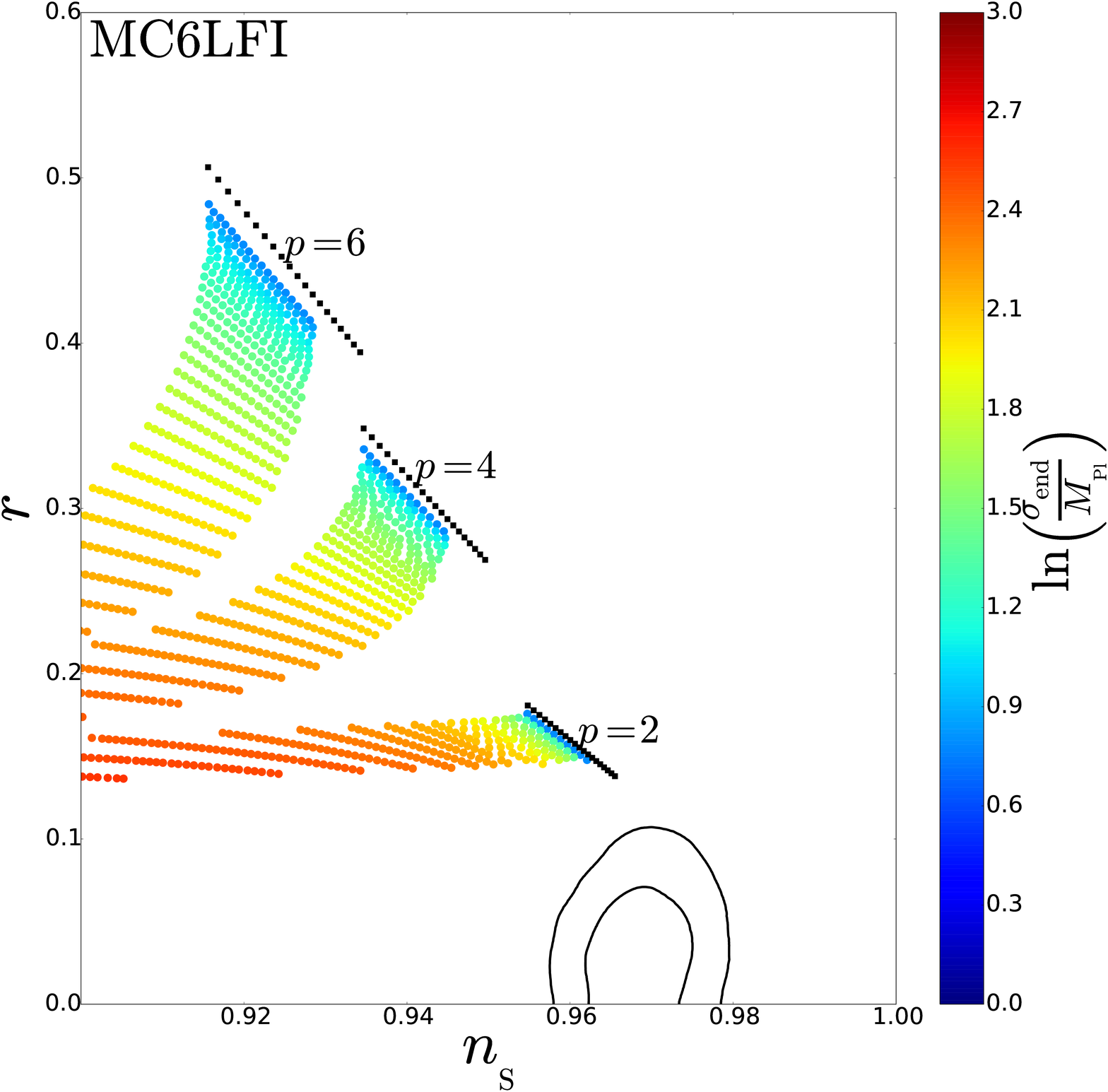}
\includegraphics[width=\wappfig,clip=true]{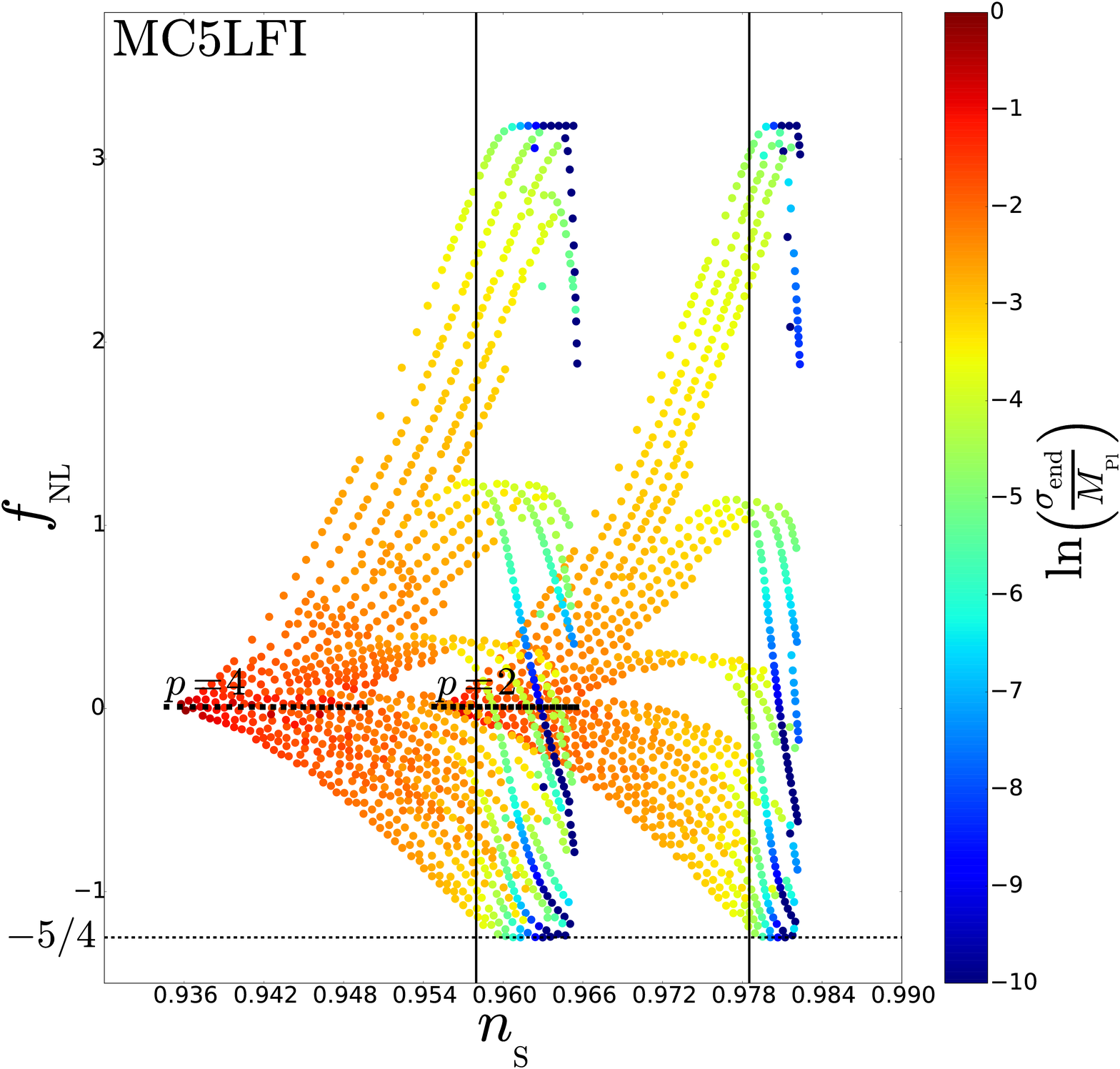}
\includegraphics[width=\wappfig,clip=true]{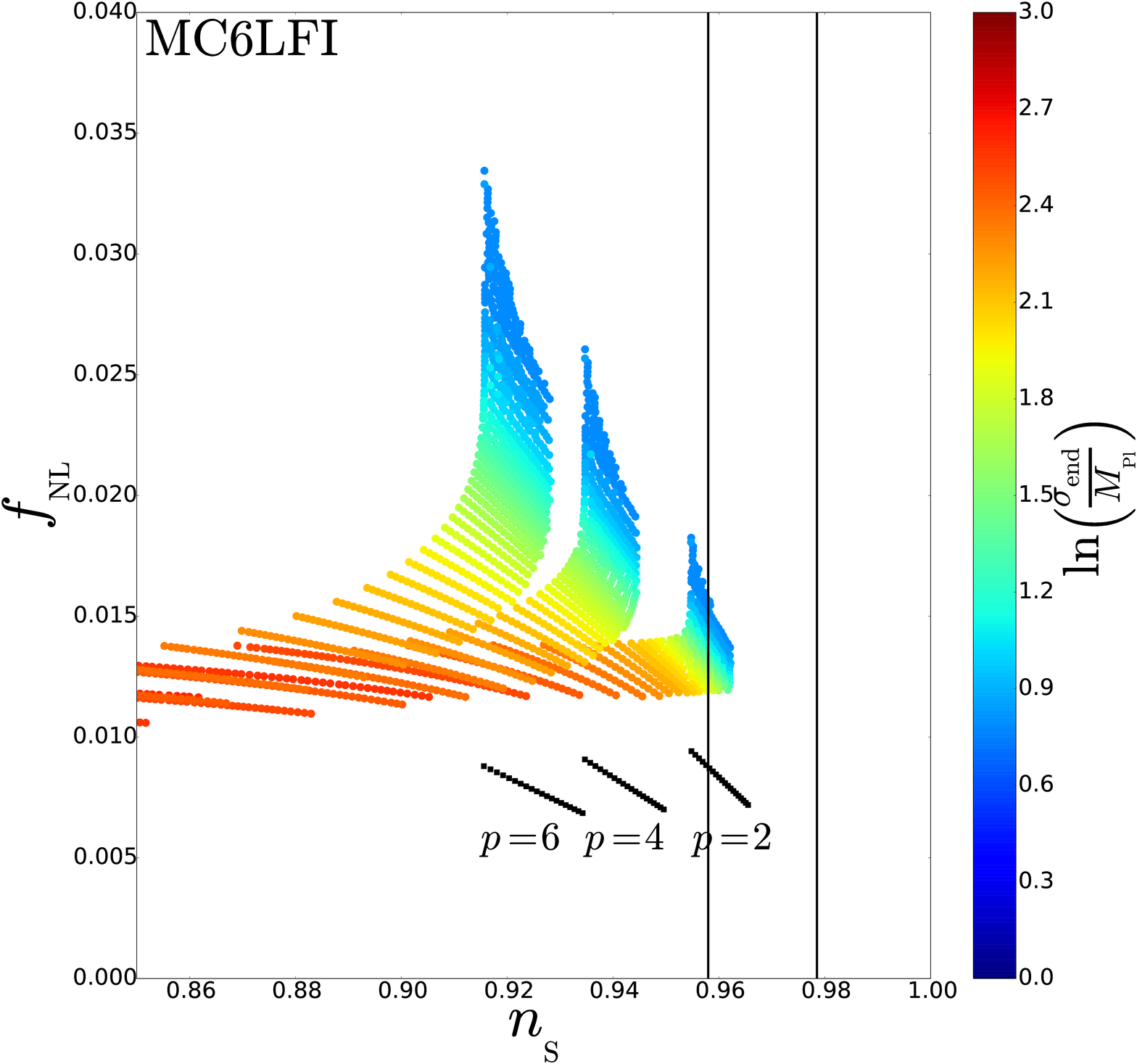}
\includegraphics[width=\wappfig,clip=true]{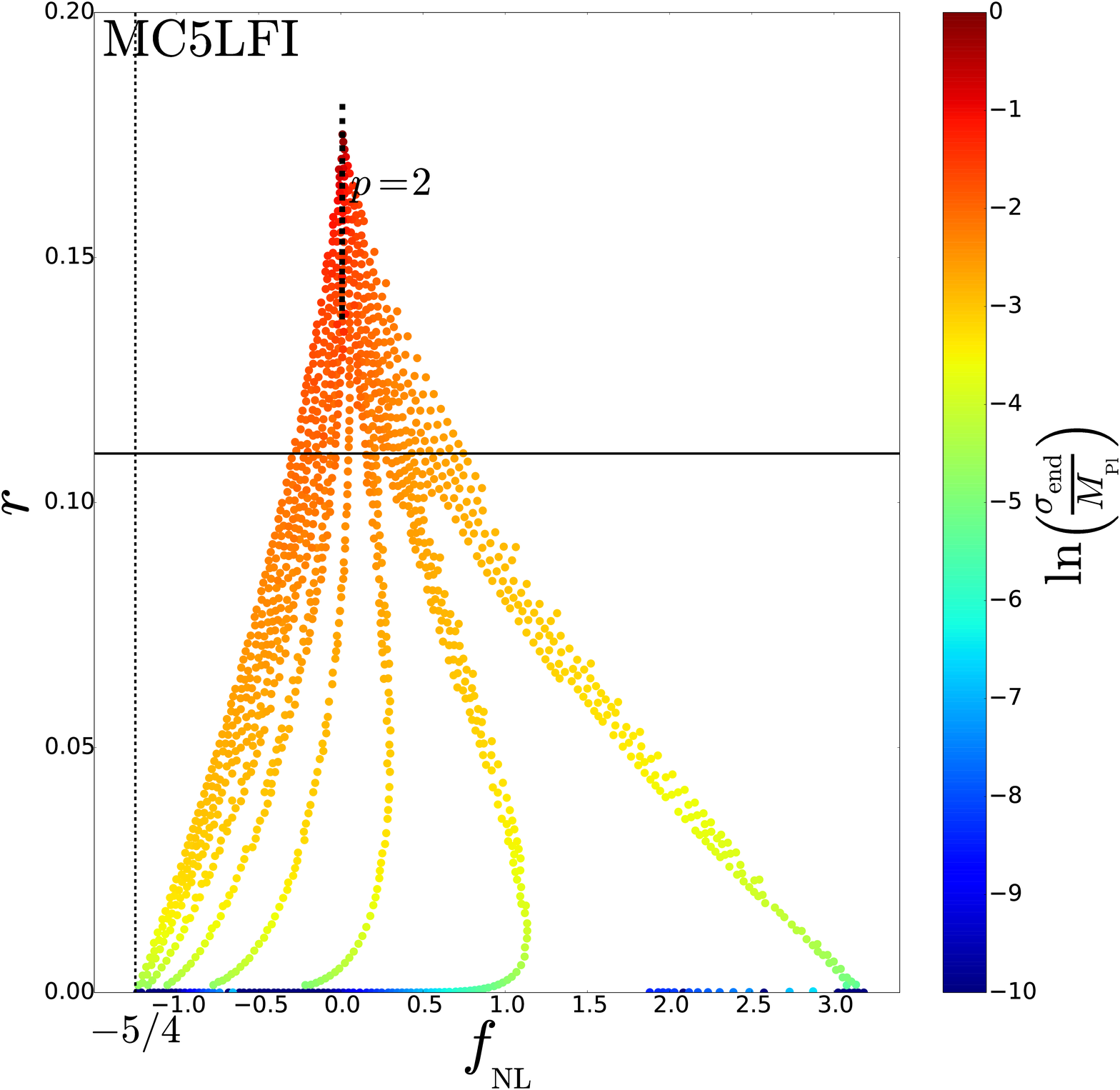}
\includegraphics[width=\wappfig,clip=true]{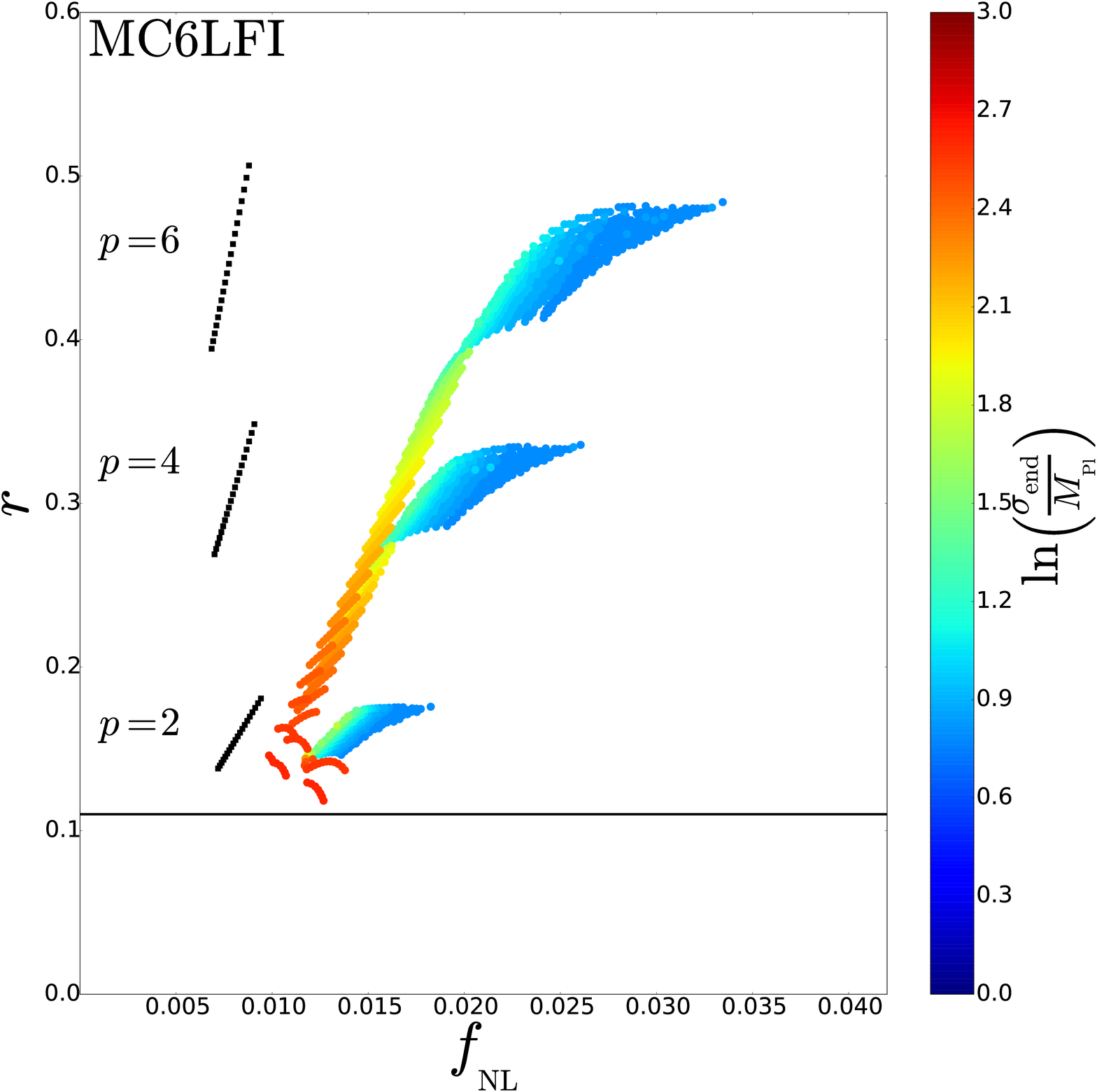}
\caption{Reheating consistent slow-roll predictions for the large field inflation models with a massive curvaton field, when reheating scenario is of the fifth (left panels) and sixth (right panels) type.}
\label{fig:CMBMCLFI56}
\end{center}
\end{figure}
\begin{figure}[!ht]
\begin{center}
\includegraphics[width=\wappfig,clip=true]{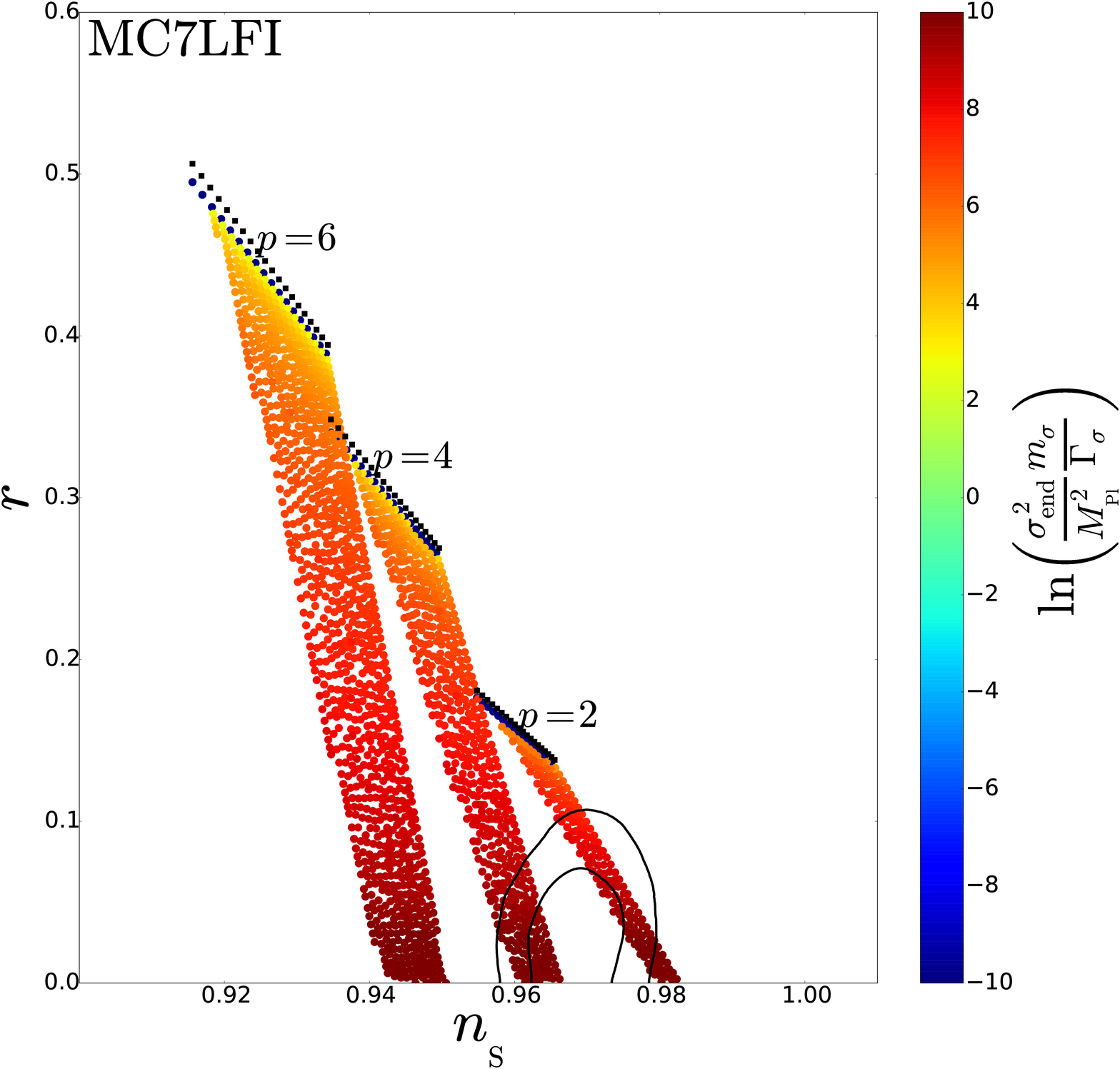}
\includegraphics[width=\wappfig,clip=true]{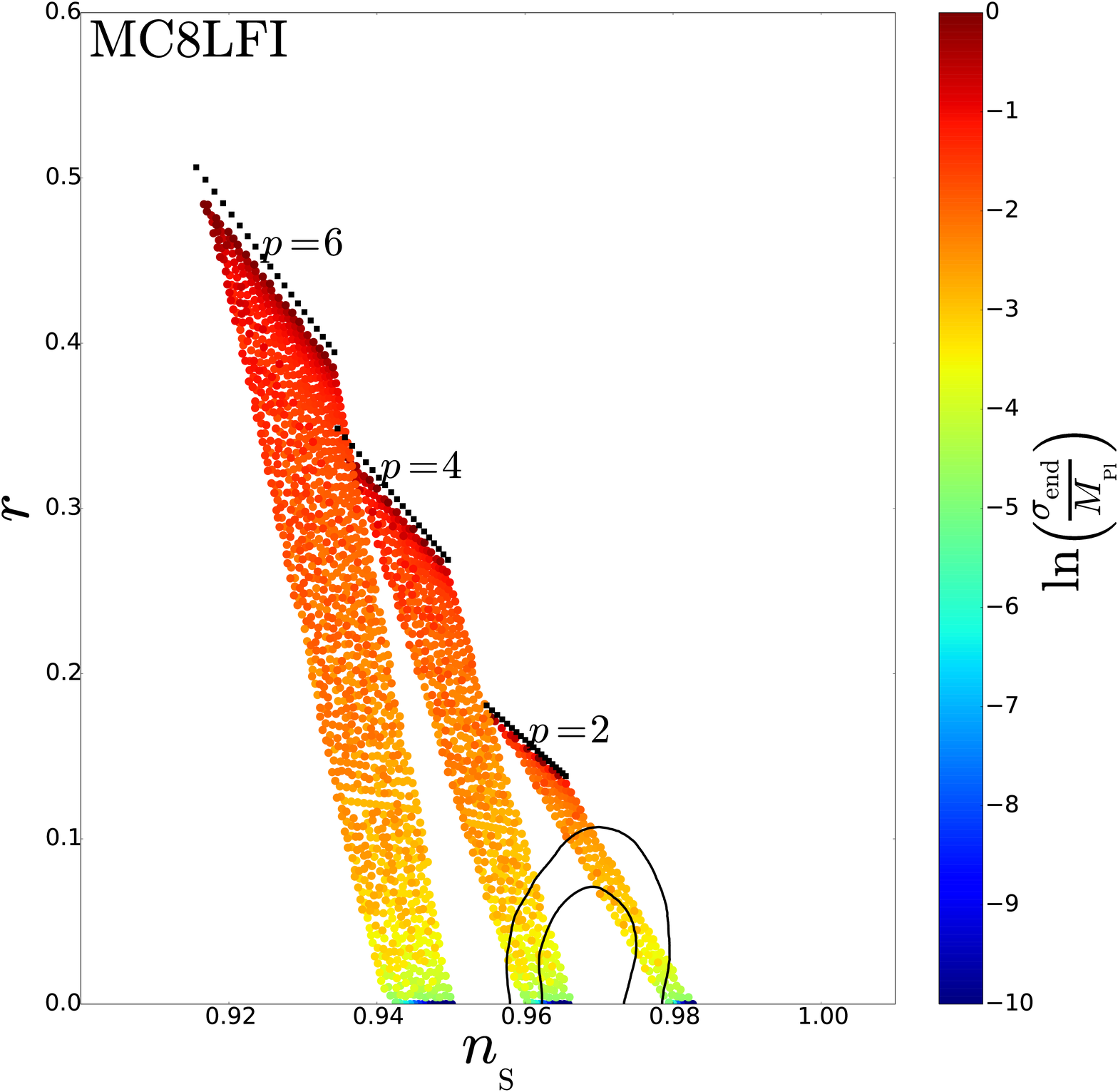}
\includegraphics[width=\wappfig,clip=true]{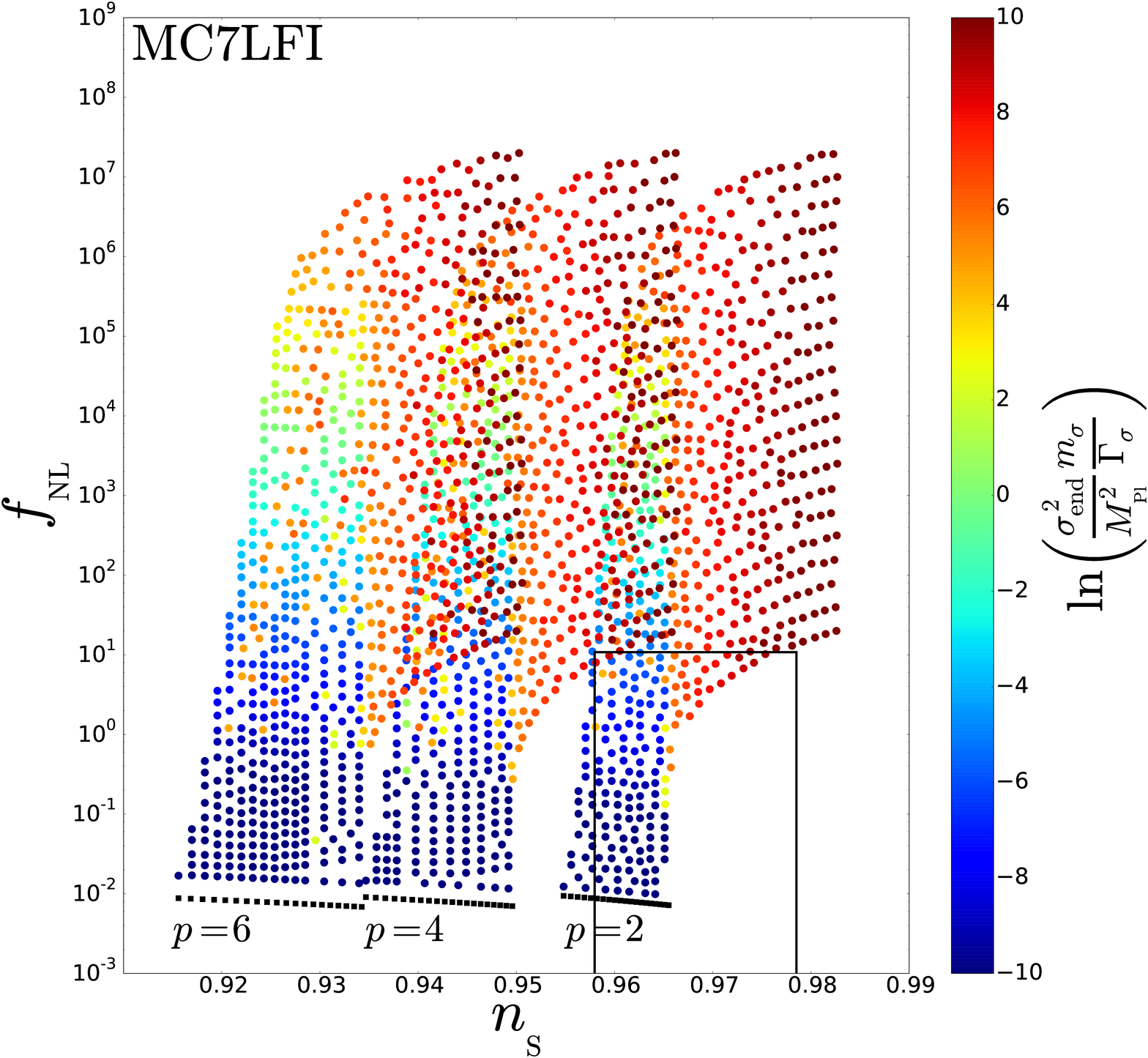}
\includegraphics[width=\wappfig,clip=true]{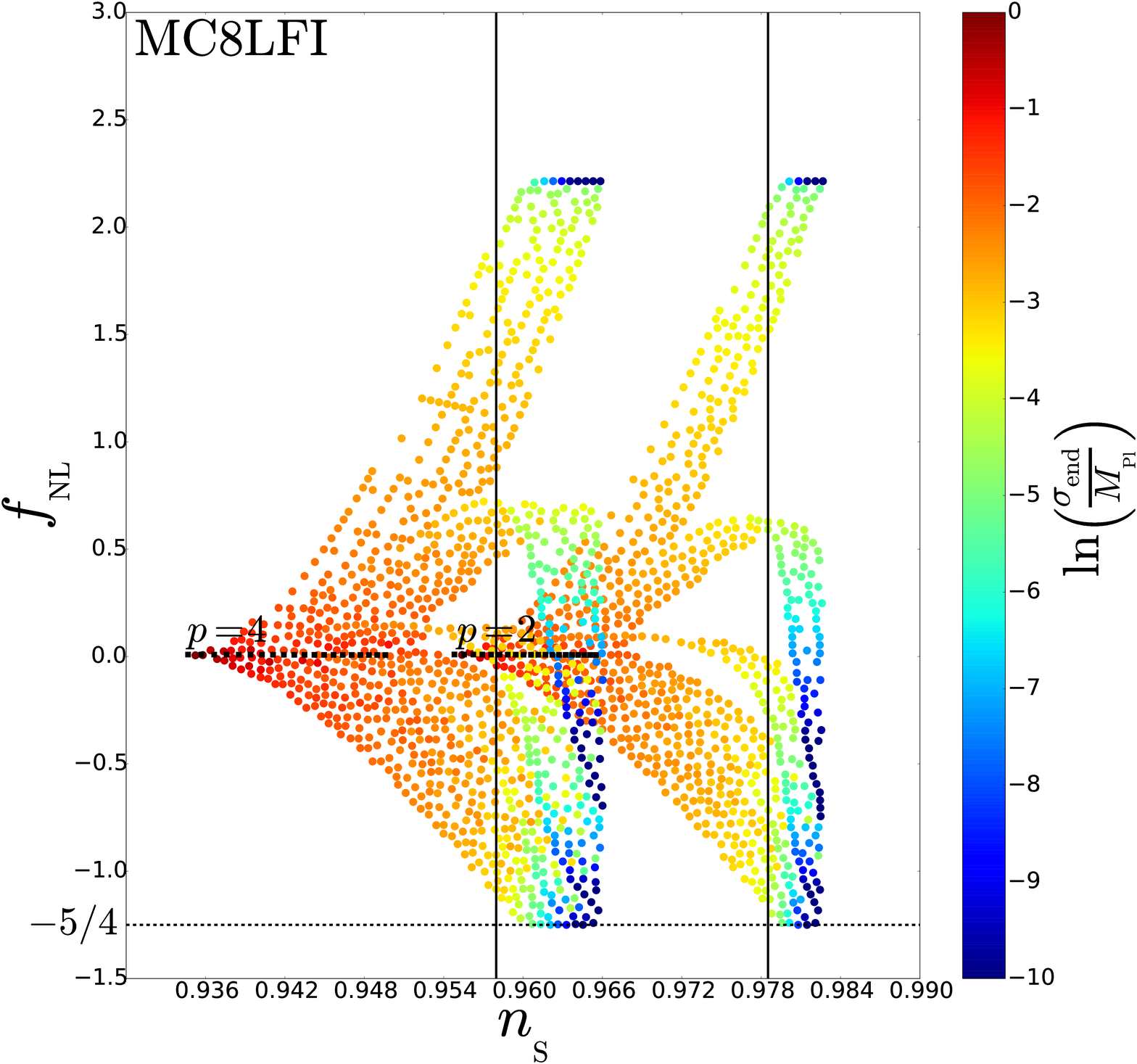}
\includegraphics[width=\wappfig,clip=true]{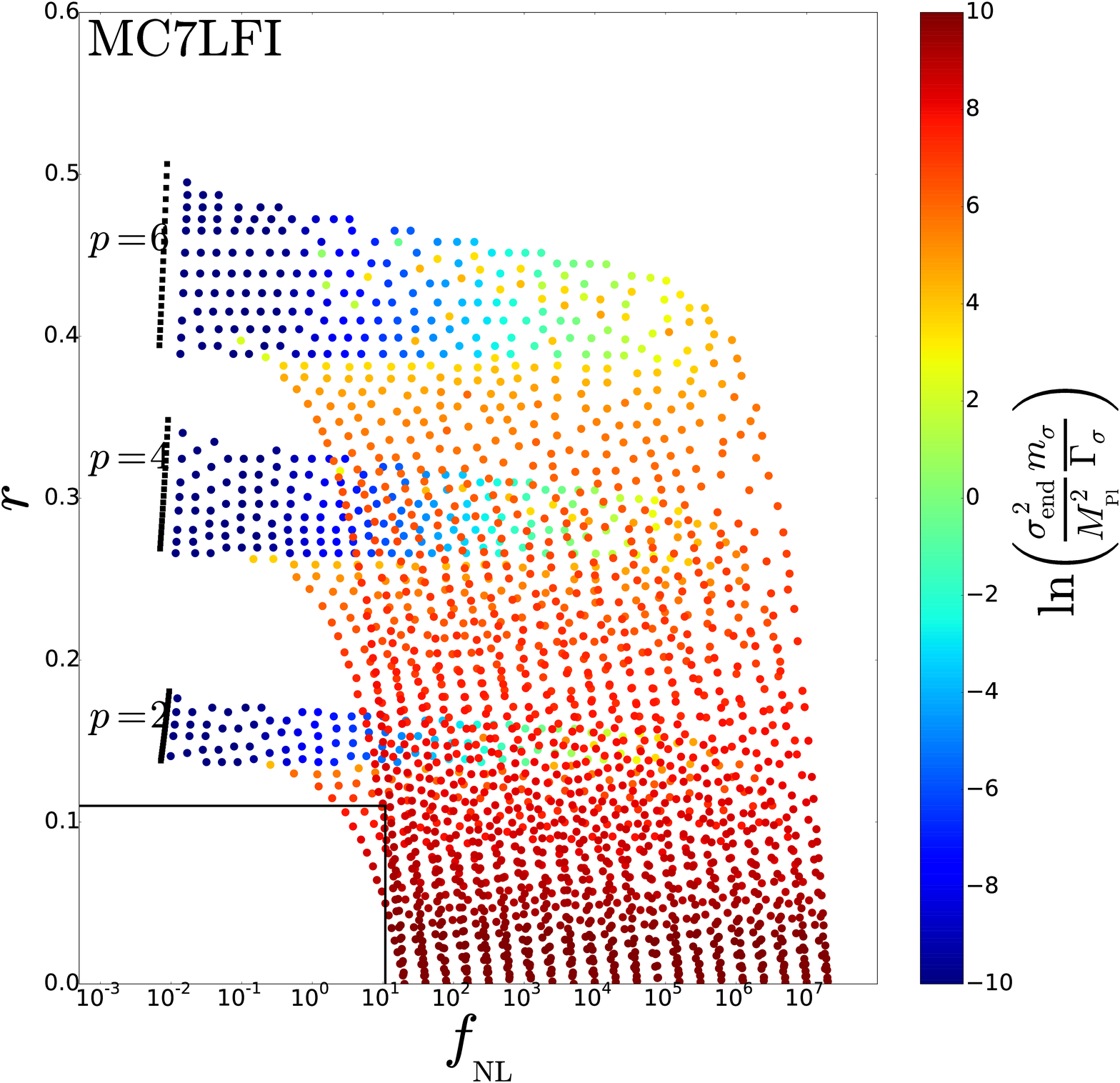}
\includegraphics[width=\wappfig,clip=true]{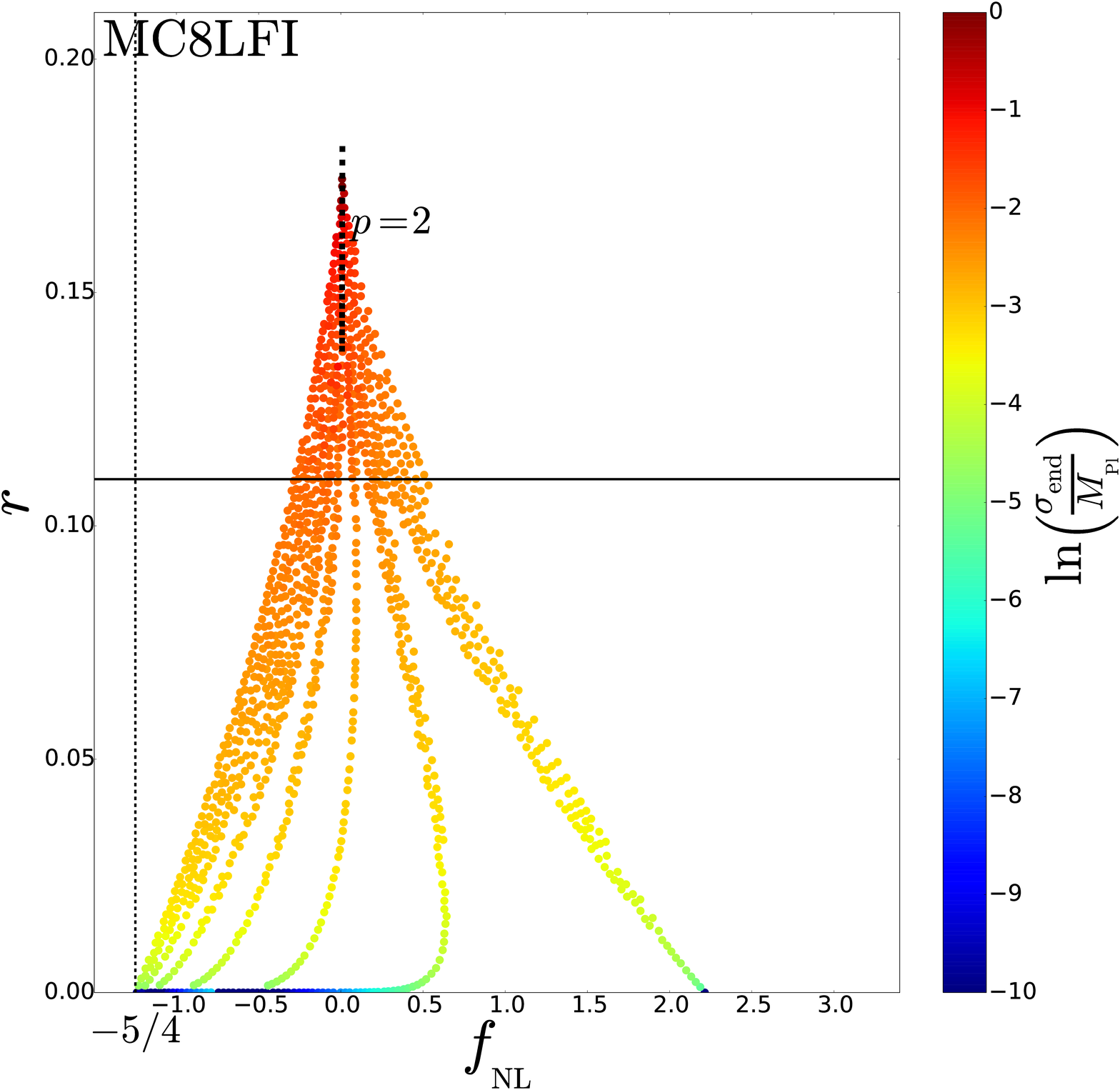}
\caption{Reheating consistent slow-roll predictions for the large field inflation models with a massive curvaton field, when reheating scenario is of the seventh (left panels) and eighth (right panels) type.}
\label{fig:CMBMCLFI78}
\end{center}
\end{figure}
\begin{figure}[!ht]
\begin{center}
\includegraphics[width=\wappfig,clip=true]{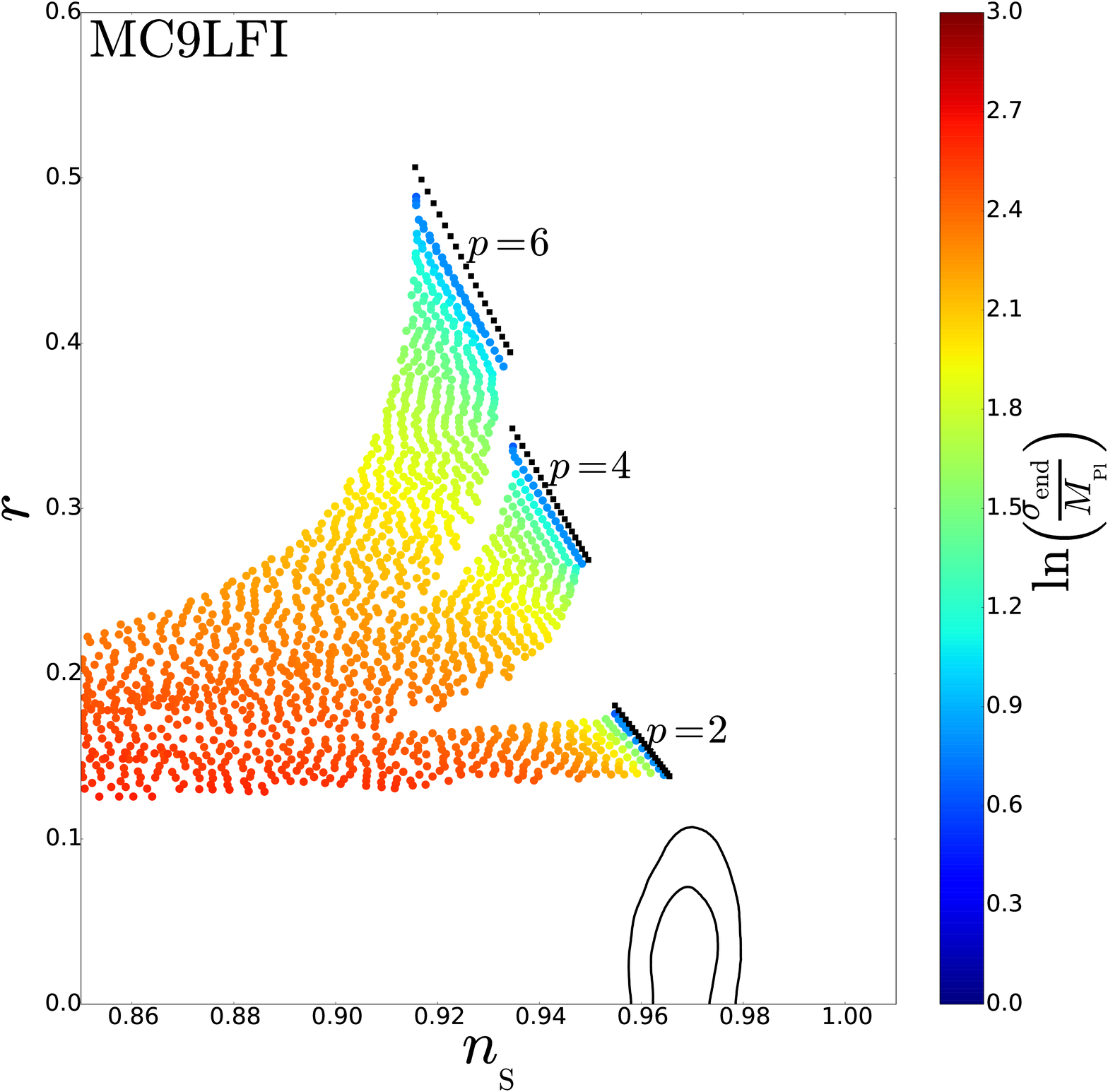}
\includegraphics[width=\wappfig,clip=true]{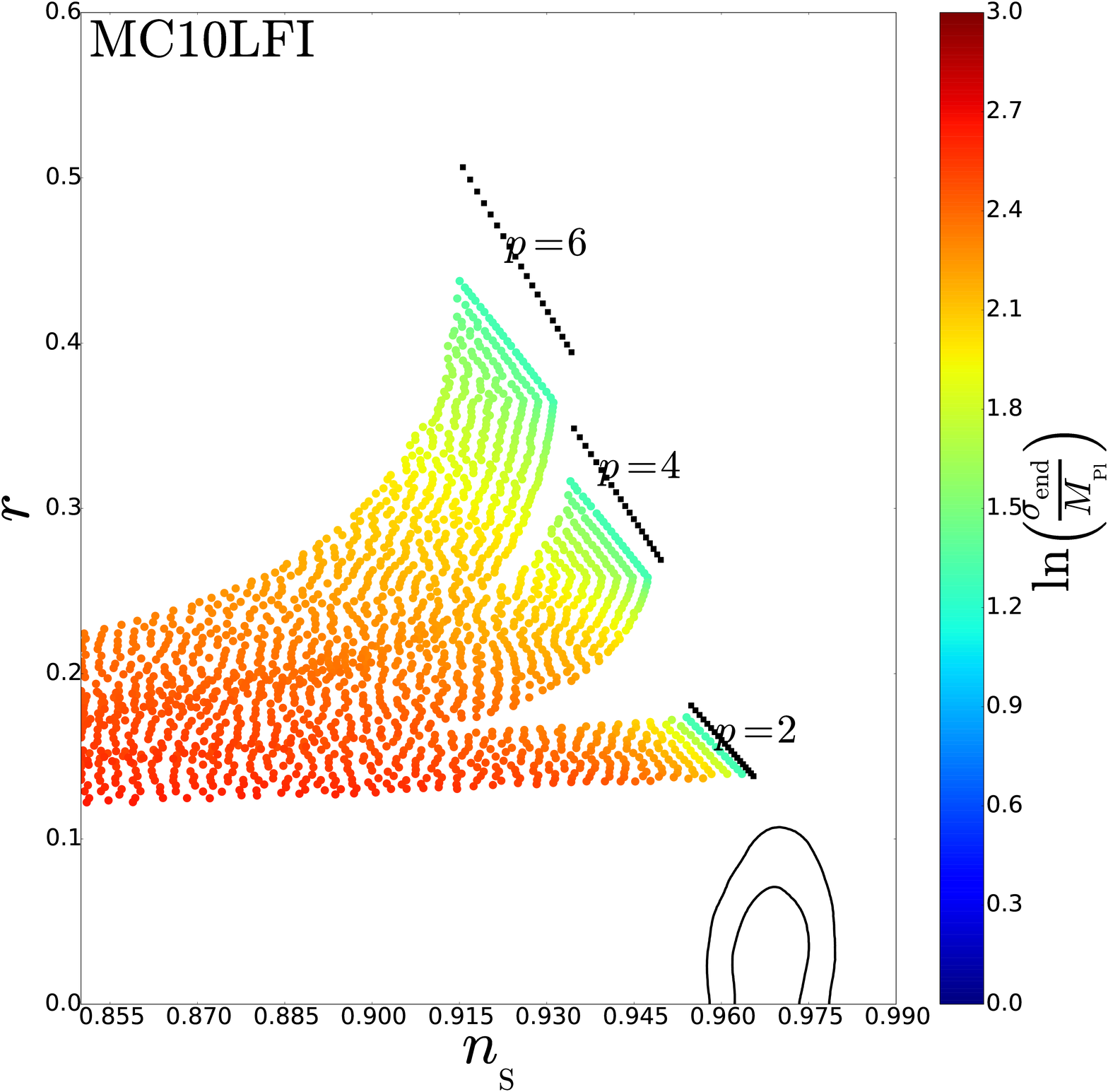}
\includegraphics[width=\wappfig,clip=true]{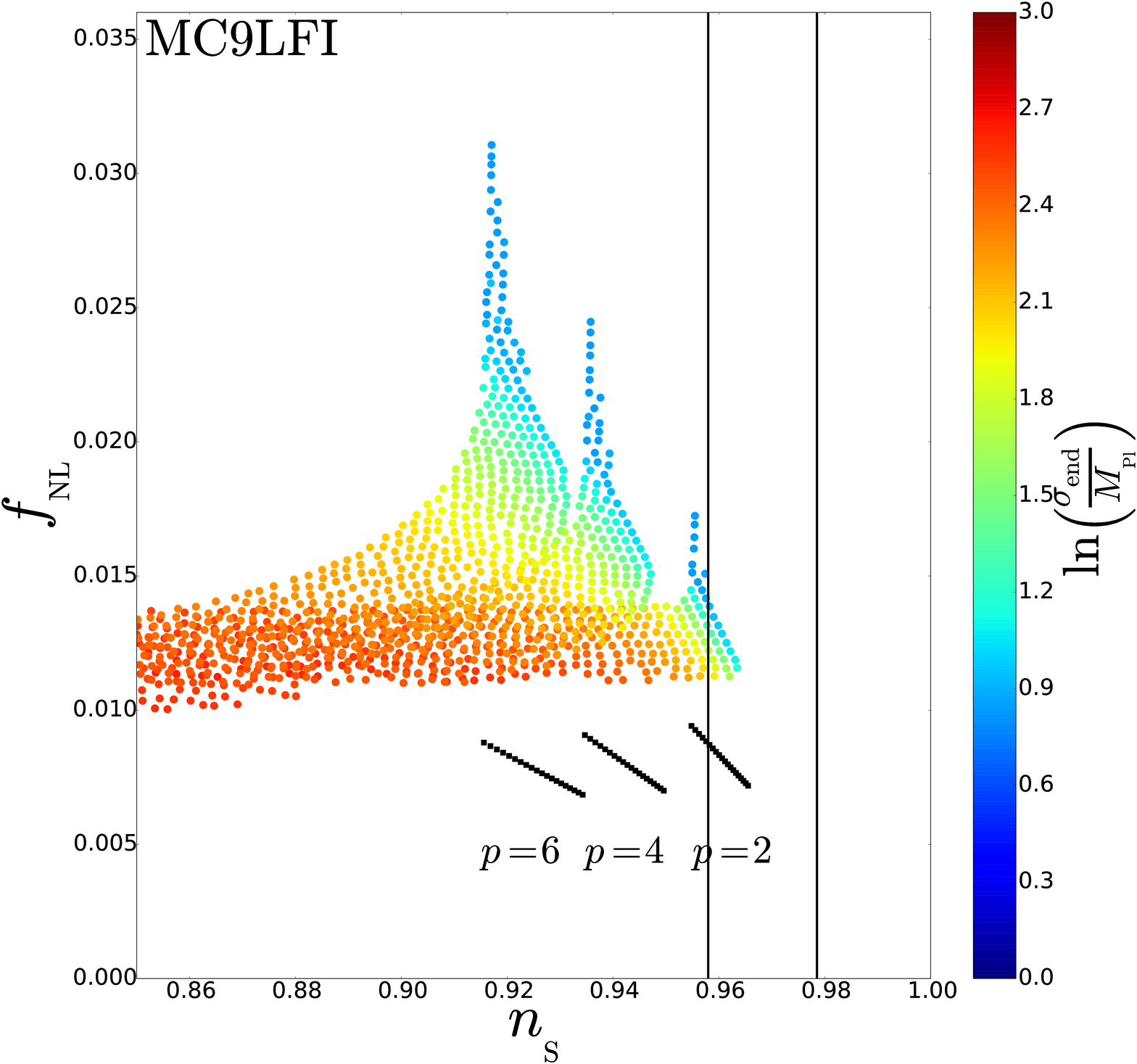}
\includegraphics[width=\wappfig,clip=true]{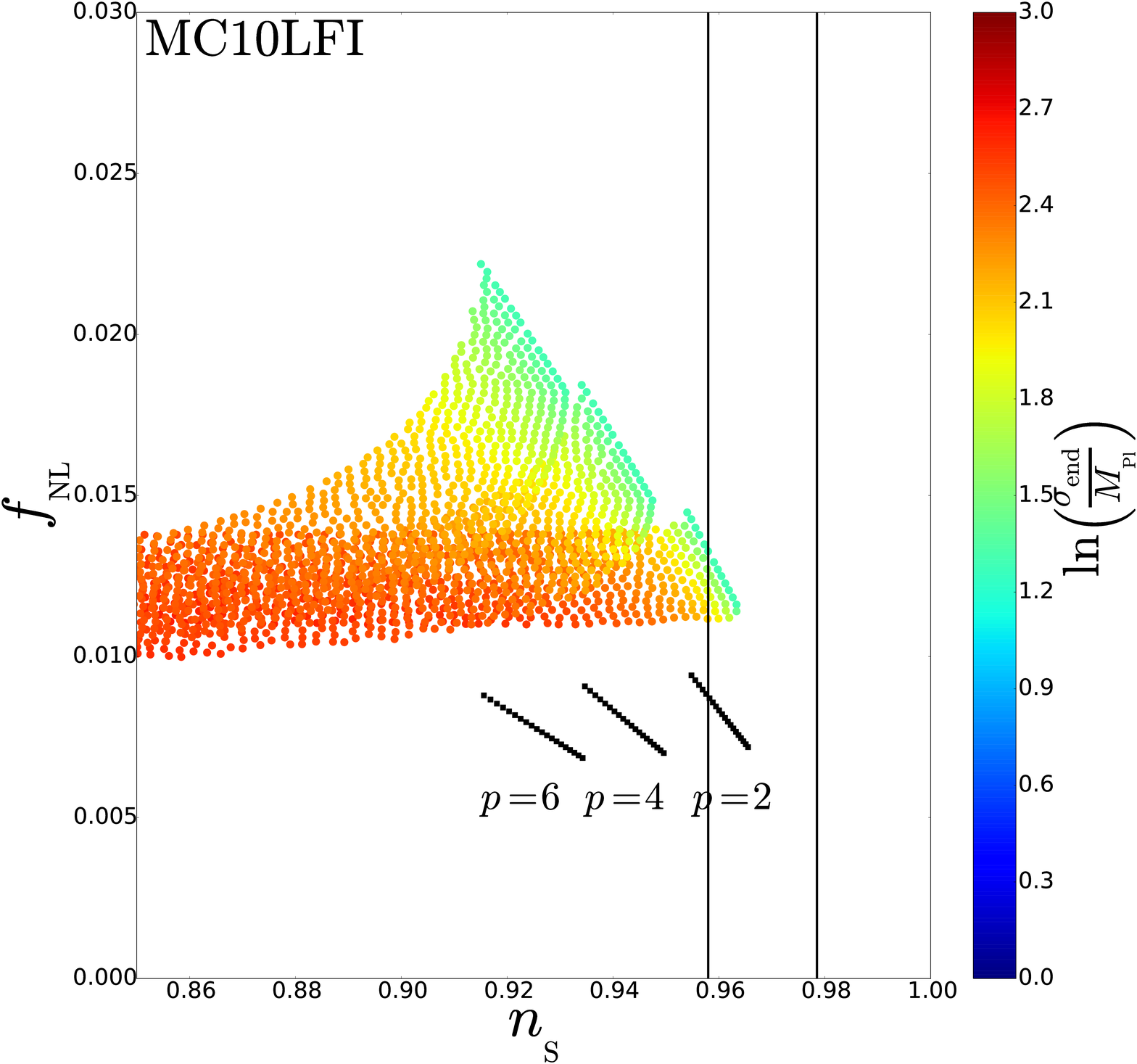}
\includegraphics[width=\wappfig,clip=true]{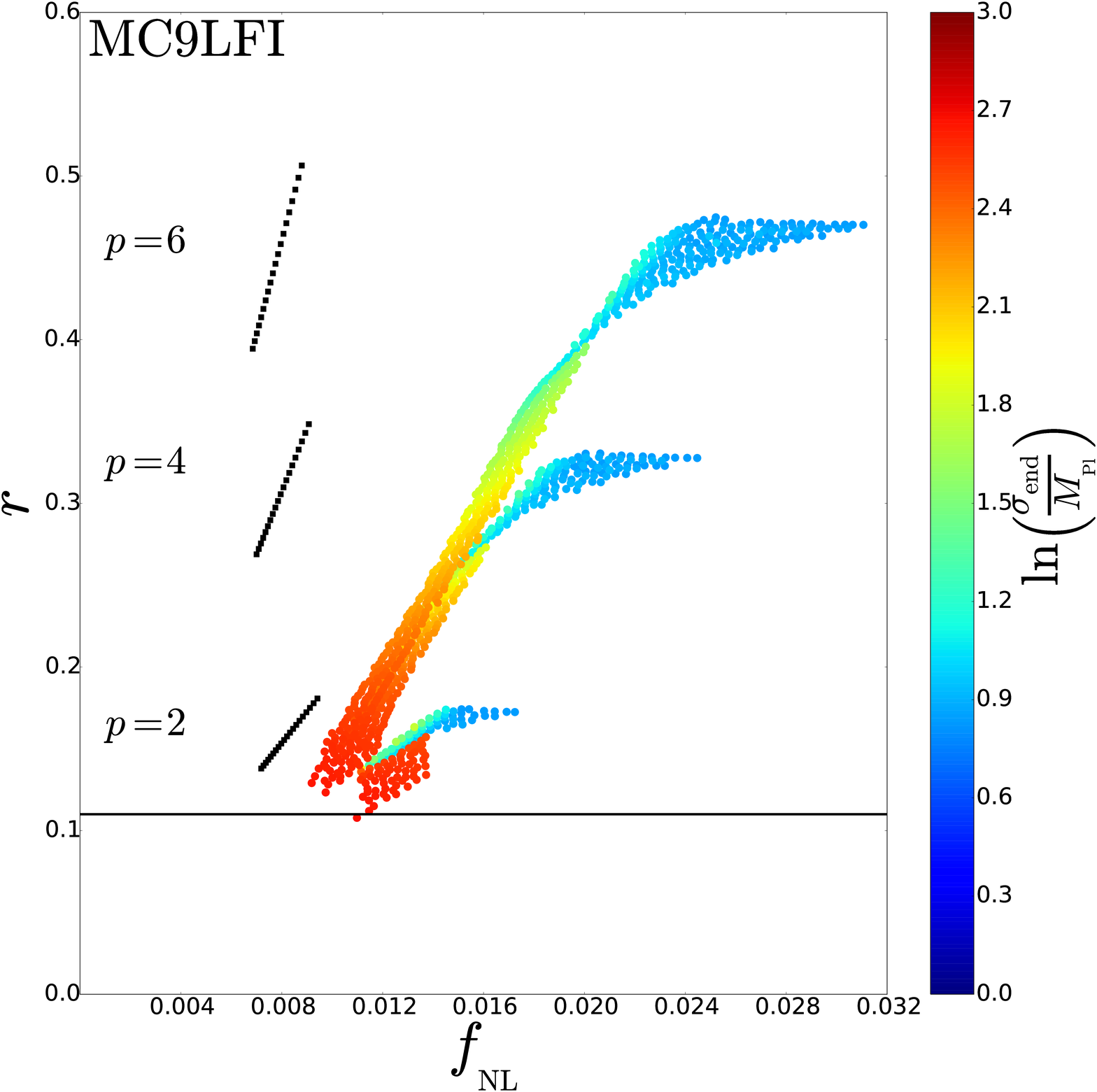}
\includegraphics[width=\wappfig,clip=true]{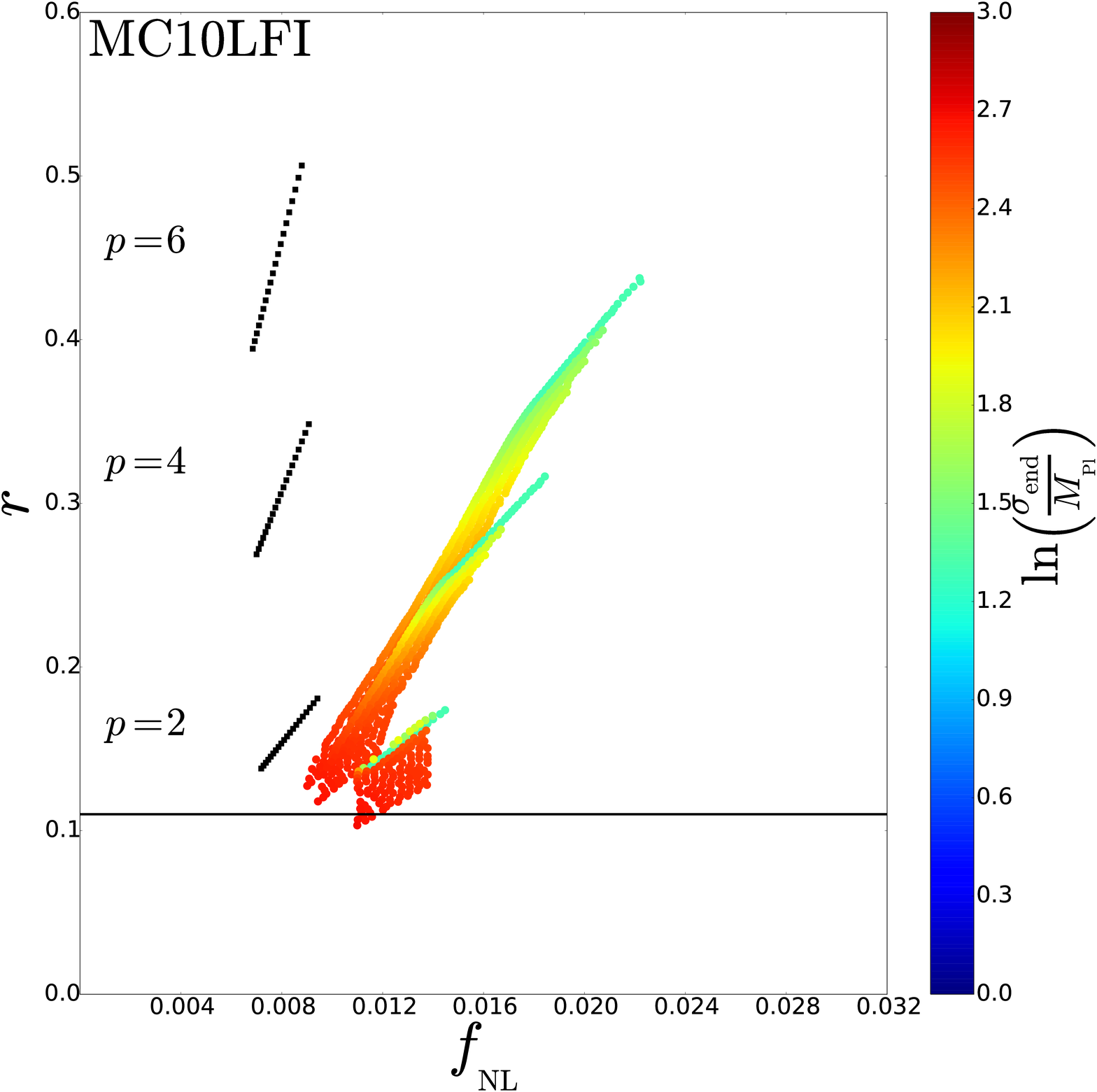}
\caption{Reheating consistent slow-roll predictions for the large field inflation models with a massive curvaton field, when reheating scenario is of the ninth (left panels) and tenth (right panels) type.}
\label{fig:CMBMCLFI910}
\end{center}
\end{figure}
\clearpage
\subsection{Higgs Inflation + Massive Scalar Field (MCHI)}
\label{sec:plots:hi}
\begin{figure}[!ht]
\begin{center}
\includegraphics[width=\wappfig,clip=true]{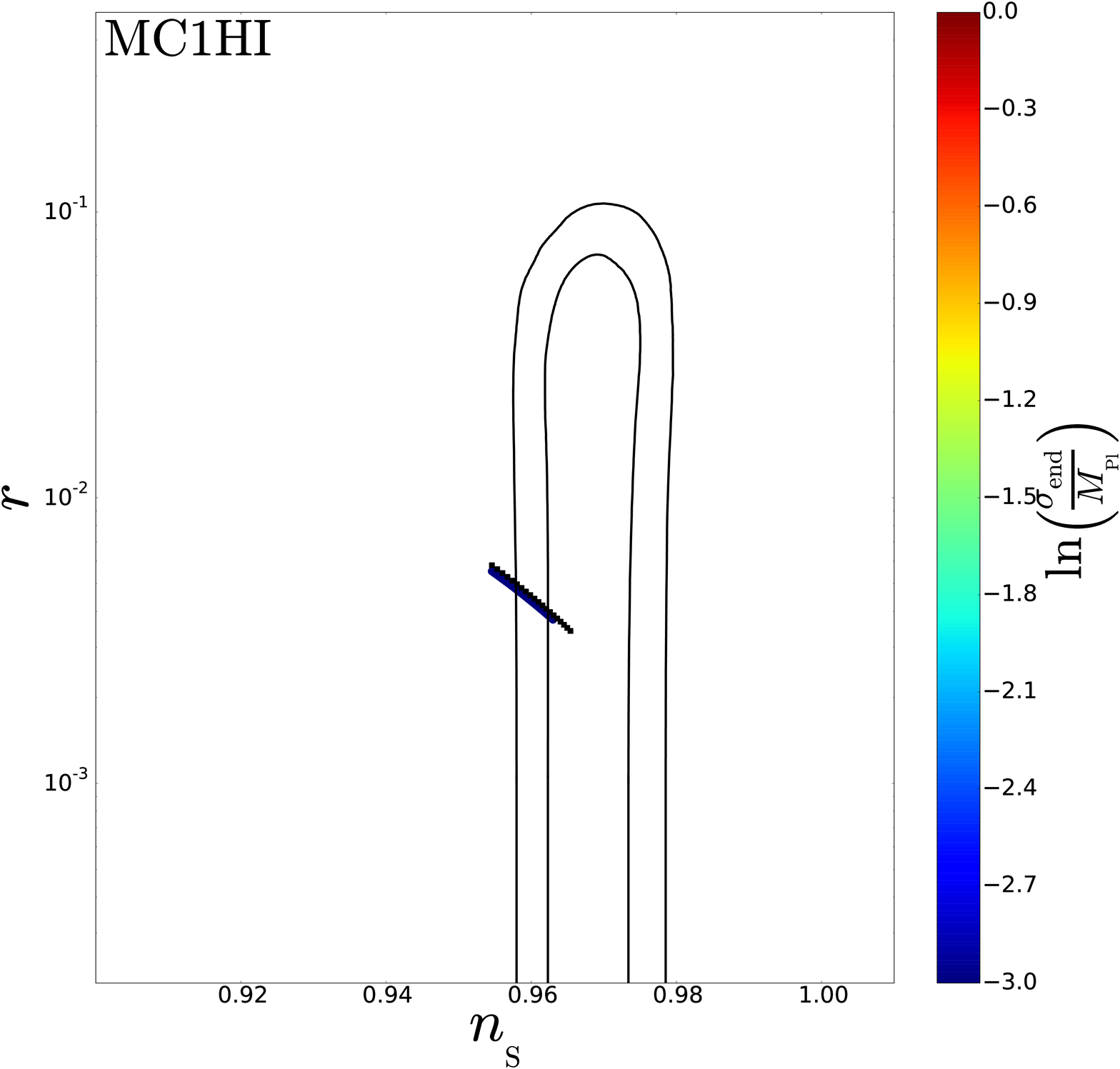}
\includegraphics[width=\wappfig,clip=true]{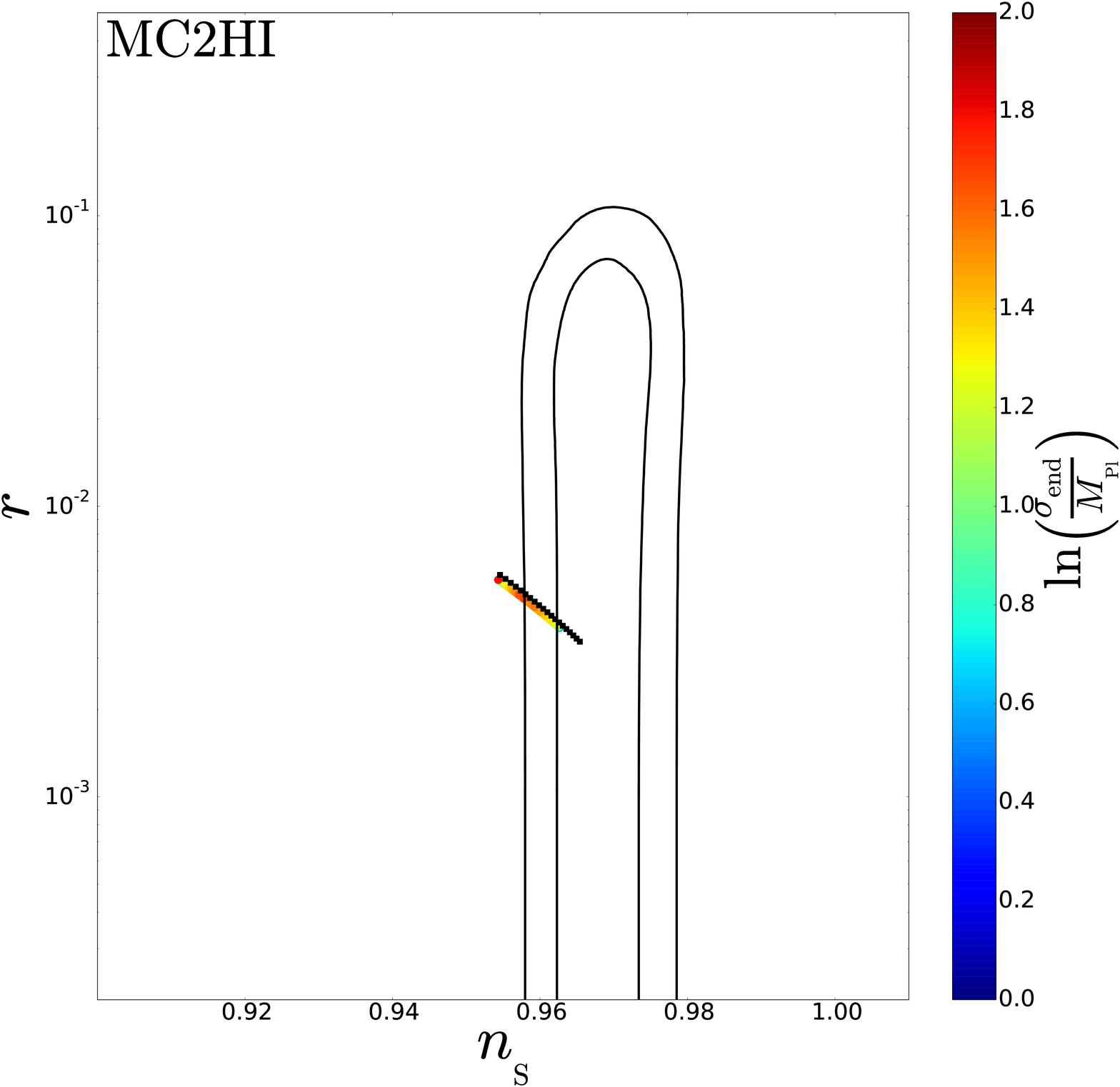}
\includegraphics[width=\wappfig,clip=true]{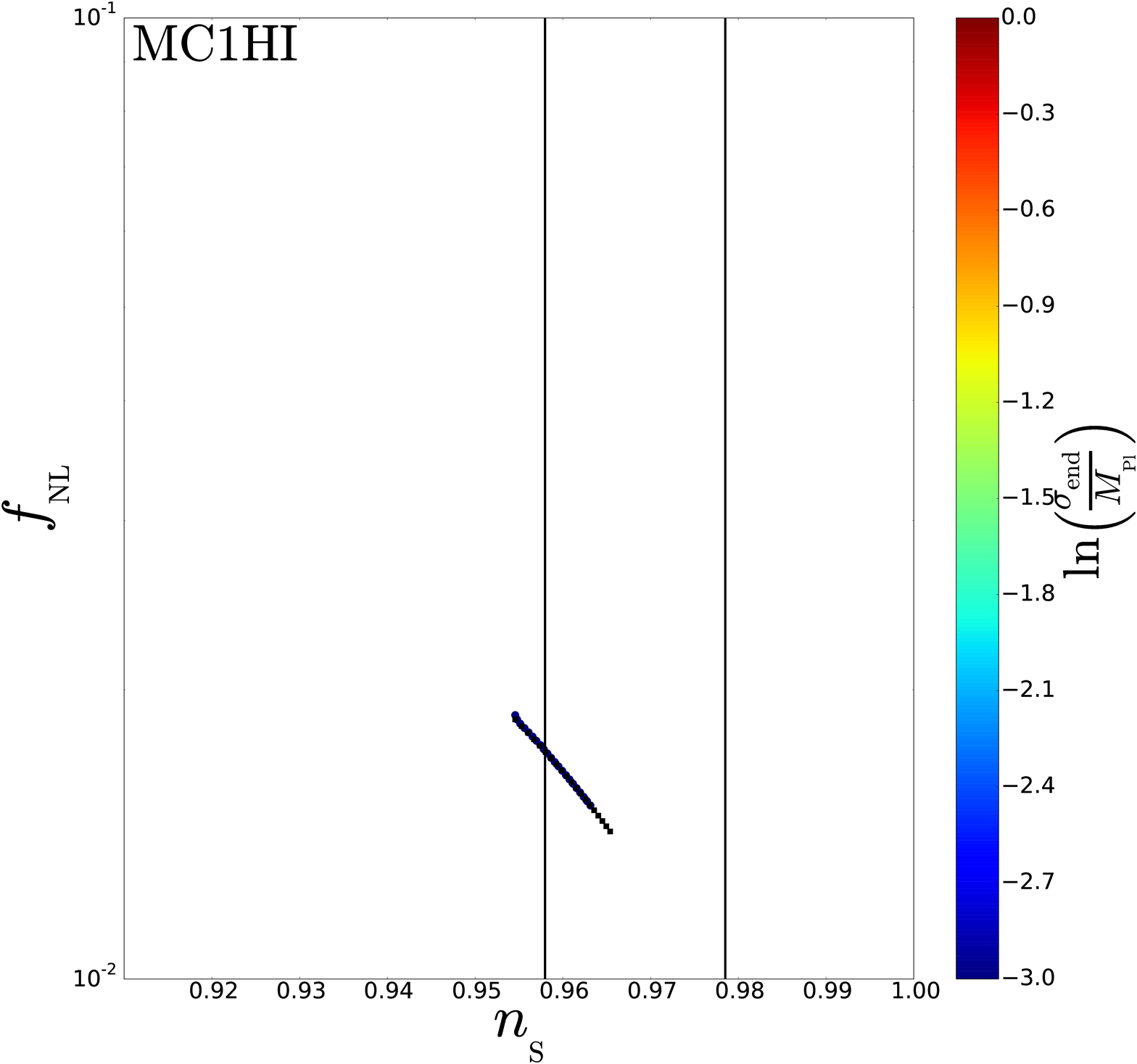}
\includegraphics[width=\wappfig,clip=true]{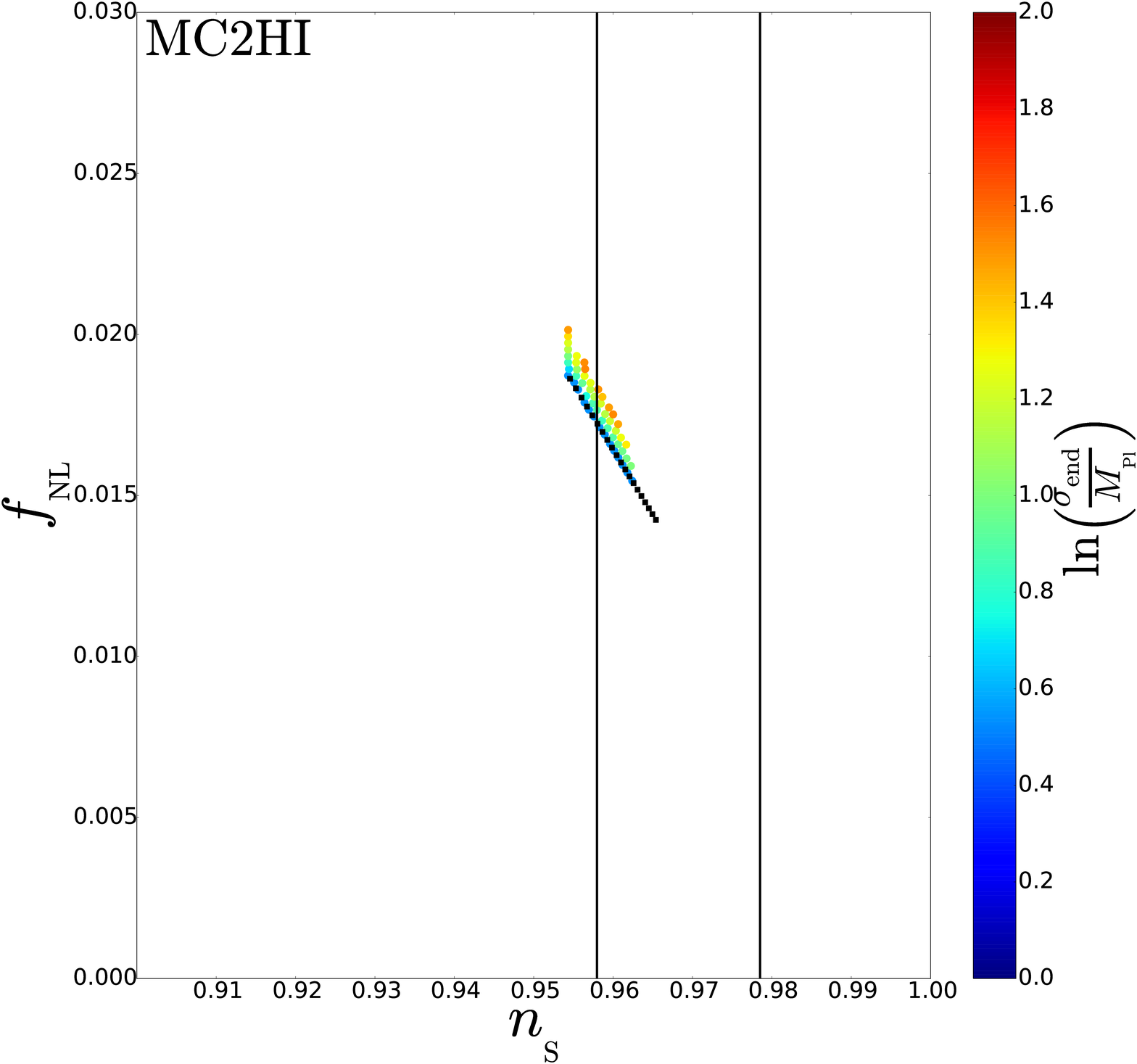}
\includegraphics[width=\wappfig,clip=true]{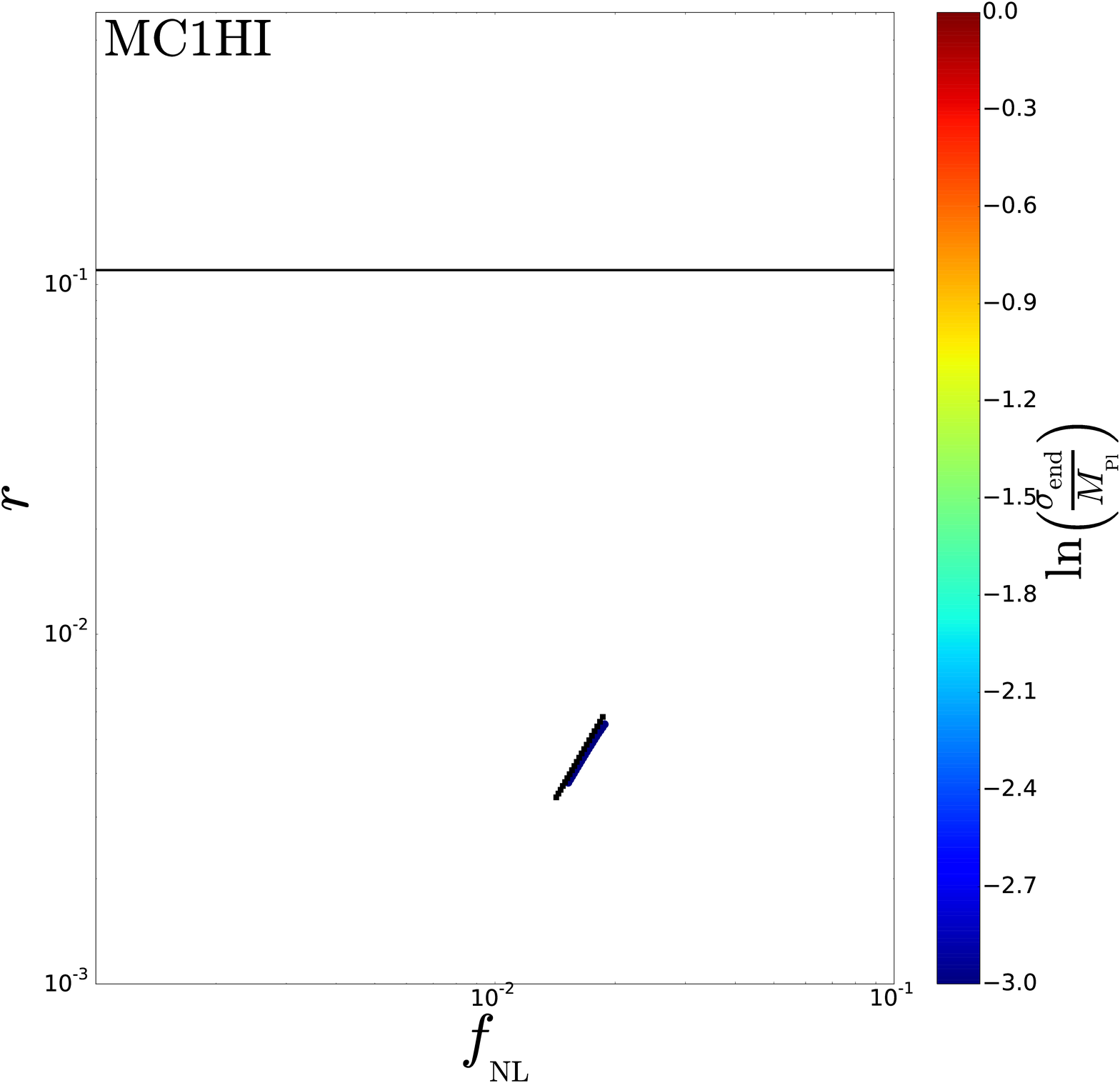}
\includegraphics[width=\wappfig,clip=true]{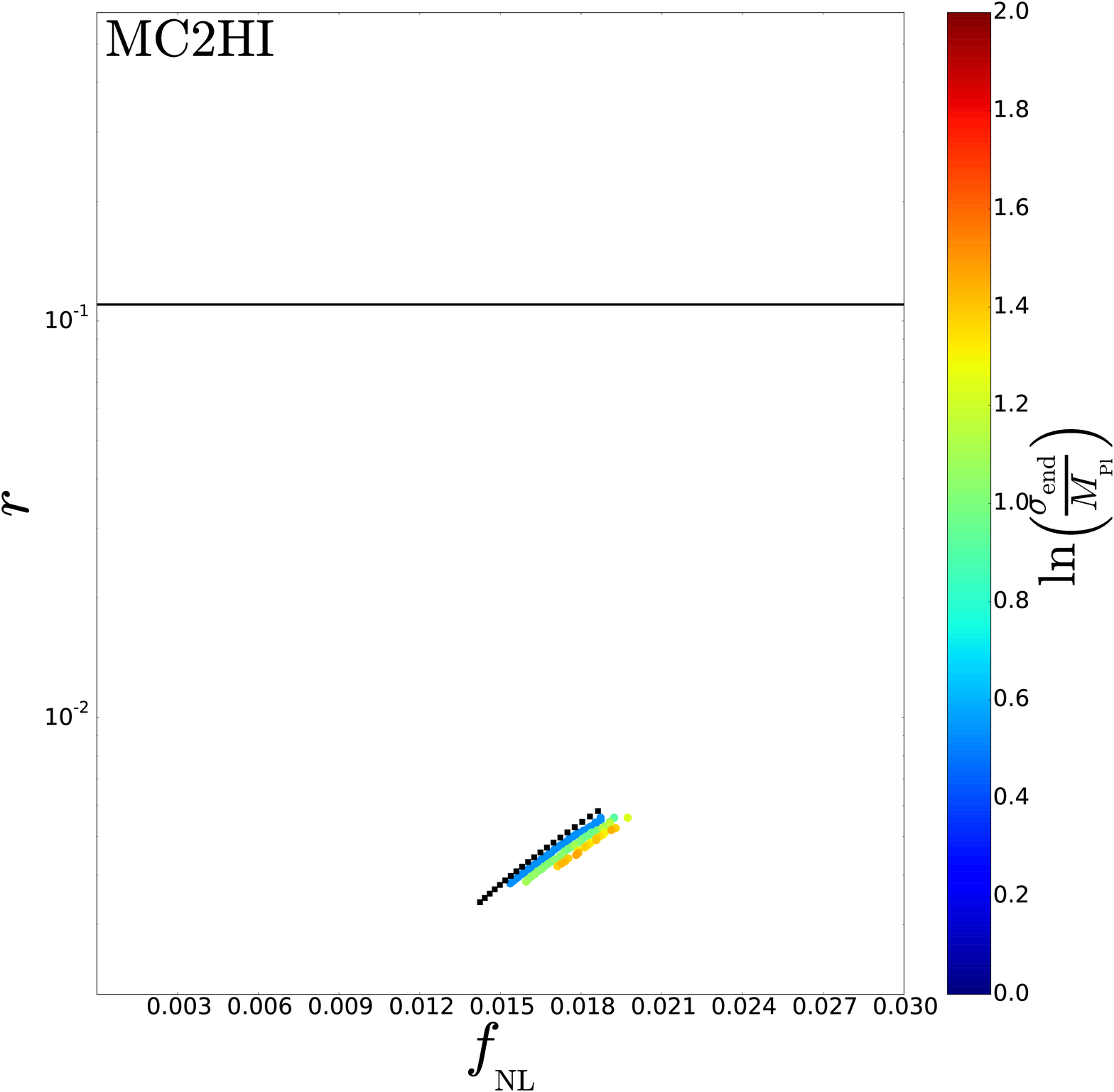}
\caption{Reheating consistent slow-roll predictions for the Higgs inflation models with a massive curvaton field, when reheating scenario is of the first (left panels) and second (right panels) type.}
\label{fig:CMBMCHI12}
\end{center}
\end{figure}
\begin{figure}[!ht]
\begin{center}
\includegraphics[width=\wappfig,clip=true]{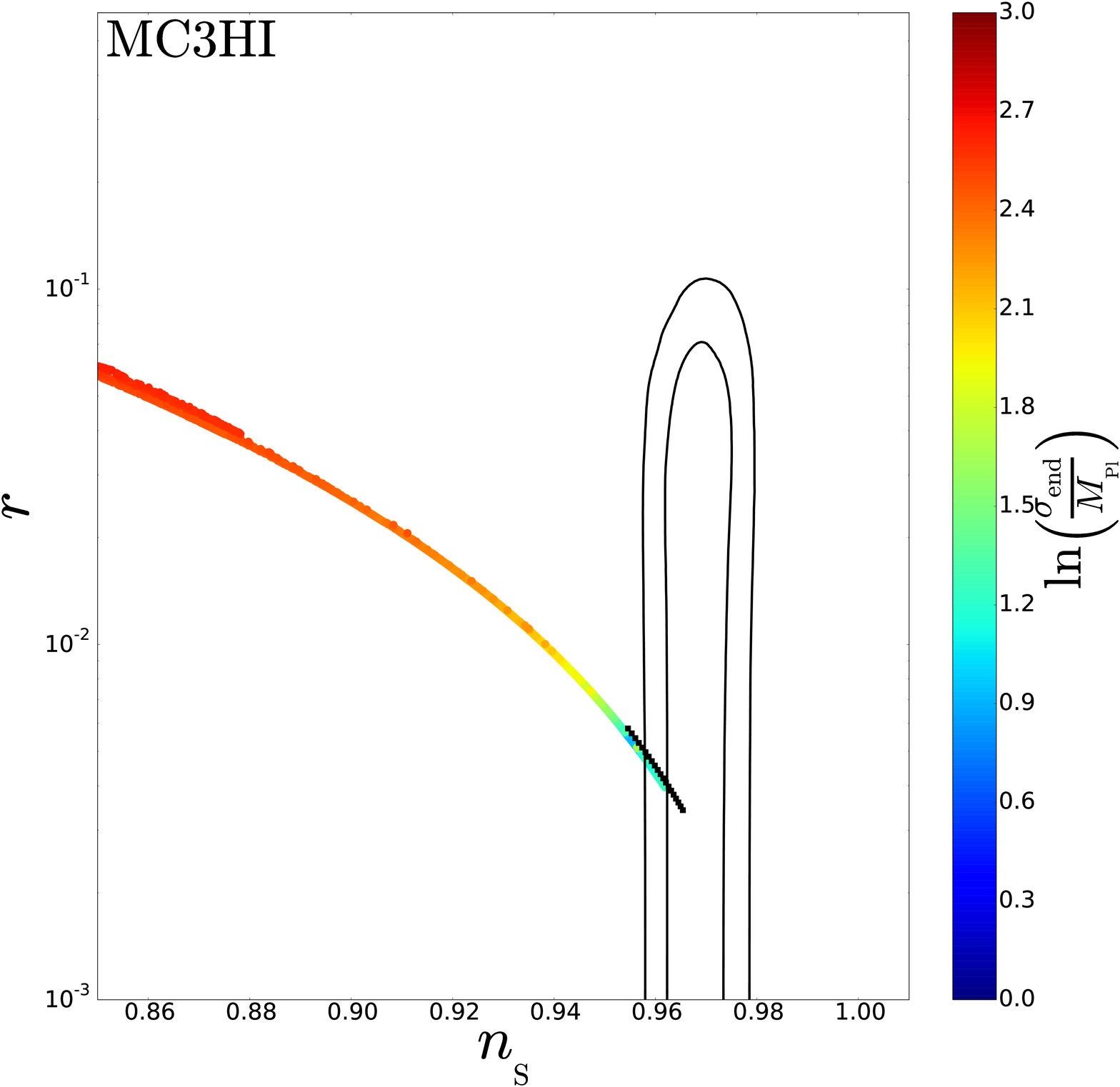}
\includegraphics[width=\wappfig,clip=true]{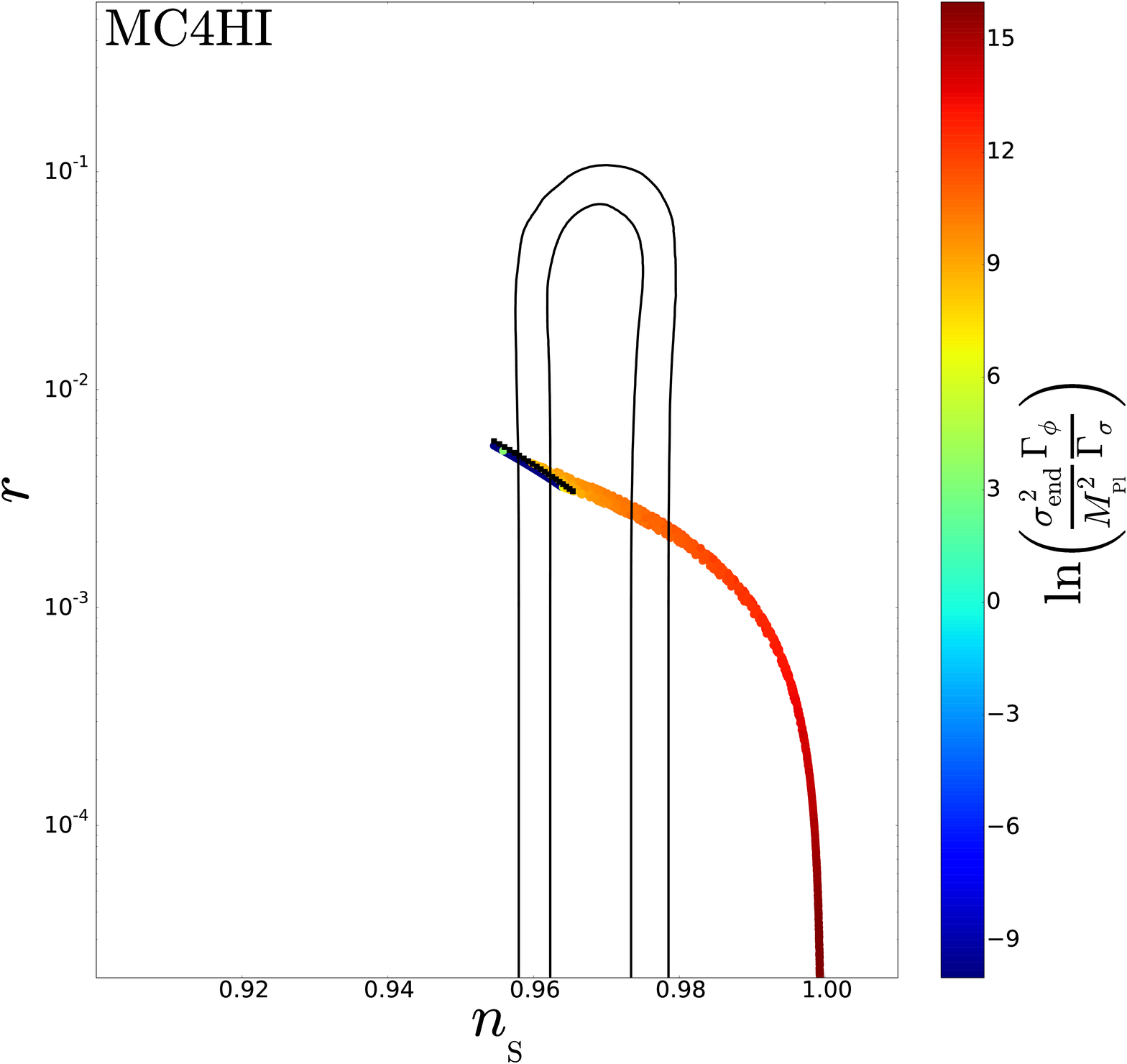}
\includegraphics[width=\wappfig,clip=true]{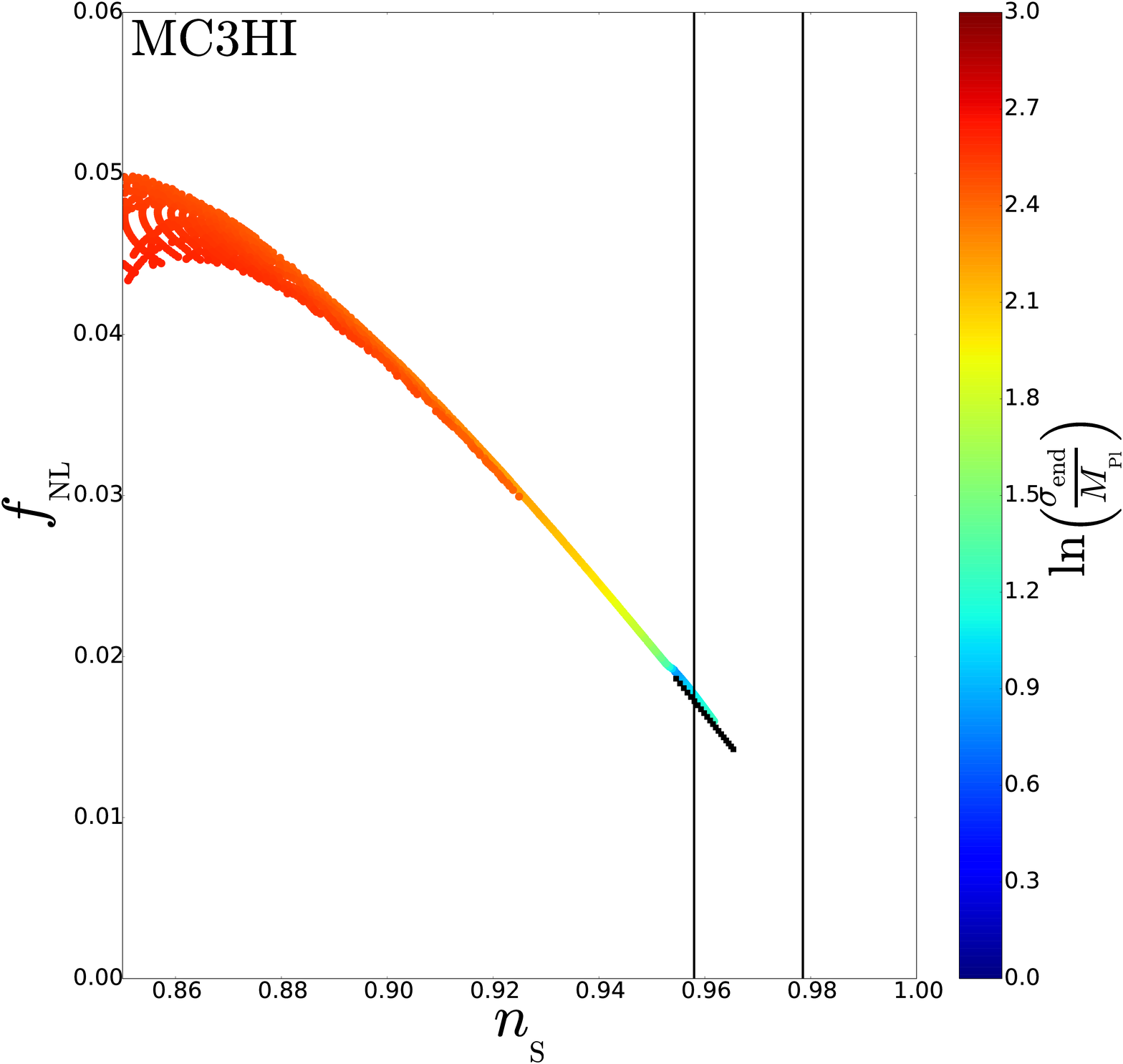}
\includegraphics[width=\wappfig,clip=true]{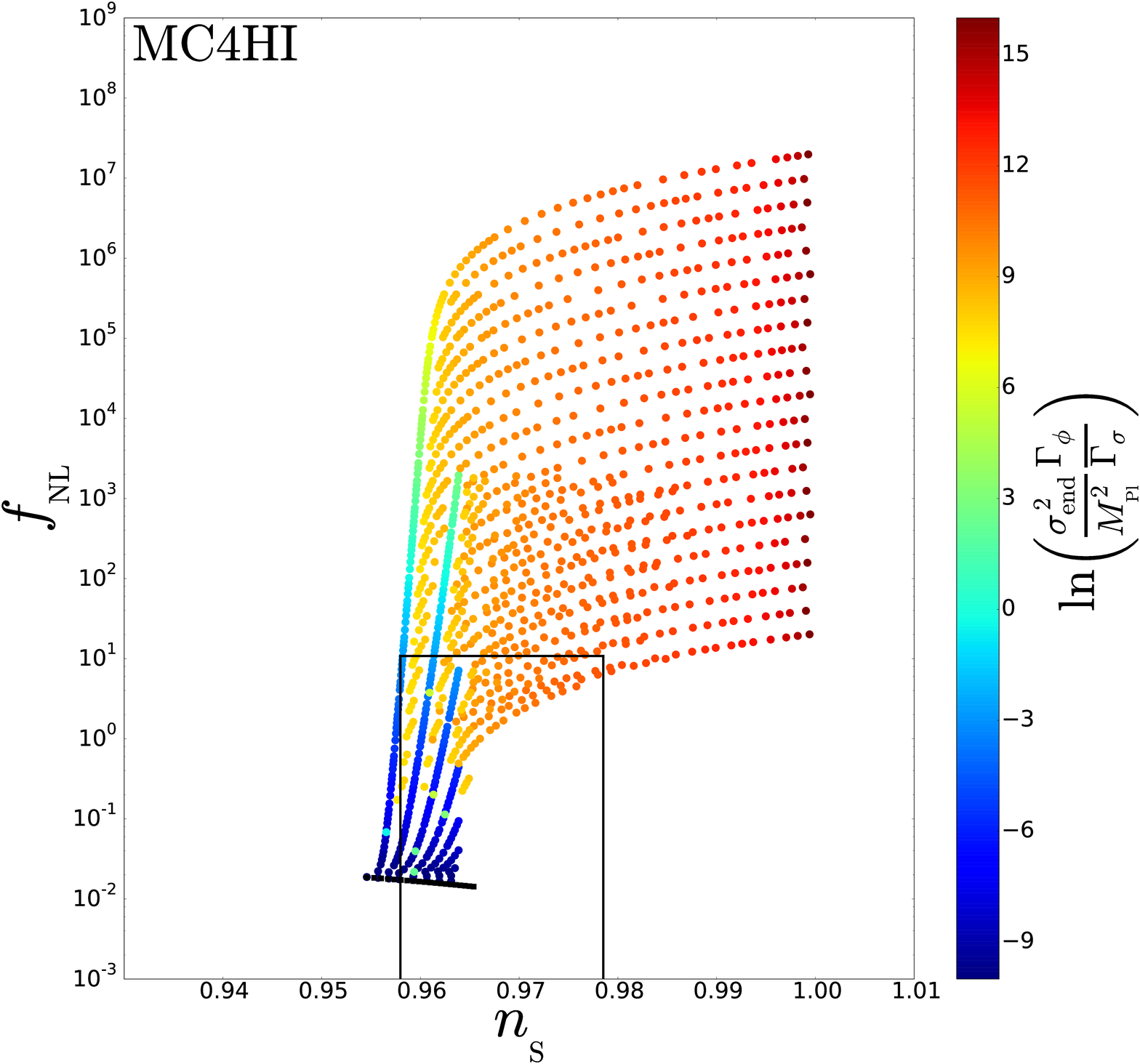}
\includegraphics[width=\wappfig,clip=true]{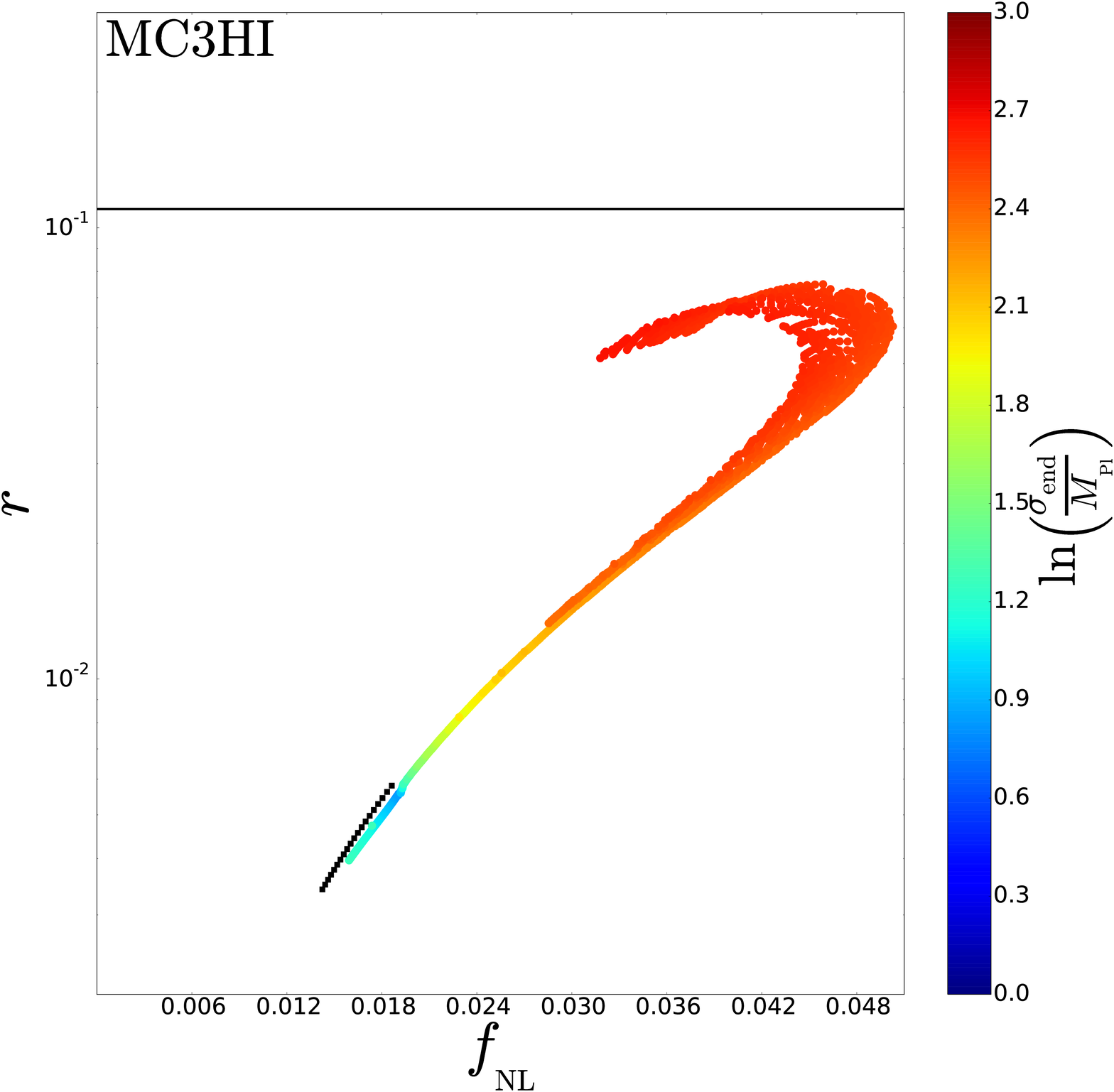}
\includegraphics[width=\wappfig,clip=true]{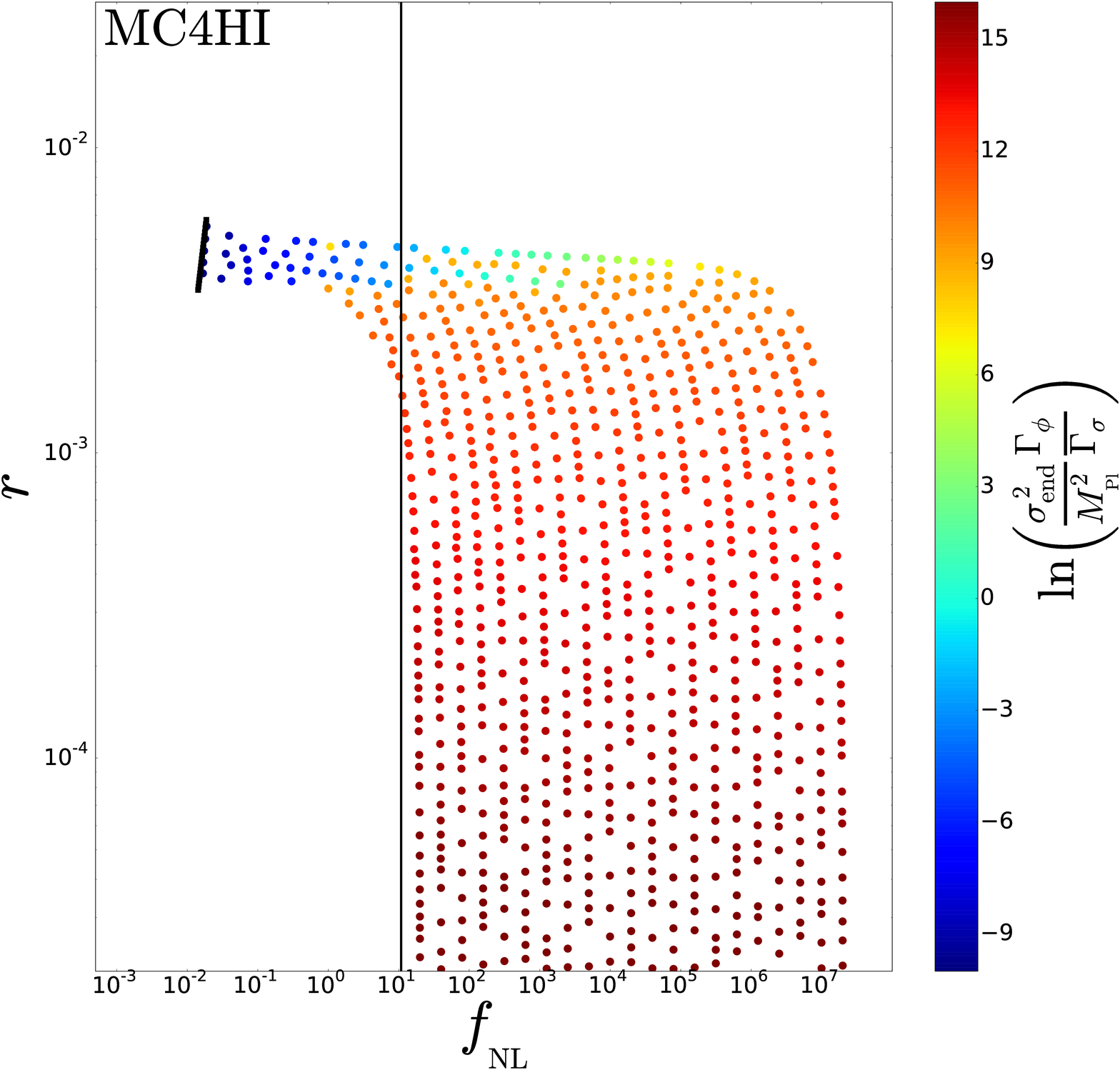}
\caption{Reheating consistent slow-roll predictions for the Higgs inflation models with a massive curvaton field, when reheating scenario is of the third (left panels) and fourth (right panels) type.}
\label{fig:CMBMCHI34}
\end{center}
\end{figure}
\begin{figure}[!ht]
\begin{center}
\includegraphics[width=\wappfig,clip=true]{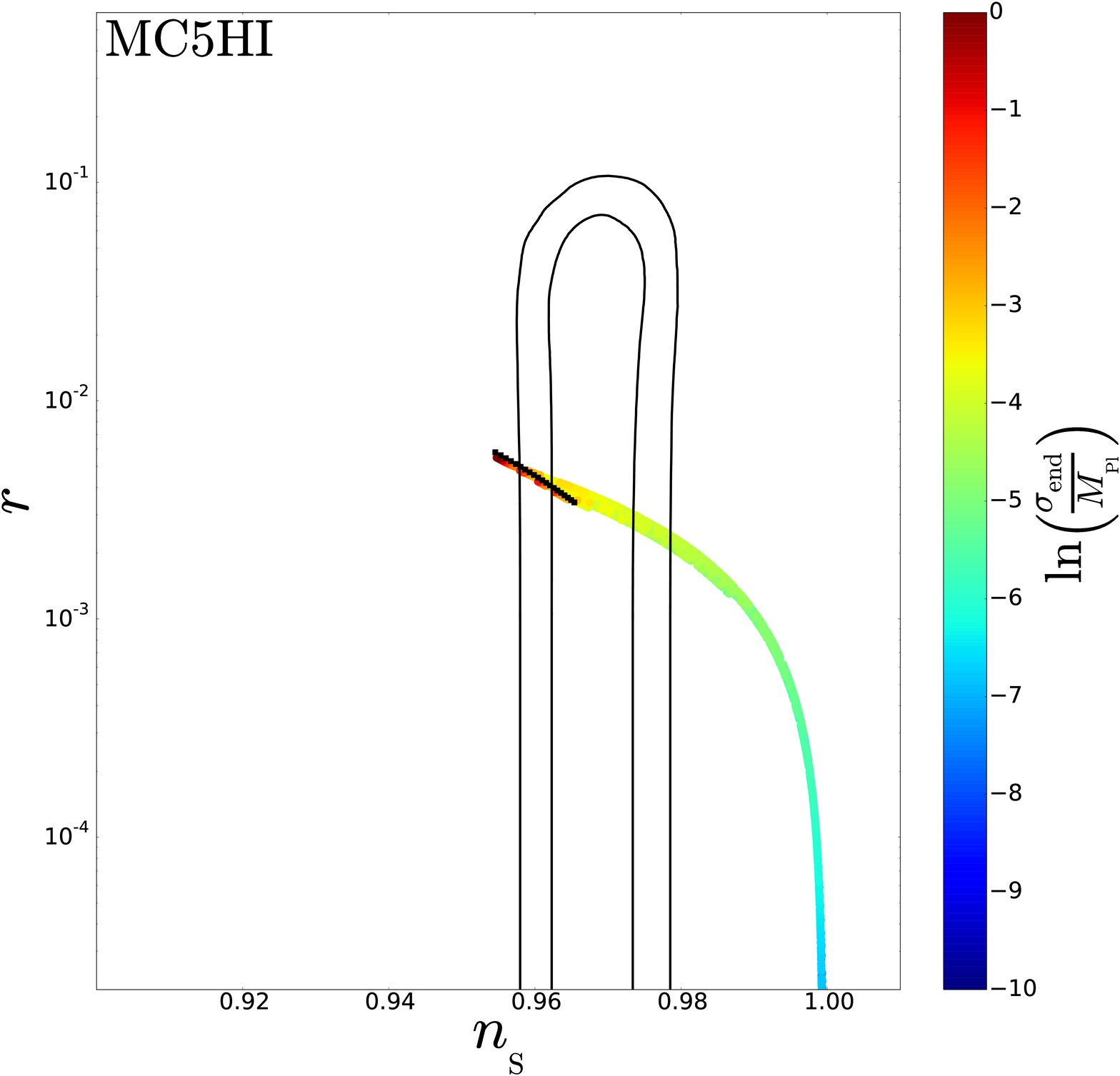}
\includegraphics[width=\wappfig,clip=true]{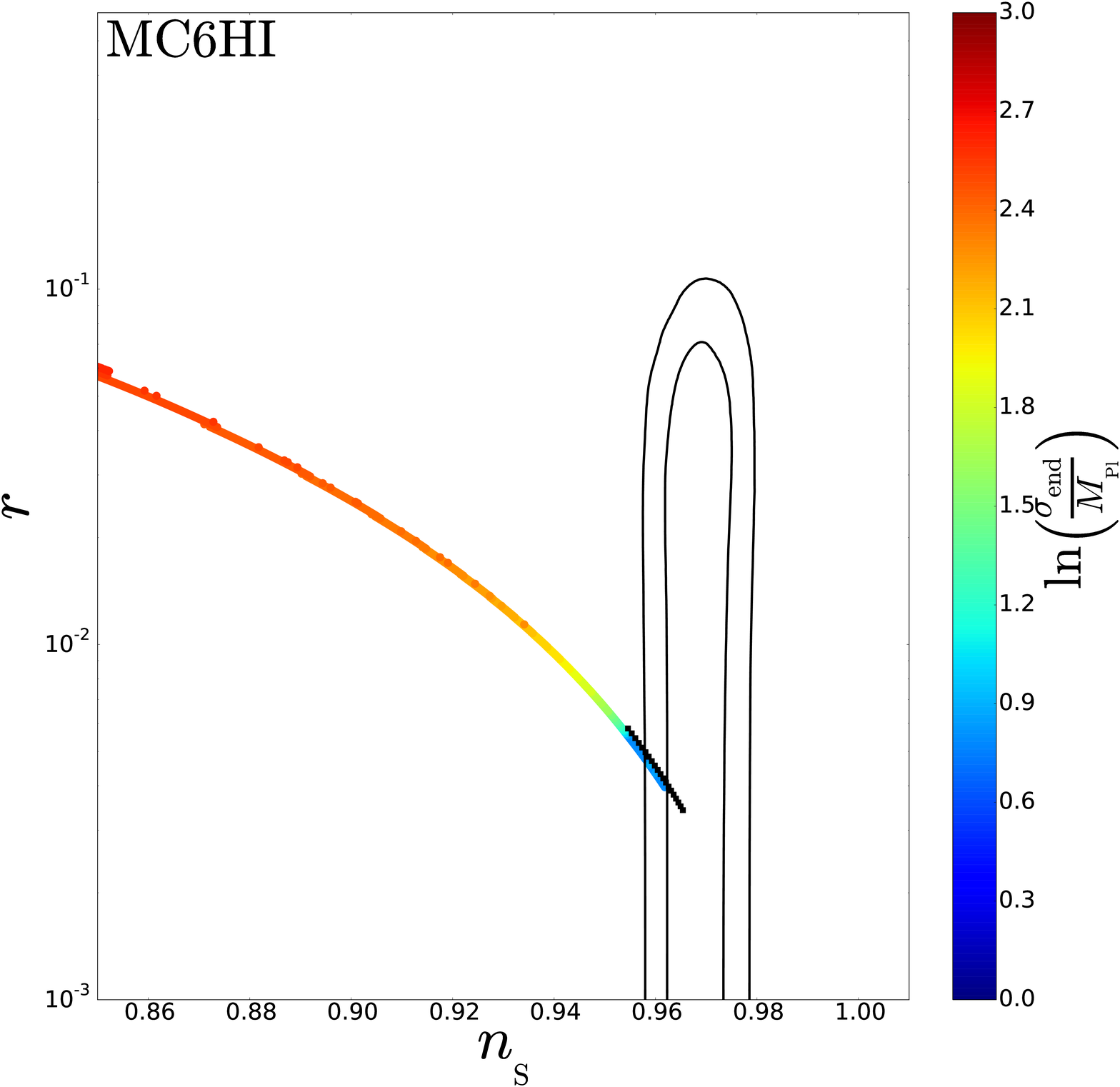}
\includegraphics[width=\wappfig,clip=true]{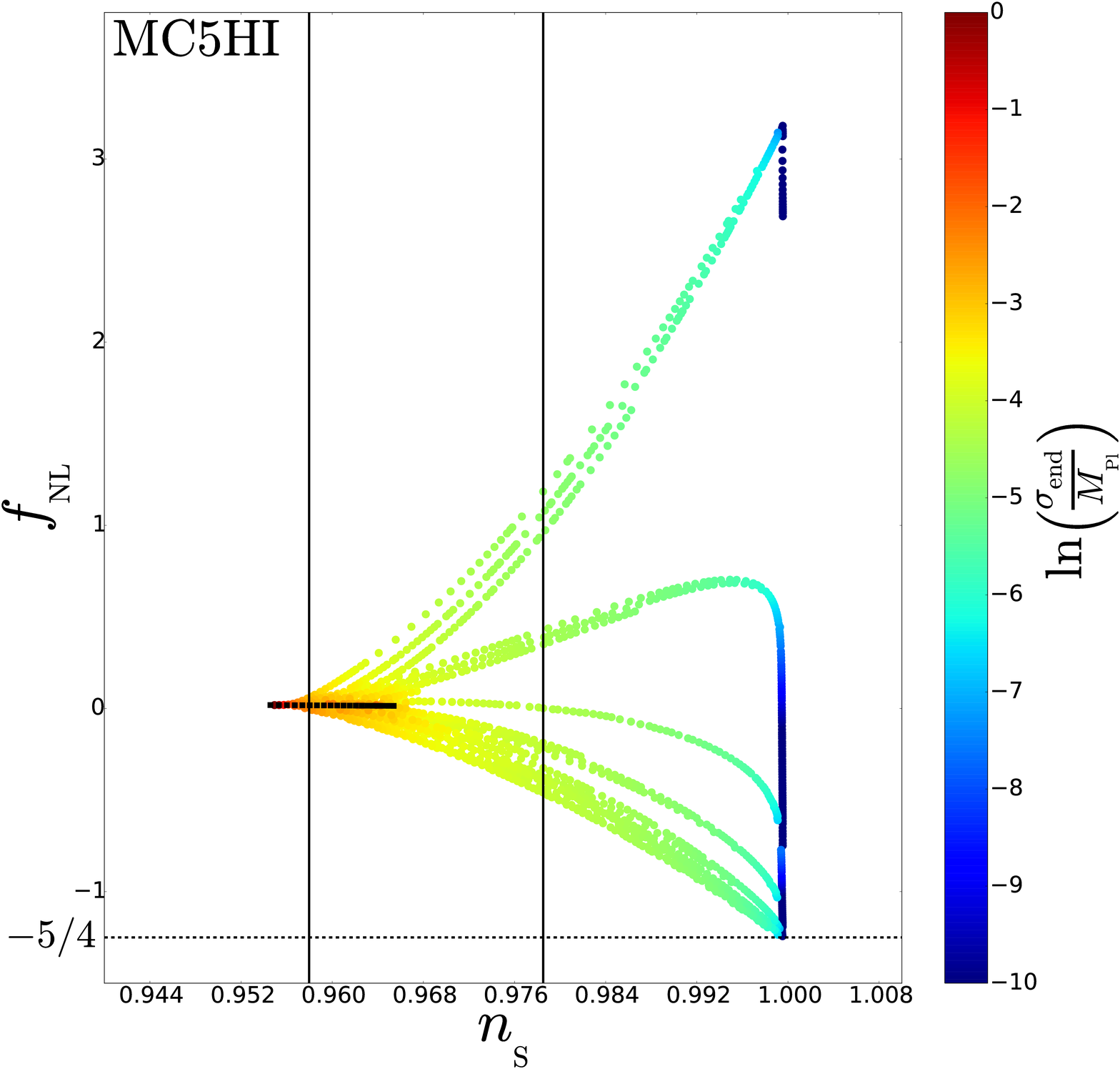}
\includegraphics[width=\wappfig,clip=true]{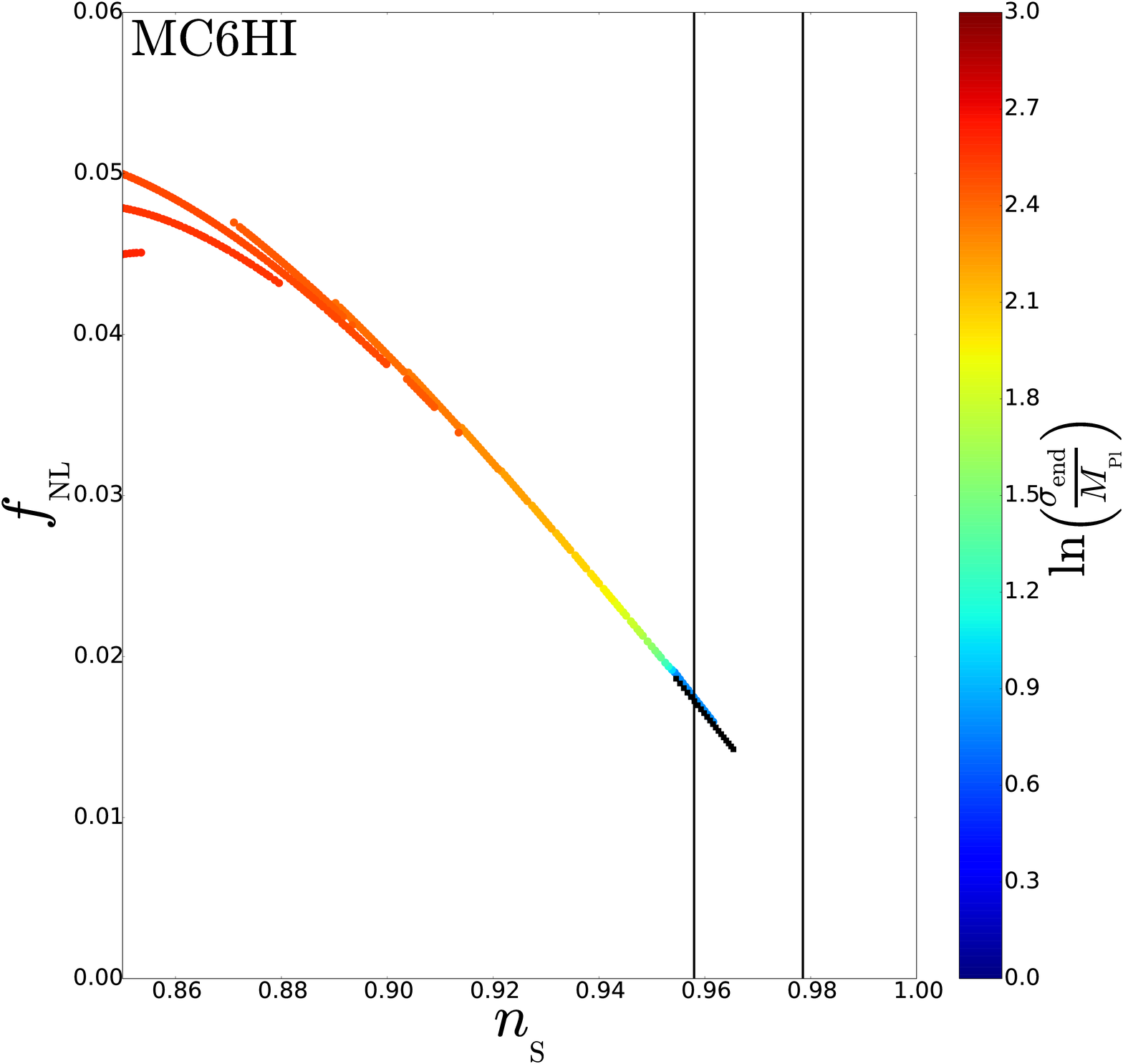}
\includegraphics[width=\wappfig,clip=true]{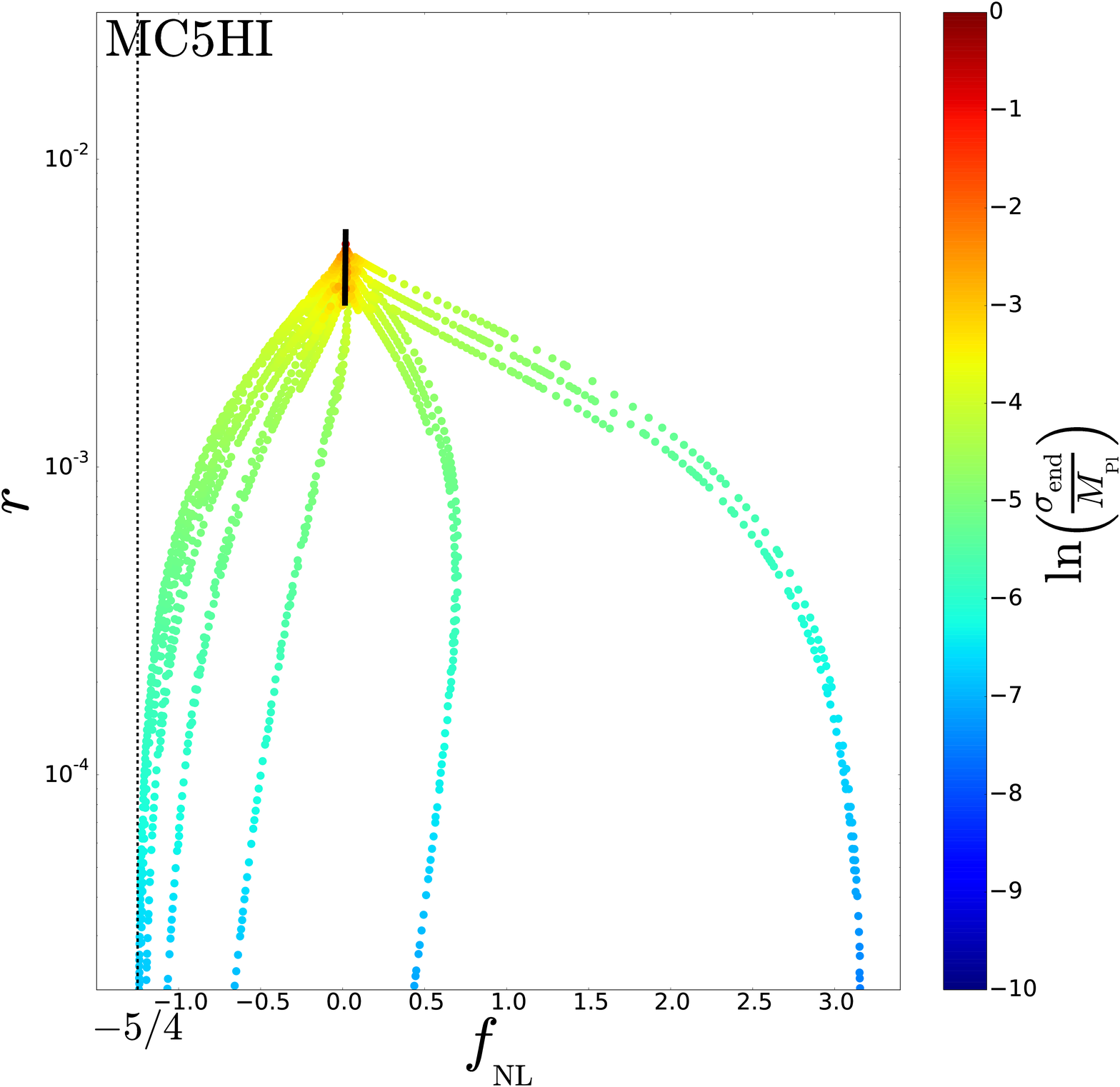}
\includegraphics[width=\wappfig,clip=true]{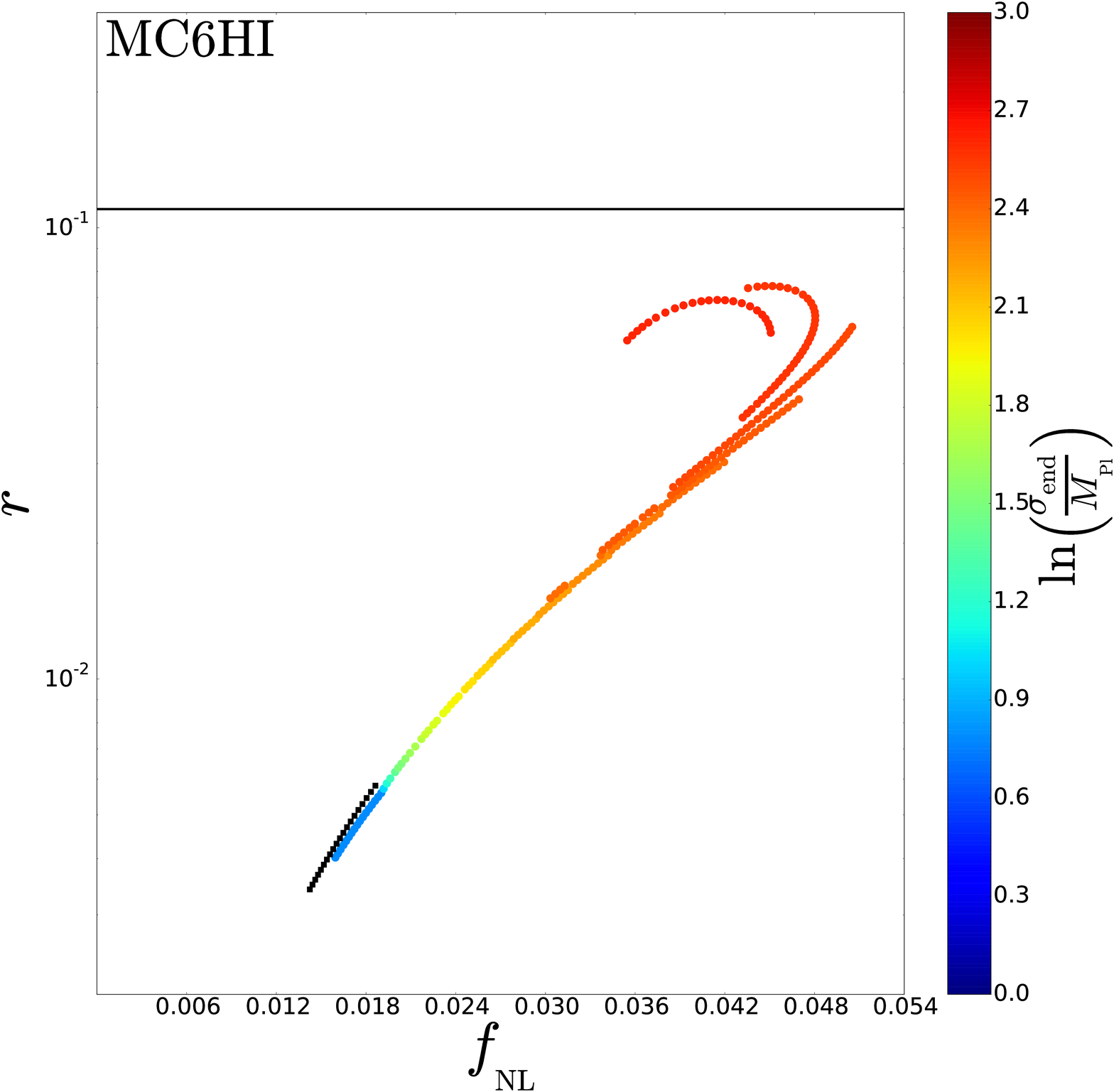}
\caption{Reheating consistent slow-roll predictions for the Higgs inflation models with a massive curvaton field, when reheating scenario is of the fifth (left panels) and sixth (right panels) type.}
\label{fig:CMBMCHI56}
\end{center}
\end{figure}
\begin{figure}[!ht]
\begin{center}
\includegraphics[width=\wappfig,clip=true]{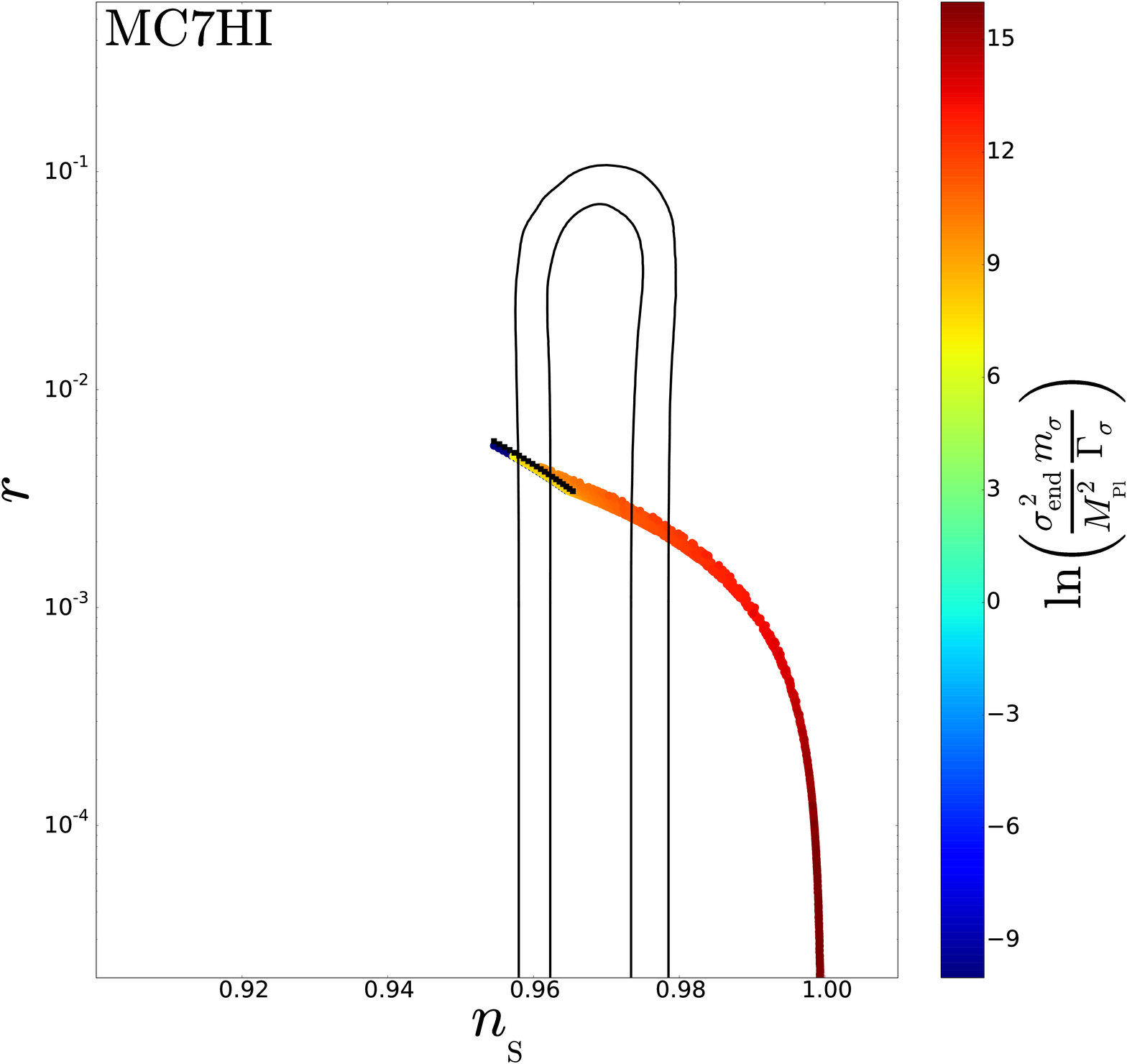}
\includegraphics[width=\wappfig,clip=true]{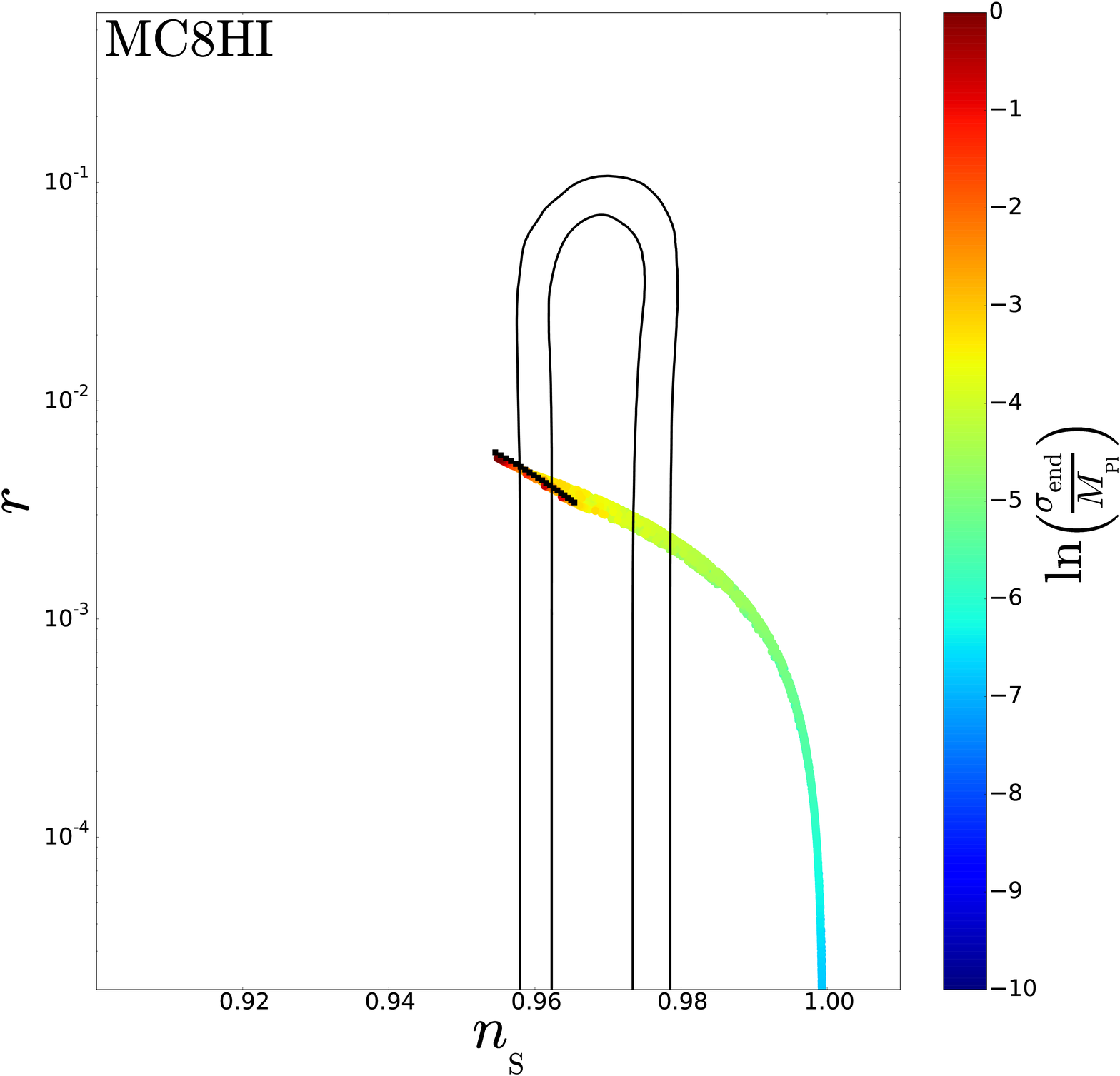}
\includegraphics[width=\wappfig,clip=true]{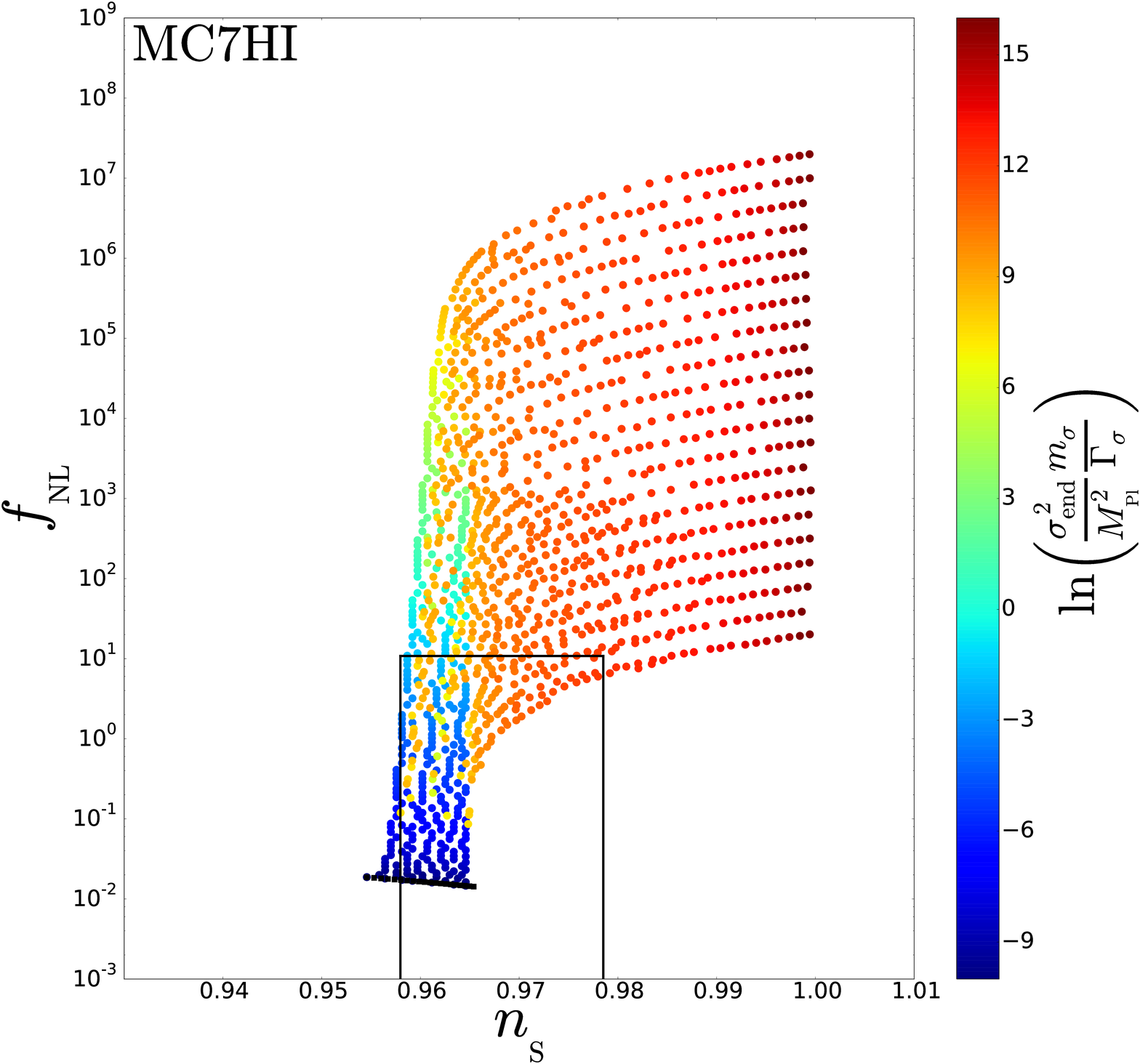}
\includegraphics[width=\wappfig,clip=true]{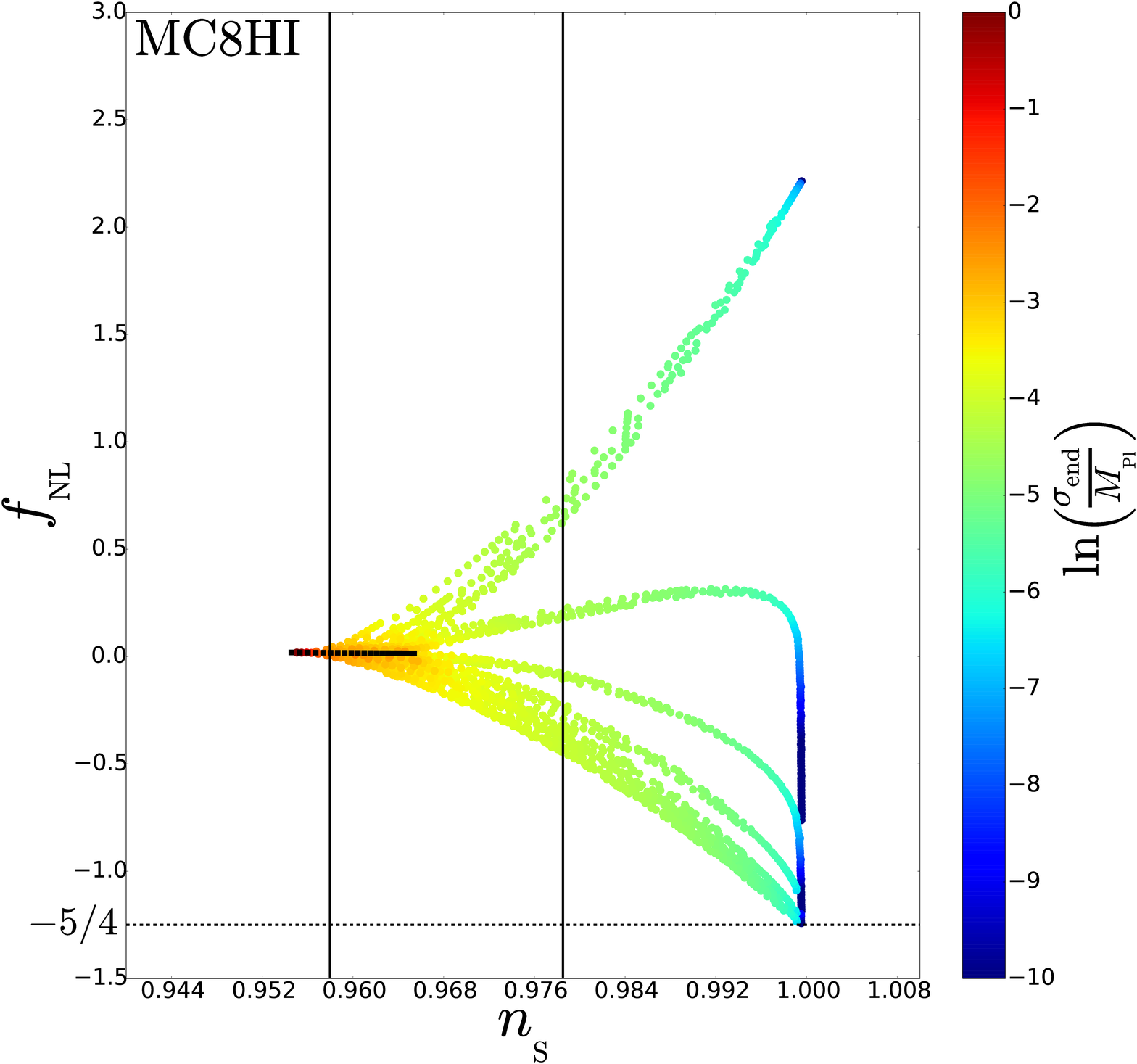}
\includegraphics[width=\wappfig,clip=true]{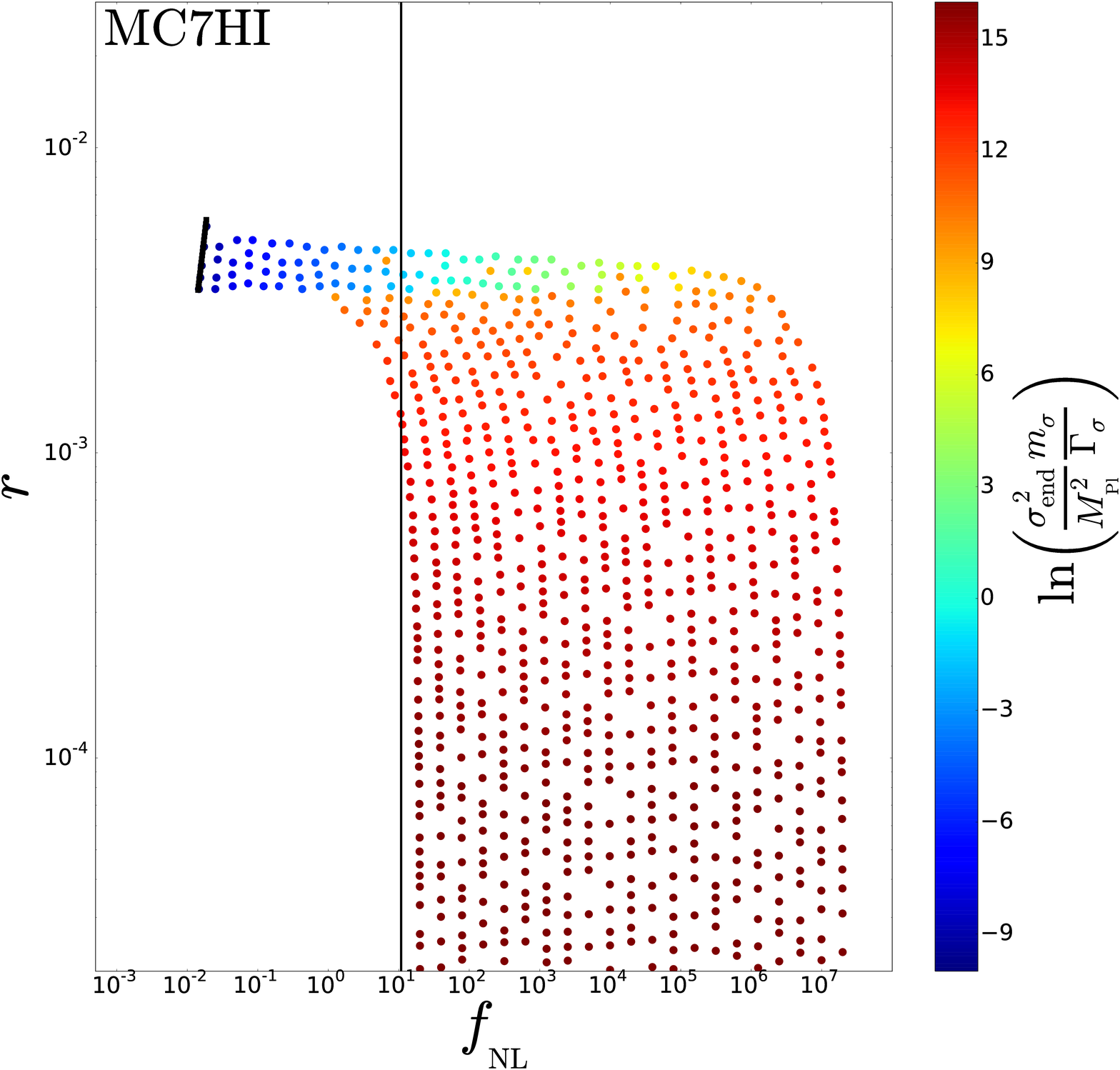}
\includegraphics[width=\wappfig,clip=true]{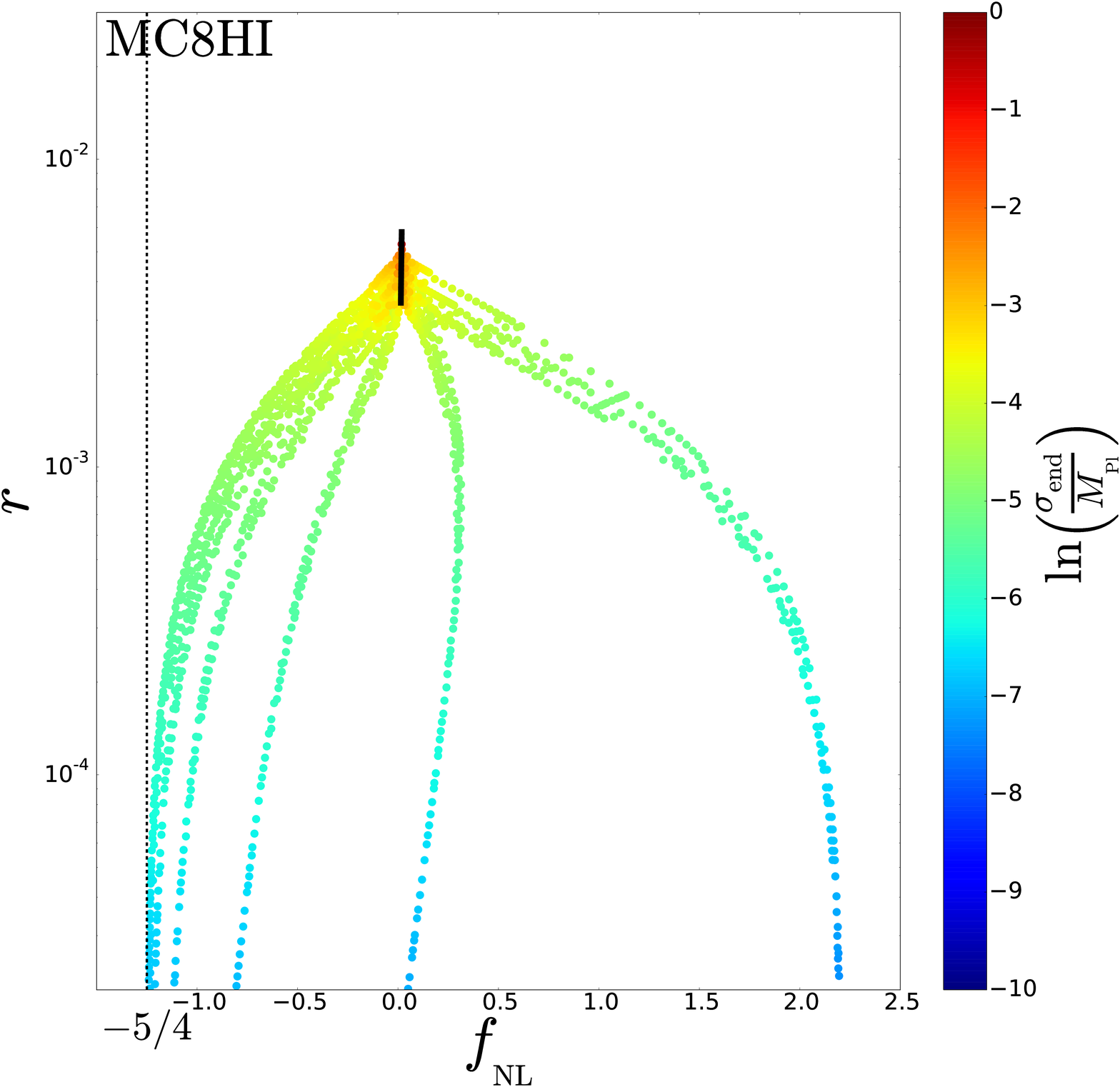}
\caption{Reheating consistent slow-roll predictions for the Higgs inflation models with a massive curvaton field, when reheating scenario is of the seventh (left panels) and eighth (right panels) type.}
\label{fig:CMBMCHI78}
\end{center}
\end{figure}
\begin{figure}[!ht]
\begin{center}
\includegraphics[width=\wappfig,clip=true]{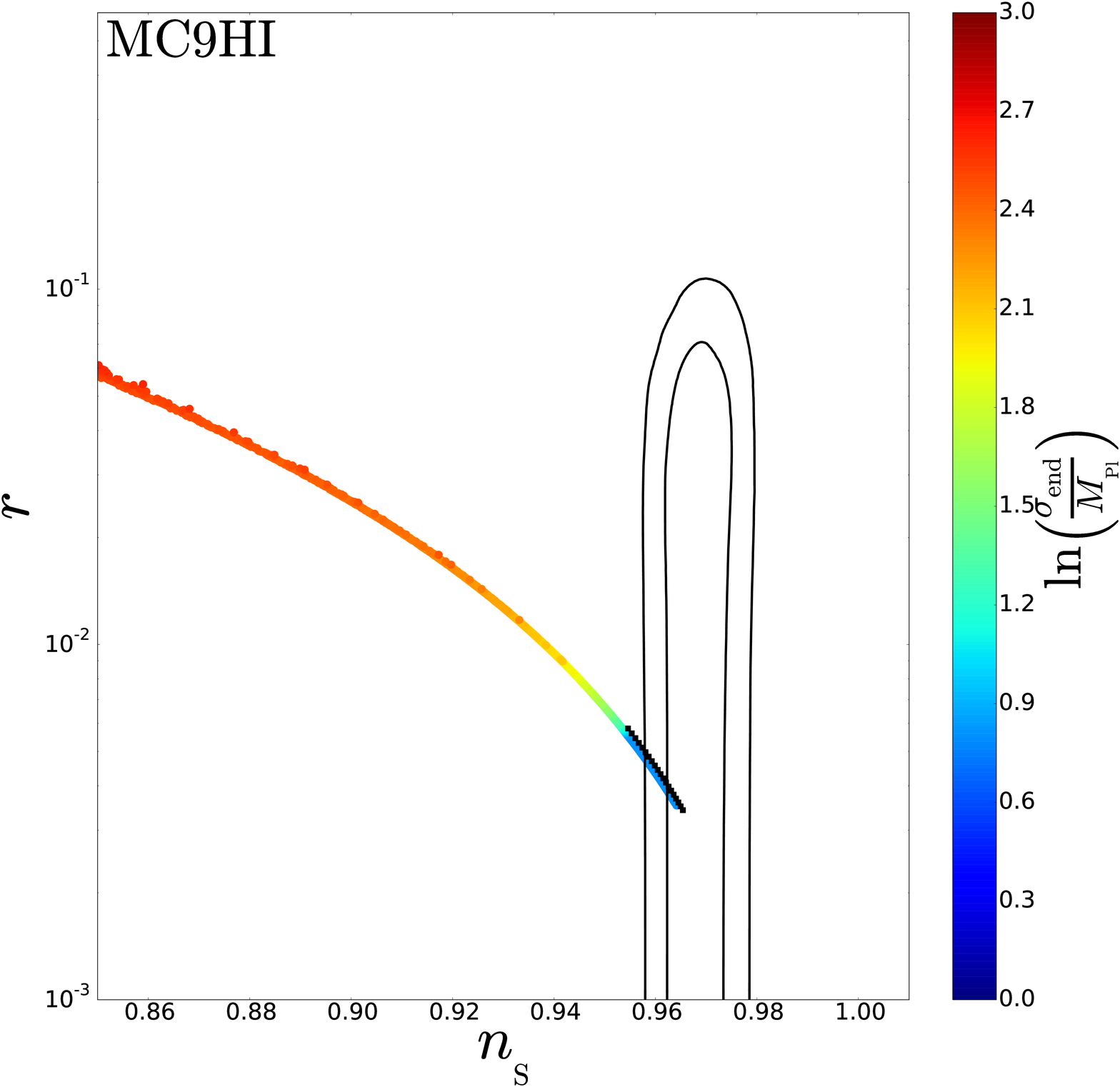}
\includegraphics[width=\wappfig,clip=true]{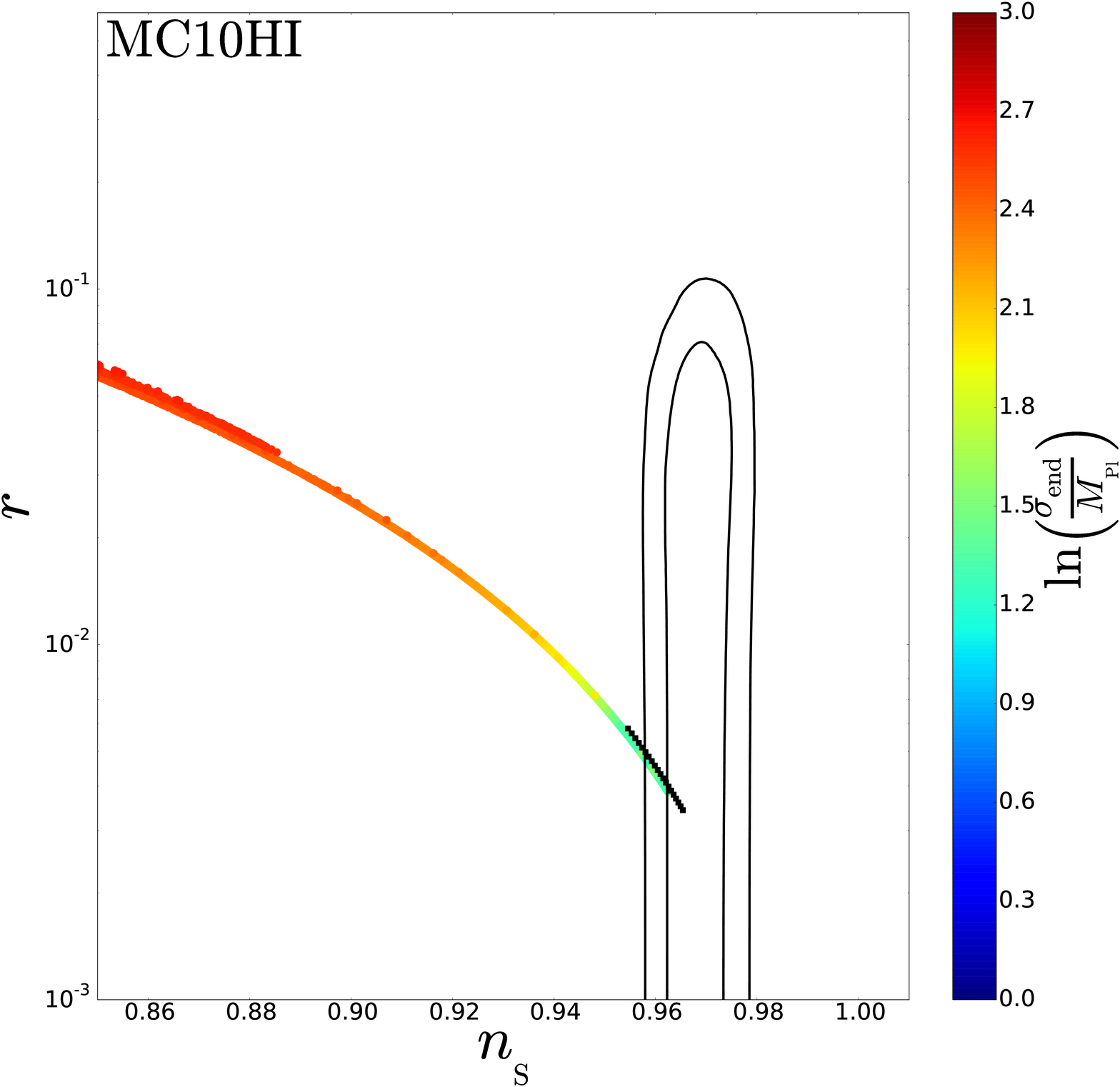}
\includegraphics[width=\wappfig,clip=true]{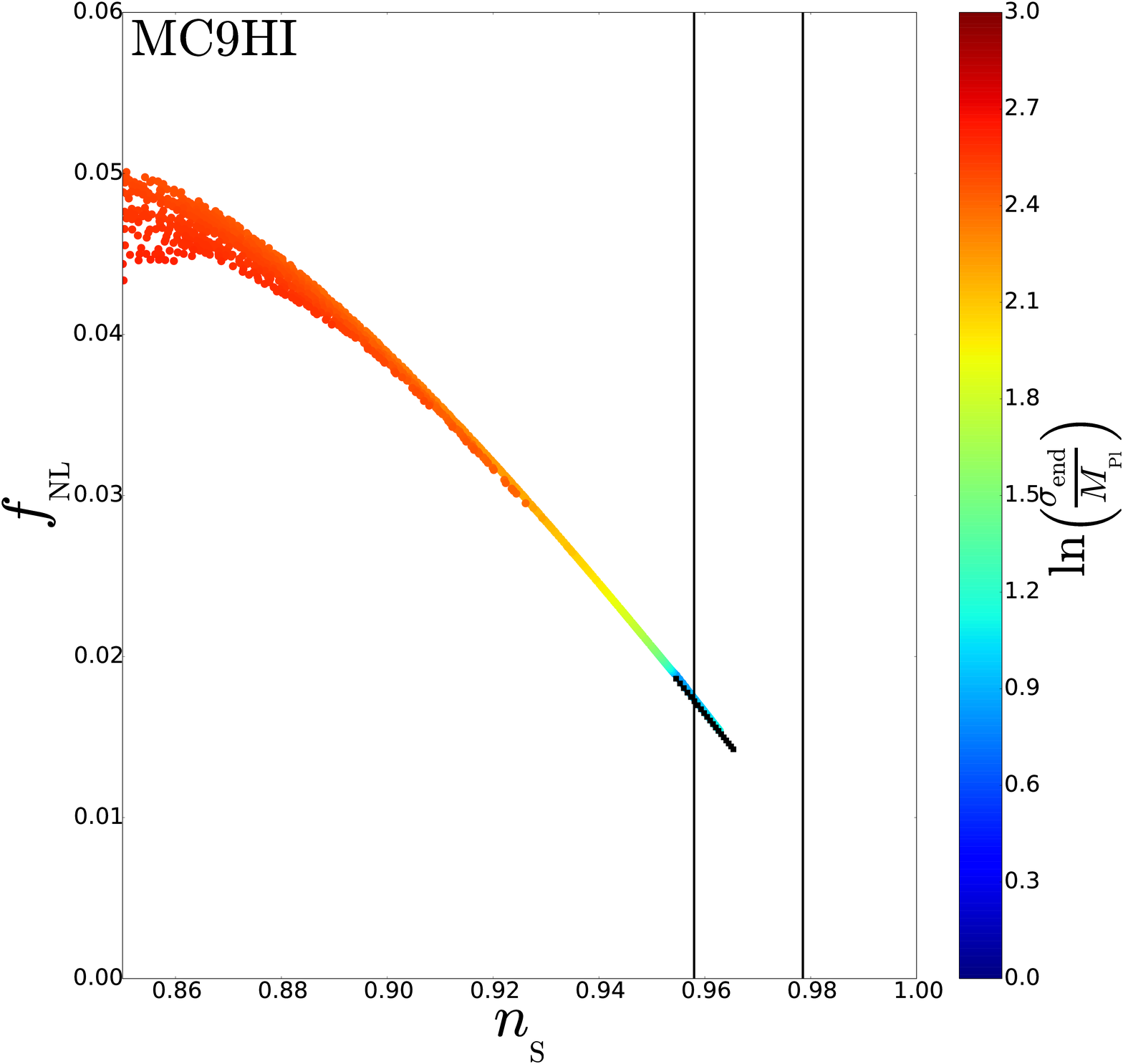}
\includegraphics[width=\wappfig,clip=true]{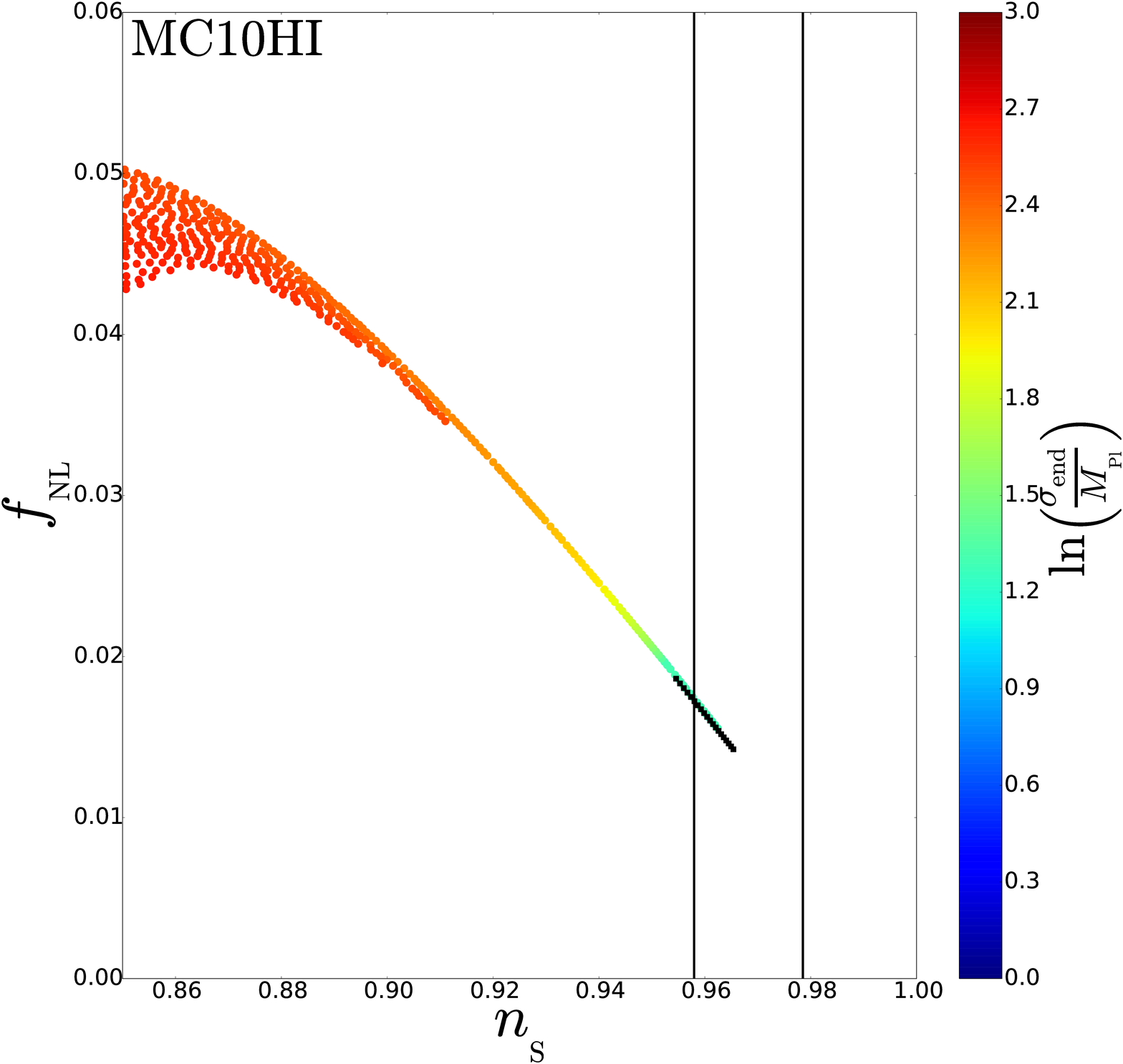}
\includegraphics[width=\wappfig,clip=true]{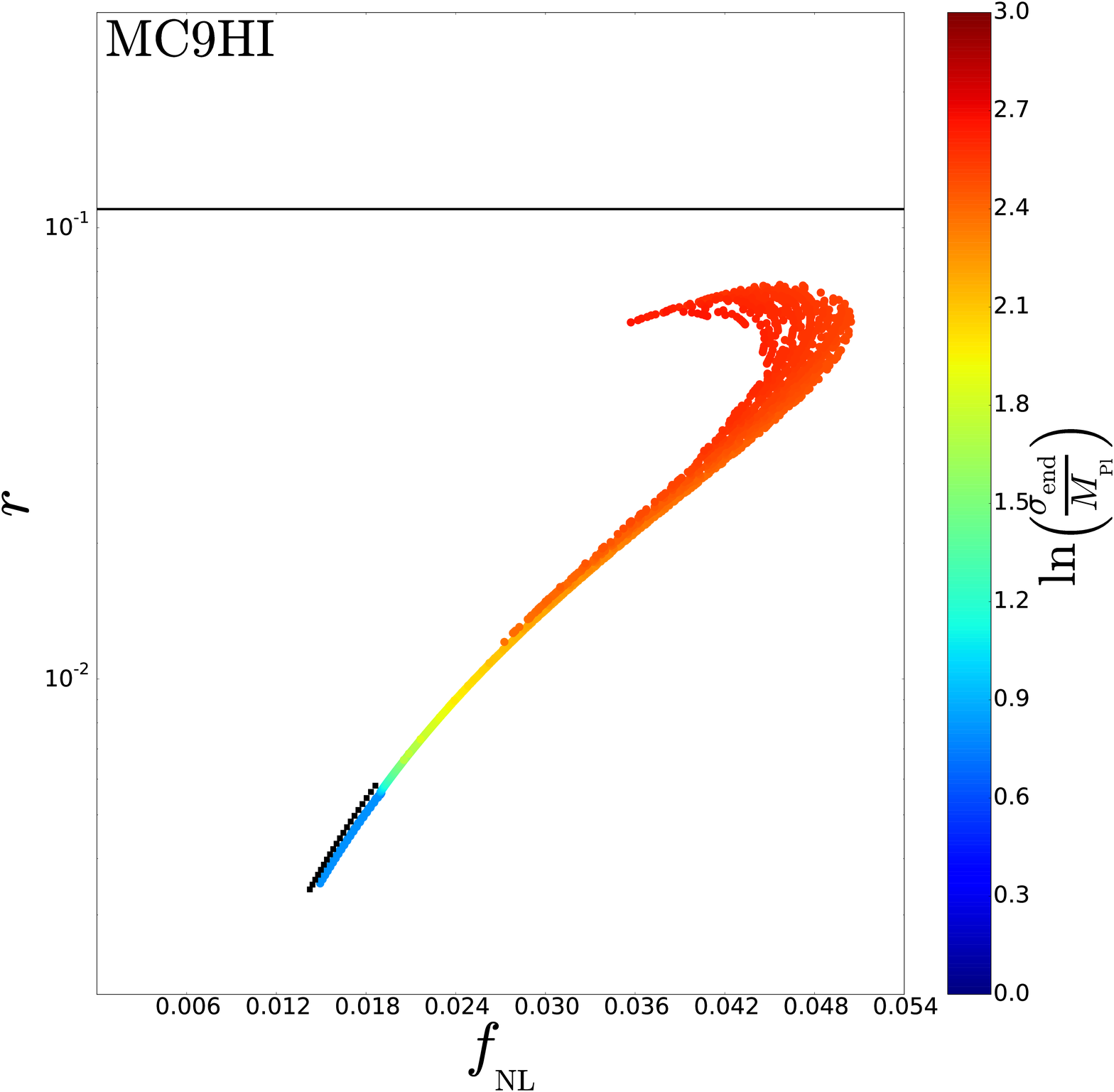}
\includegraphics[width=\wappfig,clip=true]{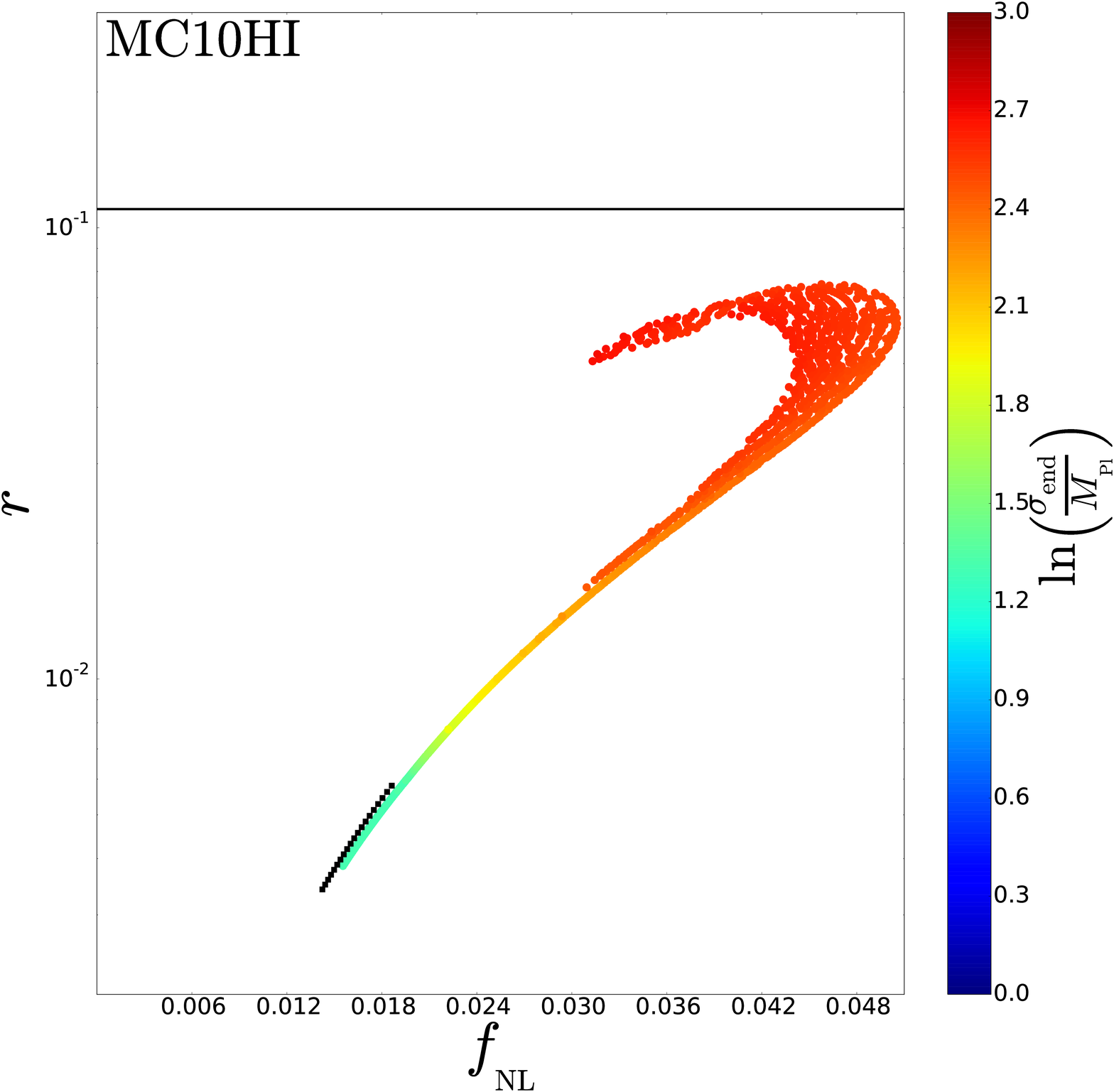}
\caption{Reheating consistent slow-roll predictions for the Higgs inflation models with a massive curvaton field, when reheating scenario is of the ninth (left panels) and tenth (right panels) type.}
\label{fig:CMBMCHI910}
\end{center}
\end{figure}
\clearpage
\subsection{Natural Inflation + Massive Scalar Field (MCNI)}
\label{sec:plots:ni}
\begin{figure}[!ht]
\begin{center}
\includegraphics[width=\wappfig,clip=true]{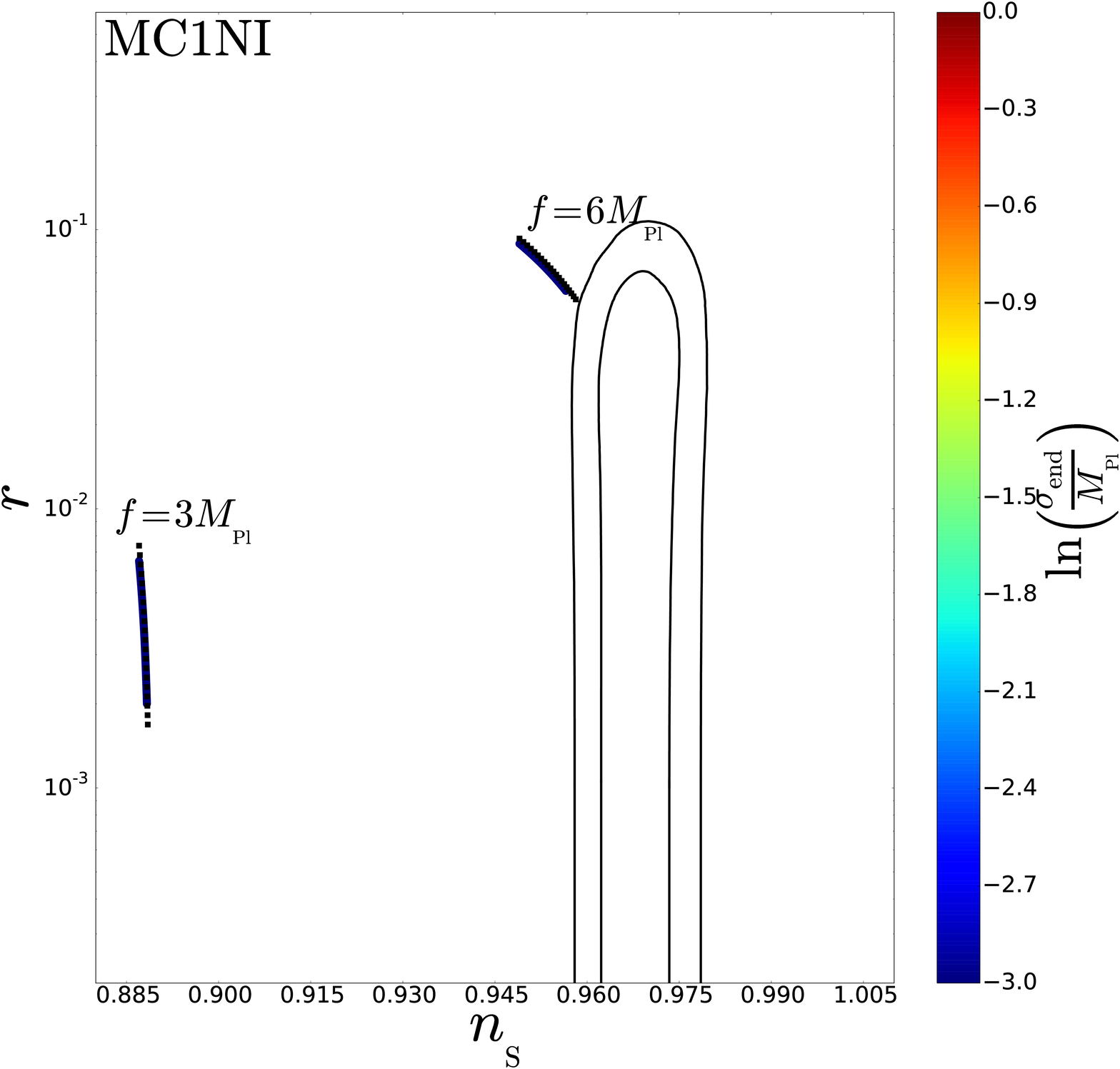}
\includegraphics[width=\wappfig,clip=true]{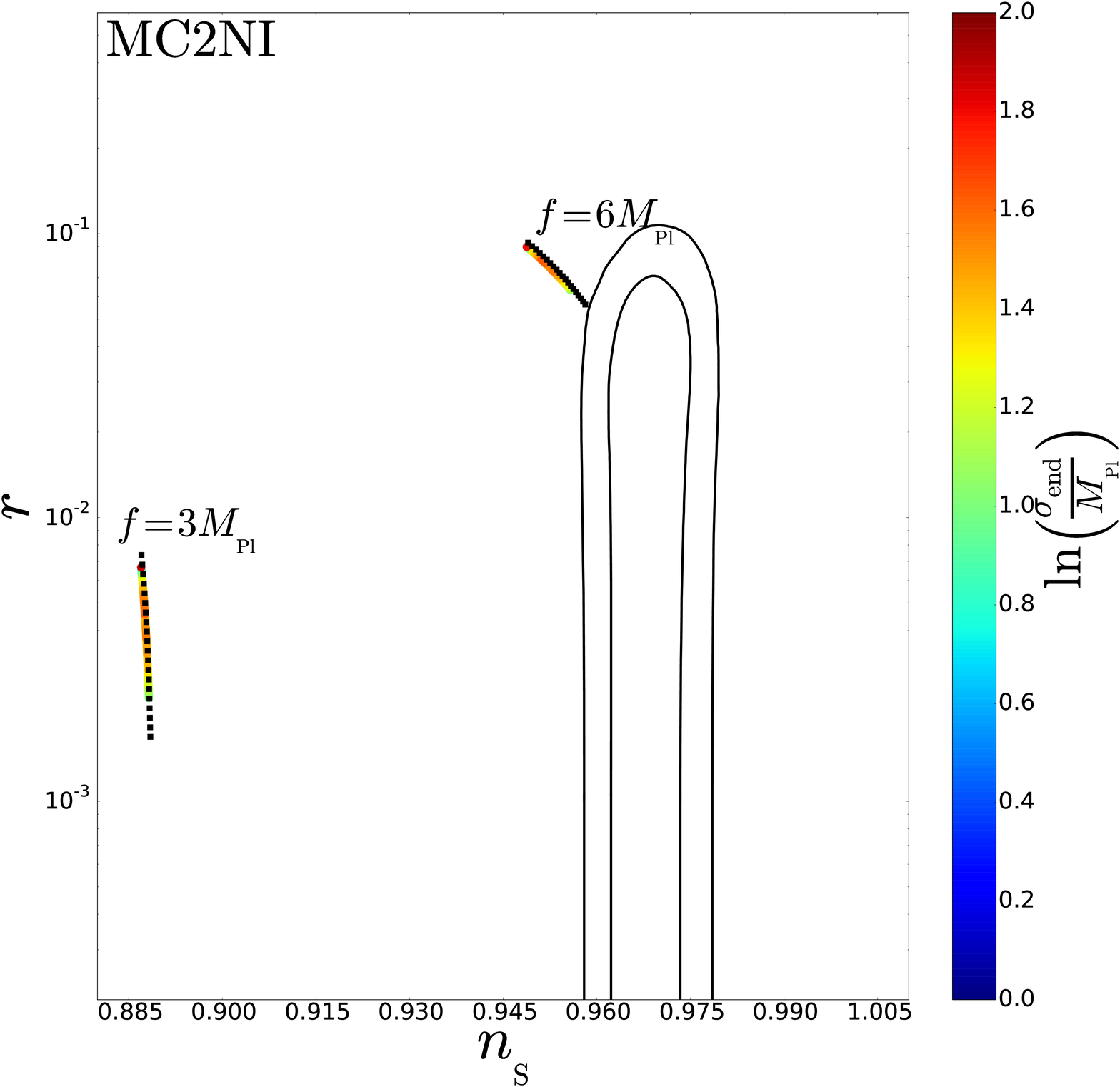}
\includegraphics[width=\wappfig,clip=true]{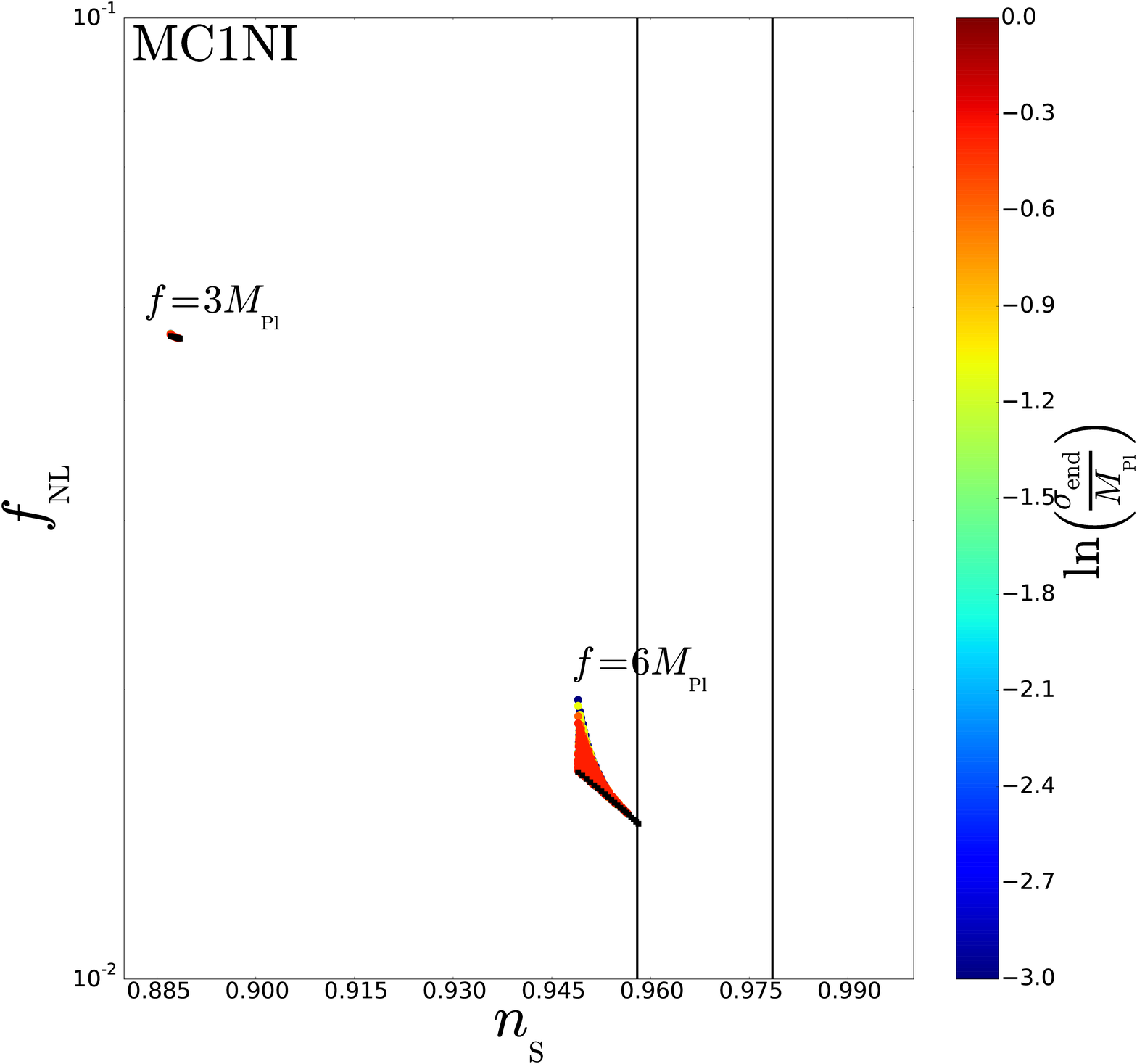}
\includegraphics[width=\wappfig,clip=true]{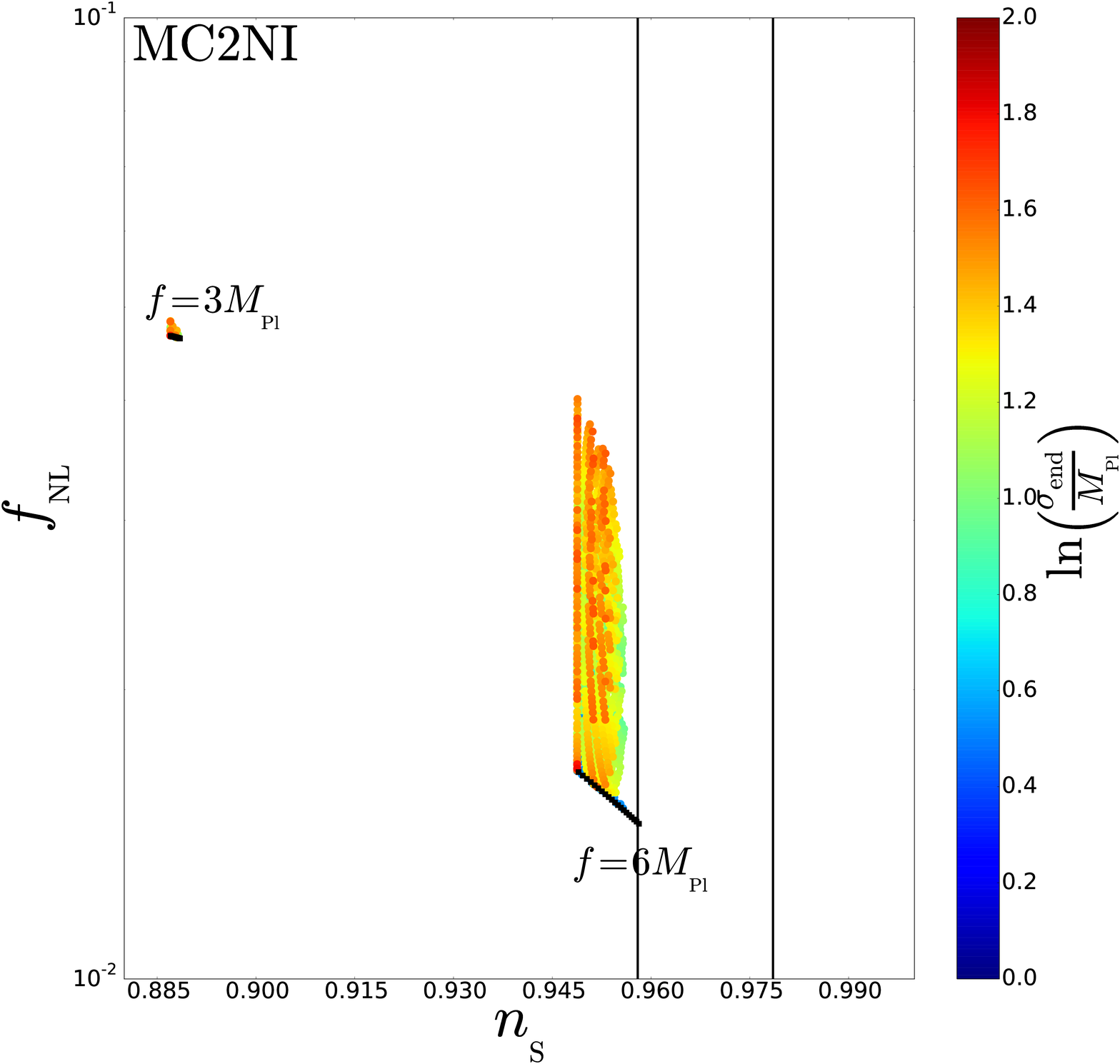}
\includegraphics[width=\wappfig,clip=true]{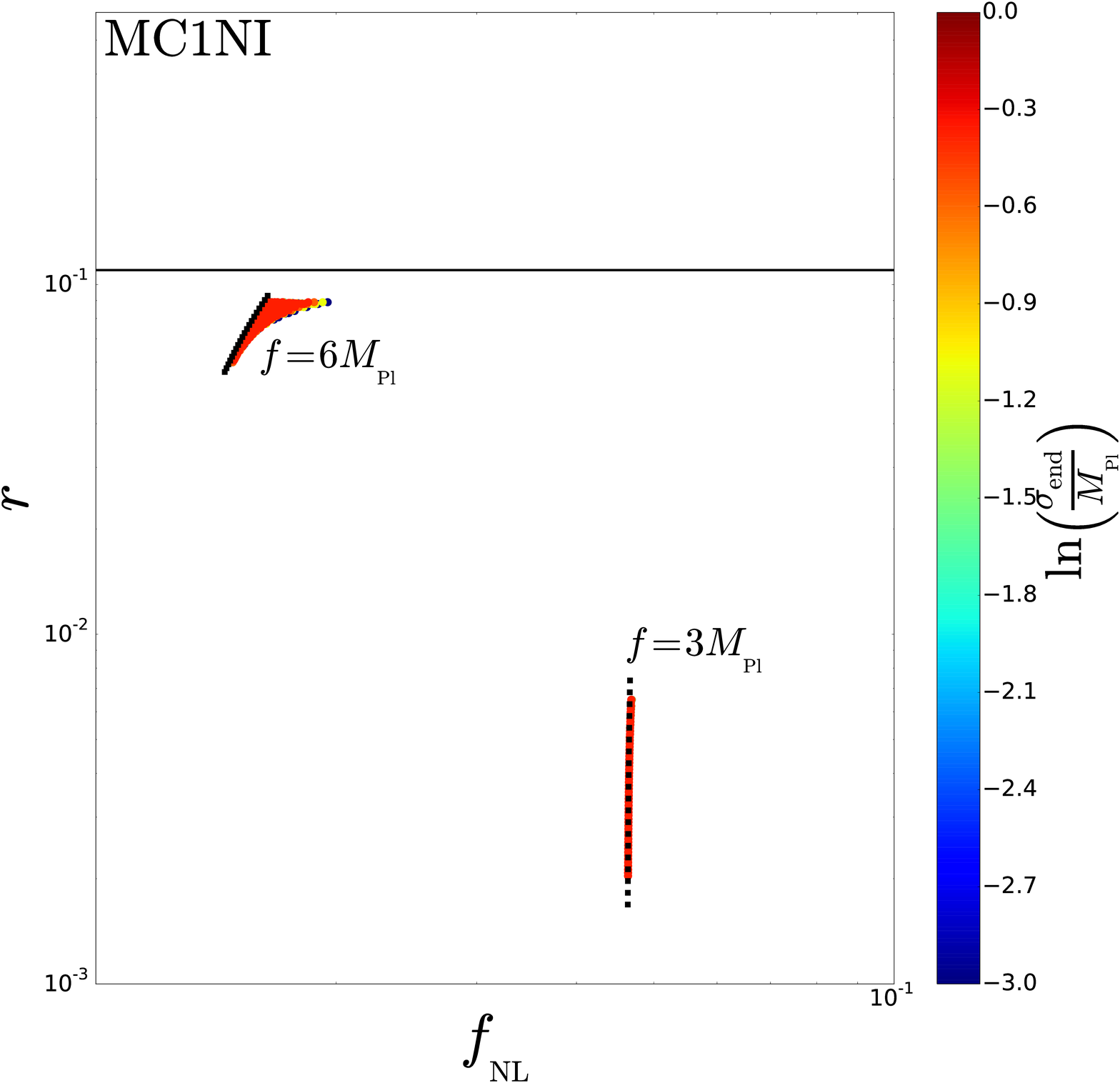}
\includegraphics[width=\wappfig,clip=true]{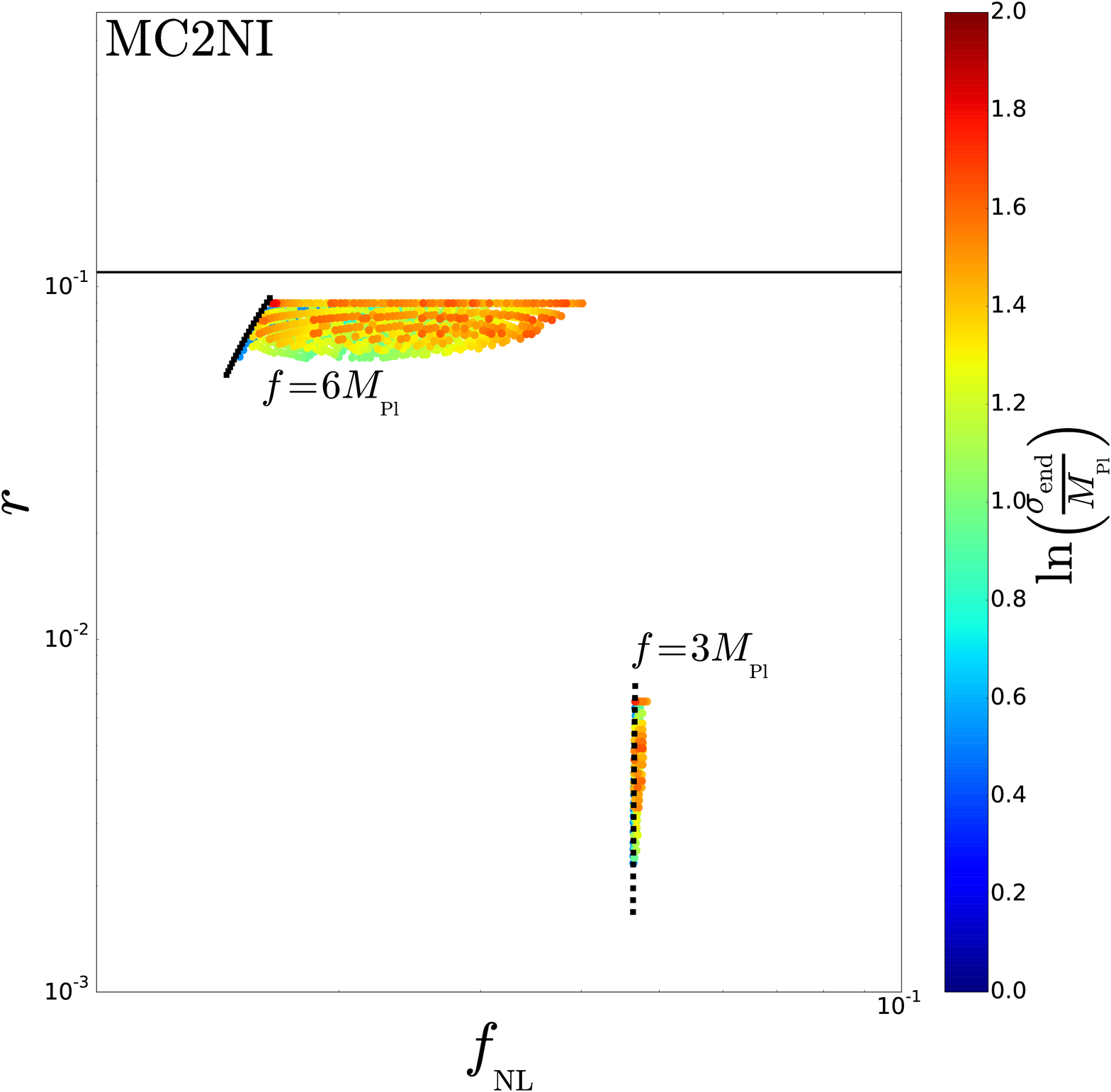}
\caption{Reheating consistent slow-roll predictions for the natural inflation models with a massive curvaton field, when reheating scenario is of the first (left panels) and second (right panels) type.}
\label{fig:CMBMCNI12}
\end{center}
\end{figure}
\begin{figure}[!ht]
\begin{center}
\includegraphics[width=\wappfig,clip=true]{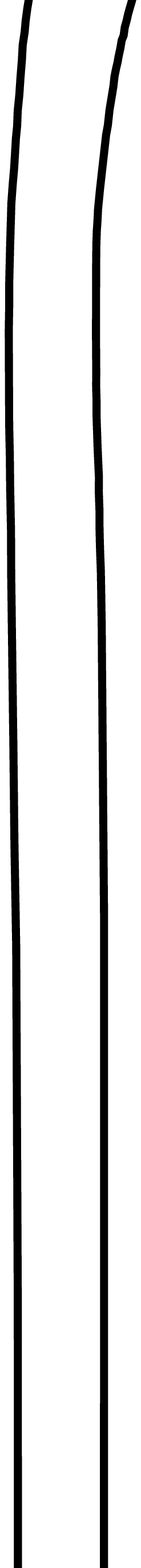}
\includegraphics[width=\wappfig,clip=true]{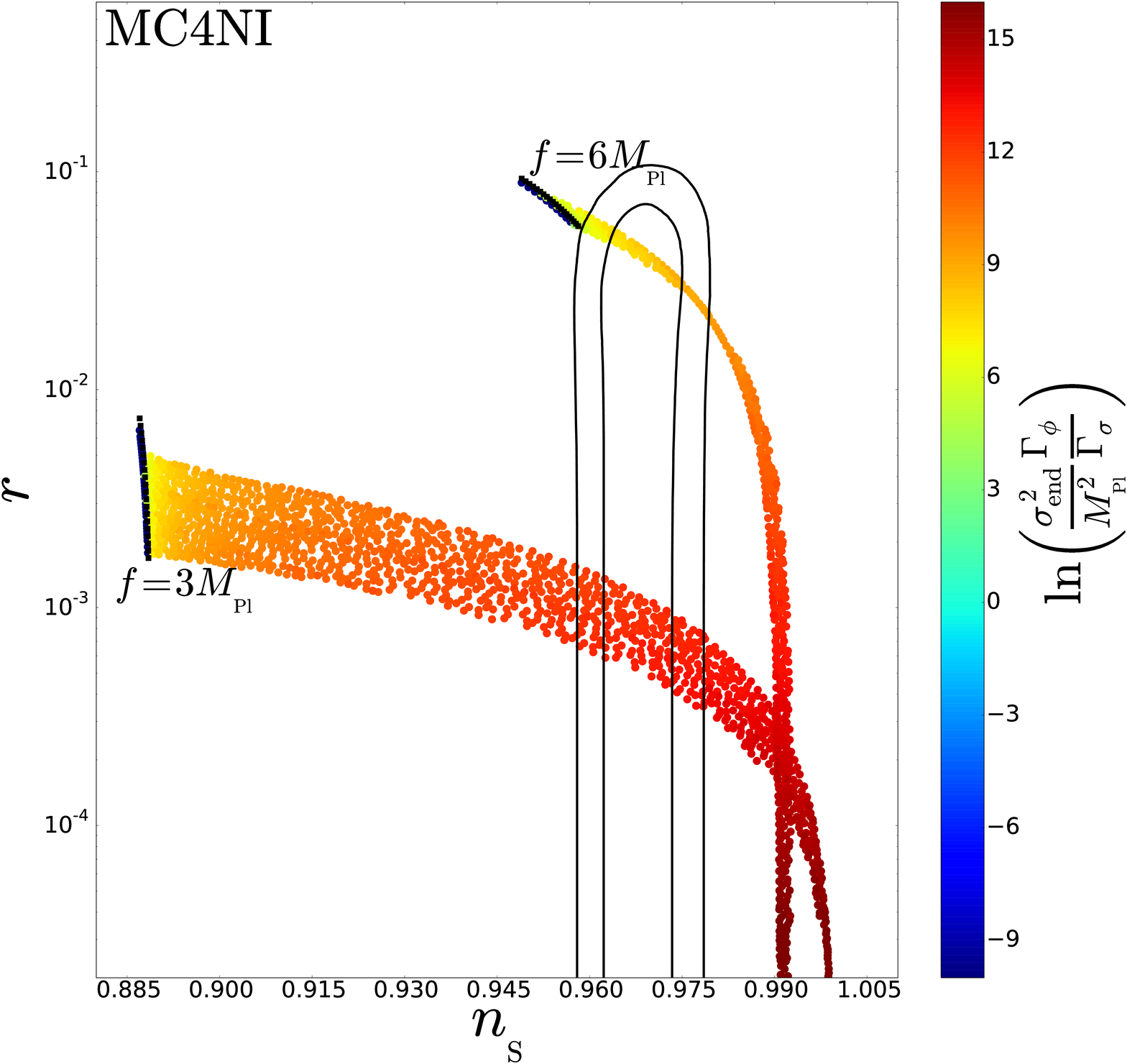}
\includegraphics[width=\wappfig,clip=true]{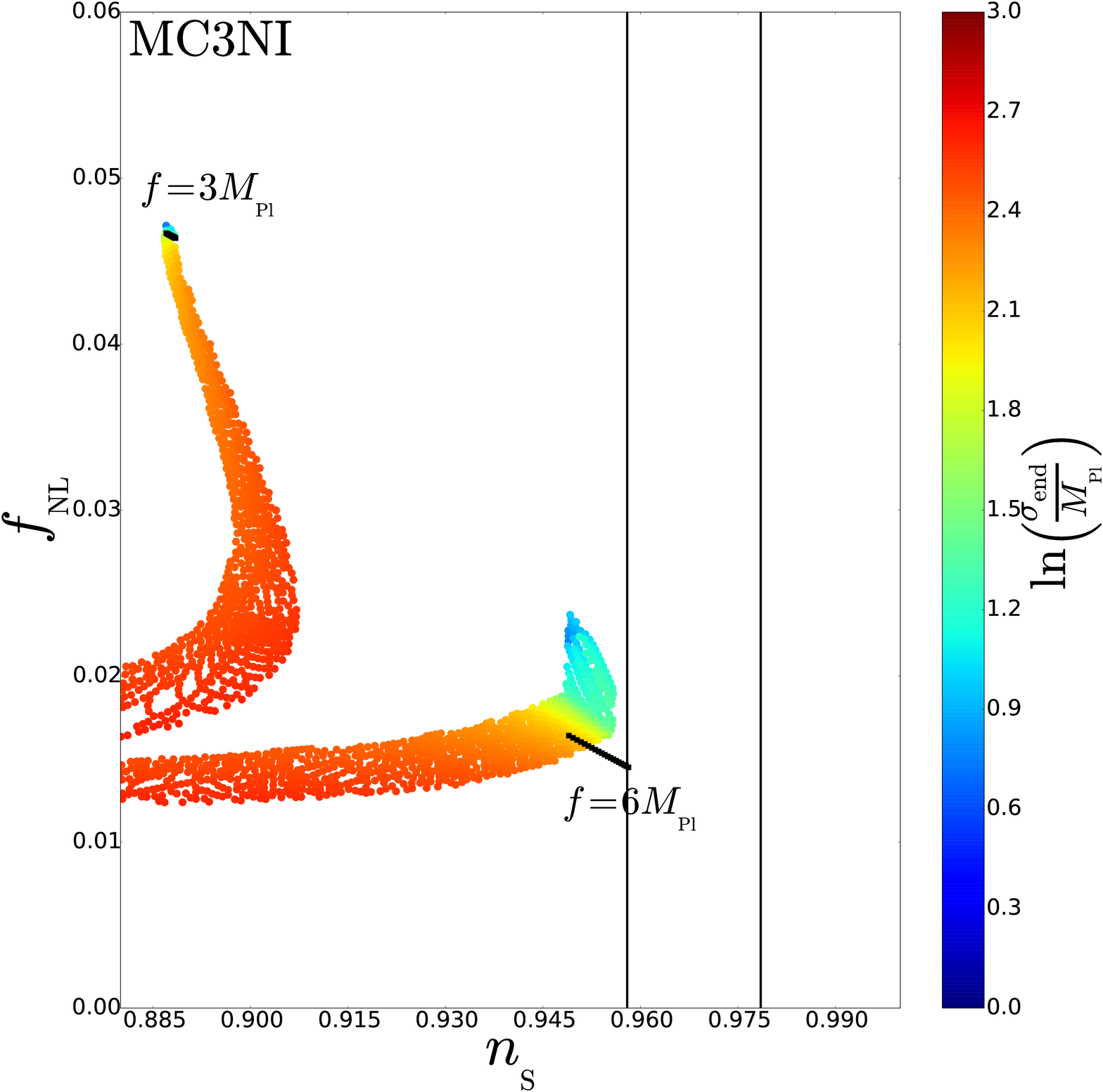}
\includegraphics[width=\wappfig,clip=true]{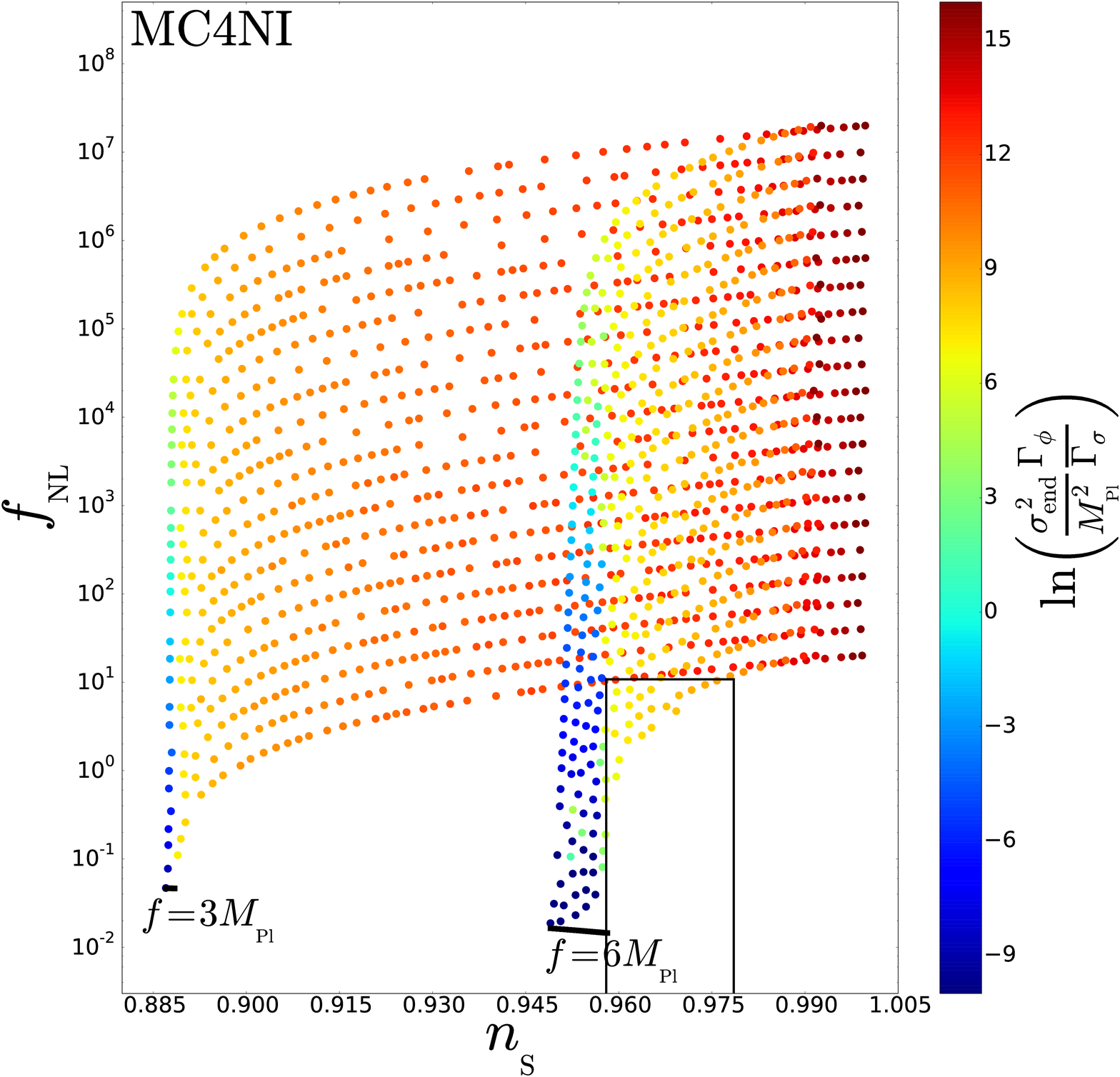}
\includegraphics[width=\wappfig,clip=true]{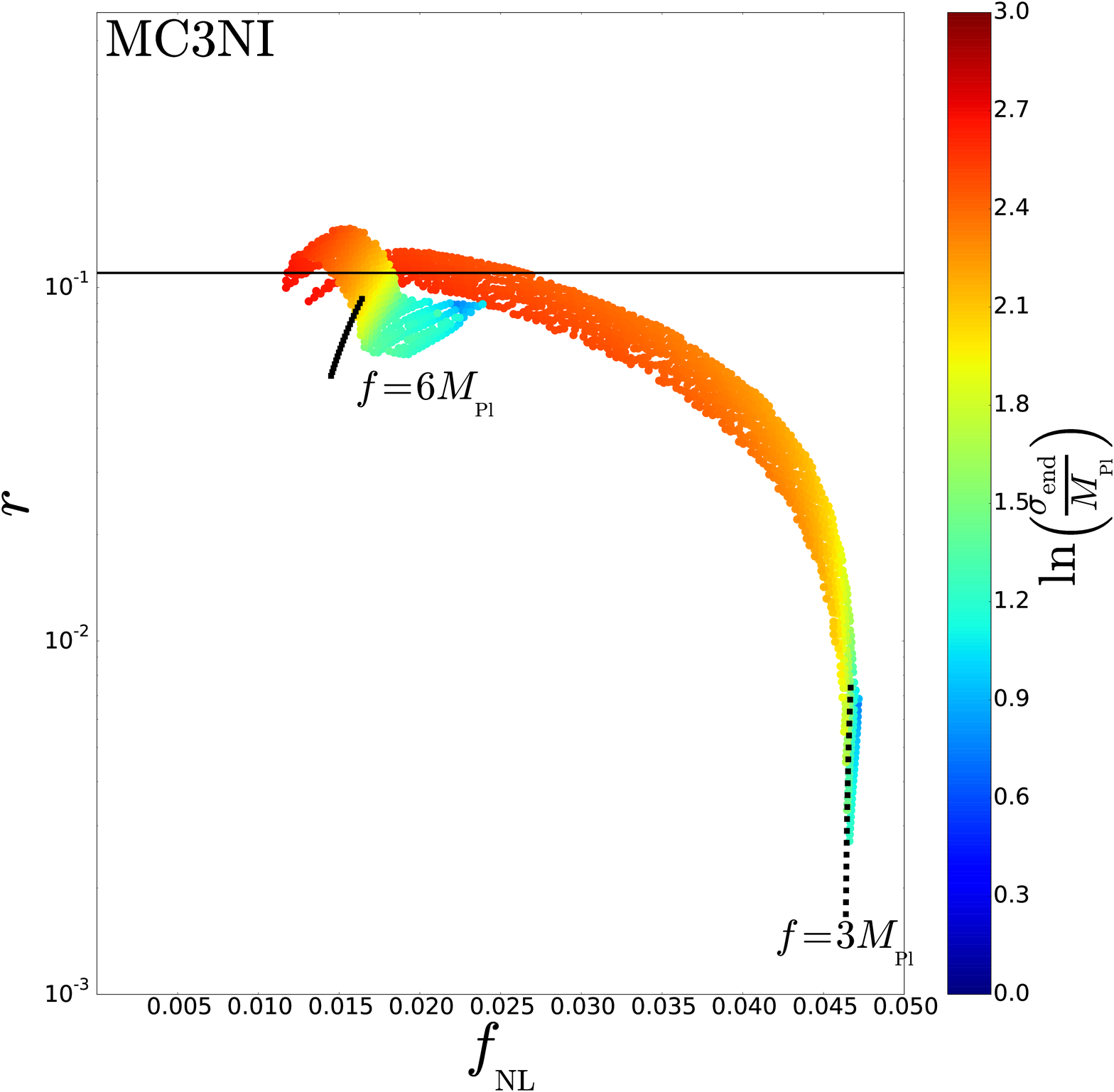}
\includegraphics[width=\wappfig,clip=true]{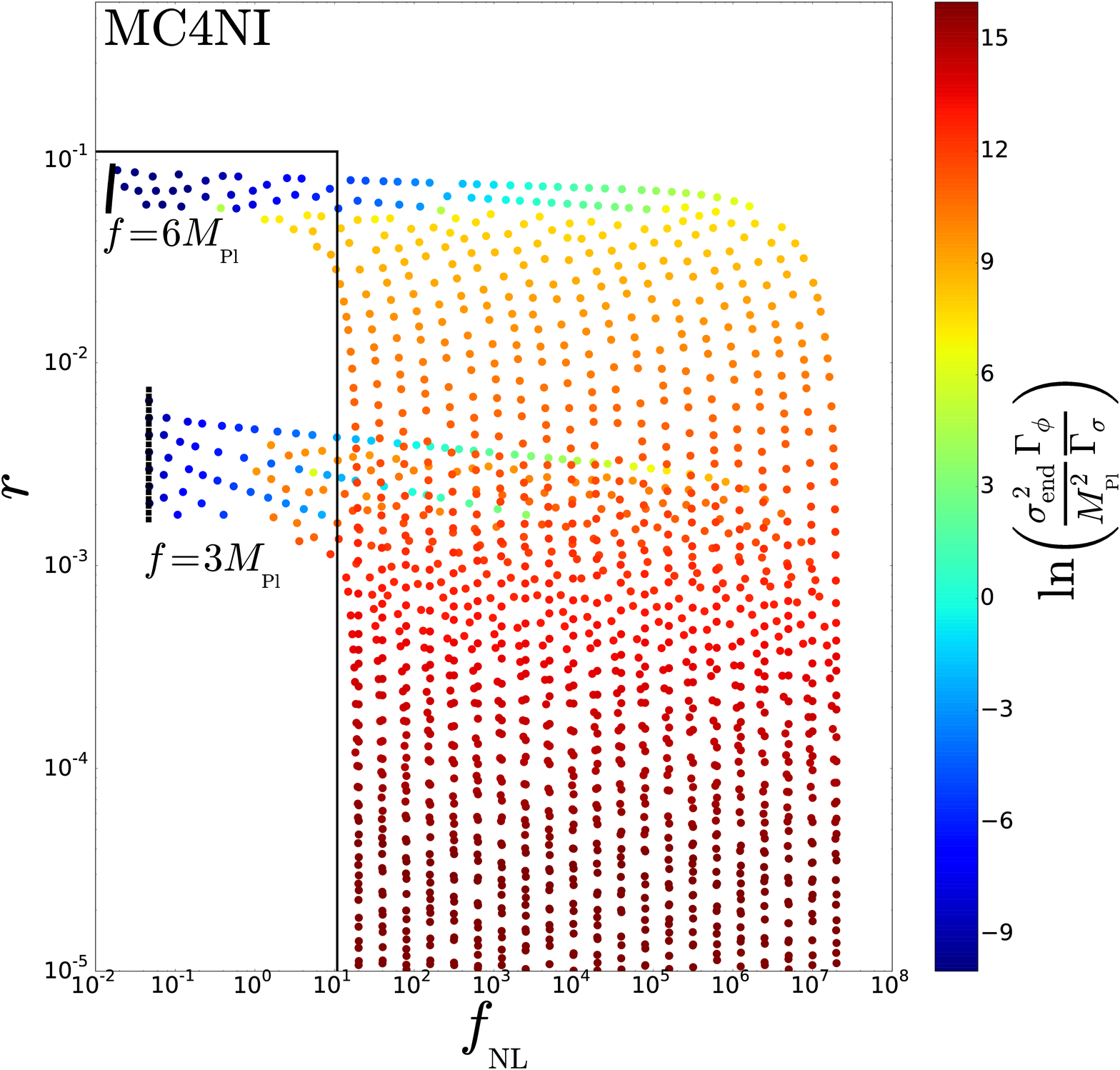}
\caption{Reheating consistent slow-roll predictions for the natural inflation models with a massive curvaton field, when reheating scenario is of the third (left panels) and fourth (right panels) type.}
\label{fig:CMBMCNI34}
\end{center}
\end{figure}
\begin{figure}[!ht]
\begin{center}
\includegraphics[width=\wappfig,clip=true]{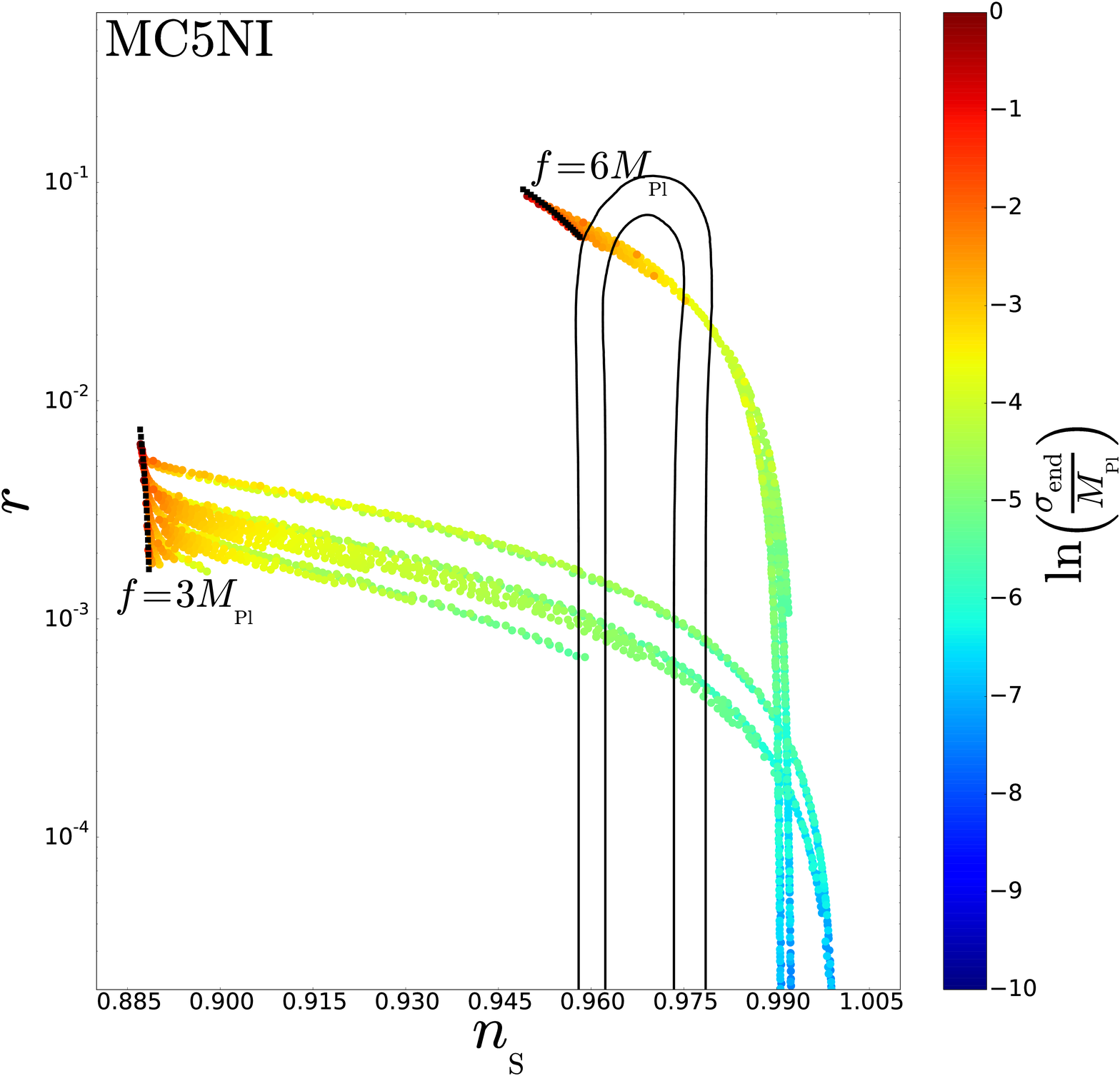}
\includegraphics[width=\wappfig,clip=true]{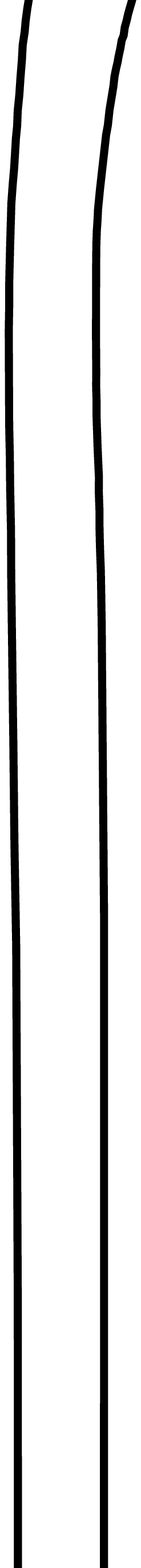}
\includegraphics[width=\wappfig,clip=true]{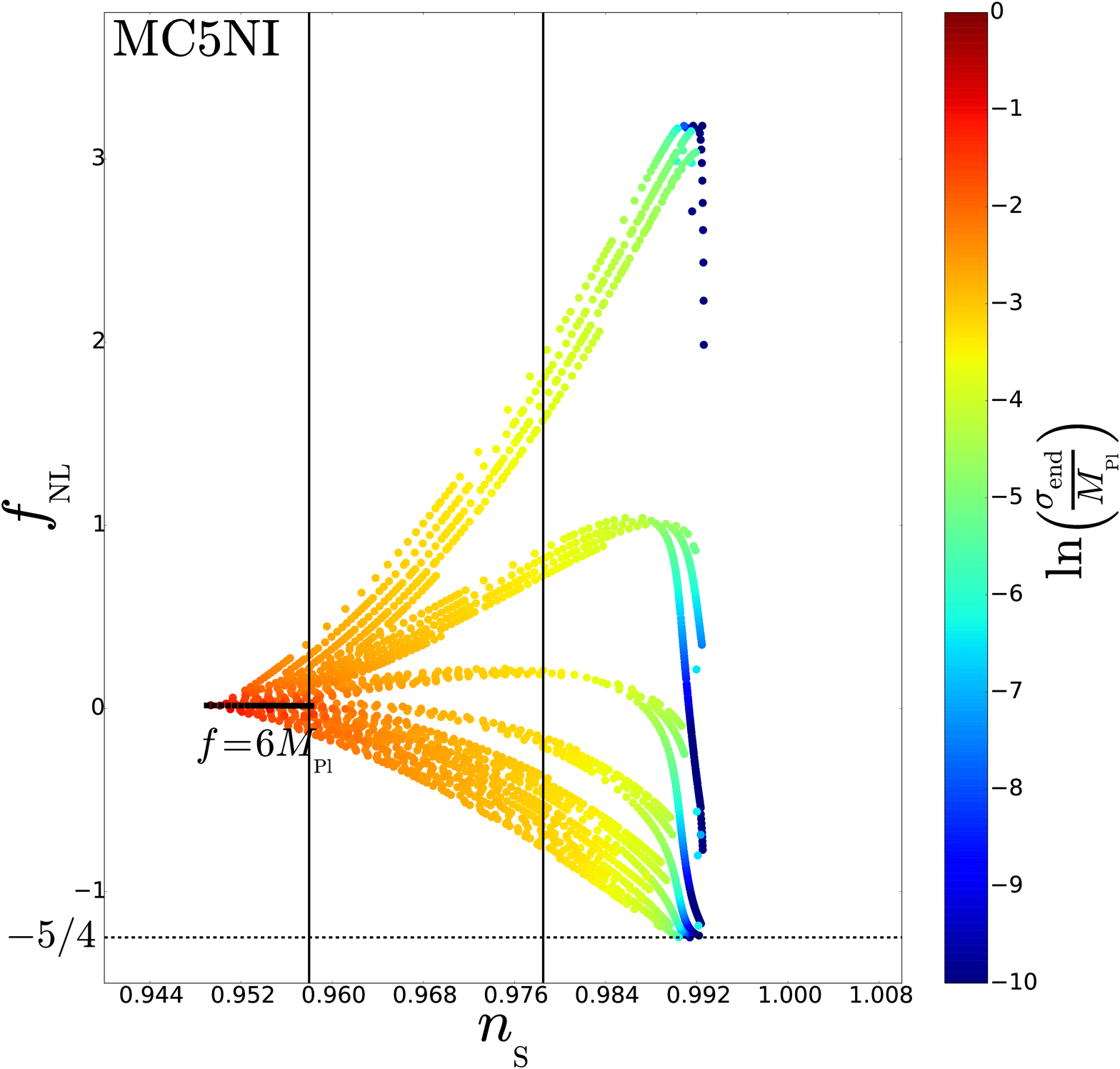}
\includegraphics[width=\wappfig,clip=true]{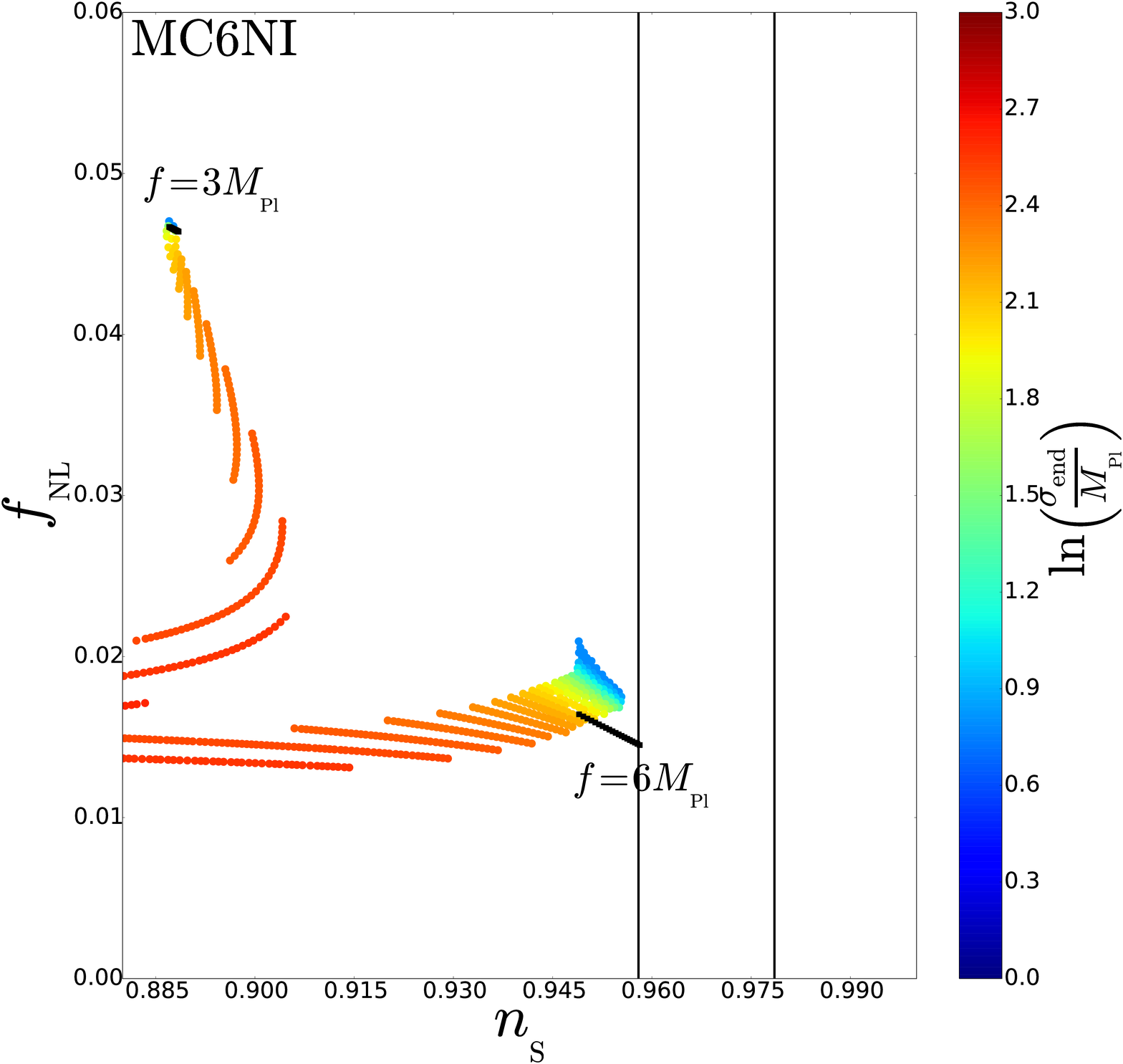}
\includegraphics[width=\wappfig,clip=true]{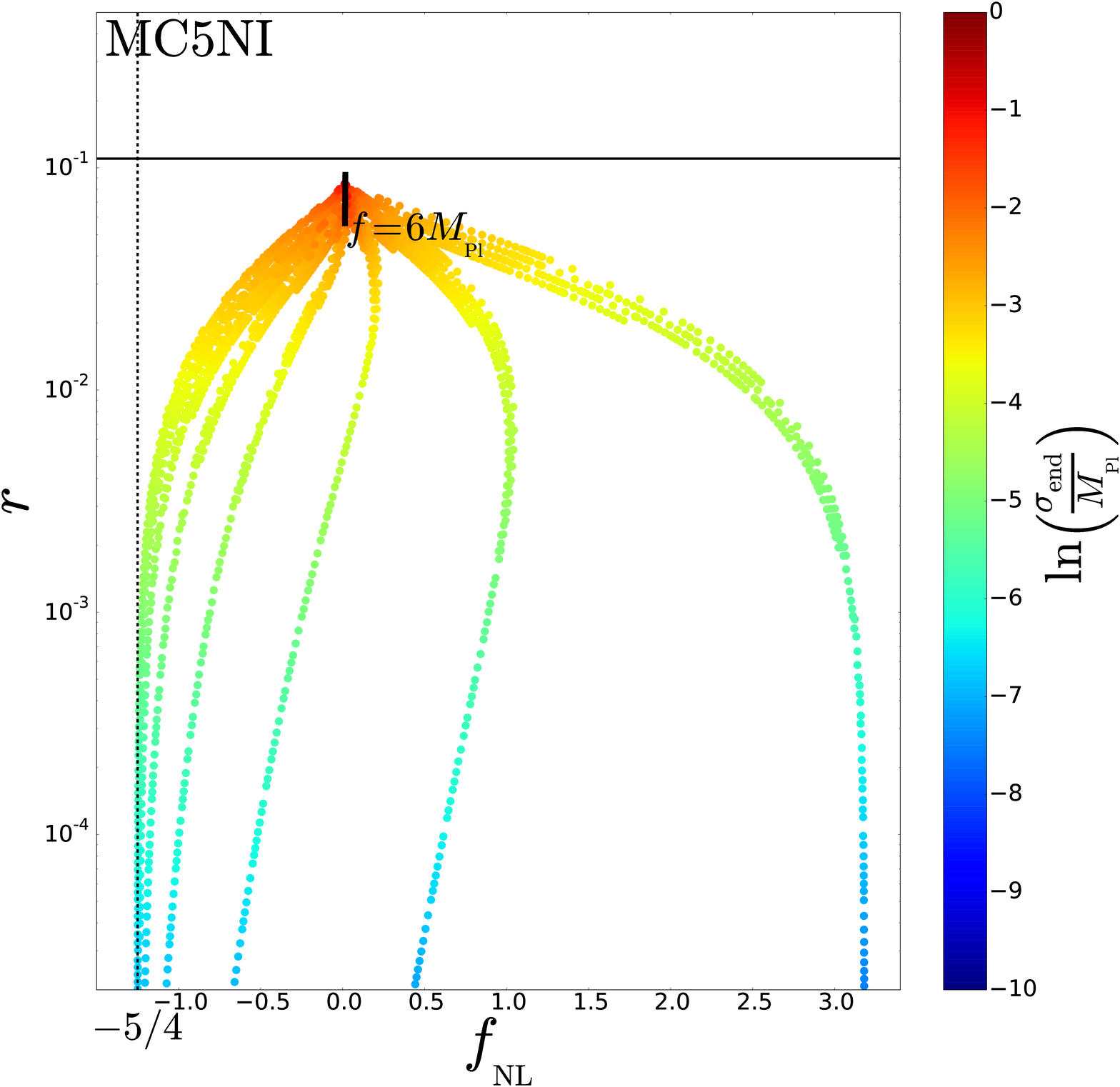}
\includegraphics[width=\wappfig,clip=true]{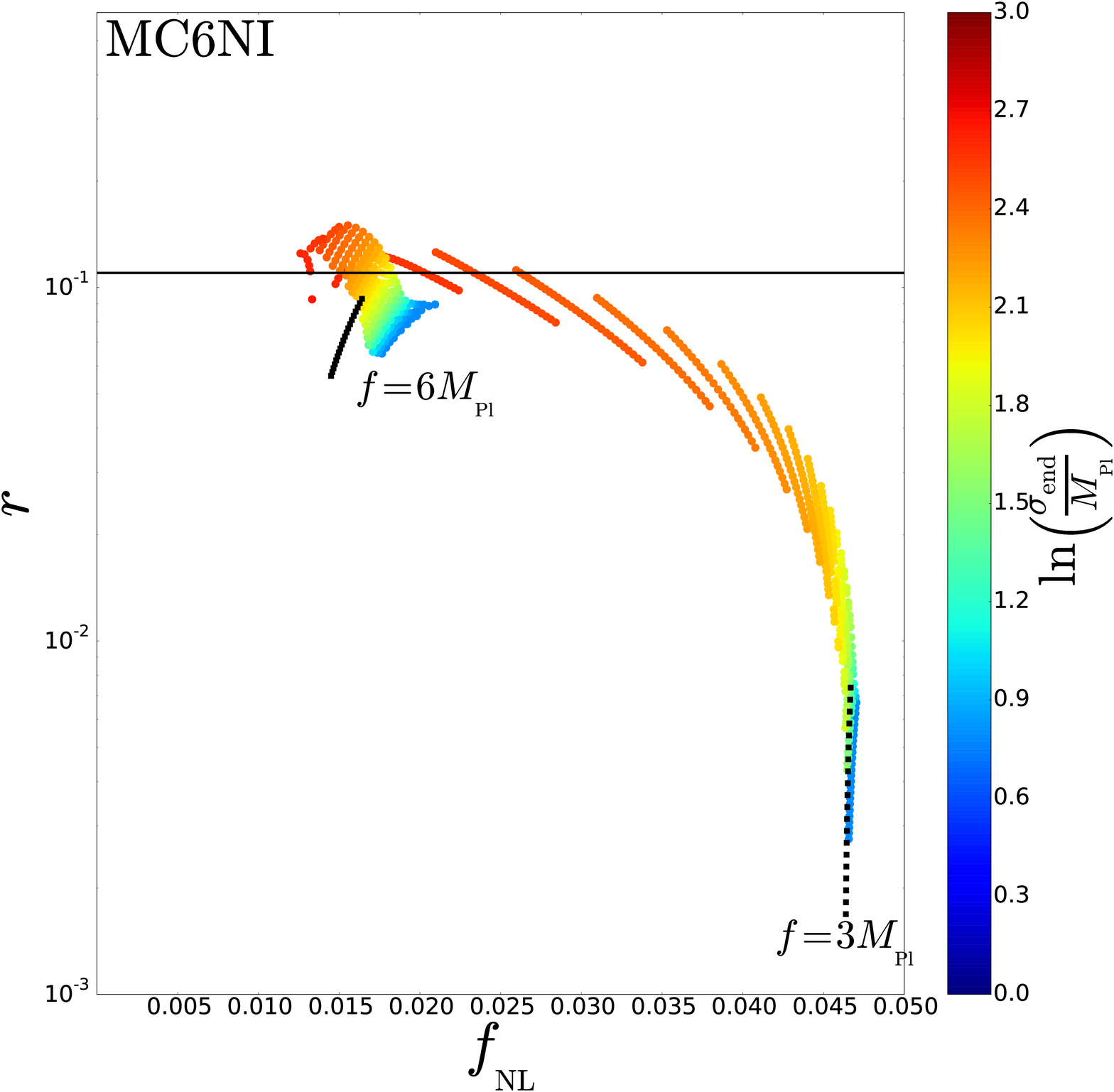}
\caption{Reheating consistent slow-roll predictions for the natural inflation models with a massive curvaton field, when reheating scenario is of the fifth (left panels) and sixth (right panels) type.}
\label{fig:CMBMCNI56}
\end{center}
\end{figure}
\begin{figure}[!ht]
\begin{center}
\includegraphics[width=\wappfig,clip=true]{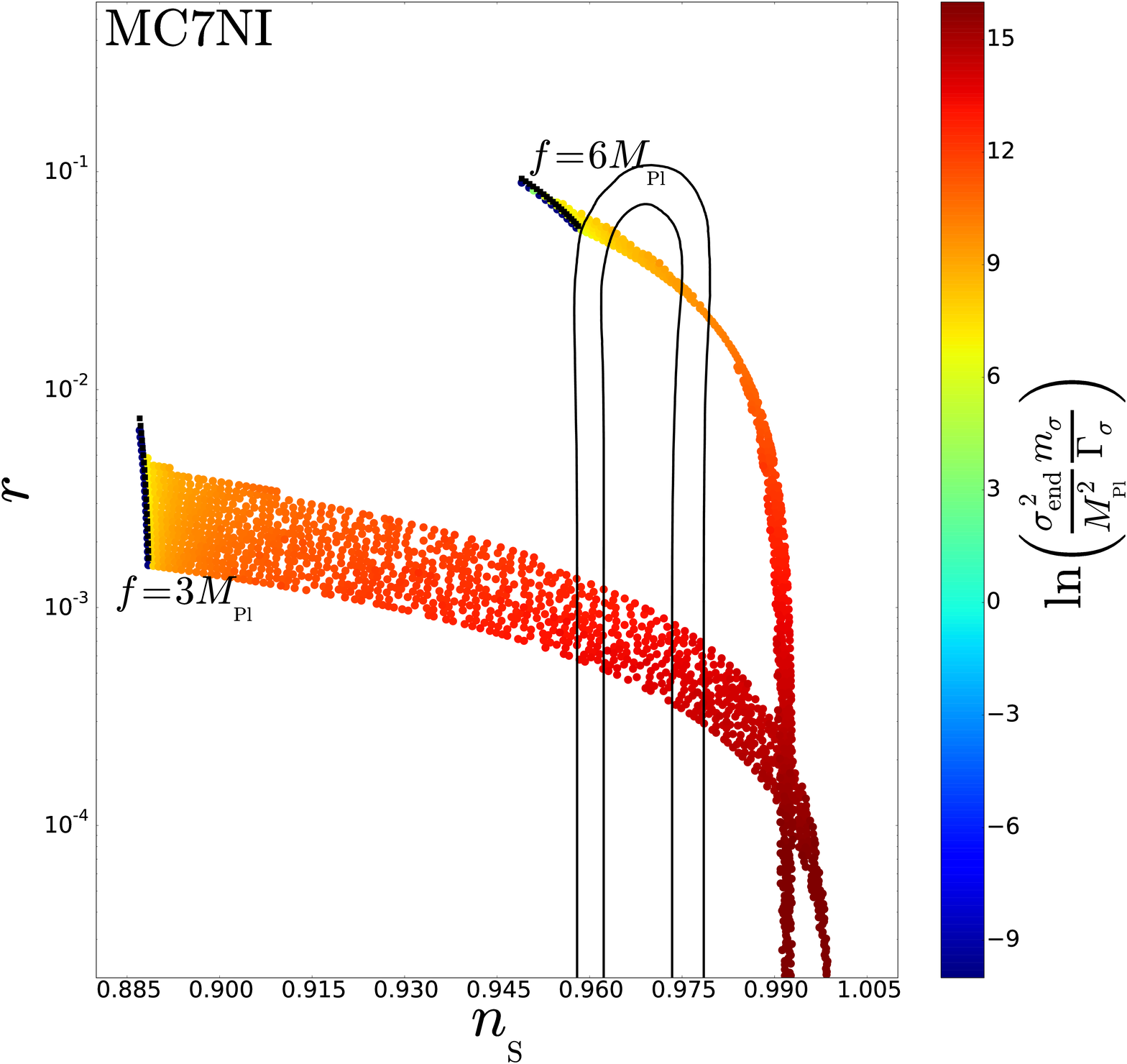}
\includegraphics[width=\wappfig,clip=true]{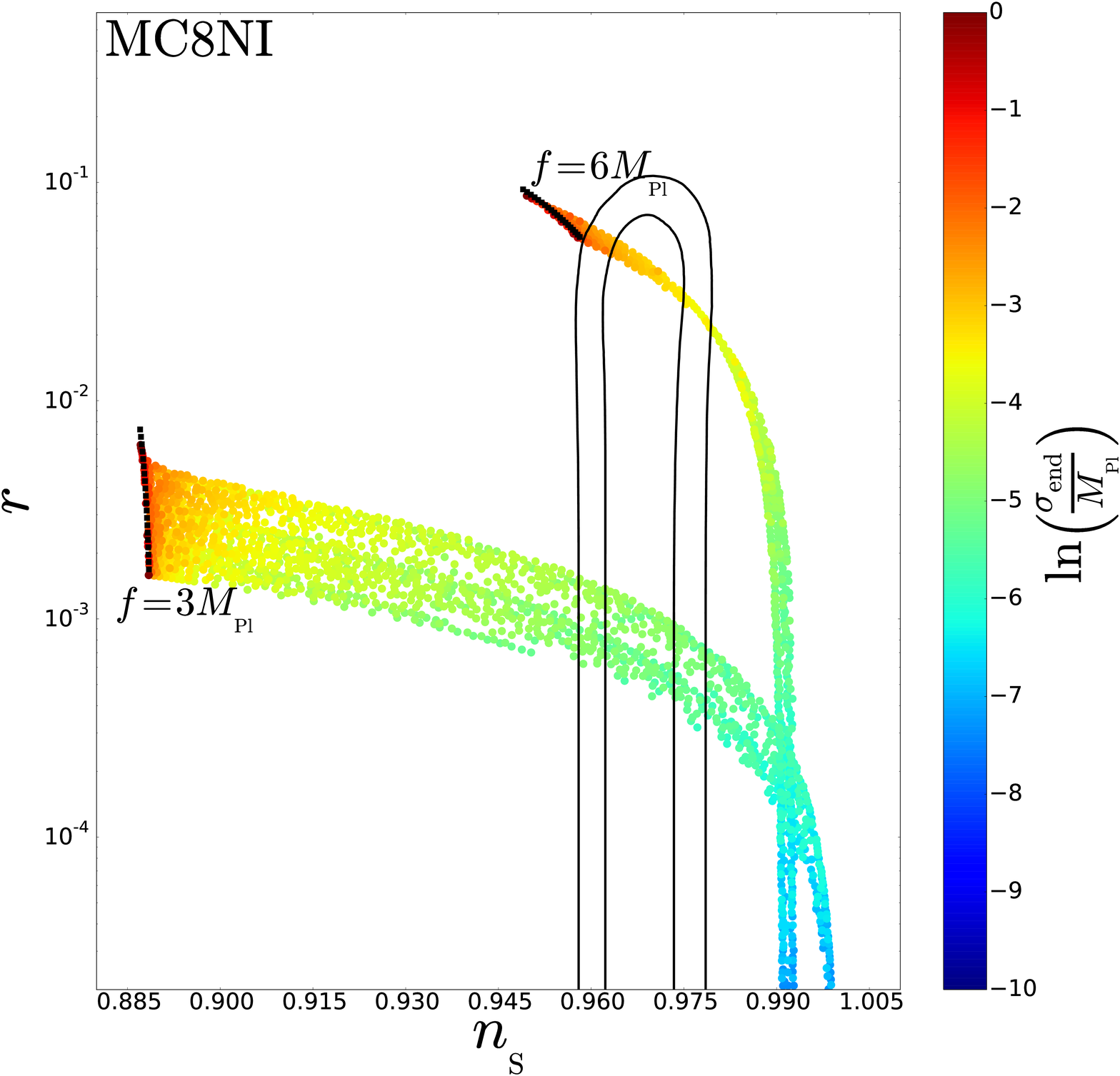}
\includegraphics[width=\wappfig,clip=true]{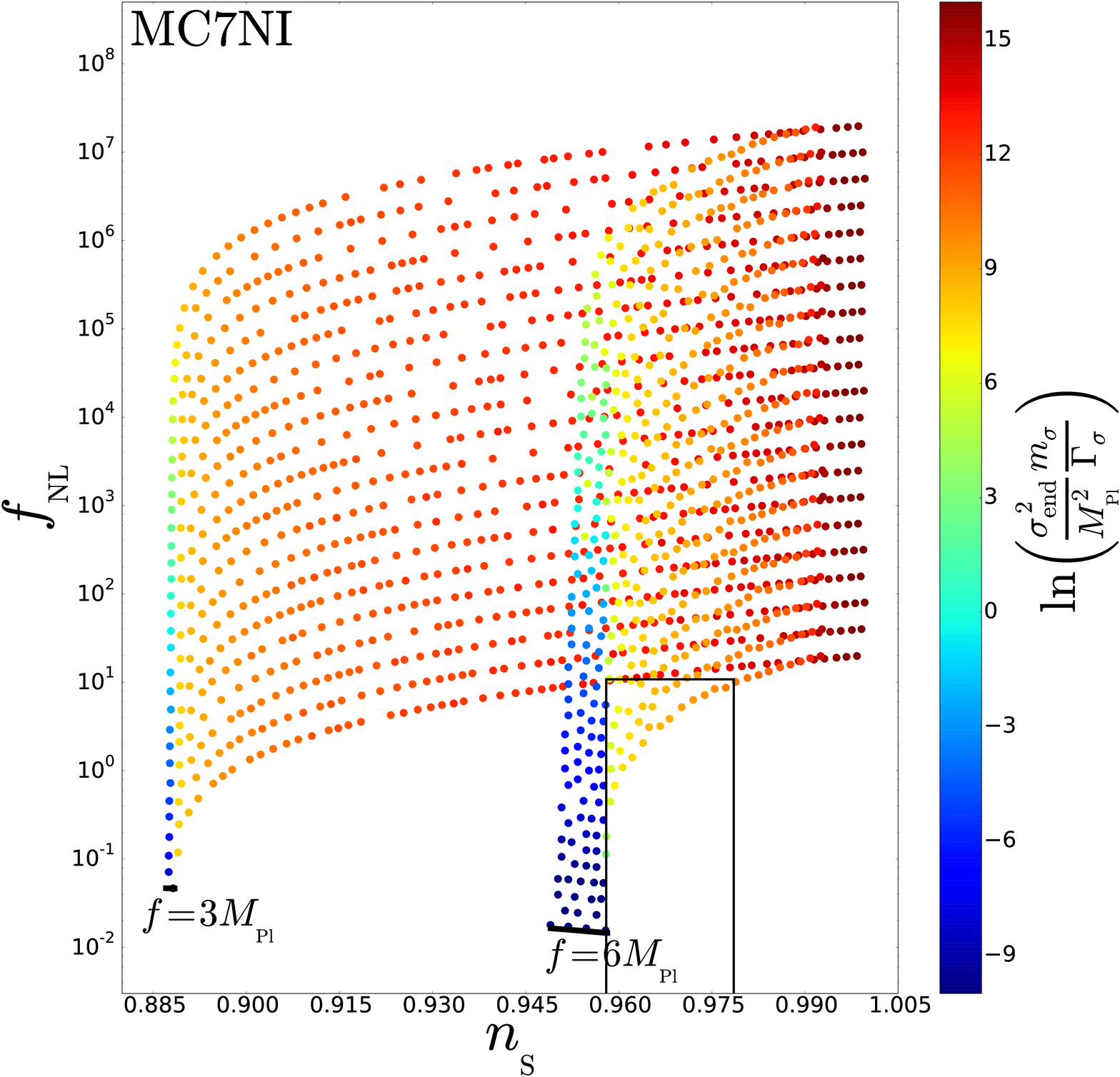}
\includegraphics[width=\wappfig,clip=true]{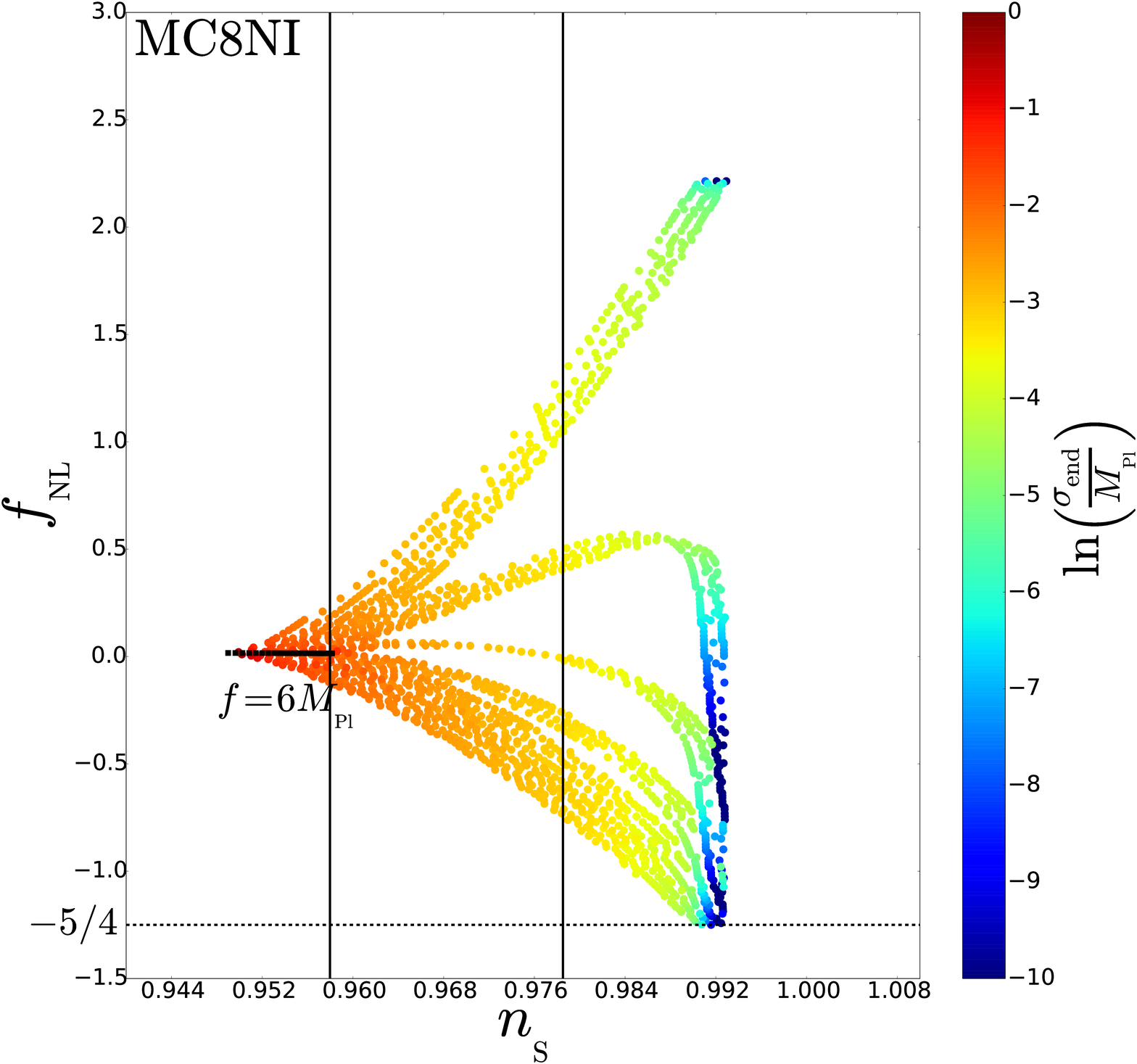}
\includegraphics[width=\wappfig,clip=true]{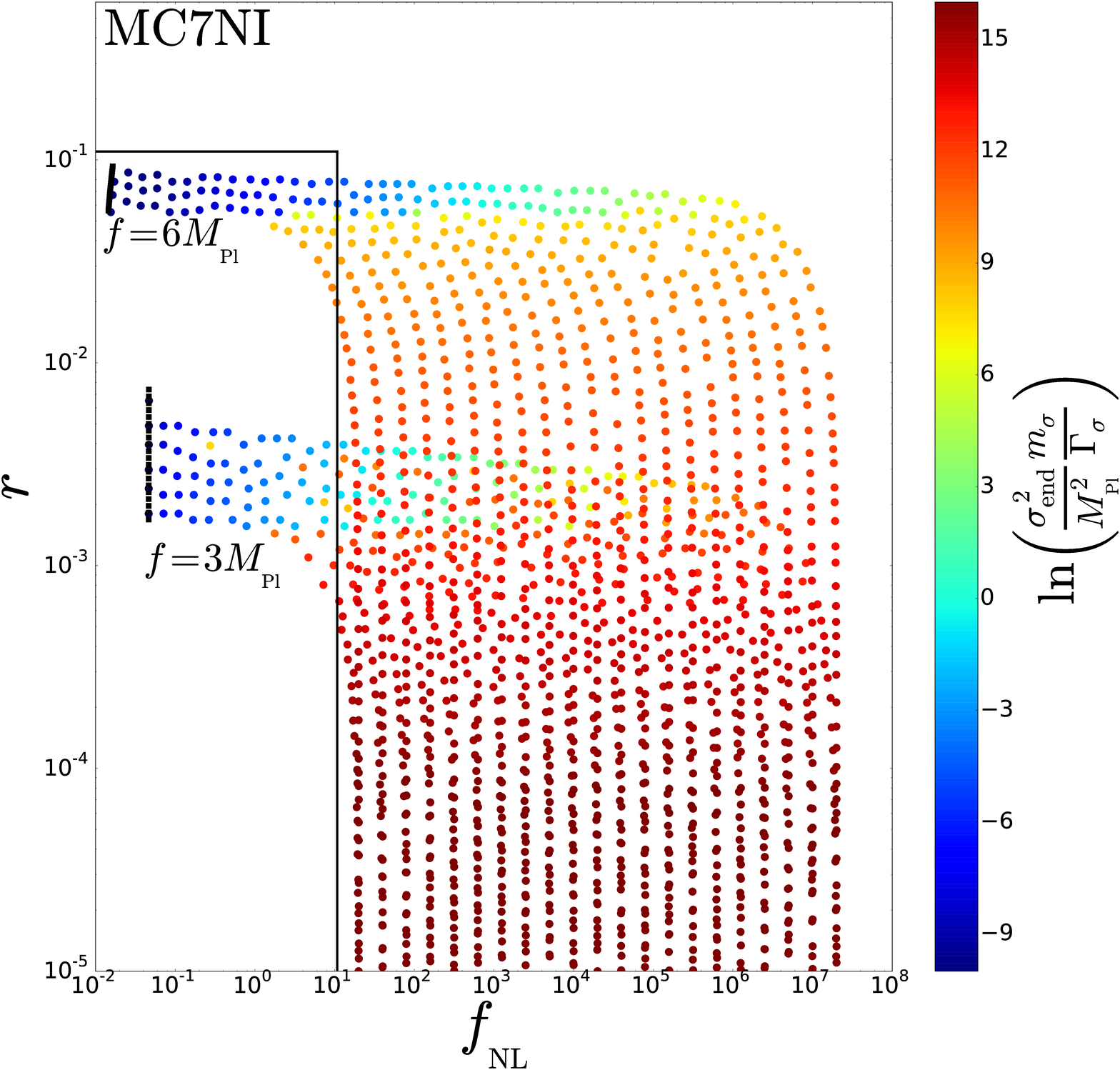}
\includegraphics[width=\wappfig,clip=true]{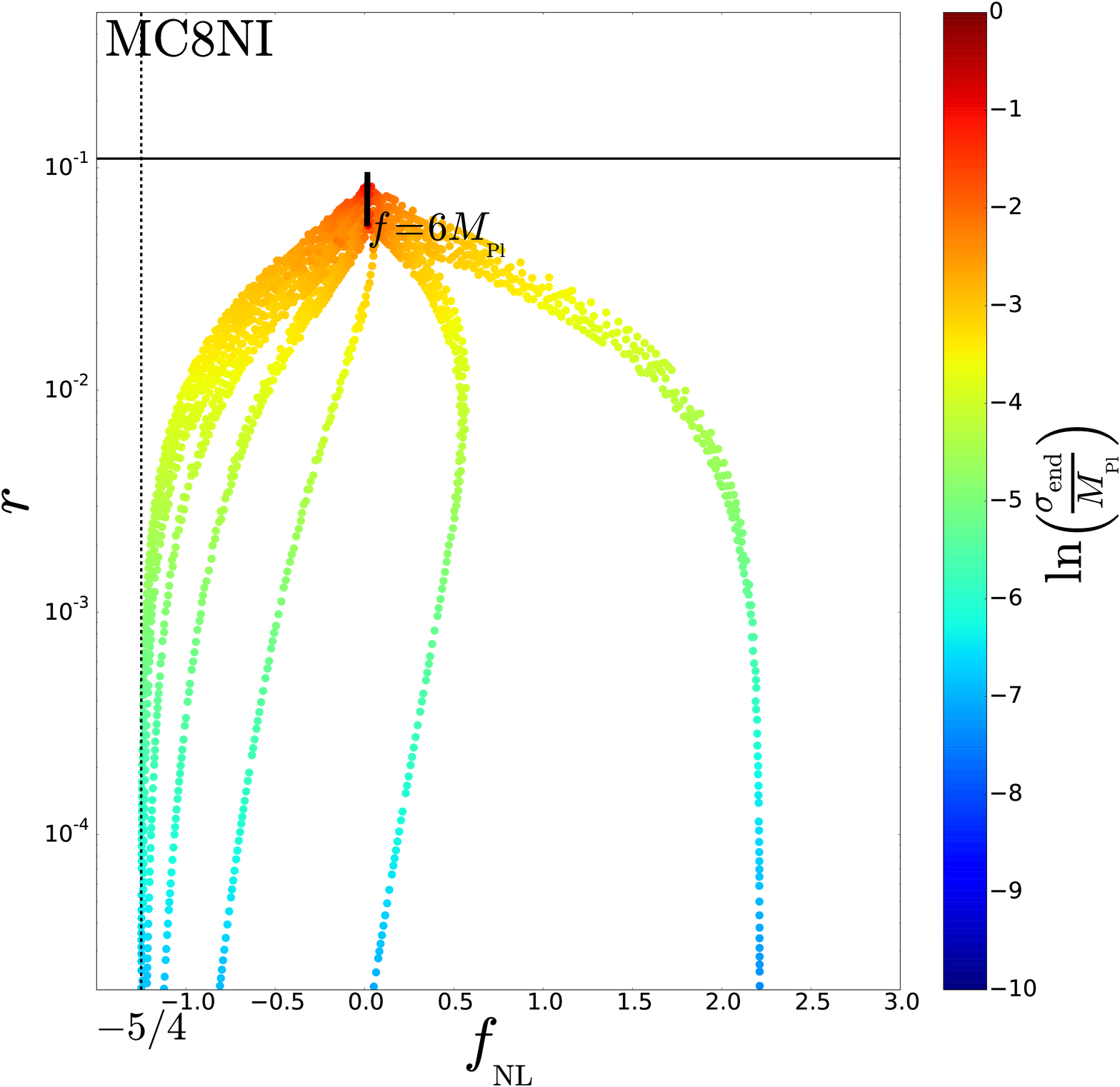}
\caption{Reheating consistent slow-roll predictions for the natural inflation models with a massive curvaton field, when reheating scenario is of the seventh (left panels) and eighth (right panels) type.}
\label{fig:CMBMCNI78}
\end{center}
\end{figure}
\begin{figure}[!ht]
\begin{center}
\includegraphics[width=\wappfig,clip=true]{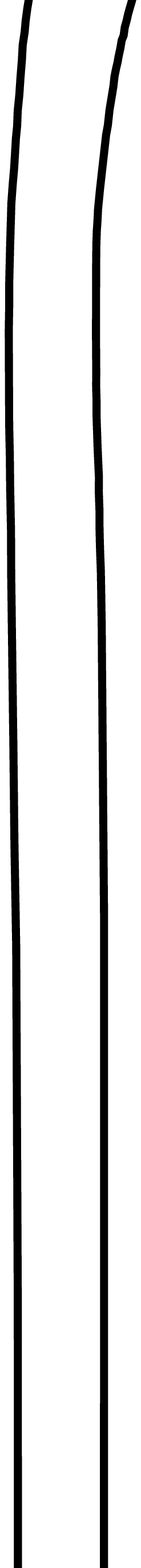}
\includegraphics[width=\wappfig,clip=true]{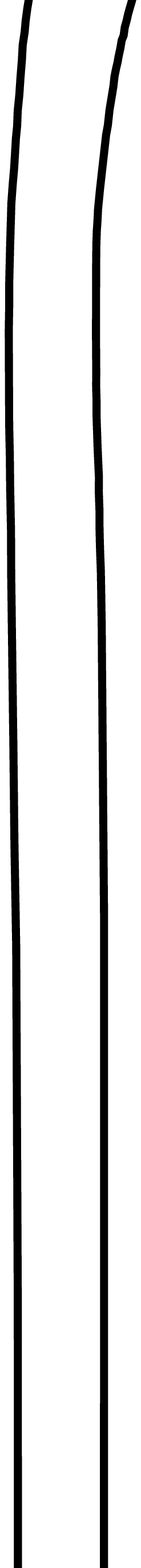}
\includegraphics[width=\wappfig,clip=true]{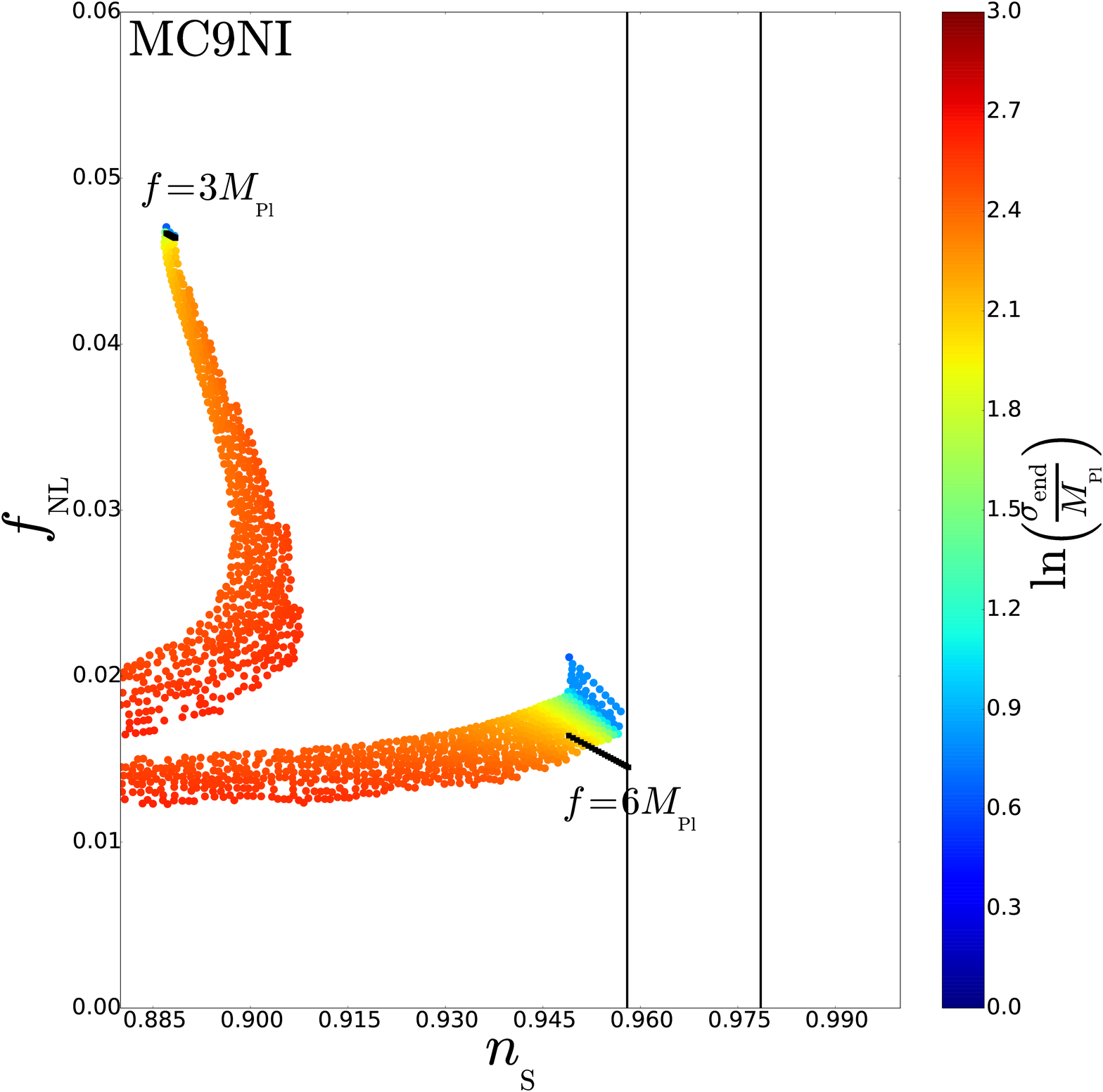}
\includegraphics[width=\wappfig,clip=true]{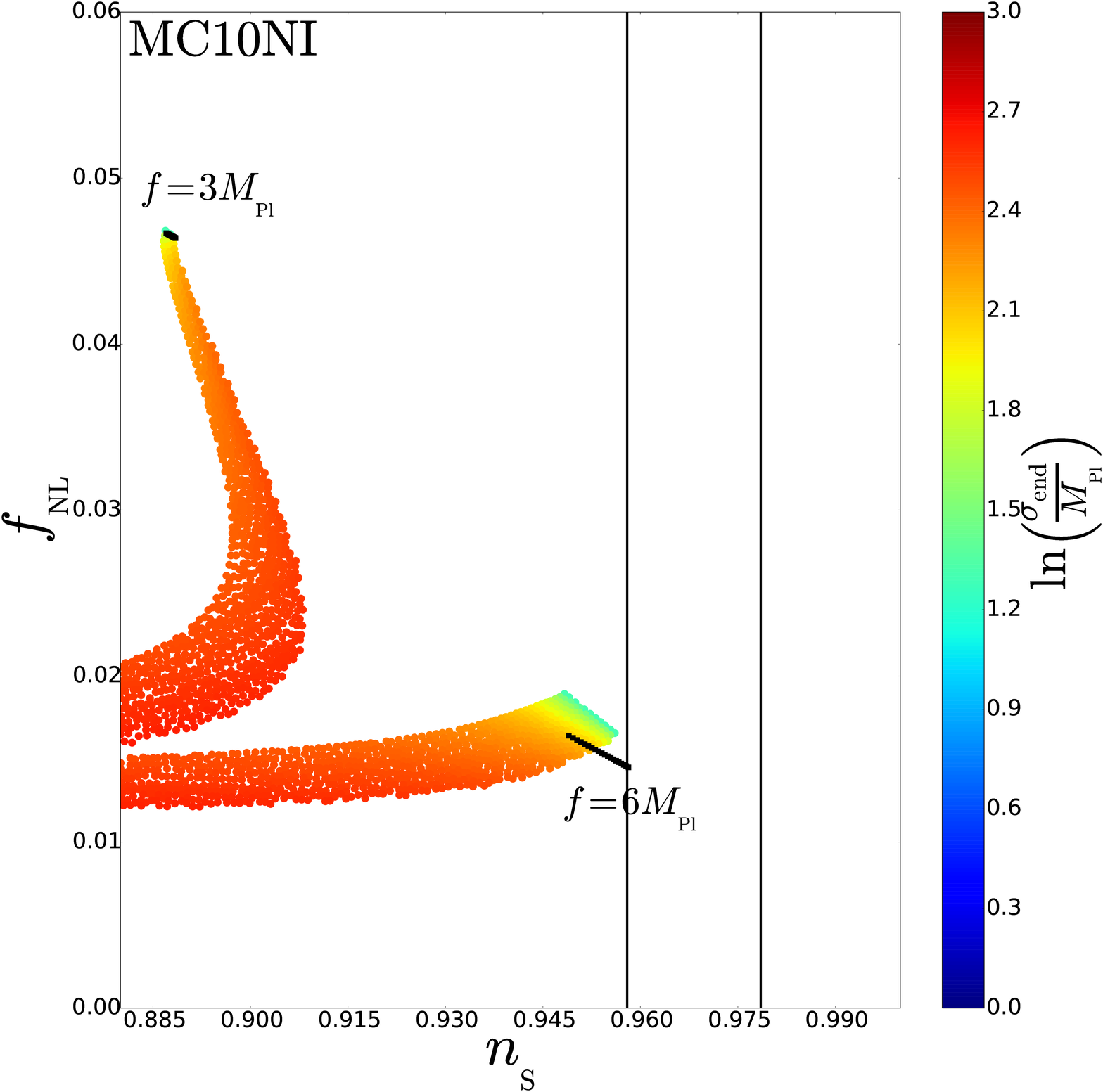}
\includegraphics[width=\wappfig,clip=true]{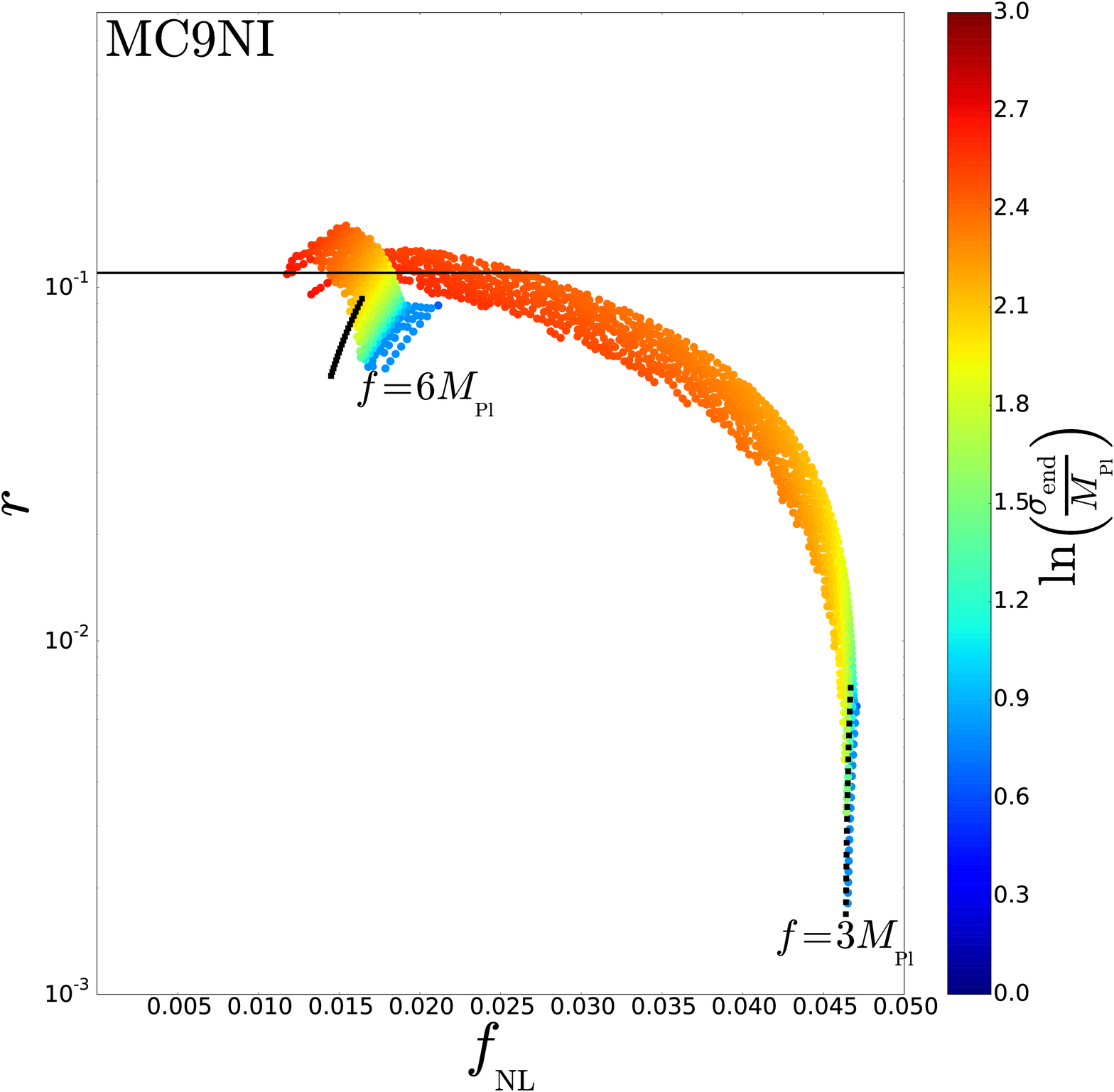}
\includegraphics[width=\wappfig,clip=true]{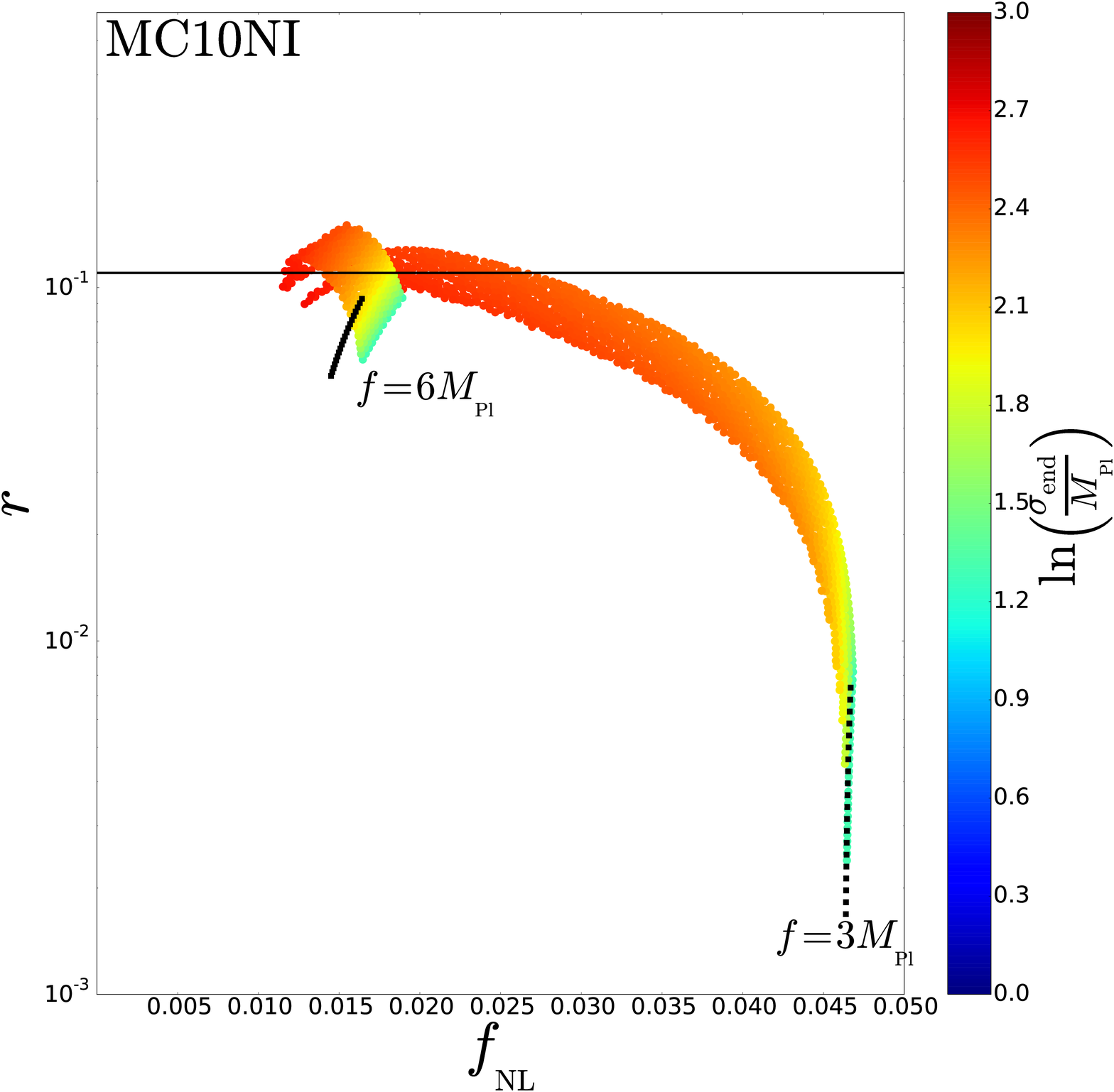}
\caption{Reheating consistent slow-roll predictions for the natural inflation models with a massive curvaton field, when reheating scenario is of the ninth (left panels) and tenth (right panels) type.}
\label{fig:CMBMCNI910}
\end{center}
\end{figure}

\clearpage
\bibliographystyle{JHEP}
\bibliography{EncyclopaediaCurvatonis}
\end{document}